\documentclass[apjl,ip,twocolumn]{aastex61}
\usepackage{comment}
\usepackage{ifthen}
\usepackage[switch]{lineno}
\usepackage{amsmath}
\newcommand{\forloop}[5][1]
{
\setcounter{#2}{#3}
\ifthenelse{#4}
	{
	#5
	\addtocounter{#2}{#1}
	\forloop[#1]{#2}{\value{#2}}{#4}{#5}
	}
	{
	}
}

\newcommand{\ctbd}[1]{}

\newcommand{\lc}{light curve}
\newcommand{\lcs}{light curves}
\newcommand{\Lc}{Light curve}

\newcommand{\masy}{\ensuremath{\rm mas\,yr^{-1}}}
\newcommand{\kms}{\ensuremath{\rm km\,s^{-1}}}
\newcommand{\ms}{\ensuremath{\rm m\,s^{-1}}}

\newcommand{\gcmc}{\ensuremath{\rm g\,cm^{-3}}}

\newcommand{\mplsini}{\ensuremath{\mpl\sin i}}
\newcommand{\teff}{\ensuremath{T_{\rm eff}}}
\newcommand{\logg}{\ensuremath{\log{g}}}
\newcommand{\vsini}{\ensuremath{v \sin{i}}}
\newcommand{\feh}{\ensuremath{\rm [Fe/H]}}

\newcommand{\vmac}{\ensuremath{v_{\rm mac}}}
\newcommand{\vmic}{\ensuremath{v_{\rm mic}}}

\newcommand{\rsun}{\ensuremath{R_\sun}}
\newcommand{\msun}{\ensuremath{M_\sun}}
\newcommand{\lsun}{\ensuremath{L_\sun}}

\newcommand{\rstar}{\ensuremath{R_\star}}
\newcommand{\mstar}{\ensuremath{M_\star}}
\newcommand{\lstar}{\ensuremath{L_\star}}

\newcommand{\teffstar}{\ensuremath{T_{\rm eff\star}}}
\newcommand{\rhostar}{\ensuremath{\rho_\star}}
\newcommand{\loggstar}{\ensuremath{\log{g_{\star}}}}

\newcommand{\mearth}{\ensuremath{M_\earth}}

\newcommand{\rpl}{\ensuremath{R_{p}}}
\newcommand{\mpl}{\ensuremath{M_{p}}}

\newcommand{\rhopl}{\ensuremath{\rho_{p}}}

\newcommand{\arstar}{\ensuremath{a/\rstar}}
\newcommand{\zrstar}{\ensuremath{\zeta/\rstar}}

\newcommand{\rjup}{\ensuremath{R_{\rm J}}}
\newcommand{\mjup}{\ensuremath{M_{\rm J}}}

\newcommand{\reffigl}[1]{Figure~\ref{fig:#1}}
\newcommand{\refsecl}[1]{\mbox{Section \ref{sec:#1}}}

\newcommand{\reftabl}[1]{Table~\ref{tab:#1}}

\newcommand{\hjs}{hot Jupiters}

\newcommand{\loopand}{\ifnum\value{planetcounter}=2 and \else\fi}
\newcommand{\loopcomma}{\ifnum\value{planetcounter}<2 ,\else. \fi}
\newcommand{\loopcommanoperiod}{\ifnum\value{planetcounter}<2 ,\else \space\fi}
\newcommand{\loopcommanospace}{\ifnum\value{planetcounter}<2 ,\else \fi}

\newcommand{\hatcurhtrxxxxA}{HATS602-035}                      
\newcommand{\hatcurfieldxxxxA}{\ensuremath{string}}            
\newcommand{\hatcurCCraxxxxA}{\ensuremath{07^{\mathrm h}29^{\mathrm m}40.63{\mathrm s}}}                     
\newcommand{\hatcurCCdecxxxxA}{\ensuremath{-29{\arcdeg}56{\arcmin}16.4{\arcsec}}}                    
\newcommand{\hatcurCCmagxxxxA}{12.746}                         
\newcommand{\hatcurCCtwomassxxxxA}{2MASS~07294061-2956163}     
\newcommand{\hatcurCCgscxxxxA}{GSC~6550-00341}                 
\newcommand{\hatcurCCtassmvxxxxA}{\ensuremath{12.746\pm0.020}} 
\newcommand{\hatcurCCtassmvshortxxxxA}{\ensuremath{12.7}}      
\newcommand{\hatcurCCtassmBxxxxA}{\ensuremath{13.232\pm0.010}} 
\newcommand{\hatcurCCtassmBshortxxxxA}{\ensuremath{13.2}}      
\newcommand{\hatcurCCtassmIxxxxA}{\ensuremath{nff\pmnff}}      
\newcommand{\hatcurCCtassmIshortxxxxA}{\ensuremath{0.0}}       
\newcommand{\hatcurCCtassmgxxxxA}{\ensuremath{12.907\pm0.020}} 
\newcommand{\hatcurCCtassmgshortxxxxA}{\ensuremath{12.9}}      
\newcommand{\hatcurCCtassmrxxxxA}{\ensuremath{12.578\pm0.050}} 
\newcommand{\hatcurCCtassmrshortxxxxA}{\ensuremath{12.6}}      
\newcommand{\hatcurCCtassmixxxxA}{\ensuremath{12.542\pm0.060}} 
\newcommand{\hatcurCCtassmishortxxxxA}{\ensuremath{12.5}}      
\newcommand{\hatcurCCtwomassJmagxxxxA}{\ensuremath{11.833\pm0.024}} 
\newcommand{\hatcurCCtwomassHmagxxxxA}{\ensuremath{11.620\pm0.024}} 
\newcommand{\hatcurCCtwomassKmagxxxxA}{\ensuremath{11.518\pm0.023}} 
\newcommand{\hatcurCCcitJmagxxxxA}{\ensuremath{11.852\pm0.024}} 
\newcommand{\hatcurCCcitHmagxxxxA}{\ensuremath{11.614\pm0.025}} 
\newcommand{\hatcurCCcitKmagxxxxA}{\ensuremath{11.542\pm0.023}} 
\newcommand{\hatcurCCbbJmagxxxxA}{\ensuremath{11.897\pm0.026}} 
\newcommand{\hatcurCCbbHmagxxxxA}{\ensuremath{11.636\pm0.025}} 
\newcommand{\hatcurCCbbKmagxxxxA}{\ensuremath{11.562\pm0.023}} 
\newcommand{\hatcurCCesoJmagxxxxA}{\ensuremath{11.899\pm0.027}} 
\newcommand{\hatcurCCesoHmagxxxxA}{\ensuremath{11.632\pm0.029}} 
\newcommand{\hatcurCCesoKmagxxxxA}{\ensuremath{11.561\pm0.023}} 
\newcommand{\hatcurCCesoJHmagxxxxA}{\ensuremath{0.267\pm0.013}} 
\newcommand{\hatcurCCesoJKmagxxxxA}{\ensuremath{0.338\pm0.036}} 
\newcommand{\hatcurCCesoHKmagxxxxA}{\ensuremath{0.071\pm0.038}} 
\newcommand{\hatcurLCdipxxxxA}{\ensuremath{10.2}}              
\newcommand{\hatcurLCrprstarxxxxA}{\ensuremath{0.0993\pm0.0029}} 
\newcommand{\hatcurLCbsqxxxxA}{\ensuremath{0.485_{-0.054}^{+0.042}}} 
\newcommand{\hatcurLCimpxxxxA}{\ensuremath{0.696_{-0.040}^{+0.030}}} 
\newcommand{\hatcurLCzetaxxxxA}{\ensuremath{15.25\pm0.14}}     
\newcommand{\hatcurLCdurxxxxA}{\ensuremath{0.1555\pm0.0029}}   
\newcommand{\hatcurLCdurshortxxxxA}{\ensuremath{0.1555}}       
\newcommand{\hatcurLCdurhrxxxxA}{\ensuremath{3.732\pm0.069}}   
\newcommand{\hatcurLCdurhrshortxxxxA}{\ensuremath{3.732}}      
\newcommand{\hatcurLCqxxxxA}{\ensuremath{0.03400\pm0.00063}}   
\newcommand{\hatcurLCqshortxxxxA}{\ensuremath{0.034}}          
\newcommand{\hatcurLCingdurxxxxA}{\ensuremath{0.0256\pm0.0029}} 
\newcommand{\hatcurLCPxxxxA}{\ensuremath{4.5776348\pm0.0000073}} 
\newcommand{\hatcurLCPprecxxxxA}{\ensuremath{4.5776348}}       
\newcommand{\hatcurLCPshortxxxxA}{\ensuremath{4.5776}}         
\newcommand{\hatcurLCTxxxxA}{\ensuremath{2457315.28338\pm0.00055}} 
\newcommand{\hatcurLCTAxxxxA}{\ensuremath{2455800.0863\pm0.0024}} 
\newcommand{\hatcurLCTBxxxxA}{\ensuremath{2457397.68079\pm0.00058}} 
\newcommand{\hatcurLChatnetmAxxxxA}{\ensuremath{12.767650\pm0.000097}} 
\newcommand{\hatcurLCiblendAxxxxA}{\ensuremath{0.911\pm0.058}} 
\newcommand{\hatcurLChatnetmBxxxxA}{\ensuremath{12.76761\pm0.00011}} 
\newcommand{\hatcurLCiblendBxxxxA}{\ensuremath{0.782\pm0.077}} 
\newcommand{\hatcurLCrhoxxxxA}{\ensuremath{0.458_{-0.057}^{+0.076}}} 
\newcommand{\hatcurSMEiteffxxxxA}{\ensuremath{6653\pm90}}      
\newcommand{\hatcurSMEizfehxxxxA}{\ensuremath{0.040\pm0.050}}  
\newcommand{\hatcurSMEizfehshortxxxxA}{\ensuremath{0.04}}      
\newcommand{\hatcurSMEiloggxxxxA}{\ensuremath{4.33\pm0.16}}    
\newcommand{\hatcurSMEivsinxxxxA}{\ensuremath{7.70\pm0.20}}    
\newcommand{\hatcurSMEivmacxxxxA}{\ensuremath{5.34\pm0.14}}    
\newcommand{\hatcurSMEivmicxxxxA}{\ensuremath{1.99\pm0.15}}    
\newcommand{\hatcurSMEiiteffxxxxA}{\ensuremath{6572\pm83}}     
\newcommand{\hatcurSMEiizfehxxxxA}{\ensuremath{0.000\pm0.044}} 
\newcommand{\hatcurSMEiizfehshortxxxxA}{\ensuremath{0.00}}     
\newcommand{\hatcurSMEiiloggxxxxA}{\ensuremath{4.167\pm0.040}} 
\newcommand{\hatcurSMEiivsinxxxxA}{\ensuremath{7.75\pm0.17}}   
\newcommand{\hatcurSMEiivmacxxxxA}{\ensuremath{5.21\pm0.13}}   
\newcommand{\hatcurSMEiivmicxxxxA}{\ensuremath{1.86\pm0.12}}   
\newcommand{\hatcurLBizxxxxA}{\ensuremath{0.1158}}             
\newcommand{\hatcurLBiizxxxxA}{\ensuremath{0.3681}}            
\newcommand{\hatcurLBiixxxxA}{\ensuremath{0.1654}}             
\newcommand{\hatcurLBiiixxxxA}{\ensuremath{0.3772}}            
\newcommand{\hatcurLBiIxxxxA}{\ensuremath{0.1472}}             
\newcommand{\hatcurLBiiIxxxxA}{\ensuremath{0.3755}}            
\newcommand{\hatcurLBigxxxxA}{\ensuremath{0.3938}}             
\newcommand{\hatcurLBiigxxxxA}{\ensuremath{0.3444}}            
\newcommand{\hatcurLBirxxxxA}{\ensuremath{0.2375}}             
\newcommand{\hatcurLBiirxxxxA}{\ensuremath{0.3875}}            
\newcommand{\hatcurLBiRxxxxA}{\ensuremath{0.2168}}             
\newcommand{\hatcurLBiiRxxxxA}{\ensuremath{0.3864}}            
\newcommand{\hatcurLBikepxxxxA}{\ensuremath{0.1000}}           
\newcommand{\hatcurLBiikepxxxxA}{\ensuremath{0.1000}}          
\newcommand{\hatcurISOmxxxxA}{\ensuremath{1.379\pm0.040}}      
\newcommand{\hatcurISOmshortxxxxA}{\ensuremath{1.38}}          
\newcommand{\hatcurISOmlongxxxxA}{\ensuremath{1.379\pm0.040}}  
\newcommand{\hatcurISOrxxxxA}{\ensuremath{1.621\pm0.085}}      
\newcommand{\hatcurISOrshortxxxxA}{\ensuremath{1.62}}          
\newcommand{\hatcurISOrlongxxxxA}{\ensuremath{1.621\pm0.085}}  
\newcommand{\hatcurISOrhoxxxxA}{\ensuremath{0.457_{-0.056}^{+0.075}}} 
\newcommand{\hatcurISOrholongxxxxA}{\ensuremath{0.457_{-0.056}^{+0.075}}} 
\newcommand{\hatcurISOloggxxxxA}{\ensuremath{4.158\pm0.038}}   
\newcommand{\hatcurISOlumxxxxA}{\ensuremath{4.37\pm0.53}}      
\newcommand{\hatcurISOlumshortxxxxA}{\ensuremath{4.37}}        
\newcommand{\hatcurISOmvxxxxA}{\ensuremath{3.13\pm0.14}}       
\newcommand{\hatcurISOvixxxxA}{\ensuremath{0.481\pm0.020}}     
\newcommand{\hatcurISOagexxxxA}{\ensuremath{2.06\pm0.30}}      
\newcommand{\hatcurISOsigmaxxxxA}{\ensuremath{0.000300\pm0.000085}} 
\newcommand{\hatcurISOMJxxxxA}{\ensuremath{2.37\pm0.12}}       
\newcommand{\hatcurISOMHxxxxA}{\ensuremath{2.16\pm0.12}}       
\newcommand{\hatcurISOMKxxxxA}{\ensuremath{2.12\pm0.12}}       
\newcommand{\hatcurISOJKxxxxA}{\ensuremath{0.240\pm0.020}}     
\newcommand{\hatcurISOspecxxxxA}{G}                            
\newcommand{\hatcurRVKxxxxA}{\ensuremath{62\pm13}}             
\newcommand{\hatcurRVrkxxxxA}{\ensuremath{0\pm0}}              
\newcommand{\hatcurRVrhxxxxA}{\ensuremath{0\pm0}}              
\newcommand{\hatcurRVkxxxxA}{\ensuremath{0\pm0}}               
\newcommand{\hatcurRVhxxxxA}{\ensuremath{0\pm0}}               
\newcommand{\hatcurRVtronexxxxA}{\ensuremath{0\pm0}}           
\newcommand{\hatcurRVtrtwoxxxxA}{\ensuremath{0\pm0}}           
\newcommand{\hatcurRVgammaAxxxxA}{\ensuremath{2916.2\pm8.7}}   
\newcommand{\hatcurRVjitterAxxxxA}{\ensuremath{26\pm12}}       
\newcommand{\hatcurRVjittertwosiglimAxxxxA}{\ensuremath{<44.6}} 
\newcommand{\hatcurRVfitrmsAxxxxA}{\ensuremath{0.0}}           
\newcommand{\hatcurRVgammaBxxxxA}{\ensuremath{3092\pm78}}      
\newcommand{\hatcurRVjitterBxxxxA}{\ensuremath{98\pm96}}       
\newcommand{\hatcurRVjittertwosiglimBxxxxA}{\ensuremath{<312.8}} 
\newcommand{\hatcurRVfitrmsBxxxxA}{\ensuremath{0.0}}           
\newcommand{\hatcurRVeccenxxxxA}{\ensuremath{0\pm0}}           
\newcommand{\hatcurRVeccentwosiglimxxxxA}{\ensuremath{<0.000}} 
\newcommand{\hatcurRVomegaxxxxA}{\ensuremath{0\pm0}}           
\newcommand{\hatcurPPixxxxA}{\ensuremath{84.98\pm0.49}}        
\newcommand{\hatcurPPgxxxxA}{\ensuremath{6.4\pm1.7}}           
\newcommand{\hatcurPPloggxxxxA}{\ensuremath{2.81\pm0.12}}      
\newcommand{\hatcurPParxxxxA}{\ensuremath{7.97\pm0.38}}        
\newcommand{\hatcurPParelxxxxA}{\ensuremath{0.06007\pm0.00058}} 
\newcommand{\hatcurPPrhoxxxxA}{\ensuremath{0.202_{-0.051}^{+0.072}}} 
\newcommand{\hatcurPPmxxxxA}{\ensuremath{0.63\pm0.13}}         
\newcommand{\hatcurPPmshortxxxxA}{\ensuremath{0.63}}           
\newcommand{\hatcurPPmlongxxxxA}{\ensuremath{0.63\pm0.13}}     
\newcommand{\hatcurPPmexxxxA}{\ensuremath{201\pm42}}           
\newcommand{\hatcurPPmeshortxxxxA}{\ensuremath{201.5}}         
\newcommand{\hatcurPPmelongxxxxA}{\ensuremath{201\pm42}}       
\newcommand{\hatcurPPrxxxxA}{\ensuremath{1.57\pm0.12}}         
\newcommand{\hatcurPPrshortxxxxA}{\ensuremath{1.57}}           
\newcommand{\hatcurPPrlongxxxxA}{\ensuremath{1.57\pm0.12}}     
\newcommand{\hatcurPPrexxxxA}{\ensuremath{17.6\pm1.3}}         
\newcommand{\hatcurPPreshortxxxxA}{\ensuremath{17.6}}          
\newcommand{\hatcurPPrelongxxxxA}{\ensuremath{17.6\pm1.3}}     
\newcommand{\hatcurPPmrcorrxxxxA}{\ensuremath{0.05}}           
\newcommand{\hatcurPPteffxxxxA}{\ensuremath{1645\pm43}}        
\newcommand{\hatcurPPthetaxxxxA}{\ensuremath{0.0351\pm0.0079}} 
\newcommand{\hatcurPPfluxperixxxxA}{\ensuremath{1.65\pm0.17}}  
\newcommand{\hatcurPPfluxperidimxxxxA}{\ensuremath{9}}         
\newcommand{\hatcurPPfluxapxxxxA}{\ensuremath{1.65\pm0.17}}    
\newcommand{\hatcurPPfluxapdimxxxxA}{\ensuremath{9}}           
\newcommand{\hatcurPPfluxavgxxxxA}{\ensuremath{1.65\pm0.17}}   
\newcommand{\hatcurPPfluxavgdimxxxxA}{\ensuremath{9}}          
\newcommand{\hatcurPPfluxavglogxxxxA}{\ensuremath{9.218\pm0.046}} 
\newcommand{\hatcurXsecphasexxxxA}{\ensuremath{0\pm0}}         
\newcommand{\hatcurXsecondaryxxxxA}{\ensuremath{2457317.57220\pm0.00056}} 
\newcommand{\hatcurXsecdurxxxxA}{\ensuremath{0.1555\pm0.0029}} 
\newcommand{\hatcurXsecingdurxxxxA}{\ensuremath{0.0256\pm0.0029}} 
\newcommand{\hatcurPPphiconjxxxxA}{\ensuremath{0\pm0}}         
\newcommand{\hatcurPPperixxxxA}{\ensuremath{2457314.13897\pm0.00055}} 
\newcommand{\hatcurPPaequivxxxxA}{\ensuremath{0.0287\pm0.0015}} 
\newcommand{\hatcurPPtcircxxxxA}{\ensuremath{182_{-58}^{+98}}} 
\newcommand{\hatcurPPtinfallxxxxA}{\ensuremath{3110_{-790}^{+1170}}} 
\newcommand{\hatcurXdistxxxxA}{\ensuremath{772\pm42}}          
\newcommand{\hatcurXAvxxxxA}{\ensuremath{0.186\pm0.069}}       
\newcommand{\hatcurXdistredxxxxA}{\ensuremath{773\pm41}}       
\newcommand{\hatcurXEBVxxxxA}{\ensuremath{0.060\pm0.022}}      
\newcommand{\hatcurXmvisoredxxxxA}{\ensuremath{12.750\pm0.020}} 
\newcommand{\hatcurXmiisoredxxxxA}{\ensuremath{12.173\pm0.017}} 
\newcommand{\hatcurXmjisoredxxxxA}{\ensuremath{11.857\pm0.014}} 
\newcommand{\hatcurXmhisoredxxxxA}{\ensuremath{11.636\pm0.015}} 
\newcommand{\hatcurXmkisoredxxxxA}{\ensuremath{11.583\pm0.016}} 
\newcommand{\hatcurXviisoredxxxxA}{\ensuremath{0.578\pm0.020}} 
\newcommand{\hatcurXvkisoredxxxxA}{\ensuremath{1.167\pm0.027}} 
\newcommand{\hatcurXjhisoredxxxxA}{\ensuremath{0.2210\pm0.0097}} 
\newcommand{\hatcurXjkisoredxxxxA}{\ensuremath{0.2740\pm0.0080}} 
\newcommand{\hatcurCCpmraxxxxA}{\ensuremath{-0.2\pm1.0}}       
\newcommand{\hatcurCCpmdecxxxxA}{\ensuremath{-5.9\pm1.7}}      
\newcommand{\hatcurCCpmxxxxA}{\ensuremath{5.9\pm2.0}}          
 \newcommand{\hatcurhtrxxxxB}{HATS601-056}                      
\newcommand{\hatcurfieldxxxxB}{\ensuremath{string}}            
\newcommand{\hatcurCCraxxxxB}{\ensuremath{06^{\mathrm h}42^{\mathrm m}17.10{\mathrm s}}}                     
\newcommand{\hatcurCCdecxxxxB}{\ensuremath{-29{\arcdeg}46{\arcmin}36.5{\arcsec}}}                    
\newcommand{\hatcurCCmagxxxxB}{13.377}                         
\newcommand{\hatcurCCtwomassxxxxB}{2MASS~06421710-2946365}     
\newcommand{\hatcurCCgscxxxxB}{GSC~6533-01514}                 
\newcommand{\hatcurCCtassmvxxxxB}{\ensuremath{13.377\pm0.030}} 
\newcommand{\hatcurCCtassmvshortxxxxB}{\ensuremath{13.4}}      
\newcommand{\hatcurCCtassmBxxxxB}{\ensuremath{13.866\pm0.020}} 
\newcommand{\hatcurCCtassmBshortxxxxB}{\ensuremath{13.9}}      
\newcommand{\hatcurCCtassmIxxxxB}{\ensuremath{nff\pmnff}}      
\newcommand{\hatcurCCtassmIshortxxxxB}{\ensuremath{0.0}}       
\newcommand{\hatcurCCtassmgxxxxB}{\ensuremath{13.557\pm0.020}} 
\newcommand{\hatcurCCtassmgshortxxxxB}{\ensuremath{13.6}}      
\newcommand{\hatcurCCtassmrxxxxB}{\ensuremath{13.272\pm0.040}} 
\newcommand{\hatcurCCtassmrshortxxxxB}{\ensuremath{13.3}}      
\newcommand{\hatcurCCtassmixxxxB}{\ensuremath{13.197\pm0.040}} 
\newcommand{\hatcurCCtassmishortxxxxB}{\ensuremath{13.2}}      
\newcommand{\hatcurCCtwomassJmagxxxxB}{\ensuremath{12.439\pm0.026}} 
\newcommand{\hatcurCCtwomassHmagxxxxB}{\ensuremath{12.242\pm0.032}} 
\newcommand{\hatcurCCtwomassKmagxxxxB}{\ensuremath{12.147\pm0.027}} 
\newcommand{\hatcurCCcitJmagxxxxB}{\ensuremath{12.460\pm0.026}} 
\newcommand{\hatcurCCcitHmagxxxxB}{\ensuremath{12.236\pm0.032}} 
\newcommand{\hatcurCCcitKmagxxxxB}{\ensuremath{12.171\pm0.027}} 
\newcommand{\hatcurCCbbJmagxxxxB}{\ensuremath{12.503\pm0.028}} 
\newcommand{\hatcurCCbbHmagxxxxB}{\ensuremath{12.258\pm0.033}} 
\newcommand{\hatcurCCbbKmagxxxxB}{\ensuremath{12.191\pm0.027}} 
\newcommand{\hatcurCCesoJmagxxxxB}{\ensuremath{12.504\pm0.029}} 
\newcommand{\hatcurCCesoHmagxxxxB}{\ensuremath{12.255\pm0.037}} 
\newcommand{\hatcurCCesoKmagxxxxB}{\ensuremath{12.190\pm0.028}} 
\newcommand{\hatcurCCesoJHmagxxxxB}{\ensuremath{0.249\pm0.045}} 
\newcommand{\hatcurCCesoJKmagxxxxB}{\ensuremath{0.314\pm0.040}} 
\newcommand{\hatcurCCesoHKmagxxxxB}{\ensuremath{0.065\pm0.047}} 
\newcommand{\hatcurLCdipxxxxB}{\ensuremath{4.7}}               
\newcommand{\hatcurLCrprstarxxxxB}{\ensuremath{0.0716\pm0.0034}} 
\newcommand{\hatcurLCbsqxxxxB}{\ensuremath{0.122_{-0.087}^{+0.104}}} 
\newcommand{\hatcurLCimpxxxxB}{\ensuremath{0.35_{-0.16}^{+0.13}}} 
\newcommand{\hatcurLCzetaxxxxB}{\ensuremath{9.738\pm0.100}}    
\newcommand{\hatcurLCdurxxxxB}{\ensuremath{0.2221\pm0.0031}}   
\newcommand{\hatcurLCdurshortxxxxB}{\ensuremath{0.2221}}       
\newcommand{\hatcurLCdurhrxxxxB}{\ensuremath{5.330\pm0.074}}   
\newcommand{\hatcurLCdurhrshortxxxxB}{\ensuremath{5.330}}      
\newcommand{\hatcurLCqxxxxB}{\ensuremath{0.06800\pm0.00095}}   
\newcommand{\hatcurLCqshortxxxxB}{\ensuremath{0.068}}          
\newcommand{\hatcurLCingdurxxxxB}{\ensuremath{0.0167\pm0.0023}} 
\newcommand{\hatcurLCPxxxxB}{\ensuremath{3.2642736\pm0.0000058}} 
\newcommand{\hatcurLCPprecxxxxB}{\ensuremath{3.2642736}}       
\newcommand{\hatcurLCPshortxxxxB}{\ensuremath{3.2643}}         
\newcommand{\hatcurLCTxxxxB}{\ensuremath{2456962.6760\pm0.0010}} 
\newcommand{\hatcurLCTAxxxxB}{\ensuremath{2455794.0660\pm0.0022}} 
\newcommand{\hatcurLCTBxxxxB}{\ensuremath{2457393.5599\pm0.0014}} 
\newcommand{\hatcurLChatnetmAxxxxB}{\ensuremath{13.317590\pm0.000078}} 
\newcommand{\hatcurLCiblendAxxxxB}{\ensuremath{0.888\pm0.078}} 
\newcommand{\hatcurLChatnetmBxxxxB}{\ensuremath{13.31754\pm0.00010}} 
\newcommand{\hatcurLCiblendBxxxxB}{\ensuremath{0.545\pm0.087}} 
\newcommand{\hatcurLChatnetmCxxxxB}{\ensuremath{13.31749\pm0.00026}} 
\newcommand{\hatcurLCiblendCxxxxB}{\ensuremath{0.83\pm0.14}}   
\newcommand{\hatcurLCrhoxxxxB}{\ensuremath{0.190\pm0.027}}     
\newcommand{\hatcurSMEiteffxxxxB}{\ensuremath{6460\pm130}}     
\newcommand{\hatcurSMEizfehxxxxB}{\ensuremath{0.010\pm0.077}}  
\newcommand{\hatcurSMEizfehshortxxxxB}{\ensuremath{0.01}}      
\newcommand{\hatcurSMEiloggxxxxB}{\ensuremath{3.90\pm0.24}}    
\newcommand{\hatcurSMEivsinxxxxB}{\ensuremath{9.52\pm0.25}}    
\newcommand{\hatcurSMEivmacxxxxB}{\ensuremath{5.04\pm0.20}}    
\newcommand{\hatcurSMEivmicxxxxB}{\ensuremath{1.70\pm0.17}}    
\newcommand{\hatcurLBizxxxxB}{\ensuremath{0.1203}}             
\newcommand{\hatcurLBiizxxxxB}{\ensuremath{0.3694}}            
\newcommand{\hatcurLBiixxxxB}{\ensuremath{0.1695}}             
\newcommand{\hatcurLBiiixxxxB}{\ensuremath{0.3788}}            
\newcommand{\hatcurLBiIxxxxB}{\ensuremath{0.1507}}             
\newcommand{\hatcurLBiiIxxxxB}{\ensuremath{0.3774}}            
\newcommand{\hatcurLBigxxxxB}{\ensuremath{0.4123}}             
\newcommand{\hatcurLBiigxxxxB}{\ensuremath{0.3323}}            
\newcommand{\hatcurLBirxxxxB}{\ensuremath{0.2457}}             
\newcommand{\hatcurLBiirxxxxB}{\ensuremath{0.3857}}            
\newcommand{\hatcurLBiRxxxxB}{\ensuremath{0.2239}}             
\newcommand{\hatcurLBiiRxxxxB}{\ensuremath{0.3855}}            
\newcommand{\hatcurLBikepxxxxB}{\ensuremath{0.1000}}           
\newcommand{\hatcurLBiikepxxxxB}{\ensuremath{0.1000}}          
\newcommand{\hatcurISOmxxxxB}{\ensuremath{1.561\pm0.069}}      
\newcommand{\hatcurISOmshortxxxxB}{\ensuremath{1.56}}          
\newcommand{\hatcurISOmlongxxxxB}{\ensuremath{1.561\pm0.069}}  
\newcommand{\hatcurISOrxxxxB}{\ensuremath{2.26_{-0.11}^{+0.18}}} 
\newcommand{\hatcurISOrshortxxxxB}{\ensuremath{2.26}}          
\newcommand{\hatcurISOrlongxxxxB}{\ensuremath{2.26_{-0.11}^{+0.18}}} 
\newcommand{\hatcurISOrhoxxxxB}{\ensuremath{0.189\pm0.027}}    
\newcommand{\hatcurISOrholongxxxxB}{\ensuremath{0.189\pm0.027}} 
\newcommand{\hatcurISOloggxxxxB}{\ensuremath{3.921\pm0.041}}   
\newcommand{\hatcurISOlumxxxxB}{\ensuremath{8.0_{-1.1}^{+1.5}}} 
\newcommand{\hatcurISOlumshortxxxxB}{\ensuremath{8.04}}        
\newcommand{\hatcurISOmvxxxxB}{\ensuremath{2.47\pm0.18}}       
\newcommand{\hatcurISOvixxxxB}{\ensuremath{0.504\pm0.033}}     
\newcommand{\hatcurISOagexxxxB}{\ensuremath{2.07\pm0.34}}      
\newcommand{\hatcurISOsigmaxxxxB}{\ensuremath{0.000400\pm0.000076}} 
\newcommand{\hatcurISOMJxxxxB}{\ensuremath{1.67\pm0.15}}       
\newcommand{\hatcurISOMHxxxxB}{\ensuremath{1.45\pm0.14}}       
\newcommand{\hatcurISOMKxxxxB}{\ensuremath{1.41\pm0.14}}       
\newcommand{\hatcurISOJKxxxxB}{\ensuremath{0.260\pm0.030}}     
\newcommand{\hatcurISOspecxxxxB}{F}                            
\newcommand{\hatcurRVKxxxxB}{\ensuremath{162\pm23}}            
\newcommand{\hatcurRVrkxxxxB}{\ensuremath{0\pm0}}              
\newcommand{\hatcurRVrhxxxxB}{\ensuremath{0\pm0}}              
\newcommand{\hatcurRVkxxxxB}{\ensuremath{0\pm0}}               
\newcommand{\hatcurRVhxxxxB}{\ensuremath{0\pm0}}               
\newcommand{\hatcurRVtronexxxxB}{\ensuremath{0\pm0}}           
\newcommand{\hatcurRVtrtwoxxxxB}{\ensuremath{0\pm0}}           
\newcommand{\hatcurRVgammaxxxxB}{\ensuremath{9194\pm17}}       
\newcommand{\hatcurRVjitterxxxxB}{\ensuremath{0\pm15}}         
\newcommand{\hatcurRVjittertwosiglimxxxxB}{\ensuremath{<33.7}} 
\newcommand{\hatcurRVfitrmsxxxxB}{\ensuremath{.1fym}}          
\newcommand{\hatcurRVeccenxxxxB}{\ensuremath{0\pm0}}           
\newcommand{\hatcurRVeccentwosiglimxxxxB}{\ensuremath{<0.000}} 
\newcommand{\hatcurRVomegaxxxxB}{\ensuremath{0\pm0}}           
\newcommand{\hatcurPPixxxxB}{\ensuremath{85.8\pm1.8}}          
\newcommand{\hatcurPPgxxxxB}{\ensuremath{15.5\pm3.2}}          
\newcommand{\hatcurPPloggxxxxB}{\ensuremath{3.191\pm0.091}}    
\newcommand{\hatcurPParxxxxB}{\ensuremath{4.74_{-0.31}^{+0.22}}} 
\newcommand{\hatcurPParelxxxxB}{\ensuremath{0.04997\pm0.00074}} 
\newcommand{\hatcurPPrhoxxxxB}{\ensuremath{0.49\pm0.13}}       
\newcommand{\hatcurPPmxxxxB}{\ensuremath{1.59\pm0.24}}         
\newcommand{\hatcurPPmshortxxxxB}{\ensuremath{1.59}}           
\newcommand{\hatcurPPmlongxxxxB}{\ensuremath{1.59\pm0.24}}     
\newcommand{\hatcurPPmexxxxB}{\ensuremath{505\pm75}}           
\newcommand{\hatcurPPmeshortxxxxB}{\ensuremath{505.3}}         
\newcommand{\hatcurPPmelongxxxxB}{\ensuremath{505\pm75}}       
\newcommand{\hatcurPPrxxxxB}{\ensuremath{1.58_{-0.12}^{+0.16}}} 
\newcommand{\hatcurPPrshortxxxxB}{\ensuremath{1.58}}           
\newcommand{\hatcurPPrlongxxxxB}{\ensuremath{1.58_{-0.12}^{+0.16}}} 
\newcommand{\hatcurPPrexxxxB}{\ensuremath{17.7_{-1.4}^{+1.8}}} 
\newcommand{\hatcurPPreshortxxxxB}{\ensuremath{17.7}}          
\newcommand{\hatcurPPrelongxxxxB}{\ensuremath{17.7_{-1.4}^{+1.8}}} 
\newcommand{\hatcurPPmrcorrxxxxB}{\ensuremath{0.20}}           
\newcommand{\hatcurPPteffxxxxB}{\ensuremath{2101\pm69}}        
\newcommand{\hatcurPPthetaxxxxB}{\ensuremath{0.063\pm0.010}}   
\newcommand{\hatcurPPfluxperixxxxB}{\ensuremath{4.39_{-0.50}^{+0.68}}} 
\newcommand{\hatcurPPfluxperidimxxxxB}{\ensuremath{9}}         
\newcommand{\hatcurPPfluxapxxxxB}{\ensuremath{4.39_{-0.50}^{+0.68}}} 
\newcommand{\hatcurPPfluxapdimxxxxB}{\ensuremath{9}}           
\newcommand{\hatcurPPfluxavgxxxxB}{\ensuremath{4.39_{-0.50}^{+0.68}}} 
\newcommand{\hatcurPPfluxavgdimxxxxB}{\ensuremath{9}}          
\newcommand{\hatcurPPfluxavglogxxxxB}{\ensuremath{9.643\pm0.057}} 
\newcommand{\hatcurXsecphasexxxxB}{\ensuremath{0\pm0}}         
\newcommand{\hatcurXsecondaryxxxxB}{\ensuremath{2456964.3081\pm0.0010}} 
\newcommand{\hatcurXsecdurxxxxB}{\ensuremath{0.2221\pm0.0031}} 
\newcommand{\hatcurXsecingdurxxxxB}{\ensuremath{0.0167\pm0.0023}} 
\newcommand{\hatcurPPphiconjxxxxB}{\ensuremath{0\pm0}}         
\newcommand{\hatcurPPperixxxxB}{\ensuremath{2456961.8599\pm0.0010}} 
\newcommand{\hatcurPPaequivxxxxB}{\ensuremath{0.0176\pm0.0011}} 
\newcommand{\hatcurPPtcircxxxxB}{\ensuremath{109\pm46}}        
\newcommand{\hatcurPPtinfallxxxxB}{\ensuremath{74\pm21}}       
\newcommand{\hatcurXdistxxxxB}{\ensuremath{1434_{-76}^{+116}}} 
\newcommand{\hatcurXAvxxxxB}{\ensuremath{0.120\pm0.096}}       
\newcommand{\hatcurXdistredxxxxB}{\ensuremath{1431_{-75}^{+116}}} 
\newcommand{\hatcurXEBVxxxxB}{\ensuremath{0.039\pm0.031}}      
\newcommand{\hatcurXmvisoredxxxxB}{\ensuremath{13.385\pm0.031}} 
\newcommand{\hatcurXmiisoredxxxxB}{\ensuremath{12.814\pm0.025}} 
\newcommand{\hatcurXmjisoredxxxxB}{\ensuremath{12.485\pm0.017}} 
\newcommand{\hatcurXmhisoredxxxxB}{\ensuremath{12.254\pm0.021}} 
\newcommand{\hatcurXmkisoredxxxxB}{\ensuremath{12.201\pm0.021}} 
\newcommand{\hatcurXviisoredxxxxB}{\ensuremath{0.569\pm0.025}} 
\newcommand{\hatcurXvkisoredxxxxB}{\ensuremath{1.183\pm0.041}} 
\newcommand{\hatcurXjhisoredxxxxB}{\ensuremath{0.231\pm0.017}} 
\newcommand{\hatcurXjkisoredxxxxB}{\ensuremath{0.283\pm0.014}} 
\newcommand{\hatcurCCpmraxxxxB}{\ensuremath{-4.6\pm2.1}}       
\newcommand{\hatcurCCpmdecxxxxB}{\ensuremath{5.6\pm2.3}}       
\newcommand{\hatcurCCpmxxxxB}{\ensuremath{7.2\pm3.1}}          
 \newcommand{\hatcurhtrxxxxC}{HATS601-061}                      
\newcommand{\hatcurfieldxxxxC}{\ensuremath{string}}            
\newcommand{\hatcurCCraxxxxC}{\ensuremath{06^{\mathrm h}54^{\mathrm m}04.18{\mathrm s}}}                     
\newcommand{\hatcurCCdecxxxxC}{\ensuremath{-27{\arcdeg}03{\arcmin}01.4{\arcsec}}}                    
\newcommand{\hatcurCCmagxxxxC}{12.681}                         
\newcommand{\hatcurCCtwomassxxxxC}{2MASS~06540416-2703013}     
\newcommand{\hatcurCCgscxxxxC}{GSC~6530-01596}                 
\newcommand{\hatcurCCtassmvxxxxC}{\ensuremath{12.681\pm0.030}} 
\newcommand{\hatcurCCtassmvshortxxxxC}{\ensuremath{12.7}}      
\newcommand{\hatcurCCtassmBxxxxC}{\ensuremath{13.175\pm0.040}} 
\newcommand{\hatcurCCtassmBshortxxxxC}{\ensuremath{13.2}}      
\newcommand{\hatcurCCtassmIxxxxC}{\ensuremath{nff\pmnff}}      
\newcommand{\hatcurCCtassmIshortxxxxC}{\ensuremath{0.0}}       
\newcommand{\hatcurCCtassmgxxxxC}{\ensuremath{12.893\pm0.040}} 
\newcommand{\hatcurCCtassmgshortxxxxC}{\ensuremath{12.9}}      
\newcommand{\hatcurCCtassmrxxxxC}{\ensuremath{12.555\pm0.020}} 
\newcommand{\hatcurCCtassmrshortxxxxC}{\ensuremath{12.6}}      
\newcommand{\hatcurCCtassmixxxxC}{\ensuremath{12.484\pm0.060}} 
\newcommand{\hatcurCCtassmishortxxxxC}{\ensuremath{12.5}}      
\newcommand{\hatcurCCtwomassJmagxxxxC}{\ensuremath{11.765\pm0.024}} 
\newcommand{\hatcurCCtwomassHmagxxxxC}{\ensuremath{11.521\pm0.023}} 
\newcommand{\hatcurCCtwomassKmagxxxxC}{\ensuremath{11.498\pm0.023}} 
\newcommand{\hatcurCCcitJmagxxxxC}{\ensuremath{11.787\pm0.024}} 
\newcommand{\hatcurCCcitHmagxxxxC}{\ensuremath{11.517\pm0.024}} 
\newcommand{\hatcurCCcitKmagxxxxC}{\ensuremath{11.522\pm0.023}} 
\newcommand{\hatcurCCbbJmagxxxxC}{\ensuremath{11.828\pm0.026}} 
\newcommand{\hatcurCCbbHmagxxxxC}{\ensuremath{11.537\pm0.024}} 
\newcommand{\hatcurCCbbKmagxxxxC}{\ensuremath{11.542\pm0.023}} 
\newcommand{\hatcurCCesoJmagxxxxC}{\ensuremath{11.829\pm0.027}} 
\newcommand{\hatcurCCesoHmagxxxxC}{\ensuremath{11.530\pm0.025}} 
\newcommand{\hatcurCCesoKmagxxxxC}{\ensuremath{11.542\pm0.024}} 
\newcommand{\hatcurCCesoJHmagxxxxC}{\ensuremath{0.299\pm0.035}} 
\newcommand{\hatcurCCesoJKmagxxxxC}{\ensuremath{0.287\pm0.036}} 
\newcommand{\hatcurCCesoHKmagxxxxC}{\ensuremath{-0.0120\pm0.0070}} 
\newcommand{\hatcurLCdipxxxxC}{\ensuremath{7.2}}               
\newcommand{\hatcurLCrprstarxxxxC}{\ensuremath{0.0823\pm0.0031}} 
\newcommand{\hatcurLCbsqxxxxC}{\ensuremath{0.767_{-0.020}^{+0.021}}} 
\newcommand{\hatcurLCimpxxxxC}{\ensuremath{0.876_{-0.011}^{+0.012}}} 
\newcommand{\hatcurLCzetaxxxxC}{\ensuremath{26.87\pm0.73}}     
\newcommand{\hatcurLCdurxxxxC}{\ensuremath{0.0982\pm0.0029}}   
\newcommand{\hatcurLCdurshortxxxxC}{\ensuremath{0.0982}}       
\newcommand{\hatcurLCdurhrxxxxC}{\ensuremath{2.356\pm0.071}}   
\newcommand{\hatcurLCdurhrshortxxxxC}{\ensuremath{2.356}}      
\newcommand{\hatcurLCqxxxxC}{\ensuremath{0.02340\pm0.00070}}   
\newcommand{\hatcurLCqshortxxxxC}{\ensuremath{0.023}}          
\newcommand{\hatcurLCingdurxxxxC}{\ensuremath{0.0277\pm0.0069}} 
\newcommand{\hatcurLCPxxxxC}{\ensuremath{4.193654\pm0.000011}} 
\newcommand{\hatcurLCPprecxxxxC}{\ensuremath{4.1936537}}       
\newcommand{\hatcurLCPshortxxxxC}{\ensuremath{4.1937}}         
\newcommand{\hatcurLCTxxxxC}{\ensuremath{2457034.7641\pm0.0014}} 
\newcommand{\hatcurLCTAxxxxC}{\ensuremath{2455797.6358\pm0.0031}} 
\newcommand{\hatcurLCTBxxxxC}{\ensuremath{2457319.9325\pm0.0018}} 
\newcommand{\hatcurLChatnetmxxxxC}{\ensuremath{12.674930\pm0.000055}} 
\newcommand{\hatcurLCiblendxxxxC}{\ensuremath{0.870\pm0.080}}  
\newcommand{\hatcurLCrhoxxxxC}{\ensuremath{1.10_{-0.34}^{+0.85}}} 
\newcommand{\hatcurSMEiteffxxxxC}{\ensuremath{6424\pm91}}      
\newcommand{\hatcurSMEizfehxxxxC}{\ensuremath{0.470\pm0.041}}  
\newcommand{\hatcurSMEizfehshortxxxxC}{\ensuremath{0.47}}      
\newcommand{\hatcurSMEiloggxxxxC}{\ensuremath{4.10\pm0.22}}    
\newcommand{\hatcurSMEivsinxxxxC}{\ensuremath{19.21\pm0.23}}   
\newcommand{\hatcurSMEivmacxxxxC}{\ensuremath{4.99\pm0.14}}    
\newcommand{\hatcurSMEivmicxxxxC}{\ensuremath{1.66\pm0.12}}    
\newcommand{\hatcurLBizxxxxC}{\ensuremath{0.1258}}             
\newcommand{\hatcurLBiizxxxxC}{\ensuremath{0.3871}}            
\newcommand{\hatcurLBiixxxxC}{\ensuremath{0.1795}}             
\newcommand{\hatcurLBiiixxxxC}{\ensuremath{0.3965}}            
\newcommand{\hatcurLBiIxxxxC}{\ensuremath{0.1592}}             
\newcommand{\hatcurLBiiIxxxxC}{\ensuremath{0.3955}}            
\newcommand{\hatcurLBigxxxxC}{\ensuremath{0.4468}}             
\newcommand{\hatcurLBiigxxxxC}{\ensuremath{0.3221}}            
\newcommand{\hatcurLBirxxxxC}{\ensuremath{0.2634}}             
\newcommand{\hatcurLBiirxxxxC}{\ensuremath{0.3954}}            
\newcommand{\hatcurLBiRxxxxC}{\ensuremath{0.2393}}             
\newcommand{\hatcurLBiiRxxxxC}{\ensuremath{0.3976}}            
\newcommand{\hatcurLBikepxxxxC}{\ensuremath{0.1000}}           
\newcommand{\hatcurLBiikepxxxxC}{\ensuremath{0.1000}}          
\newcommand{\hatcurISOmxxxxC}{\ensuremath{1.402\pm0.041}}      
\newcommand{\hatcurISOmshortxxxxC}{\ensuremath{1.40}}          
\newcommand{\hatcurISOmlongxxxxC}{\ensuremath{1.402\pm0.041}}  
\newcommand{\hatcurISOrxxxxC}{\ensuremath{1.416\pm0.074}}      
\newcommand{\hatcurISOrshortxxxxC}{\ensuremath{1.42}}          
\newcommand{\hatcurISOrlongxxxxC}{\ensuremath{1.416\pm0.074}}  
\newcommand{\hatcurISOrhoxxxxC}{\ensuremath{0.697\pm0.088}}    
\newcommand{\hatcurISOrholongxxxxC}{\ensuremath{0.697\pm0.088}} 
\newcommand{\hatcurISOloggxxxxC}{\ensuremath{4.284\pm0.036}}   
\newcommand{\hatcurISOlumxxxxC}{\ensuremath{3.01\pm0.43}}      
\newcommand{\hatcurISOlumshortxxxxC}{\ensuremath{3.01}}        
\newcommand{\hatcurISOmvxxxxC}{\ensuremath{3.52\pm0.16}}       
\newcommand{\hatcurISOvixxxxC}{\ensuremath{0.532\pm0.023}}     
\newcommand{\hatcurISOagexxxxC}{\ensuremath{0.64_{-0.28}^{+0.44}}} 
\newcommand{\hatcurISOsigmaxxxxC}{\ensuremath{0.0074\pm0.0042}} 
\newcommand{\hatcurISOMJxxxxC}{\ensuremath{2.68\pm0.13}}       
\newcommand{\hatcurISOMHxxxxC}{\ensuremath{2.46\pm0.12}}       
\newcommand{\hatcurISOMKxxxxC}{\ensuremath{2.42\pm0.12}}       
\newcommand{\hatcurISOJKxxxxC}{\ensuremath{0.260\pm0.020}}     
\newcommand{\hatcurISOspecxxxxC}{F}                            
\newcommand{\hatcurRVKxxxxC}{\ensuremath{820\pm170}}           
\newcommand{\hatcurRVrkxxxxC}{\ensuremath{0\pm0}}              
\newcommand{\hatcurRVrhxxxxC}{\ensuremath{0\pm0}}              
\newcommand{\hatcurRVkxxxxC}{\ensuremath{0\pm0}}               
\newcommand{\hatcurRVhxxxxC}{\ensuremath{0\pm0}}               
\newcommand{\hatcurRVtronexxxxC}{\ensuremath{0\pm0}}           
\newcommand{\hatcurRVtrtwoxxxxC}{\ensuremath{0\pm0}}           
\newcommand{\hatcurRVgammaAxxxxC}{\ensuremath{37140\pm230}}    
\newcommand{\hatcurRVjitterAxxxxC}{\ensuremath{640\pm190}}     
\newcommand{\hatcurRVjittertwosiglimAxxxxC}{\ensuremath{<1017.4}} 
\newcommand{\hatcurRVfitrmsAxxxxC}{\ensuremath{0.0}}           
\newcommand{\hatcurRVgammaBxxxxC}{\ensuremath{37150\pm330}}    
\newcommand{\hatcurRVjitterBxxxxC}{\ensuremath{610\pm310}}     
\newcommand{\hatcurRVjittertwosiglimBxxxxC}{\ensuremath{<1339.8}} 
\newcommand{\hatcurRVfitrmsBxxxxC}{\ensuremath{0.0}}           
\newcommand{\hatcurRVgammaCxxxxC}{\ensuremath{38150\pm130}}    
\newcommand{\hatcurRVjitterCxxxxC}{\ensuremath{260\pm100}}     
\newcommand{\hatcurRVjittertwosiglimCxxxxC}{\ensuremath{<488.3}} 
\newcommand{\hatcurRVfitrmsCxxxxC}{\ensuremath{0.0}}           
\newcommand{\hatcurRVeccenxxxxC}{\ensuremath{0\pm0}}           
\newcommand{\hatcurRVeccentwosiglimxxxxC}{\ensuremath{<0.000}} 
\newcommand{\hatcurRVomegaxxxxC}{\ensuremath{0\pm0}}           
\newcommand{\hatcurPPixxxxC}{\ensuremath{84.20_{-0.37}^{+0.27}}} 
\newcommand{\hatcurPPgxxxxC}{\ensuremath{153\pm39}}            
\newcommand{\hatcurPPloggxxxxC}{\ensuremath{4.18\pm0.12}}      
\newcommand{\hatcurPParxxxxC}{\ensuremath{8.66\pm0.37}}        
\newcommand{\hatcurPParelxxxxC}{\ensuremath{0.05707\pm0.00056}} 
\newcommand{\hatcurPPrhoxxxxC}{\ensuremath{6.8\pm2.0}}         
\newcommand{\hatcurPPmxxxxC}{\ensuremath{8.2\pm1.7}}           
\newcommand{\hatcurPPmshortxxxxC}{\ensuremath{8.16}}           
\newcommand{\hatcurPPmlongxxxxC}{\ensuremath{8.2\pm1.7}}       
\newcommand{\hatcurPPmexxxxC}{\ensuremath{2590\pm530}}         
\newcommand{\hatcurPPmeshortxxxxC}{\ensuremath{2593.9}}        
\newcommand{\hatcurPPmelongxxxxC}{\ensuremath{2590\pm530}}     
\newcommand{\hatcurPPrxxxxC}{\ensuremath{1.137\pm0.081}}       
\newcommand{\hatcurPPrshortxxxxC}{\ensuremath{1.14}}           
\newcommand{\hatcurPPrlongxxxxC}{\ensuremath{1.137\pm0.081}}   
\newcommand{\hatcurPPrexxxxC}{\ensuremath{12.74\pm0.90}}       
\newcommand{\hatcurPPreshortxxxxC}{\ensuremath{12.7}}          
\newcommand{\hatcurPPrelongxxxxC}{\ensuremath{12.74\pm0.90}}   
\newcommand{\hatcurPPmrcorrxxxxC}{\ensuremath{0.01}}           
\newcommand{\hatcurPPteffxxxxC}{\ensuremath{1538\pm46}}        
\newcommand{\hatcurPPthetaxxxxC}{\ensuremath{0.57\pm0.13}}     
\newcommand{\hatcurPPfluxperixxxxC}{\ensuremath{1.26\pm0.16}}  
\newcommand{\hatcurPPfluxperidimxxxxC}{\ensuremath{9}}         
\newcommand{\hatcurPPfluxapxxxxC}{\ensuremath{1.26\pm0.16}}    
\newcommand{\hatcurPPfluxapdimxxxxC}{\ensuremath{9}}           
\newcommand{\hatcurPPfluxavgxxxxC}{\ensuremath{1.26\pm0.16}}   
\newcommand{\hatcurPPfluxavgdimxxxxC}{\ensuremath{9}}          
\newcommand{\hatcurPPfluxavglogxxxxC}{\ensuremath{9.101\pm0.052}} 
\newcommand{\hatcurXsecphasexxxxC}{\ensuremath{0\pm0}}         
\newcommand{\hatcurXsecondaryxxxxC}{\ensuremath{2457036.8609\pm0.0014}} 
\newcommand{\hatcurXsecdurxxxxC}{\ensuremath{0.0982\pm0.0030}} 
\newcommand{\hatcurXsecingdurxxxxC}{\ensuremath{0.0277\pm0.0039}} 
\newcommand{\hatcurPPphiconjxxxxC}{\ensuremath{0\pm0}}         
\newcommand{\hatcurPPperixxxxC}{\ensuremath{2457033.7157\pm0.0014}} 
\newcommand{\hatcurPPaequivxxxxC}{\ensuremath{0.0329\pm0.0020}} 
\newcommand{\hatcurPPtcircxxxxC}{\ensuremath{8000_{-2800}^{+3800}}} 
\newcommand{\hatcurPPtinfallxxxxC}{\ensuremath{340\pm110}}     
\newcommand{\hatcurXdistxxxxC}{\ensuremath{669\pm37}}          
\newcommand{\hatcurXAvxxxxC}{\ensuremath{0.051_{-0.051}^{+0.082}}} 
\newcommand{\hatcurXdistredxxxxC}{\ensuremath{661\pm36}}       
\newcommand{\hatcurXEBVxxxxC}{\ensuremath{0.016_{-0.016}^{+0.027}}} 
\newcommand{\hatcurXmvisoredxxxxC}{\ensuremath{12.692\pm0.030}} 
\newcommand{\hatcurXmiisoredxxxxC}{\ensuremath{12.125\pm0.021}} 
\newcommand{\hatcurXmjisoredxxxxC}{\ensuremath{11.798\pm0.014}} 
\newcommand{\hatcurXmhisoredxxxxC}{\ensuremath{11.570\pm0.016}} 
\newcommand{\hatcurXmkisoredxxxxC}{\ensuremath{11.522\pm0.016}} 
\newcommand{\hatcurXviisoredxxxxC}{\ensuremath{0.564\pm0.020}} 
\newcommand{\hatcurXvkisoredxxxxC}{\ensuremath{1.169\pm0.036}} 
\newcommand{\hatcurXjhisoredxxxxC}{\ensuremath{0.227\pm0.011}} 
\newcommand{\hatcurXjkisoredxxxxC}{\ensuremath{0.275\pm0.011}} 
\newcommand{\hatcurCCpmraxxxxC}{\ensuremath{0.60\pm0.90}}      
\newcommand{\hatcurCCpmdecxxxxC}{\ensuremath{-7.0\pm1.0}}      
\newcommand{\hatcurCCpmxxxxC}{\ensuremath{7.0\pm1.3}}          
 \newcommand{\hatcurhtrxxxxD}{HATS602-002}                      
\newcommand{\hatcurfieldxxxxD}{\ensuremath{string}}            
\newcommand{\hatcurCCraxxxxD}{\ensuremath{07^{\mathrm h}13^{\mathrm m}48.58{\mathrm s}}}                     
\newcommand{\hatcurCCdecxxxxD}{\ensuremath{-33{\arcdeg}26{\arcmin}14.4{\arcsec}}}                    
\newcommand{\hatcurCCmagxxxxD}{13.617}                         
\newcommand{\hatcurCCtwomassxxxxD}{2MASS~07134857-3326143}     
\newcommand{\hatcurCCgscxxxxD}{GSC~7107-03973}                 
\newcommand{\hatcurCCtassmvxxxxD}{\ensuremath{13.617\pm0.010}} 
\newcommand{\hatcurCCtassmvshortxxxxD}{\ensuremath{13.6}}      
\newcommand{\hatcurCCtassmBxxxxD}{\ensuremath{14.243\pm0.010}} 
\newcommand{\hatcurCCtassmBshortxxxxD}{\ensuremath{14.2}}      
\newcommand{\hatcurCCtassmIxxxxD}{\ensuremath{nff\pmnff}}      
\newcommand{\hatcurCCtassmIshortxxxxD}{\ensuremath{0.0}}       
\newcommand{\hatcurCCtassmgxxxxD}{\ensuremath{13.922\pm0.050}} 
\newcommand{\hatcurCCtassmgshortxxxxD}{\ensuremath{13.9}}      
\newcommand{\hatcurCCtassmrxxxxD}{\ensuremath{13.498\pm0.010}} 
\newcommand{\hatcurCCtassmrshortxxxxD}{\ensuremath{13.5}}      
\newcommand{\hatcurCCtassmixxxxD}{\ensuremath{13.396\pm0.050}} 
\newcommand{\hatcurCCtassmishortxxxxD}{\ensuremath{13.4}}      
\newcommand{\hatcurCCtwomassJmagxxxxD}{\ensuremath{12.543\pm0.024}} 
\newcommand{\hatcurCCtwomassHmagxxxxD}{\ensuremath{12.281\pm0.026}} 
\newcommand{\hatcurCCtwomassKmagxxxxD}{\ensuremath{12.245\pm0.029}} 
\newcommand{\hatcurCCcitJmagxxxxD}{\ensuremath{12.564\pm0.024}} 
\newcommand{\hatcurCCcitHmagxxxxD}{\ensuremath{12.277\pm0.026}} 
\newcommand{\hatcurCCcitKmagxxxxD}{\ensuremath{12.269\pm0.029}} 
\newcommand{\hatcurCCbbJmagxxxxD}{\ensuremath{12.607\pm0.026}} 
\newcommand{\hatcurCCbbHmagxxxxD}{\ensuremath{12.297\pm0.027}} 
\newcommand{\hatcurCCbbKmagxxxxD}{\ensuremath{12.289\pm0.029}} 
\newcommand{\hatcurCCesoJmagxxxxD}{\ensuremath{12.608\pm0.027}} 
\newcommand{\hatcurCCesoHmagxxxxD}{\ensuremath{12.291\pm0.030}} 
\newcommand{\hatcurCCesoKmagxxxxD}{\ensuremath{12.288\pm0.030}} 
\newcommand{\hatcurCCesoJHmagxxxxD}{\ensuremath{0.317\pm0.038}} 
\newcommand{\hatcurCCesoJKmagxxxxD}{\ensuremath{0.321\pm0.040}} 
\newcommand{\hatcurCCesoHKmagxxxxD}{\ensuremath{0.003\pm0.043}} 
\newcommand{\hatcurLCdipxxxxD}{\ensuremath{10.5}}              
\newcommand{\hatcurLCrprstarxxxxD}{\ensuremath{0.0976\pm0.0040}} 
\newcommand{\hatcurLCbsqxxxxD}{\ensuremath{0.21_{-0.13}^{+0.16}}} 
\newcommand{\hatcurLCimpxxxxD}{\ensuremath{0.46_{-0.18}^{+0.15}}} 
\newcommand{\hatcurLCzetaxxxxD}{\ensuremath{16.51\pm0.20}}     
\newcommand{\hatcurLCdurxxxxD}{\ensuremath{0.1361\pm0.0032}}   
\newcommand{\hatcurLCdurshortxxxxD}{\ensuremath{0.1361}}       
\newcommand{\hatcurLCdurhrxxxxD}{\ensuremath{3.265\pm0.076}}   
\newcommand{\hatcurLCdurhrshortxxxxD}{\ensuremath{3.265}}      
\newcommand{\hatcurLCqxxxxD}{\ensuremath{0.0594\pm0.0014}}     
\newcommand{\hatcurLCqshortxxxxD}{\ensuremath{0.059}}          
\newcommand{\hatcurLCingdurxxxxD}{\ensuremath{0.0149\pm0.0036}} 
\newcommand{\hatcurLCPxxxxD}{\ensuremath{2.2921020\pm0.0000021}} 
\newcommand{\hatcurLCPprecxxxxD}{\ensuremath{2.2921020}}       
\newcommand{\hatcurLCPshortxxxxD}{\ensuremath{2.2921}}         
\newcommand{\hatcurLCTxxxxD}{\ensuremath{2456768.60734\pm0.00069}} 
\newcommand{\hatcurLCTAxxxxD}{\ensuremath{2455794.4640\pm0.0011}} 
\newcommand{\hatcurLCTBxxxxD}{\ensuremath{2457396.64325\pm0.00090}} 
\newcommand{\hatcurLChatnetmAxxxxD}{\ensuremath{13.483320\pm0.000093}} 
\newcommand{\hatcurLCiblendAxxxxD}{\ensuremath{0.928\pm0.058}} 
\newcommand{\hatcurLChatnetmBxxxxD}{\ensuremath{13.48309\pm0.00010}} 
\newcommand{\hatcurLCiblendBxxxxD}{\ensuremath{0.885\pm0.064}} 
\newcommand{\hatcurLCrhoxxxxD}{\ensuremath{0.56\pm0.13}}       
\newcommand{\hatcurSMEiteffxxxxD}{\ensuremath{6320\pm170}}     
\newcommand{\hatcurSMEizfehxxxxD}{\ensuremath{0.320\pm0.076}}  
\newcommand{\hatcurSMEizfehshortxxxxD}{\ensuremath{0.32}}      
\newcommand{\hatcurSMEiloggxxxxD}{\ensuremath{4.55\pm0.21}}    
\newcommand{\hatcurSMEivsinxxxxD}{\ensuremath{5.86\pm0.49}}    
\newcommand{\hatcurSMEivmacxxxxD}{\ensuremath{4.82\pm0.26}}    
\newcommand{\hatcurSMEivmicxxxxD}{\ensuremath{1.53\pm0.19}}    
\newcommand{\hatcurSMEiiteffxxxxD}{\ensuremath{6060\pm120}}    
\newcommand{\hatcurSMEiizfehxxxxD}{\ensuremath{0.220\pm0.070}} 
\newcommand{\hatcurSMEiizfehshortxxxxD}{\ensuremath{0.22}}     
\newcommand{\hatcurSMEiiloggxxxxD}{\ensuremath{4.199\pm0.070}} 
\newcommand{\hatcurSMEiivsinxxxxD}{\ensuremath{6.04\pm0.36}}   
\newcommand{\hatcurSMEiivmacxxxxD}{\ensuremath{4.42\pm0.19}}   
\newcommand{\hatcurSMEiivmicxxxxD}{\ensuremath{1.28\pm0.10}}   
\newcommand{\hatcurLBizxxxxD}{\ensuremath{0.1719}}             
\newcommand{\hatcurLBiizxxxxD}{\ensuremath{0.3525}}            
\newcommand{\hatcurLBiixxxxD}{\ensuremath{0.2292}}             
\newcommand{\hatcurLBiiixxxxD}{\ensuremath{0.3581}}            
\newcommand{\hatcurLBiIxxxxD}{\ensuremath{0.2090}}             
\newcommand{\hatcurLBiiIxxxxD}{\ensuremath{0.3575}}            
\newcommand{\hatcurLBigxxxxD}{\ensuremath{0.5019}}             
\newcommand{\hatcurLBiigxxxxD}{\ensuremath{0.2762}}            
\newcommand{\hatcurLBirxxxxD}{\ensuremath{0.3138}}             
\newcommand{\hatcurLBiirxxxxD}{\ensuremath{0.3565}}            
\newcommand{\hatcurLBiRxxxxD}{\ensuremath{0.2901}}             
\newcommand{\hatcurLBiiRxxxxD}{\ensuremath{0.3582}}            
\newcommand{\hatcurLBikepxxxxD}{\ensuremath{0.1000}}           
\newcommand{\hatcurLBiikepxxxxD}{\ensuremath{0.1000}}          
\newcommand{\hatcurISOmxxxxD}{\ensuremath{1.273\pm0.067}}      
\newcommand{\hatcurISOmshortxxxxD}{\ensuremath{1.27}}          
\newcommand{\hatcurISOmlongxxxxD}{\ensuremath{1.273\pm0.067}}  
\newcommand{\hatcurISOrxxxxD}{\ensuremath{1.48_{-0.12}^{+0.20}}} 
\newcommand{\hatcurISOrshortxxxxD}{\ensuremath{1.48}}          
\newcommand{\hatcurISOrlongxxxxD}{\ensuremath{1.48_{-0.12}^{+0.20}}} 
\newcommand{\hatcurISOrhoxxxxD}{\ensuremath{0.55\pm0.13}}      
\newcommand{\hatcurISOrholongxxxxD}{\ensuremath{0.55\pm0.13}}  
\newcommand{\hatcurISOloggxxxxD}{\ensuremath{4.201\pm0.070}}   
\newcommand{\hatcurISOlumxxxxD}{\ensuremath{2.66_{-0.47}^{+0.77}}} 
\newcommand{\hatcurISOlumshortxxxxD}{\ensuremath{2.66}}        
\newcommand{\hatcurISOmvxxxxD}{\ensuremath{3.71\pm0.25}}       
\newcommand{\hatcurISOvixxxxD}{\ensuremath{0.616\pm0.036}}     
\newcommand{\hatcurISOagexxxxD}{\ensuremath{3.26\pm0.83}}      
\newcommand{\hatcurISOsigmaxxxxD}{\ensuremath{0.00120\pm0.00021}} 
\newcommand{\hatcurISOMJxxxxD}{\ensuremath{2.71\pm0.23}}       
\newcommand{\hatcurISOMHxxxxD}{\ensuremath{2.42\pm0.22}}       
\newcommand{\hatcurISOMKxxxxD}{\ensuremath{2.37\pm0.22}}       
\newcommand{\hatcurISOJKxxxxD}{\ensuremath{0.340\pm0.020}}     
\newcommand{\hatcurISOspecxxxxD}{F}                            
\newcommand{\hatcurRVKxxxxD}{\ensuremath{246\pm17}}            
\newcommand{\hatcurRVrkxxxxD}{\ensuremath{0\pm0}}              
\newcommand{\hatcurRVrhxxxxD}{\ensuremath{0\pm0}}              
\newcommand{\hatcurRVkxxxxD}{\ensuremath{0\pm0}}               
\newcommand{\hatcurRVhxxxxD}{\ensuremath{0\pm0}}               
\newcommand{\hatcurRVtronexxxxD}{\ensuremath{0\pm0}}           
\newcommand{\hatcurRVtrtwoxxxxD}{\ensuremath{0\pm0}}           
\newcommand{\hatcurRVgammaAxxxxD}{\ensuremath{8131\pm67}}      
\newcommand{\hatcurRVjitterAxxxxD}{\ensuremath{100\pm90}}      
\newcommand{\hatcurRVjittertwosiglimAxxxxD}{\ensuremath{<272.9}} 
\newcommand{\hatcurRVfitrmsAxxxxD}{\ensuremath{0.0}}           
\newcommand{\hatcurRVgammaBxxxxD}{\ensuremath{8163\pm13}}      
\newcommand{\hatcurRVjitterBxxxxD}{\ensuremath{0\pm12}}        
\newcommand{\hatcurRVjittertwosiglimBxxxxD}{\ensuremath{<20.6}} 
\newcommand{\hatcurRVfitrmsBxxxxD}{\ensuremath{0.0}}           
\newcommand{\hatcurRVeccenxxxxD}{\ensuremath{0\pm0}}           
\newcommand{\hatcurRVeccentwosiglimxxxxD}{\ensuremath{<0.000}} 
\newcommand{\hatcurRVomegaxxxxD}{\ensuremath{0\pm0}}           
\newcommand{\hatcurPPixxxxD}{\ensuremath{85.1\pm2.1}}          
\newcommand{\hatcurPPgxxxxD}{\ensuremath{23.4\pm5.2}}          
\newcommand{\hatcurPPloggxxxxD}{\ensuremath{3.369_{-0.137}^{+0.092}}} 
\newcommand{\hatcurPParxxxxD}{\ensuremath{5.36_{-0.56}^{+0.39}}} 
\newcommand{\hatcurPParelxxxxD}{\ensuremath{0.03689\pm0.00065}} 
\newcommand{\hatcurPPrhoxxxxD}{\ensuremath{0.83\pm0.28}}       
\newcommand{\hatcurPPmxxxxD}{\ensuremath{1.88\pm0.15}}         
\newcommand{\hatcurPPmshortxxxxD}{\ensuremath{1.88}}           
\newcommand{\hatcurPPmlongxxxxD}{\ensuremath{1.88\pm0.15}}     
\newcommand{\hatcurPPmexxxxD}{\ensuremath{598\pm46}}           
\newcommand{\hatcurPPmeshortxxxxD}{\ensuremath{597.8}}         
\newcommand{\hatcurPPmelongxxxxD}{\ensuremath{598\pm46}}       
\newcommand{\hatcurPPrxxxxD}{\ensuremath{1.40_{-0.14}^{+0.25}}} 
\newcommand{\hatcurPPrshortxxxxD}{\ensuremath{1.40}}           
\newcommand{\hatcurPPrlongxxxxD}{\ensuremath{1.40_{-0.14}^{+0.25}}} 
\newcommand{\hatcurPPrexxxxD}{\ensuremath{15.7_{-1.6}^{+2.8}}} 
\newcommand{\hatcurPPreshortxxxxD}{\ensuremath{15.7}}          
\newcommand{\hatcurPPrelongxxxxD}{\ensuremath{15.7_{-1.6}^{+2.8}}} 
\newcommand{\hatcurPPmrcorrxxxxD}{\ensuremath{0.39}}           
\newcommand{\hatcurPPteffxxxxD}{\ensuremath{1856_{-76}^{+105}}} 
\newcommand{\hatcurPPthetaxxxxD}{\ensuremath{0.076\pm0.011}}   
\newcommand{\hatcurPPfluxperixxxxD}{\ensuremath{2.67_{-0.41}^{+0.66}}} 
\newcommand{\hatcurPPfluxperidimxxxxD}{\ensuremath{9}}         
\newcommand{\hatcurPPfluxapxxxxD}{\ensuremath{2.67_{-0.41}^{+0.66}}} 
\newcommand{\hatcurPPfluxapdimxxxxD}{\ensuremath{9}}           
\newcommand{\hatcurPPfluxavgxxxxD}{\ensuremath{2.67_{-0.41}^{+0.66}}} 
\newcommand{\hatcurPPfluxavgdimxxxxD}{\ensuremath{9}}          
\newcommand{\hatcurPPfluxavglogxxxxD}{\ensuremath{9.427_{-0.073}^{+0.096}}} 
\newcommand{\hatcurXsecphasexxxxD}{\ensuremath{0\pm0}}         
\newcommand{\hatcurXsecondaryxxxxD}{\ensuremath{2456769.75339\pm0.00069}} 
\newcommand{\hatcurXsecdurxxxxD}{\ensuremath{0.1361\pm0.0032}} 
\newcommand{\hatcurXsecingdurxxxxD}{\ensuremath{0.0149\pm0.0036}} 
\newcommand{\hatcurPPphiconjxxxxD}{\ensuremath{0\pm0}}         
\newcommand{\hatcurPPperixxxxD}{\ensuremath{2456768.03431\pm0.00069}} 
\newcommand{\hatcurPPaequivxxxxD}{\ensuremath{0.0226\pm0.0021}} 
\newcommand{\hatcurPPtcircxxxxD}{\ensuremath{44\pm24}}         
\newcommand{\hatcurPPtinfallxxxxD}{\ensuremath{65\pm25}}       
\newcommand{\hatcurXdistxxxxD}{\ensuremath{964_{-79}^{+128}}}  
\newcommand{\hatcurXAvxxxxD}{\ensuremath{0.000\pm0.065}}       
\newcommand{\hatcurXdistredxxxxD}{\ensuremath{942_{-80}^{+126}}} 
\newcommand{\hatcurXEBVxxxxD}{\ensuremath{0.000\pm0.021}}      
\newcommand{\hatcurXmvisoredxxxxD}{\ensuremath{13.624\pm0.014}} 
\newcommand{\hatcurXmiisoredxxxxD}{\ensuremath{12.992\pm0.015}} 
\newcommand{\hatcurXmjisoredxxxxD}{\ensuremath{12.589_{-0.028}^{+0.019}}} 
\newcommand{\hatcurXmhisoredxxxxD}{\ensuremath{12.301_{-0.048}^{+0.026}}} 
\newcommand{\hatcurXmkisoredxxxxD}{\ensuremath{12.248_{-0.047}^{+0.026}}} 
\newcommand{\hatcurXviisoredxxxxD}{\ensuremath{0.630_{-0.015}^{+0.025}}} 
\newcommand{\hatcurXvkisoredxxxxD}{\ensuremath{1.374_{-0.031}^{+0.062}}} 
\newcommand{\hatcurXjhisoredxxxxD}{\ensuremath{0.291\pm0.016}} 
\newcommand{\hatcurXjkisoredxxxxD}{\ensuremath{0.343_{-0.012}^{+0.018}}} 
\newcommand{\hatcurCCpmraxxxxD}{\ensuremath{0.9\pm1.3}}        
\newcommand{\hatcurCCpmdecxxxxD}{\ensuremath{1.7\pm1.3}}       
\newcommand{\hatcurCCpmxxxxD}{\ensuremath{1.9\pm1.8}}          
 \newcommand{\hatcurCCbbHmag}[1]{\ifnum#1=39 
\hatcurCCbbHmagxxxxA
\else
\ifnum#1=40 
\hatcurCCbbHmagxxxxB
\else
\ifnum#1=41 
\hatcurCCbbHmagxxxxC
\else
\ifnum#1=42 
\hatcurCCbbHmagxxxxD
\else
??????\fi
\fi
\fi
\fi
}
\newcommand{\hatcurCCbbJmag}[1]{\ifnum#1=39 
\hatcurCCbbJmagxxxxA
\else
\ifnum#1=40 
\hatcurCCbbJmagxxxxB
\else
\ifnum#1=41 
\hatcurCCbbJmagxxxxC
\else
\ifnum#1=42 
\hatcurCCbbJmagxxxxD
\else
??????\fi
\fi
\fi
\fi
}
\newcommand{\hatcurCCbbKmag}[1]{\ifnum#1=39 
\hatcurCCbbKmagxxxxA
\else
\ifnum#1=40 
\hatcurCCbbKmagxxxxB
\else
\ifnum#1=41 
\hatcurCCbbKmagxxxxC
\else
\ifnum#1=42 
\hatcurCCbbKmagxxxxD
\else
??????\fi
\fi
\fi
\fi
}
\newcommand{\hatcurCCcitHmag}[1]{\ifnum#1=39 
\hatcurCCcitHmagxxxxA
\else
\ifnum#1=40 
\hatcurCCcitHmagxxxxB
\else
\ifnum#1=41 
\hatcurCCcitHmagxxxxC
\else
\ifnum#1=42 
\hatcurCCcitHmagxxxxD
\else
??????\fi
\fi
\fi
\fi
}
\newcommand{\hatcurCCcitJmag}[1]{\ifnum#1=39 
\hatcurCCcitJmagxxxxA
\else
\ifnum#1=40 
\hatcurCCcitJmagxxxxB
\else
\ifnum#1=41 
\hatcurCCcitJmagxxxxC
\else
\ifnum#1=42 
\hatcurCCcitJmagxxxxD
\else
??????\fi
\fi
\fi
\fi
}
\newcommand{\hatcurCCcitKmag}[1]{\ifnum#1=39 
\hatcurCCcitKmagxxxxA
\else
\ifnum#1=40 
\hatcurCCcitKmagxxxxB
\else
\ifnum#1=41 
\hatcurCCcitKmagxxxxC
\else
\ifnum#1=42 
\hatcurCCcitKmagxxxxD
\else
??????\fi
\fi
\fi
\fi
}
\newcommand{\hatcurCCdec}[1]{\ifnum#1=39 
\hatcurCCdecxxxxA
\else
\ifnum#1=40 
\hatcurCCdecxxxxB
\else
\ifnum#1=41 
\hatcurCCdecxxxxC
\else
\ifnum#1=42 
\hatcurCCdecxxxxD
\else
??????\fi
\fi
\fi
\fi
}
\newcommand{\hatcurCCesoHKmag}[1]{\ifnum#1=39 
\hatcurCCesoHKmagxxxxA
\else
\ifnum#1=40 
\hatcurCCesoHKmagxxxxB
\else
\ifnum#1=41 
\hatcurCCesoHKmagxxxxC
\else
\ifnum#1=42 
\hatcurCCesoHKmagxxxxD
\else
??????\fi
\fi
\fi
\fi
}
\newcommand{\hatcurCCesoHmag}[1]{\ifnum#1=39 
\hatcurCCesoHmagxxxxA
\else
\ifnum#1=40 
\hatcurCCesoHmagxxxxB
\else
\ifnum#1=41 
\hatcurCCesoHmagxxxxC
\else
\ifnum#1=42 
\hatcurCCesoHmagxxxxD
\else
??????\fi
\fi
\fi
\fi
}
\newcommand{\hatcurCCesoJHmag}[1]{\ifnum#1=39 
\hatcurCCesoJHmagxxxxA
\else
\ifnum#1=40 
\hatcurCCesoJHmagxxxxB
\else
\ifnum#1=41 
\hatcurCCesoJHmagxxxxC
\else
\ifnum#1=42 
\hatcurCCesoJHmagxxxxD
\else
??????\fi
\fi
\fi
\fi
}
\newcommand{\hatcurCCesoJKmag}[1]{\ifnum#1=39 
\hatcurCCesoJKmagxxxxA
\else
\ifnum#1=40 
\hatcurCCesoJKmagxxxxB
\else
\ifnum#1=41 
\hatcurCCesoJKmagxxxxC
\else
\ifnum#1=42 
\hatcurCCesoJKmagxxxxD
\else
??????\fi
\fi
\fi
\fi
}
\newcommand{\hatcurCCesoJmag}[1]{\ifnum#1=39 
\hatcurCCesoJmagxxxxA
\else
\ifnum#1=40 
\hatcurCCesoJmagxxxxB
\else
\ifnum#1=41 
\hatcurCCesoJmagxxxxC
\else
\ifnum#1=42 
\hatcurCCesoJmagxxxxD
\else
??????\fi
\fi
\fi
\fi
}
\newcommand{\hatcurCCesoKmag}[1]{\ifnum#1=39 
\hatcurCCesoKmagxxxxA
\else
\ifnum#1=40 
\hatcurCCesoKmagxxxxB
\else
\ifnum#1=41 
\hatcurCCesoKmagxxxxC
\else
\ifnum#1=42 
\hatcurCCesoKmagxxxxD
\else
??????\fi
\fi
\fi
\fi
}
\newcommand{\hatcurCCgsc}[1]{\ifnum#1=39 
\hatcurCCgscxxxxA
\else
\ifnum#1=40 
\hatcurCCgscxxxxB
\else
\ifnum#1=41 
\hatcurCCgscxxxxC
\else
\ifnum#1=42 
\hatcurCCgscxxxxD
\else
??????\fi
\fi
\fi
\fi
}
\newcommand{\hatcurCCmag}[1]{\ifnum#1=39 
\hatcurCCmagxxxxA
\else
\ifnum#1=40 
\hatcurCCmagxxxxB
\else
\ifnum#1=41 
\hatcurCCmagxxxxC
\else
\ifnum#1=42 
\hatcurCCmagxxxxD
\else
??????\fi
\fi
\fi
\fi
}
\newcommand{\hatcurCCpm}[1]{\ifnum#1=39 
\hatcurCCpmxxxxA
\else
\ifnum#1=40 
\hatcurCCpmxxxxB
\else
\ifnum#1=41 
\hatcurCCpmxxxxC
\else
\ifnum#1=42 
\hatcurCCpmxxxxD
\else
??????\fi
\fi
\fi
\fi
}
\newcommand{\hatcurCCpmdec}[1]{\ifnum#1=39 
\hatcurCCpmdecxxxxA
\else
\ifnum#1=40 
\hatcurCCpmdecxxxxB
\else
\ifnum#1=41 
\hatcurCCpmdecxxxxC
\else
\ifnum#1=42 
\hatcurCCpmdecxxxxD
\else
??????\fi
\fi
\fi
\fi
}
\newcommand{\hatcurCCpmra}[1]{\ifnum#1=39 
\hatcurCCpmraxxxxA
\else
\ifnum#1=40 
\hatcurCCpmraxxxxB
\else
\ifnum#1=41 
\hatcurCCpmraxxxxC
\else
\ifnum#1=42 
\hatcurCCpmraxxxxD
\else
??????\fi
\fi
\fi
\fi
}
\newcommand{\hatcurCCra}[1]{\ifnum#1=39 
\hatcurCCraxxxxA
\else
\ifnum#1=40 
\hatcurCCraxxxxB
\else
\ifnum#1=41 
\hatcurCCraxxxxC
\else
\ifnum#1=42 
\hatcurCCraxxxxD
\else
??????\fi
\fi
\fi
\fi
}
\newcommand{\hatcurCCtassmB}[1]{\ifnum#1=39 
\hatcurCCtassmBxxxxA
\else
\ifnum#1=40 
\hatcurCCtassmBxxxxB
\else
\ifnum#1=41 
\hatcurCCtassmBxxxxC
\else
\ifnum#1=42 
\hatcurCCtassmBxxxxD
\else
??????\fi
\fi
\fi
\fi
}
\newcommand{\hatcurCCtassmBshort}[1]{\ifnum#1=39 
\hatcurCCtassmBshortxxxxA
\else
\ifnum#1=40 
\hatcurCCtassmBshortxxxxB
\else
\ifnum#1=41 
\hatcurCCtassmBshortxxxxC
\else
\ifnum#1=42 
\hatcurCCtassmBshortxxxxD
\else
??????\fi
\fi
\fi
\fi
}
\newcommand{\hatcurCCtassmg}[1]{\ifnum#1=39 
\hatcurCCtassmgxxxxA
\else
\ifnum#1=40 
\hatcurCCtassmgxxxxB
\else
\ifnum#1=41 
\hatcurCCtassmgxxxxC
\else
\ifnum#1=42 
\hatcurCCtassmgxxxxD
\else
??????\fi
\fi
\fi
\fi
}
\newcommand{\hatcurCCtassmgshort}[1]{\ifnum#1=39 
\hatcurCCtassmgshortxxxxA
\else
\ifnum#1=40 
\hatcurCCtassmgshortxxxxB
\else
\ifnum#1=41 
\hatcurCCtassmgshortxxxxC
\else
\ifnum#1=42 
\hatcurCCtassmgshortxxxxD
\else
??????\fi
\fi
\fi
\fi
}
\newcommand{\hatcurCCtassmi}[1]{\ifnum#1=39 
\hatcurCCtassmixxxxA
\else
\ifnum#1=40 
\hatcurCCtassmixxxxB
\else
\ifnum#1=41 
\hatcurCCtassmixxxxC
\else
\ifnum#1=42 
\hatcurCCtassmixxxxD
\else
??????\fi
\fi
\fi
\fi
}
\newcommand{\hatcurCCtassmI}[1]{\ifnum#1=39 
\hatcurCCtassmIxxxxA
\else
\ifnum#1=40 
\hatcurCCtassmIxxxxB
\else
\ifnum#1=41 
\hatcurCCtassmIxxxxC
\else
\ifnum#1=42 
\hatcurCCtassmIxxxxD
\else
??????\fi
\fi
\fi
\fi
}
\newcommand{\hatcurCCtassmishort}[1]{\ifnum#1=39 
\hatcurCCtassmishortxxxxA
\else
\ifnum#1=40 
\hatcurCCtassmishortxxxxB
\else
\ifnum#1=41 
\hatcurCCtassmishortxxxxC
\else
\ifnum#1=42 
\hatcurCCtassmishortxxxxD
\else
??????\fi
\fi
\fi
\fi
}
\newcommand{\hatcurCCtassmIshort}[1]{\ifnum#1=39 
\hatcurCCtassmIshortxxxxA
\else
\ifnum#1=40 
\hatcurCCtassmIshortxxxxB
\else
\ifnum#1=41 
\hatcurCCtassmIshortxxxxC
\else
\ifnum#1=42 
\hatcurCCtassmIshortxxxxD
\else
??????\fi
\fi
\fi
\fi
}
\newcommand{\hatcurCCtassmr}[1]{\ifnum#1=39 
\hatcurCCtassmrxxxxA
\else
\ifnum#1=40 
\hatcurCCtassmrxxxxB
\else
\ifnum#1=41 
\hatcurCCtassmrxxxxC
\else
\ifnum#1=42 
\hatcurCCtassmrxxxxD
\else
??????\fi
\fi
\fi
\fi
}
\newcommand{\hatcurCCtassmrshort}[1]{\ifnum#1=39 
\hatcurCCtassmrshortxxxxA
\else
\ifnum#1=40 
\hatcurCCtassmrshortxxxxB
\else
\ifnum#1=41 
\hatcurCCtassmrshortxxxxC
\else
\ifnum#1=42 
\hatcurCCtassmrshortxxxxD
\else
??????\fi
\fi
\fi
\fi
}
\newcommand{\hatcurCCtassmv}[1]{\ifnum#1=39 
\hatcurCCtassmvxxxxA
\else
\ifnum#1=40 
\hatcurCCtassmvxxxxB
\else
\ifnum#1=41 
\hatcurCCtassmvxxxxC
\else
\ifnum#1=42 
\hatcurCCtassmvxxxxD
\else
??????\fi
\fi
\fi
\fi
}
\newcommand{\hatcurCCtassmvshort}[1]{\ifnum#1=39 
\hatcurCCtassmvshortxxxxA
\else
\ifnum#1=40 
\hatcurCCtassmvshortxxxxB
\else
\ifnum#1=41 
\hatcurCCtassmvshortxxxxC
\else
\ifnum#1=42 
\hatcurCCtassmvshortxxxxD
\else
??????\fi
\fi
\fi
\fi
}
\newcommand{\hatcurCCtwomass}[1]{\ifnum#1=39 
\hatcurCCtwomassxxxxA
\else
\ifnum#1=40 
\hatcurCCtwomassxxxxB
\else
\ifnum#1=41 
\hatcurCCtwomassxxxxC
\else
\ifnum#1=42 
\hatcurCCtwomassxxxxD
\else
??????\fi
\fi
\fi
\fi
}
\newcommand{\hatcurCCtwomassHmag}[1]{\ifnum#1=39 
\hatcurCCtwomassHmagxxxxA
\else
\ifnum#1=40 
\hatcurCCtwomassHmagxxxxB
\else
\ifnum#1=41 
\hatcurCCtwomassHmagxxxxC
\else
\ifnum#1=42 
\hatcurCCtwomassHmagxxxxD
\else
??????\fi
\fi
\fi
\fi
}
\newcommand{\hatcurCCtwomassJmag}[1]{\ifnum#1=39 
\hatcurCCtwomassJmagxxxxA
\else
\ifnum#1=40 
\hatcurCCtwomassJmagxxxxB
\else
\ifnum#1=41 
\hatcurCCtwomassJmagxxxxC
\else
\ifnum#1=42 
\hatcurCCtwomassJmagxxxxD
\else
??????\fi
\fi
\fi
\fi
}
\newcommand{\hatcurCCtwomassKmag}[1]{\ifnum#1=39 
\hatcurCCtwomassKmagxxxxA
\else
\ifnum#1=40 
\hatcurCCtwomassKmagxxxxB
\else
\ifnum#1=41 
\hatcurCCtwomassKmagxxxxC
\else
\ifnum#1=42 
\hatcurCCtwomassKmagxxxxD
\else
??????\fi
\fi
\fi
\fi
}
\newcommand{\hatcurfield}[1]{\ifnum#1=39 
\hatcurfieldxxxxA
\else
\ifnum#1=40 
\hatcurfieldxxxxB
\else
\ifnum#1=41 
\hatcurfieldxxxxC
\else
\ifnum#1=42 
\hatcurfieldxxxxD
\else
??????\fi
\fi
\fi
\fi
}
\newcommand{\hatcurhtr}[1]{\ifnum#1=39 
\hatcurhtrxxxxA
\else
\ifnum#1=40 
\hatcurhtrxxxxB
\else
\ifnum#1=41 
\hatcurhtrxxxxC
\else
\ifnum#1=42 
\hatcurhtrxxxxD
\else
??????\fi
\fi
\fi
\fi
}
\newcommand{\hatcurISOage}[1]{\ifnum#1=39 
\hatcurISOagexxxxA
\else
\ifnum#1=40 
\hatcurISOagexxxxB
\else
\ifnum#1=41 
\hatcurISOagexxxxC
\else
\ifnum#1=42 
\hatcurISOagexxxxD
\else
??????\fi
\fi
\fi
\fi
}
\newcommand{\hatcurISOJK}[1]{\ifnum#1=39 
\hatcurISOJKxxxxA
\else
\ifnum#1=40 
\hatcurISOJKxxxxB
\else
\ifnum#1=41 
\hatcurISOJKxxxxC
\else
\ifnum#1=42 
\hatcurISOJKxxxxD
\else
??????\fi
\fi
\fi
\fi
}
\newcommand{\hatcurISOlogg}[1]{\ifnum#1=39 
\hatcurISOloggxxxxA
\else
\ifnum#1=40 
\hatcurISOloggxxxxB
\else
\ifnum#1=41 
\hatcurISOloggxxxxC
\else
\ifnum#1=42 
\hatcurISOloggxxxxD
\else
??????\fi
\fi
\fi
\fi
}
\newcommand{\hatcurISOlum}[1]{\ifnum#1=39 
\hatcurISOlumxxxxA
\else
\ifnum#1=40 
\hatcurISOlumxxxxB
\else
\ifnum#1=41 
\hatcurISOlumxxxxC
\else
\ifnum#1=42 
\hatcurISOlumxxxxD
\else
??????\fi
\fi
\fi
\fi
}
\newcommand{\hatcurISOlumshort}[1]{\ifnum#1=39 
\hatcurISOlumshortxxxxA
\else
\ifnum#1=40 
\hatcurISOlumshortxxxxB
\else
\ifnum#1=41 
\hatcurISOlumshortxxxxC
\else
\ifnum#1=42 
\hatcurISOlumshortxxxxD
\else
??????\fi
\fi
\fi
\fi
}
\newcommand{\hatcurISOm}[1]{\ifnum#1=39 
\hatcurISOmxxxxA
\else
\ifnum#1=40 
\hatcurISOmxxxxB
\else
\ifnum#1=41 
\hatcurISOmxxxxC
\else
\ifnum#1=42 
\hatcurISOmxxxxD
\else
??????\fi
\fi
\fi
\fi
}
\newcommand{\hatcurISOMH}[1]{\ifnum#1=39 
\hatcurISOMHxxxxA
\else
\ifnum#1=40 
\hatcurISOMHxxxxB
\else
\ifnum#1=41 
\hatcurISOMHxxxxC
\else
\ifnum#1=42 
\hatcurISOMHxxxxD
\else
??????\fi
\fi
\fi
\fi
}
\newcommand{\hatcurISOMJ}[1]{\ifnum#1=39 
\hatcurISOMJxxxxA
\else
\ifnum#1=40 
\hatcurISOMJxxxxB
\else
\ifnum#1=41 
\hatcurISOMJxxxxC
\else
\ifnum#1=42 
\hatcurISOMJxxxxD
\else
??????\fi
\fi
\fi
\fi
}
\newcommand{\hatcurISOMK}[1]{\ifnum#1=39 
\hatcurISOMKxxxxA
\else
\ifnum#1=40 
\hatcurISOMKxxxxB
\else
\ifnum#1=41 
\hatcurISOMKxxxxC
\else
\ifnum#1=42 
\hatcurISOMKxxxxD
\else
??????\fi
\fi
\fi
\fi
}
\newcommand{\hatcurISOmlong}[1]{\ifnum#1=39 
\hatcurISOmlongxxxxA
\else
\ifnum#1=40 
\hatcurISOmlongxxxxB
\else
\ifnum#1=41 
\hatcurISOmlongxxxxC
\else
\ifnum#1=42 
\hatcurISOmlongxxxxD
\else
??????\fi
\fi
\fi
\fi
}
\newcommand{\hatcurISOmshort}[1]{\ifnum#1=39 
\hatcurISOmshortxxxxA
\else
\ifnum#1=40 
\hatcurISOmshortxxxxB
\else
\ifnum#1=41 
\hatcurISOmshortxxxxC
\else
\ifnum#1=42 
\hatcurISOmshortxxxxD
\else
??????\fi
\fi
\fi
\fi
}
\newcommand{\hatcurISOmv}[1]{\ifnum#1=39 
\hatcurISOmvxxxxA
\else
\ifnum#1=40 
\hatcurISOmvxxxxB
\else
\ifnum#1=41 
\hatcurISOmvxxxxC
\else
\ifnum#1=42 
\hatcurISOmvxxxxD
\else
??????\fi
\fi
\fi
\fi
}
\newcommand{\hatcurISOr}[1]{\ifnum#1=39 
\hatcurISOrxxxxA
\else
\ifnum#1=40 
\hatcurISOrxxxxB
\else
\ifnum#1=41 
\hatcurISOrxxxxC
\else
\ifnum#1=42 
\hatcurISOrxxxxD
\else
??????\fi
\fi
\fi
\fi
}
\newcommand{\hatcurISOrho}[1]{\ifnum#1=39 
\hatcurISOrhoxxxxA
\else
\ifnum#1=40 
\hatcurISOrhoxxxxB
\else
\ifnum#1=41 
\hatcurISOrhoxxxxC
\else
\ifnum#1=42 
\hatcurISOrhoxxxxD
\else
??????\fi
\fi
\fi
\fi
}
\newcommand{\hatcurISOrholong}[1]{\ifnum#1=39 
\hatcurISOrholongxxxxA
\else
\ifnum#1=40 
\hatcurISOrholongxxxxB
\else
\ifnum#1=41 
\hatcurISOrholongxxxxC
\else
\ifnum#1=42 
\hatcurISOrholongxxxxD
\else
??????\fi
\fi
\fi
\fi
}
\newcommand{\hatcurISOrlong}[1]{\ifnum#1=39 
\hatcurISOrlongxxxxA
\else
\ifnum#1=40 
\hatcurISOrlongxxxxB
\else
\ifnum#1=41 
\hatcurISOrlongxxxxC
\else
\ifnum#1=42 
\hatcurISOrlongxxxxD
\else
??????\fi
\fi
\fi
\fi
}
\newcommand{\hatcurISOrshort}[1]{\ifnum#1=39 
\hatcurISOrshortxxxxA
\else
\ifnum#1=40 
\hatcurISOrshortxxxxB
\else
\ifnum#1=41 
\hatcurISOrshortxxxxC
\else
\ifnum#1=42 
\hatcurISOrshortxxxxD
\else
??????\fi
\fi
\fi
\fi
}
\newcommand{\hatcurISOsigma}[1]{\ifnum#1=39 
\hatcurISOsigmaxxxxA
\else
\ifnum#1=40 
\hatcurISOsigmaxxxxB
\else
\ifnum#1=41 
\hatcurISOsigmaxxxxC
\else
\ifnum#1=42 
\hatcurISOsigmaxxxxD
\else
??????\fi
\fi
\fi
\fi
}
\newcommand{\hatcurISOspec}[1]{\ifnum#1=39 
\hatcurISOspecxxxxA
\else
\ifnum#1=40 
\hatcurISOspecxxxxB
\else
\ifnum#1=41 
\hatcurISOspecxxxxC
\else
\ifnum#1=42 
\hatcurISOspecxxxxD
\else
??????\fi
\fi
\fi
\fi
}
\newcommand{\hatcurISOvi}[1]{\ifnum#1=39 
\hatcurISOvixxxxA
\else
\ifnum#1=40 
\hatcurISOvixxxxB
\else
\ifnum#1=41 
\hatcurISOvixxxxC
\else
\ifnum#1=42 
\hatcurISOvixxxxD
\else
??????\fi
\fi
\fi
\fi
}
\newcommand{\hatcurLBig}[1]{\ifnum#1=39 
\hatcurLBigxxxxA
\else
\ifnum#1=40 
\hatcurLBigxxxxB
\else
\ifnum#1=41 
\hatcurLBigxxxxC
\else
\ifnum#1=42 
\hatcurLBigxxxxD
\else
??????\fi
\fi
\fi
\fi
}
\newcommand{\hatcurLBii}[1]{\ifnum#1=39 
\hatcurLBiixxxxA
\else
\ifnum#1=40 
\hatcurLBiixxxxB
\else
\ifnum#1=41 
\hatcurLBiixxxxC
\else
\ifnum#1=42 
\hatcurLBiixxxxD
\else
??????\fi
\fi
\fi
\fi
}
\newcommand{\hatcurLBiI}[1]{\ifnum#1=39 
\hatcurLBiIxxxxA
\else
\ifnum#1=40 
\hatcurLBiIxxxxB
\else
\ifnum#1=41 
\hatcurLBiIxxxxC
\else
\ifnum#1=42 
\hatcurLBiIxxxxD
\else
??????\fi
\fi
\fi
\fi
}
\newcommand{\hatcurLBiig}[1]{\ifnum#1=39 
\hatcurLBiigxxxxA
\else
\ifnum#1=40 
\hatcurLBiigxxxxB
\else
\ifnum#1=41 
\hatcurLBiigxxxxC
\else
\ifnum#1=42 
\hatcurLBiigxxxxD
\else
??????\fi
\fi
\fi
\fi
}
\newcommand{\hatcurLBiii}[1]{\ifnum#1=39 
\hatcurLBiiixxxxA
\else
\ifnum#1=40 
\hatcurLBiiixxxxB
\else
\ifnum#1=41 
\hatcurLBiiixxxxC
\else
\ifnum#1=42 
\hatcurLBiiixxxxD
\else
??????\fi
\fi
\fi
\fi
}
\newcommand{\hatcurLBiiI}[1]{\ifnum#1=39 
\hatcurLBiiIxxxxA
\else
\ifnum#1=40 
\hatcurLBiiIxxxxB
\else
\ifnum#1=41 
\hatcurLBiiIxxxxC
\else
\ifnum#1=42 
\hatcurLBiiIxxxxD
\else
??????\fi
\fi
\fi
\fi
}
\newcommand{\hatcurLBiikep}[1]{\ifnum#1=39 
\hatcurLBiikepxxxxA
\else
\ifnum#1=40 
\hatcurLBiikepxxxxB
\else
\ifnum#1=41 
\hatcurLBiikepxxxxC
\else
\ifnum#1=42 
\hatcurLBiikepxxxxD
\else
??????\fi
\fi
\fi
\fi
}
\newcommand{\hatcurLBiir}[1]{\ifnum#1=39 
\hatcurLBiirxxxxA
\else
\ifnum#1=40 
\hatcurLBiirxxxxB
\else
\ifnum#1=41 
\hatcurLBiirxxxxC
\else
\ifnum#1=42 
\hatcurLBiirxxxxD
\else
??????\fi
\fi
\fi
\fi
}
\newcommand{\hatcurLBiiR}[1]{\ifnum#1=39 
\hatcurLBiiRxxxxA
\else
\ifnum#1=40 
\hatcurLBiiRxxxxB
\else
\ifnum#1=41 
\hatcurLBiiRxxxxC
\else
\ifnum#1=42 
\hatcurLBiiRxxxxD
\else
??????\fi
\fi
\fi
\fi
}
\newcommand{\hatcurLBiiz}[1]{\ifnum#1=39 
\hatcurLBiizxxxxA
\else
\ifnum#1=40 
\hatcurLBiizxxxxB
\else
\ifnum#1=41 
\hatcurLBiizxxxxC
\else
\ifnum#1=42 
\hatcurLBiizxxxxD
\else
??????\fi
\fi
\fi
\fi
}
\newcommand{\hatcurLBikep}[1]{\ifnum#1=39 
\hatcurLBikepxxxxA
\else
\ifnum#1=40 
\hatcurLBikepxxxxB
\else
\ifnum#1=41 
\hatcurLBikepxxxxC
\else
\ifnum#1=42 
\hatcurLBikepxxxxD
\else
??????\fi
\fi
\fi
\fi
}
\newcommand{\hatcurLBir}[1]{\ifnum#1=39 
\hatcurLBirxxxxA
\else
\ifnum#1=40 
\hatcurLBirxxxxB
\else
\ifnum#1=41 
\hatcurLBirxxxxC
\else
\ifnum#1=42 
\hatcurLBirxxxxD
\else
??????\fi
\fi
\fi
\fi
}
\newcommand{\hatcurLBiR}[1]{\ifnum#1=39 
\hatcurLBiRxxxxA
\else
\ifnum#1=40 
\hatcurLBiRxxxxB
\else
\ifnum#1=41 
\hatcurLBiRxxxxC
\else
\ifnum#1=42 
\hatcurLBiRxxxxD
\else
??????\fi
\fi
\fi
\fi
}
\newcommand{\hatcurLBiz}[1]{\ifnum#1=39 
\hatcurLBizxxxxA
\else
\ifnum#1=40 
\hatcurLBizxxxxB
\else
\ifnum#1=41 
\hatcurLBizxxxxC
\else
\ifnum#1=42 
\hatcurLBizxxxxD
\else
??????\fi
\fi
\fi
\fi
}
\newcommand{\hatcurLCbsq}[1]{\ifnum#1=39 
\hatcurLCbsqxxxxA
\else
\ifnum#1=40 
\hatcurLCbsqxxxxB
\else
\ifnum#1=41 
\hatcurLCbsqxxxxC
\else
\ifnum#1=42 
\hatcurLCbsqxxxxD
\else
??????\fi
\fi
\fi
\fi
}
\newcommand{\hatcurLCdip}[1]{\ifnum#1=39 
\hatcurLCdipxxxxA
\else
\ifnum#1=40 
\hatcurLCdipxxxxB
\else
\ifnum#1=41 
\hatcurLCdipxxxxC
\else
\ifnum#1=42 
\hatcurLCdipxxxxD
\else
??????\fi
\fi
\fi
\fi
}
\newcommand{\hatcurLCdur}[1]{\ifnum#1=39 
\hatcurLCdurxxxxA
\else
\ifnum#1=40 
\hatcurLCdurxxxxB
\else
\ifnum#1=41 
\hatcurLCdurxxxxC
\else
\ifnum#1=42 
\hatcurLCdurxxxxD
\else
??????\fi
\fi
\fi
\fi
}
\newcommand{\hatcurLCdurhr}[1]{\ifnum#1=39 
\hatcurLCdurhrxxxxA
\else
\ifnum#1=40 
\hatcurLCdurhrxxxxB
\else
\ifnum#1=41 
\hatcurLCdurhrxxxxC
\else
\ifnum#1=42 
\hatcurLCdurhrxxxxD
\else
??????\fi
\fi
\fi
\fi
}
\newcommand{\hatcurLCdurhrshort}[1]{\ifnum#1=39 
\hatcurLCdurhrshortxxxxA
\else
\ifnum#1=40 
\hatcurLCdurhrshortxxxxB
\else
\ifnum#1=41 
\hatcurLCdurhrshortxxxxC
\else
\ifnum#1=42 
\hatcurLCdurhrshortxxxxD
\else
??????\fi
\fi
\fi
\fi
}
\newcommand{\hatcurLCdurshort}[1]{\ifnum#1=39 
\hatcurLCdurshortxxxxA
\else
\ifnum#1=40 
\hatcurLCdurshortxxxxB
\else
\ifnum#1=41 
\hatcurLCdurshortxxxxC
\else
\ifnum#1=42 
\hatcurLCdurshortxxxxD
\else
??????\fi
\fi
\fi
\fi
}
\newcommand{\hatcurLChatnetm}[1]{\ifnum#1=41 
\hatcurLChatnetmxxxxC
\else
??????\fi
}
\newcommand{\hatcurLChatnetmA}[1]{\ifnum#1=39 
\hatcurLChatnetmAxxxxA
\else
\ifnum#1=40 
\hatcurLChatnetmAxxxxB
\else
\ifnum#1=42 
\hatcurLChatnetmAxxxxD
\else
??????\fi
\fi
\fi
}
\newcommand{\hatcurLChatnetmB}[1]{\ifnum#1=39 
\hatcurLChatnetmBxxxxA
\else
\ifnum#1=40 
\hatcurLChatnetmBxxxxB
\else
\ifnum#1=42 
\hatcurLChatnetmBxxxxD
\else
??????\fi
\fi
\fi
}
\newcommand{\hatcurLChatnetmC}[1]{\ifnum#1=40 
\hatcurLChatnetmCxxxxB
\else
??????\fi
}
\newcommand{\hatcurLCiblend}[1]{\ifnum#1=41 
\hatcurLCiblendxxxxC
\else
??????\fi
}
\newcommand{\hatcurLCiblendA}[1]{\ifnum#1=39 
\hatcurLCiblendAxxxxA
\else
\ifnum#1=40 
\hatcurLCiblendAxxxxB
\else
\ifnum#1=42 
\hatcurLCiblendAxxxxD
\else
??????\fi
\fi
\fi
}
\newcommand{\hatcurLCiblendB}[1]{\ifnum#1=39 
\hatcurLCiblendBxxxxA
\else
\ifnum#1=40 
\hatcurLCiblendBxxxxB
\else
\ifnum#1=42 
\hatcurLCiblendBxxxxD
\else
??????\fi
\fi
\fi
}
\newcommand{\hatcurLCiblendC}[1]{\ifnum#1=40 
\hatcurLCiblendCxxxxB
\else
??????\fi
}
\newcommand{\hatcurLCimp}[1]{\ifnum#1=39 
\hatcurLCimpxxxxA
\else
\ifnum#1=40 
\hatcurLCimpxxxxB
\else
\ifnum#1=41 
\hatcurLCimpxxxxC
\else
\ifnum#1=42 
\hatcurLCimpxxxxD
\else
??????\fi
\fi
\fi
\fi
}
\newcommand{\hatcurLCingdur}[1]{\ifnum#1=39 
\hatcurLCingdurxxxxA
\else
\ifnum#1=40 
\hatcurLCingdurxxxxB
\else
\ifnum#1=41 
\hatcurLCingdurxxxxC
\else
\ifnum#1=42 
\hatcurLCingdurxxxxD
\else
??????\fi
\fi
\fi
\fi
}
\newcommand{\hatcurLCP}[1]{\ifnum#1=39 
\hatcurLCPxxxxA
\else
\ifnum#1=40 
\hatcurLCPxxxxB
\else
\ifnum#1=41 
\hatcurLCPxxxxC
\else
\ifnum#1=42 
\hatcurLCPxxxxD
\else
??????\fi
\fi
\fi
\fi
}
\newcommand{\hatcurLCPprec}[1]{\ifnum#1=39 
\hatcurLCPprecxxxxA
\else
\ifnum#1=40 
\hatcurLCPprecxxxxB
\else
\ifnum#1=41 
\hatcurLCPprecxxxxC
\else
\ifnum#1=42 
\hatcurLCPprecxxxxD
\else
??????\fi
\fi
\fi
\fi
}
\newcommand{\hatcurLCPshort}[1]{\ifnum#1=39 
\hatcurLCPshortxxxxA
\else
\ifnum#1=40 
\hatcurLCPshortxxxxB
\else
\ifnum#1=41 
\hatcurLCPshortxxxxC
\else
\ifnum#1=42 
\hatcurLCPshortxxxxD
\else
??????\fi
\fi
\fi
\fi
}
\newcommand{\hatcurLCq}[1]{\ifnum#1=39 
\hatcurLCqxxxxA
\else
\ifnum#1=40 
\hatcurLCqxxxxB
\else
\ifnum#1=41 
\hatcurLCqxxxxC
\else
\ifnum#1=42 
\hatcurLCqxxxxD
\else
??????\fi
\fi
\fi
\fi
}
\newcommand{\hatcurLCqshort}[1]{\ifnum#1=39 
\hatcurLCqshortxxxxA
\else
\ifnum#1=40 
\hatcurLCqshortxxxxB
\else
\ifnum#1=41 
\hatcurLCqshortxxxxC
\else
\ifnum#1=42 
\hatcurLCqshortxxxxD
\else
??????\fi
\fi
\fi
\fi
}
\newcommand{\hatcurLCrho}[1]{\ifnum#1=39 
\hatcurLCrhoxxxxA
\else
\ifnum#1=40 
\hatcurLCrhoxxxxB
\else
\ifnum#1=41 
\hatcurLCrhoxxxxC
\else
\ifnum#1=42 
\hatcurLCrhoxxxxD
\else
??????\fi
\fi
\fi
\fi
}
\newcommand{\hatcurLCrprstar}[1]{\ifnum#1=39 
\hatcurLCrprstarxxxxA
\else
\ifnum#1=40 
\hatcurLCrprstarxxxxB
\else
\ifnum#1=41 
\hatcurLCrprstarxxxxC
\else
\ifnum#1=42 
\hatcurLCrprstarxxxxD
\else
??????\fi
\fi
\fi
\fi
}
\newcommand{\hatcurLCT}[1]{\ifnum#1=39 
\hatcurLCTxxxxA
\else
\ifnum#1=40 
\hatcurLCTxxxxB
\else
\ifnum#1=41 
\hatcurLCTxxxxC
\else
\ifnum#1=42 
\hatcurLCTxxxxD
\else
??????\fi
\fi
\fi
\fi
}
\newcommand{\hatcurLCTA}[1]{\ifnum#1=39 
\hatcurLCTAxxxxA
\else
\ifnum#1=40 
\hatcurLCTAxxxxB
\else
\ifnum#1=41 
\hatcurLCTAxxxxC
\else
\ifnum#1=42 
\hatcurLCTAxxxxD
\else
??????\fi
\fi
\fi
\fi
}
\newcommand{\hatcurLCTB}[1]{\ifnum#1=39 
\hatcurLCTBxxxxA
\else
\ifnum#1=40 
\hatcurLCTBxxxxB
\else
\ifnum#1=41 
\hatcurLCTBxxxxC
\else
\ifnum#1=42 
\hatcurLCTBxxxxD
\else
??????\fi
\fi
\fi
\fi
}
\newcommand{\hatcurLCzeta}[1]{\ifnum#1=39 
\hatcurLCzetaxxxxA
\else
\ifnum#1=40 
\hatcurLCzetaxxxxB
\else
\ifnum#1=41 
\hatcurLCzetaxxxxC
\else
\ifnum#1=42 
\hatcurLCzetaxxxxD
\else
??????\fi
\fi
\fi
\fi
}
\newcommand{\hatcurPPaequiv}[1]{\ifnum#1=39 
\hatcurPPaequivxxxxA
\else
\ifnum#1=40 
\hatcurPPaequivxxxxB
\else
\ifnum#1=41 
\hatcurPPaequivxxxxC
\else
\ifnum#1=42 
\hatcurPPaequivxxxxD
\else
??????\fi
\fi
\fi
\fi
}
\newcommand{\hatcurPPar}[1]{\ifnum#1=39 
\hatcurPParxxxxA
\else
\ifnum#1=40 
\hatcurPParxxxxB
\else
\ifnum#1=41 
\hatcurPParxxxxC
\else
\ifnum#1=42 
\hatcurPParxxxxD
\else
??????\fi
\fi
\fi
\fi
}
\newcommand{\hatcurPParel}[1]{\ifnum#1=39 
\hatcurPParelxxxxA
\else
\ifnum#1=40 
\hatcurPParelxxxxB
\else
\ifnum#1=41 
\hatcurPParelxxxxC
\else
\ifnum#1=42 
\hatcurPParelxxxxD
\else
??????\fi
\fi
\fi
\fi
}
\newcommand{\hatcurPPfluxap}[1]{\ifnum#1=39 
\hatcurPPfluxapxxxxA
\else
\ifnum#1=40 
\hatcurPPfluxapxxxxB
\else
\ifnum#1=41 
\hatcurPPfluxapxxxxC
\else
\ifnum#1=42 
\hatcurPPfluxapxxxxD
\else
??????\fi
\fi
\fi
\fi
}
\newcommand{\hatcurPPfluxapdim}[1]{\ifnum#1=39 
\hatcurPPfluxapdimxxxxA
\else
\ifnum#1=40 
\hatcurPPfluxapdimxxxxB
\else
\ifnum#1=41 
\hatcurPPfluxapdimxxxxC
\else
\ifnum#1=42 
\hatcurPPfluxapdimxxxxD
\else
??????\fi
\fi
\fi
\fi
}
\newcommand{\hatcurPPfluxavg}[1]{\ifnum#1=39 
\hatcurPPfluxavgxxxxA
\else
\ifnum#1=40 
\hatcurPPfluxavgxxxxB
\else
\ifnum#1=41 
\hatcurPPfluxavgxxxxC
\else
\ifnum#1=42 
\hatcurPPfluxavgxxxxD
\else
??????\fi
\fi
\fi
\fi
}
\newcommand{\hatcurPPfluxavgdim}[1]{\ifnum#1=39 
\hatcurPPfluxavgdimxxxxA
\else
\ifnum#1=40 
\hatcurPPfluxavgdimxxxxB
\else
\ifnum#1=41 
\hatcurPPfluxavgdimxxxxC
\else
\ifnum#1=42 
\hatcurPPfluxavgdimxxxxD
\else
??????\fi
\fi
\fi
\fi
}
\newcommand{\hatcurPPfluxavglog}[1]{\ifnum#1=39 
\hatcurPPfluxavglogxxxxA
\else
\ifnum#1=40 
\hatcurPPfluxavglogxxxxB
\else
\ifnum#1=41 
\hatcurPPfluxavglogxxxxC
\else
\ifnum#1=42 
\hatcurPPfluxavglogxxxxD
\else
??????\fi
\fi
\fi
\fi
}
\newcommand{\hatcurPPfluxperi}[1]{\ifnum#1=39 
\hatcurPPfluxperixxxxA
\else
\ifnum#1=40 
\hatcurPPfluxperixxxxB
\else
\ifnum#1=41 
\hatcurPPfluxperixxxxC
\else
\ifnum#1=42 
\hatcurPPfluxperixxxxD
\else
??????\fi
\fi
\fi
\fi
}
\newcommand{\hatcurPPfluxperidim}[1]{\ifnum#1=39 
\hatcurPPfluxperidimxxxxA
\else
\ifnum#1=40 
\hatcurPPfluxperidimxxxxB
\else
\ifnum#1=41 
\hatcurPPfluxperidimxxxxC
\else
\ifnum#1=42 
\hatcurPPfluxperidimxxxxD
\else
??????\fi
\fi
\fi
\fi
}
\newcommand{\hatcurPPg}[1]{\ifnum#1=39 
\hatcurPPgxxxxA
\else
\ifnum#1=40 
\hatcurPPgxxxxB
\else
\ifnum#1=41 
\hatcurPPgxxxxC
\else
\ifnum#1=42 
\hatcurPPgxxxxD
\else
??????\fi
\fi
\fi
\fi
}
\newcommand{\hatcurPPi}[1]{\ifnum#1=39 
\hatcurPPixxxxA
\else
\ifnum#1=40 
\hatcurPPixxxxB
\else
\ifnum#1=41 
\hatcurPPixxxxC
\else
\ifnum#1=42 
\hatcurPPixxxxD
\else
??????\fi
\fi
\fi
\fi
}
\newcommand{\hatcurPPlogg}[1]{\ifnum#1=39 
\hatcurPPloggxxxxA
\else
\ifnum#1=40 
\hatcurPPloggxxxxB
\else
\ifnum#1=41 
\hatcurPPloggxxxxC
\else
\ifnum#1=42 
\hatcurPPloggxxxxD
\else
??????\fi
\fi
\fi
\fi
}
\newcommand{\hatcurPPm}[1]{\ifnum#1=39 
\hatcurPPmxxxxA
\else
\ifnum#1=40 
\hatcurPPmxxxxB
\else
\ifnum#1=41 
\hatcurPPmxxxxC
\else
\ifnum#1=42 
\hatcurPPmxxxxD
\else
??????\fi
\fi
\fi
\fi
}
\newcommand{\hatcurPPme}[1]{\ifnum#1=39 
\hatcurPPmexxxxA
\else
\ifnum#1=40 
\hatcurPPmexxxxB
\else
\ifnum#1=41 
\hatcurPPmexxxxC
\else
\ifnum#1=42 
\hatcurPPmexxxxD
\else
??????\fi
\fi
\fi
\fi
}
\newcommand{\hatcurPPmelong}[1]{\ifnum#1=39 
\hatcurPPmelongxxxxA
\else
\ifnum#1=40 
\hatcurPPmelongxxxxB
\else
\ifnum#1=41 
\hatcurPPmelongxxxxC
\else
\ifnum#1=42 
\hatcurPPmelongxxxxD
\else
??????\fi
\fi
\fi
\fi
}
\newcommand{\hatcurPPmeshort}[1]{\ifnum#1=39 
\hatcurPPmeshortxxxxA
\else
\ifnum#1=40 
\hatcurPPmeshortxxxxB
\else
\ifnum#1=41 
\hatcurPPmeshortxxxxC
\else
\ifnum#1=42 
\hatcurPPmeshortxxxxD
\else
??????\fi
\fi
\fi
\fi
}
\newcommand{\hatcurPPmlong}[1]{\ifnum#1=39 
\hatcurPPmlongxxxxA
\else
\ifnum#1=40 
\hatcurPPmlongxxxxB
\else
\ifnum#1=41 
\hatcurPPmlongxxxxC
\else
\ifnum#1=42 
\hatcurPPmlongxxxxD
\else
??????\fi
\fi
\fi
\fi
}
\newcommand{\hatcurPPmrcorr}[1]{\ifnum#1=39 
\hatcurPPmrcorrxxxxA
\else
\ifnum#1=40 
\hatcurPPmrcorrxxxxB
\else
\ifnum#1=41 
\hatcurPPmrcorrxxxxC
\else
\ifnum#1=42 
\hatcurPPmrcorrxxxxD
\else
??????\fi
\fi
\fi
\fi
}
\newcommand{\hatcurPPmshort}[1]{\ifnum#1=39 
\hatcurPPmshortxxxxA
\else
\ifnum#1=40 
\hatcurPPmshortxxxxB
\else
\ifnum#1=41 
\hatcurPPmshortxxxxC
\else
\ifnum#1=42 
\hatcurPPmshortxxxxD
\else
??????\fi
\fi
\fi
\fi
}
\newcommand{\hatcurPPperi}[1]{\ifnum#1=39 
\hatcurPPperixxxxA
\else
\ifnum#1=40 
\hatcurPPperixxxxB
\else
\ifnum#1=41 
\hatcurPPperixxxxC
\else
\ifnum#1=42 
\hatcurPPperixxxxD
\else
??????\fi
\fi
\fi
\fi
}
\newcommand{\hatcurPPphiconj}[1]{\ifnum#1=39 
\hatcurPPphiconjxxxxA
\else
\ifnum#1=40 
\hatcurPPphiconjxxxxB
\else
\ifnum#1=41 
\hatcurPPphiconjxxxxC
\else
\ifnum#1=42 
\hatcurPPphiconjxxxxD
\else
??????\fi
\fi
\fi
\fi
}
\newcommand{\hatcurPPr}[1]{\ifnum#1=39 
\hatcurPPrxxxxA
\else
\ifnum#1=40 
\hatcurPPrxxxxB
\else
\ifnum#1=41 
\hatcurPPrxxxxC
\else
\ifnum#1=42 
\hatcurPPrxxxxD
\else
??????\fi
\fi
\fi
\fi
}
\newcommand{\hatcurPPre}[1]{\ifnum#1=39 
\hatcurPPrexxxxA
\else
\ifnum#1=40 
\hatcurPPrexxxxB
\else
\ifnum#1=41 
\hatcurPPrexxxxC
\else
\ifnum#1=42 
\hatcurPPrexxxxD
\else
??????\fi
\fi
\fi
\fi
}
\newcommand{\hatcurPPrelong}[1]{\ifnum#1=39 
\hatcurPPrelongxxxxA
\else
\ifnum#1=40 
\hatcurPPrelongxxxxB
\else
\ifnum#1=41 
\hatcurPPrelongxxxxC
\else
\ifnum#1=42 
\hatcurPPrelongxxxxD
\else
??????\fi
\fi
\fi
\fi
}
\newcommand{\hatcurPPreshort}[1]{\ifnum#1=39 
\hatcurPPreshortxxxxA
\else
\ifnum#1=40 
\hatcurPPreshortxxxxB
\else
\ifnum#1=41 
\hatcurPPreshortxxxxC
\else
\ifnum#1=42 
\hatcurPPreshortxxxxD
\else
??????\fi
\fi
\fi
\fi
}
\newcommand{\hatcurPPrho}[1]{\ifnum#1=39 
\hatcurPPrhoxxxxA
\else
\ifnum#1=40 
\hatcurPPrhoxxxxB
\else
\ifnum#1=41 
\hatcurPPrhoxxxxC
\else
\ifnum#1=42 
\hatcurPPrhoxxxxD
\else
??????\fi
\fi
\fi
\fi
}
\newcommand{\hatcurPPrlong}[1]{\ifnum#1=39 
\hatcurPPrlongxxxxA
\else
\ifnum#1=40 
\hatcurPPrlongxxxxB
\else
\ifnum#1=41 
\hatcurPPrlongxxxxC
\else
\ifnum#1=42 
\hatcurPPrlongxxxxD
\else
??????\fi
\fi
\fi
\fi
}
\newcommand{\hatcurPPrshort}[1]{\ifnum#1=39 
\hatcurPPrshortxxxxA
\else
\ifnum#1=40 
\hatcurPPrshortxxxxB
\else
\ifnum#1=41 
\hatcurPPrshortxxxxC
\else
\ifnum#1=42 
\hatcurPPrshortxxxxD
\else
??????\fi
\fi
\fi
\fi
}
\newcommand{\hatcurPPtcirc}[1]{\ifnum#1=39 
\hatcurPPtcircxxxxA
\else
\ifnum#1=40 
\hatcurPPtcircxxxxB
\else
\ifnum#1=41 
\hatcurPPtcircxxxxC
\else
\ifnum#1=42 
\hatcurPPtcircxxxxD
\else
??????\fi
\fi
\fi
\fi
}
\newcommand{\hatcurPPteff}[1]{\ifnum#1=39 
\hatcurPPteffxxxxA
\else
\ifnum#1=40 
\hatcurPPteffxxxxB
\else
\ifnum#1=41 
\hatcurPPteffxxxxC
\else
\ifnum#1=42 
\hatcurPPteffxxxxD
\else
??????\fi
\fi
\fi
\fi
}
\newcommand{\hatcurPPtheta}[1]{\ifnum#1=39 
\hatcurPPthetaxxxxA
\else
\ifnum#1=40 
\hatcurPPthetaxxxxB
\else
\ifnum#1=41 
\hatcurPPthetaxxxxC
\else
\ifnum#1=42 
\hatcurPPthetaxxxxD
\else
??????\fi
\fi
\fi
\fi
}
\newcommand{\hatcurPPtinfall}[1]{\ifnum#1=39 
\hatcurPPtinfallxxxxA
\else
\ifnum#1=40 
\hatcurPPtinfallxxxxB
\else
\ifnum#1=41 
\hatcurPPtinfallxxxxC
\else
\ifnum#1=42 
\hatcurPPtinfallxxxxD
\else
??????\fi
\fi
\fi
\fi
}
\newcommand{\hatcurRVeccen}[1]{\ifnum#1=39 
\hatcurRVeccenxxxxA
\else
\ifnum#1=40 
\hatcurRVeccenxxxxB
\else
\ifnum#1=41 
\hatcurRVeccenxxxxC
\else
\ifnum#1=42 
\hatcurRVeccenxxxxD
\else
??????\fi
\fi
\fi
\fi
}
\newcommand{\hatcurRVeccentwosiglim}[1]{\ifnum#1=39 
\hatcurRVeccentwosiglimxxxxA
\else
\ifnum#1=40 
\hatcurRVeccentwosiglimxxxxB
\else
\ifnum#1=41 
\hatcurRVeccentwosiglimxxxxC
\else
\ifnum#1=42 
\hatcurRVeccentwosiglimxxxxD
\else
??????\fi
\fi
\fi
\fi
}
\newcommand{\hatcurRVfitrms}[1]{\ifnum#1=40 
\hatcurRVfitrmsxxxxB
\else
??????\fi
}
\newcommand{\hatcurRVfitrmsA}[1]{\ifnum#1=39 
\hatcurRVfitrmsAxxxxA
\else
\ifnum#1=41 
\hatcurRVfitrmsAxxxxC
\else
\ifnum#1=42 
\hatcurRVfitrmsAxxxxD
\else
??????\fi
\fi
\fi
}
\newcommand{\hatcurRVfitrmsB}[1]{\ifnum#1=39 
\hatcurRVfitrmsBxxxxA
\else
\ifnum#1=41 
\hatcurRVfitrmsBxxxxC
\else
\ifnum#1=42 
\hatcurRVfitrmsBxxxxD
\else
??????\fi
\fi
\fi
}
\newcommand{\hatcurRVfitrmsC}[1]{\ifnum#1=41 
\hatcurRVfitrmsCxxxxC
\else
??????\fi
}
\newcommand{\hatcurRVgamma}[1]{\ifnum#1=40 
\hatcurRVgammaxxxxB
\else
??????\fi
}
\newcommand{\hatcurRVgammaA}[1]{\ifnum#1=39 
\hatcurRVgammaAxxxxA
\else
\ifnum#1=41 
\hatcurRVgammaAxxxxC
\else
\ifnum#1=42 
\hatcurRVgammaAxxxxD
\else
??????\fi
\fi
\fi
}
\newcommand{\hatcurRVgammaB}[1]{\ifnum#1=39 
\hatcurRVgammaBxxxxA
\else
\ifnum#1=41 
\hatcurRVgammaBxxxxC
\else
\ifnum#1=42 
\hatcurRVgammaBxxxxD
\else
??????\fi
\fi
\fi
}
\newcommand{\hatcurRVgammaC}[1]{\ifnum#1=41 
\hatcurRVgammaCxxxxC
\else
??????\fi
}
\newcommand{\hatcurRVh}[1]{\ifnum#1=39 
\hatcurRVhxxxxA
\else
\ifnum#1=40 
\hatcurRVhxxxxB
\else
\ifnum#1=41 
\hatcurRVhxxxxC
\else
\ifnum#1=42 
\hatcurRVhxxxxD
\else
??????\fi
\fi
\fi
\fi
}
\newcommand{\hatcurRVjitter}[1]{\ifnum#1=40 
\hatcurRVjitterxxxxB
\else
??????\fi
}
\newcommand{\hatcurRVjitterA}[1]{\ifnum#1=39 
\hatcurRVjitterAxxxxA
\else
\ifnum#1=41 
\hatcurRVjitterAxxxxC
\else
\ifnum#1=42 
\hatcurRVjitterAxxxxD
\else
??????\fi
\fi
\fi
}
\newcommand{\hatcurRVjitterB}[1]{\ifnum#1=39 
\hatcurRVjitterBxxxxA
\else
\ifnum#1=41 
\hatcurRVjitterBxxxxC
\else
\ifnum#1=42 
\hatcurRVjitterBxxxxD
\else
??????\fi
\fi
\fi
}
\newcommand{\hatcurRVjitterC}[1]{\ifnum#1=41 
\hatcurRVjitterCxxxxC
\else
??????\fi
}
\newcommand{\hatcurRVjittertwosiglim}[1]{\ifnum#1=40 
\hatcurRVjittertwosiglimxxxxB
\else
??????\fi
}
\newcommand{\hatcurRVjittertwosiglimA}[1]{\ifnum#1=39 
\hatcurRVjittertwosiglimAxxxxA
\else
\ifnum#1=41 
\hatcurRVjittertwosiglimAxxxxC
\else
\ifnum#1=42 
\hatcurRVjittertwosiglimAxxxxD
\else
??????\fi
\fi
\fi
}
\newcommand{\hatcurRVjittertwosiglimB}[1]{\ifnum#1=39 
\hatcurRVjittertwosiglimBxxxxA
\else
\ifnum#1=41 
\hatcurRVjittertwosiglimBxxxxC
\else
\ifnum#1=42 
\hatcurRVjittertwosiglimBxxxxD
\else
??????\fi
\fi
\fi
}
\newcommand{\hatcurRVjittertwosiglimC}[1]{\ifnum#1=41 
\hatcurRVjittertwosiglimCxxxxC
\else
??????\fi
}
\newcommand{\hatcurRVk}[1]{\ifnum#1=39 
\hatcurRVkxxxxA
\else
\ifnum#1=40 
\hatcurRVkxxxxB
\else
\ifnum#1=41 
\hatcurRVkxxxxC
\else
\ifnum#1=42 
\hatcurRVkxxxxD
\else
??????\fi
\fi
\fi
\fi
}
\newcommand{\hatcurRVK}[1]{\ifnum#1=39 
\hatcurRVKxxxxA
\else
\ifnum#1=40 
\hatcurRVKxxxxB
\else
\ifnum#1=41 
\hatcurRVKxxxxC
\else
\ifnum#1=42 
\hatcurRVKxxxxD
\else
??????\fi
\fi
\fi
\fi
}
\newcommand{\hatcurRVomega}[1]{\ifnum#1=39 
\hatcurRVomegaxxxxA
\else
\ifnum#1=40 
\hatcurRVomegaxxxxB
\else
\ifnum#1=41 
\hatcurRVomegaxxxxC
\else
\ifnum#1=42 
\hatcurRVomegaxxxxD
\else
??????\fi
\fi
\fi
\fi
}
\newcommand{\hatcurRVrh}[1]{\ifnum#1=39 
\hatcurRVrhxxxxA
\else
\ifnum#1=40 
\hatcurRVrhxxxxB
\else
\ifnum#1=41 
\hatcurRVrhxxxxC
\else
\ifnum#1=42 
\hatcurRVrhxxxxD
\else
??????\fi
\fi
\fi
\fi
}
\newcommand{\hatcurRVrk}[1]{\ifnum#1=39 
\hatcurRVrkxxxxA
\else
\ifnum#1=40 
\hatcurRVrkxxxxB
\else
\ifnum#1=41 
\hatcurRVrkxxxxC
\else
\ifnum#1=42 
\hatcurRVrkxxxxD
\else
??????\fi
\fi
\fi
\fi
}
\newcommand{\hatcurRVtrone}[1]{\ifnum#1=39 
\hatcurRVtronexxxxA
\else
\ifnum#1=40 
\hatcurRVtronexxxxB
\else
\ifnum#1=41 
\hatcurRVtronexxxxC
\else
\ifnum#1=42 
\hatcurRVtronexxxxD
\else
??????\fi
\fi
\fi
\fi
}
\newcommand{\hatcurRVtrtwo}[1]{\ifnum#1=39 
\hatcurRVtrtwoxxxxA
\else
\ifnum#1=40 
\hatcurRVtrtwoxxxxB
\else
\ifnum#1=41 
\hatcurRVtrtwoxxxxC
\else
\ifnum#1=42 
\hatcurRVtrtwoxxxxD
\else
??????\fi
\fi
\fi
\fi
}
\newcommand{\hatcurSMEiilogg}[1]{\ifnum#1=39 
\hatcurSMEiiloggxxxxA
\else
\ifnum#1=42 
\hatcurSMEiiloggxxxxD
\else
??????\fi
\fi
}
\newcommand{\hatcurSMEiiteff}[1]{\ifnum#1=39 
\hatcurSMEiiteffxxxxA
\else
\ifnum#1=42 
\hatcurSMEiiteffxxxxD
\else
??????\fi
\fi
}
\newcommand{\hatcurSMEiivmac}[1]{\ifnum#1=39 
\hatcurSMEiivmacxxxxA
\else
\ifnum#1=42 
\hatcurSMEiivmacxxxxD
\else
??????\fi
\fi
}
\newcommand{\hatcurSMEiivmic}[1]{\ifnum#1=39 
\hatcurSMEiivmicxxxxA
\else
\ifnum#1=42 
\hatcurSMEiivmicxxxxD
\else
??????\fi
\fi
}
\newcommand{\hatcurSMEiivsin}[1]{\ifnum#1=39 
\hatcurSMEiivsinxxxxA
\else
\ifnum#1=42 
\hatcurSMEiivsinxxxxD
\else
??????\fi
\fi
}
\newcommand{\hatcurSMEiizfeh}[1]{\ifnum#1=39 
\hatcurSMEiizfehxxxxA
\else
\ifnum#1=42 
\hatcurSMEiizfehxxxxD
\else
??????\fi
\fi
}
\newcommand{\hatcurSMEiizfehshort}[1]{\ifnum#1=39 
\hatcurSMEiizfehshortxxxxA
\else
\ifnum#1=42 
\hatcurSMEiizfehshortxxxxD
\else
??????\fi
\fi
}
\newcommand{\hatcurSMEilogg}[1]{\ifnum#1=39 
\hatcurSMEiloggxxxxA
\else
\ifnum#1=40 
\hatcurSMEiloggxxxxB
\else
\ifnum#1=41 
\hatcurSMEiloggxxxxC
\else
\ifnum#1=42 
\hatcurSMEiloggxxxxD
\else
??????\fi
\fi
\fi
\fi
}
\newcommand{\hatcurSMEiteff}[1]{\ifnum#1=39 
\hatcurSMEiteffxxxxA
\else
\ifnum#1=40 
\hatcurSMEiteffxxxxB
\else
\ifnum#1=41 
\hatcurSMEiteffxxxxC
\else
\ifnum#1=42 
\hatcurSMEiteffxxxxD
\else
??????\fi
\fi
\fi
\fi
}
\newcommand{\hatcurSMEivmac}[1]{\ifnum#1=39 
\hatcurSMEivmacxxxxA
\else
\ifnum#1=40 
\hatcurSMEivmacxxxxB
\else
\ifnum#1=41 
\hatcurSMEivmacxxxxC
\else
\ifnum#1=42 
\hatcurSMEivmacxxxxD
\else
??????\fi
\fi
\fi
\fi
}
\newcommand{\hatcurSMEivmic}[1]{\ifnum#1=39 
\hatcurSMEivmicxxxxA
\else
\ifnum#1=40 
\hatcurSMEivmicxxxxB
\else
\ifnum#1=41 
\hatcurSMEivmicxxxxC
\else
\ifnum#1=42 
\hatcurSMEivmicxxxxD
\else
??????\fi
\fi
\fi
\fi
}
\newcommand{\hatcurSMEivsin}[1]{\ifnum#1=39 
\hatcurSMEivsinxxxxA
\else
\ifnum#1=40 
\hatcurSMEivsinxxxxB
\else
\ifnum#1=41 
\hatcurSMEivsinxxxxC
\else
\ifnum#1=42 
\hatcurSMEivsinxxxxD
\else
??????\fi
\fi
\fi
\fi
}
\newcommand{\hatcurSMEizfeh}[1]{\ifnum#1=39 
\hatcurSMEizfehxxxxA
\else
\ifnum#1=40 
\hatcurSMEizfehxxxxB
\else
\ifnum#1=41 
\hatcurSMEizfehxxxxC
\else
\ifnum#1=42 
\hatcurSMEizfehxxxxD
\else
??????\fi
\fi
\fi
\fi
}
\newcommand{\hatcurSMEizfehshort}[1]{\ifnum#1=39 
\hatcurSMEizfehshortxxxxA
\else
\ifnum#1=40 
\hatcurSMEizfehshortxxxxB
\else
\ifnum#1=41 
\hatcurSMEizfehshortxxxxC
\else
\ifnum#1=42 
\hatcurSMEizfehshortxxxxD
\else
??????\fi
\fi
\fi
\fi
}
\newcommand{\hatcurXAv}[1]{\ifnum#1=39 
\hatcurXAvxxxxA
\else
\ifnum#1=40 
\hatcurXAvxxxxB
\else
\ifnum#1=41 
\hatcurXAvxxxxC
\else
\ifnum#1=42 
\hatcurXAvxxxxD
\else
??????\fi
\fi
\fi
\fi
}
\newcommand{\hatcurXdist}[1]{\ifnum#1=39 
\hatcurXdistxxxxA
\else
\ifnum#1=40 
\hatcurXdistxxxxB
\else
\ifnum#1=41 
\hatcurXdistxxxxC
\else
\ifnum#1=42 
\hatcurXdistxxxxD
\else
??????\fi
\fi
\fi
\fi
}
\newcommand{\hatcurXdistred}[1]{\ifnum#1=39 
\hatcurXdistredxxxxA
\else
\ifnum#1=40 
\hatcurXdistredxxxxB
\else
\ifnum#1=41 
\hatcurXdistredxxxxC
\else
\ifnum#1=42 
\hatcurXdistredxxxxD
\else
??????\fi
\fi
\fi
\fi
}
\newcommand{\hatcurXEBV}[1]{\ifnum#1=39 
\hatcurXEBVxxxxA
\else
\ifnum#1=40 
\hatcurXEBVxxxxB
\else
\ifnum#1=41 
\hatcurXEBVxxxxC
\else
\ifnum#1=42 
\hatcurXEBVxxxxD
\else
??????\fi
\fi
\fi
\fi
}
\newcommand{\hatcurXjhisored}[1]{\ifnum#1=39 
\hatcurXjhisoredxxxxA
\else
\ifnum#1=40 
\hatcurXjhisoredxxxxB
\else
\ifnum#1=41 
\hatcurXjhisoredxxxxC
\else
\ifnum#1=42 
\hatcurXjhisoredxxxxD
\else
??????\fi
\fi
\fi
\fi
}
\newcommand{\hatcurXjkisored}[1]{\ifnum#1=39 
\hatcurXjkisoredxxxxA
\else
\ifnum#1=40 
\hatcurXjkisoredxxxxB
\else
\ifnum#1=41 
\hatcurXjkisoredxxxxC
\else
\ifnum#1=42 
\hatcurXjkisoredxxxxD
\else
??????\fi
\fi
\fi
\fi
}
\newcommand{\hatcurXmhisored}[1]{\ifnum#1=39 
\hatcurXmhisoredxxxxA
\else
\ifnum#1=40 
\hatcurXmhisoredxxxxB
\else
\ifnum#1=41 
\hatcurXmhisoredxxxxC
\else
\ifnum#1=42 
\hatcurXmhisoredxxxxD
\else
??????\fi
\fi
\fi
\fi
}
\newcommand{\hatcurXmiisored}[1]{\ifnum#1=39 
\hatcurXmiisoredxxxxA
\else
\ifnum#1=40 
\hatcurXmiisoredxxxxB
\else
\ifnum#1=41 
\hatcurXmiisoredxxxxC
\else
\ifnum#1=42 
\hatcurXmiisoredxxxxD
\else
??????\fi
\fi
\fi
\fi
}
\newcommand{\hatcurXmjisored}[1]{\ifnum#1=39 
\hatcurXmjisoredxxxxA
\else
\ifnum#1=40 
\hatcurXmjisoredxxxxB
\else
\ifnum#1=41 
\hatcurXmjisoredxxxxC
\else
\ifnum#1=42 
\hatcurXmjisoredxxxxD
\else
??????\fi
\fi
\fi
\fi
}
\newcommand{\hatcurXmkisored}[1]{\ifnum#1=39 
\hatcurXmkisoredxxxxA
\else
\ifnum#1=40 
\hatcurXmkisoredxxxxB
\else
\ifnum#1=41 
\hatcurXmkisoredxxxxC
\else
\ifnum#1=42 
\hatcurXmkisoredxxxxD
\else
??????\fi
\fi
\fi
\fi
}
\newcommand{\hatcurXmvisored}[1]{\ifnum#1=39 
\hatcurXmvisoredxxxxA
\else
\ifnum#1=40 
\hatcurXmvisoredxxxxB
\else
\ifnum#1=41 
\hatcurXmvisoredxxxxC
\else
\ifnum#1=42 
\hatcurXmvisoredxxxxD
\else
??????\fi
\fi
\fi
\fi
}
\newcommand{\hatcurXsecdur}[1]{\ifnum#1=39 
\hatcurXsecdurxxxxA
\else
\ifnum#1=40 
\hatcurXsecdurxxxxB
\else
\ifnum#1=41 
\hatcurXsecdurxxxxC
\else
\ifnum#1=42 
\hatcurXsecdurxxxxD
\else
??????\fi
\fi
\fi
\fi
}
\newcommand{\hatcurXsecingdur}[1]{\ifnum#1=39 
\hatcurXsecingdurxxxxA
\else
\ifnum#1=40 
\hatcurXsecingdurxxxxB
\else
\ifnum#1=41 
\hatcurXsecingdurxxxxC
\else
\ifnum#1=42 
\hatcurXsecingdurxxxxD
\else
??????\fi
\fi
\fi
\fi
}
\newcommand{\hatcurXsecondary}[1]{\ifnum#1=39 
\hatcurXsecondaryxxxxA
\else
\ifnum#1=40 
\hatcurXsecondaryxxxxB
\else
\ifnum#1=41 
\hatcurXsecondaryxxxxC
\else
\ifnum#1=42 
\hatcurXsecondaryxxxxD
\else
??????\fi
\fi
\fi
\fi
}
\newcommand{\hatcurXsecphase}[1]{\ifnum#1=39 
\hatcurXsecphasexxxxA
\else
\ifnum#1=40 
\hatcurXsecphasexxxxB
\else
\ifnum#1=41 
\hatcurXsecphasexxxxC
\else
\ifnum#1=42 
\hatcurXsecphasexxxxD
\else
??????\fi
\fi
\fi
\fi
}
\newcommand{\hatcurXviisored}[1]{\ifnum#1=39 
\hatcurXviisoredxxxxA
\else
\ifnum#1=40 
\hatcurXviisoredxxxxB
\else
\ifnum#1=41 
\hatcurXviisoredxxxxC
\else
\ifnum#1=42 
\hatcurXviisoredxxxxD
\else
??????\fi
\fi
\fi
\fi
}
\newcommand{\hatcurXvkisored}[1]{\ifnum#1=39 
\hatcurXvkisoredxxxxA
\else
\ifnum#1=40 
\hatcurXvkisoredxxxxB
\else
\ifnum#1=41 
\hatcurXvkisoredxxxxC
\else
\ifnum#1=42 
\hatcurXvkisoredxxxxD
\else
??????\fi
\fi
\fi
\fi
}
 \newcommand{\hatcurhtreccenxxxxA}{HATS602-035}                      
\newcommand{\hatcurfieldeccenxxxxA}{\ensuremath{string}}            
\newcommand{\hatcurCCraeccenxxxxA}{\ensuremath{07^{\mathrm h}29^{\mathrm m}40.63{\mathrm s}}}                     
\newcommand{\hatcurCCdececcenxxxxA}{\ensuremath{-29{\arcdeg}56{\arcmin}16.4{\arcsec}}}                    
\newcommand{\hatcurCCmageccenxxxxA}{12.746}                         
\newcommand{\hatcurCCtwomasseccenxxxxA}{2MASS~07294061-2956163}     
\newcommand{\hatcurCCgsceccenxxxxA}{GSC~6550-00341}                 
\newcommand{\hatcurCCtassmveccenxxxxA}{\ensuremath{12.746\pm0.020}} 
\newcommand{\hatcurCCtassmvshorteccenxxxxA}{\ensuremath{12.7}}      
\newcommand{\hatcurCCtassmBeccenxxxxA}{\ensuremath{13.232\pm0.010}} 
\newcommand{\hatcurCCtassmBshorteccenxxxxA}{\ensuremath{13.2}}      
\newcommand{\hatcurCCtassmIeccenxxxxA}{\ensuremath{nff\pmnff}}      
\newcommand{\hatcurCCtassmIshorteccenxxxxA}{\ensuremath{0.0}}       
\newcommand{\hatcurCCtassmgeccenxxxxA}{\ensuremath{12.907\pm0.020}} 
\newcommand{\hatcurCCtassmgshorteccenxxxxA}{\ensuremath{12.9}}      
\newcommand{\hatcurCCtassmreccenxxxxA}{\ensuremath{12.578\pm0.050}} 
\newcommand{\hatcurCCtassmrshorteccenxxxxA}{\ensuremath{12.6}}      
\newcommand{\hatcurCCtassmieccenxxxxA}{\ensuremath{12.542\pm0.060}} 
\newcommand{\hatcurCCtassmishorteccenxxxxA}{\ensuremath{12.5}}      
\newcommand{\hatcurCCtwomassJmageccenxxxxA}{\ensuremath{11.833\pm0.024}} 
\newcommand{\hatcurCCtwomassHmageccenxxxxA}{\ensuremath{11.620\pm0.024}} 
\newcommand{\hatcurCCtwomassKmageccenxxxxA}{\ensuremath{11.518\pm0.023}} 
\newcommand{\hatcurCCcitJmageccenxxxxA}{\ensuremath{11.852\pm0.024}} 
\newcommand{\hatcurCCcitHmageccenxxxxA}{\ensuremath{11.614\pm0.025}} 
\newcommand{\hatcurCCcitKmageccenxxxxA}{\ensuremath{11.542\pm0.023}} 
\newcommand{\hatcurCCbbJmageccenxxxxA}{\ensuremath{11.897\pm0.026}} 
\newcommand{\hatcurCCbbHmageccenxxxxA}{\ensuremath{11.636\pm0.025}} 
\newcommand{\hatcurCCbbKmageccenxxxxA}{\ensuremath{11.562\pm0.023}} 
\newcommand{\hatcurCCesoJmageccenxxxxA}{\ensuremath{11.899\pm0.027}} 
\newcommand{\hatcurCCesoHmageccenxxxxA}{\ensuremath{11.632\pm0.029}} 
\newcommand{\hatcurCCesoKmageccenxxxxA}{\ensuremath{11.561\pm0.023}} 
\newcommand{\hatcurCCesoJHmageccenxxxxA}{\ensuremath{0.267\pm0.013}} 
\newcommand{\hatcurCCesoJKmageccenxxxxA}{\ensuremath{0.338\pm0.036}} 
\newcommand{\hatcurCCesoHKmageccenxxxxA}{\ensuremath{0.071\pm0.038}} 
\newcommand{\hatcurLCdipeccenxxxxA}{\ensuremath{10.2}}              
\newcommand{\hatcurLCrprstareccenxxxxA}{\ensuremath{0.0996\pm0.0030}} 
\newcommand{\hatcurLCbsqeccenxxxxA}{\ensuremath{0.485_{-0.051}^{+0.042}}} 
\newcommand{\hatcurLCimpeccenxxxxA}{\ensuremath{0.697_{-0.037}^{+0.030}}} 
\newcommand{\hatcurLCzetaeccenxxxxA}{\ensuremath{15.25\pm0.15}}     
\newcommand{\hatcurLCdureccenxxxxA}{\ensuremath{0.1555\pm0.0030}}   
\newcommand{\hatcurLCdurshorteccenxxxxA}{\ensuremath{0.1555}}       
\newcommand{\hatcurLCdurhreccenxxxxA}{\ensuremath{3.732\pm0.071}}   
\newcommand{\hatcurLCdurhrshorteccenxxxxA}{\ensuremath{3.732}}      
\newcommand{\hatcurLCqeccenxxxxA}{\ensuremath{0.03400\pm0.00065}}   
\newcommand{\hatcurLCqshorteccenxxxxA}{\ensuremath{0.034}}          
\newcommand{\hatcurLCingdureccenxxxxA}{\ensuremath{0.0256\pm0.0029}} 
\newcommand{\hatcurLCPeccenxxxxA}{\ensuremath{4.5776350\pm0.0000073}} 
\newcommand{\hatcurLCPprececcenxxxxA}{\ensuremath{4.5776350}}       
\newcommand{\hatcurLCPshorteccenxxxxA}{\ensuremath{4.5776}}         
\newcommand{\hatcurLCTeccenxxxxA}{\ensuremath{2457242.04136\pm0.00075}} 
\newcommand{\hatcurLCTAeccenxxxxA}{\ensuremath{2455800.0864\pm0.0024}} 
\newcommand{\hatcurLCTBeccenxxxxA}{\ensuremath{2457397.68095\pm0.00080}} 
\newcommand{\hatcurLChatnetmAeccenxxxxA}{\ensuremath{12.767660\pm0.000094}} 
\newcommand{\hatcurLCiblendAeccenxxxxA}{\ensuremath{0.907\pm0.058}} 
\newcommand{\hatcurLChatnetmBeccenxxxxA}{\ensuremath{12.76761\pm0.00011}} 
\newcommand{\hatcurLCiblendBeccenxxxxA}{\ensuremath{0.775\pm0.076}} 
\newcommand{\hatcurLCrhoeccenxxxxA}{\ensuremath{0.50_{-0.14}^{+0.25}}} 
\newcommand{\hatcurSMEiteffeccenxxxxA}{\ensuremath{6653\pm90}}      
\newcommand{\hatcurSMEizfeheccenxxxxA}{\ensuremath{0.040\pm0.050}}  
\newcommand{\hatcurSMEizfehshorteccenxxxxA}{\ensuremath{0.04}}      
\newcommand{\hatcurSMEiloggeccenxxxxA}{\ensuremath{4.33\pm0.16}}    
\newcommand{\hatcurSMEivsineccenxxxxA}{\ensuremath{7.70\pm0.20}}    
\newcommand{\hatcurSMEivmaceccenxxxxA}{\ensuremath{5.34\pm0.14}}    
\newcommand{\hatcurSMEivmiceccenxxxxA}{\ensuremath{1.99\pm0.15}}    
\newcommand{\hatcurSMEiiteffeccenxxxxA}{\ensuremath{6572\pm83}}     
\newcommand{\hatcurSMEiizfeheccenxxxxA}{\ensuremath{0.000\pm0.044}} 
\newcommand{\hatcurSMEiizfehshorteccenxxxxA}{\ensuremath{0.00}}     
\newcommand{\hatcurSMEiiloggeccenxxxxA}{\ensuremath{4.167\pm0.040}} 
\newcommand{\hatcurSMEiivsineccenxxxxA}{\ensuremath{7.75\pm0.17}}   
\newcommand{\hatcurSMEiivmaceccenxxxxA}{\ensuremath{5.21\pm0.13}}   
\newcommand{\hatcurSMEiivmiceccenxxxxA}{\ensuremath{1.86\pm0.12}}   
\newcommand{\hatcurLBizeccenxxxxA}{\ensuremath{0.1158}}             
\newcommand{\hatcurLBiizeccenxxxxA}{\ensuremath{0.3681}}            
\newcommand{\hatcurLBiieccenxxxxA}{\ensuremath{0.1654}}             
\newcommand{\hatcurLBiiieccenxxxxA}{\ensuremath{0.3772}}            
\newcommand{\hatcurLBiIeccenxxxxA}{\ensuremath{0.1472}}             
\newcommand{\hatcurLBiiIeccenxxxxA}{\ensuremath{0.3755}}            
\newcommand{\hatcurLBigeccenxxxxA}{\ensuremath{0.3938}}             
\newcommand{\hatcurLBiigeccenxxxxA}{\ensuremath{0.3444}}            
\newcommand{\hatcurLBireccenxxxxA}{\ensuremath{0.2375}}             
\newcommand{\hatcurLBiireccenxxxxA}{\ensuremath{0.3875}}            
\newcommand{\hatcurLBiReccenxxxxA}{\ensuremath{0.2168}}             
\newcommand{\hatcurLBiiReccenxxxxA}{\ensuremath{0.3864}}            
\newcommand{\hatcurLBikepeccenxxxxA}{\ensuremath{0.1000}}           
\newcommand{\hatcurLBiikepeccenxxxxA}{\ensuremath{0.1000}}          
\newcommand{\hatcurISOmeccenxxxxA}{\ensuremath{1.374_{-0.057}^{+0.075}}} 
\newcommand{\hatcurISOmshorteccenxxxxA}{\ensuremath{1.37}}          
\newcommand{\hatcurISOmlongeccenxxxxA}{\ensuremath{1.374_{-0.057}^{+0.075}}} 
\newcommand{\hatcurISOreccenxxxxA}{\ensuremath{1.59\pm0.25}}        
\newcommand{\hatcurISOrshorteccenxxxxA}{\ensuremath{1.59}}          
\newcommand{\hatcurISOrlongeccenxxxxA}{\ensuremath{1.59\pm0.25}}    
\newcommand{\hatcurISOrhoeccenxxxxA}{\ensuremath{0.48_{-0.14}^{+0.18}}} 
\newcommand{\hatcurISOrholongeccenxxxxA}{\ensuremath{0.48_{-0.14}^{+0.18}}} 
\newcommand{\hatcurISOloggeccenxxxxA}{\ensuremath{4.171\pm0.100}}   
\newcommand{\hatcurISOlumeccenxxxxA}{\ensuremath{4.23_{-0.90}^{+1.29}}} 
\newcommand{\hatcurISOlumshorteccenxxxxA}{\ensuremath{4.23}}        
\newcommand{\hatcurISOmveccenxxxxA}{\ensuremath{3.16\pm0.32}}       
\newcommand{\hatcurISOvieccenxxxxA}{\ensuremath{0.481\pm0.020}}     
\newcommand{\hatcurISOageeccenxxxxA}{\ensuremath{1.97_{-0.45}^{+0.32}}} 
\newcommand{\hatcurISOsigmaeccenxxxxA}{\ensuremath{0.00030\pm0.00016}} 
\newcommand{\hatcurISOMJeccenxxxxA}{\ensuremath{2.40\pm0.31}}       
\newcommand{\hatcurISOMHeccenxxxxA}{\ensuremath{2.20\pm0.30}}       
\newcommand{\hatcurISOMKeccenxxxxA}{\ensuremath{2.16\pm0.30}}       
\newcommand{\hatcurISOJKeccenxxxxA}{\ensuremath{0.240\pm0.020}}     
\newcommand{\hatcurISOspececcenxxxxA}{G}                            
\newcommand{\hatcurRVKeccenxxxxA}{\ensuremath{62\pm14}}             
\newcommand{\hatcurRVrkeccenxxxxA}{\ensuremath{0.05\pm0.19}}        
\newcommand{\hatcurRVrheccenxxxxA}{\ensuremath{-0.05\pm0.26}}       
\newcommand{\hatcurRVkeccenxxxxA}{\ensuremath{0.012\pm0.075}}       
\newcommand{\hatcurRVheccenxxxxA}{\ensuremath{-0.01\pm0.12}}        
\newcommand{\hatcurRVtroneeccenxxxxA}{\ensuremath{0\pm0}}           
\newcommand{\hatcurRVtrtwoeccenxxxxA}{\ensuremath{0\pm0}}           
\newcommand{\hatcurRVgammaAeccenxxxxA}{\ensuremath{2916\pm10}}      
\newcommand{\hatcurRVjitterAeccenxxxxA}{\ensuremath{29\pm13}}       
\newcommand{\hatcurRVjittertwosiglimAeccenxxxxA}{\ensuremath{<49.9}} 
\newcommand{\hatcurRVfitrmsAeccenxxxxA}{\ensuremath{0.0}}           
\newcommand{\hatcurRVgammaBeccenxxxxA}{\ensuremath{3079\pm75}}      
\newcommand{\hatcurRVjitterBeccenxxxxA}{\ensuremath{93\pm72}}       
\newcommand{\hatcurRVjittertwosiglimBeccenxxxxA}{\ensuremath{<253.2}} 
\newcommand{\hatcurRVfitrmsBeccenxxxxA}{\ensuremath{0.0}}           
\newcommand{\hatcurRVecceneccenxxxxA}{\ensuremath{0.091\pm0.089}}   
\newcommand{\hatcurRVeccentwosiglimeccenxxxxA}{\ensuremath{<0.275}} 
\newcommand{\hatcurRVomegaeccenxxxxA}{\ensuremath{220\pm110}}       
\newcommand{\hatcurPPieccenxxxxA}{\ensuremath{85.14_{-1.24}^{+0.93}}} 
\newcommand{\hatcurPPgeccenxxxxA}{\ensuremath{6.4\pm2.1}}           
\newcommand{\hatcurPPloggeccenxxxxA}{\ensuremath{2.81_{-0.17}^{+0.13}}} 
\newcommand{\hatcurPPareccenxxxxA}{\ensuremath{8.10\pm0.94}}        
\newcommand{\hatcurPPareleccenxxxxA}{\ensuremath{0.0600\pm0.0011}}  
\newcommand{\hatcurPPrhoeccenxxxxA}{\ensuremath{0.205_{-0.080}^{+0.104}}} 
\newcommand{\hatcurPPmeccenxxxxA}{\ensuremath{0.62\pm0.14}}         
\newcommand{\hatcurPPmshorteccenxxxxA}{\ensuremath{0.62}}           
\newcommand{\hatcurPPmlongeccenxxxxA}{\ensuremath{0.62\pm0.14}}     
\newcommand{\hatcurPPmeeccenxxxxA}{\ensuremath{196\pm44}}           
\newcommand{\hatcurPPmeshorteccenxxxxA}{\ensuremath{195.7}}         
\newcommand{\hatcurPPmelongeccenxxxxA}{\ensuremath{196\pm44}}       
\newcommand{\hatcurPPreccenxxxxA}{\ensuremath{1.54_{-0.18}^{+0.24}}} 
\newcommand{\hatcurPPrshorteccenxxxxA}{\ensuremath{1.54}}           
\newcommand{\hatcurPPrlongeccenxxxxA}{\ensuremath{1.54_{-0.18}^{+0.24}}} 
\newcommand{\hatcurPPreeccenxxxxA}{\ensuremath{17.3_{-2.1}^{+2.7}}} 
\newcommand{\hatcurPPreshorteccenxxxxA}{\ensuremath{17.3}}          
\newcommand{\hatcurPPrelongeccenxxxxA}{\ensuremath{17.3_{-2.1}^{+2.7}}} 
\newcommand{\hatcurPPmrcorreccenxxxxA}{\ensuremath{0.12}}           
\newcommand{\hatcurPPteffeccenxxxxA}{\ensuremath{1630\pm110}}       
\newcommand{\hatcurPPthetaeccenxxxxA}{\ensuremath{0.0346\pm0.0091}} 
\newcommand{\hatcurPPfluxperieccenxxxxA}{\ensuremath{1.87_{-0.26}^{+0.74}}} 
\newcommand{\hatcurPPfluxperidimeccenxxxxA}{\ensuremath{9}}         
\newcommand{\hatcurPPfluxapeccenxxxxA}{\ensuremath{1.43_{-0.41}^{+0.28}}} 
\newcommand{\hatcurPPfluxapdimeccenxxxxA}{\ensuremath{9}}           
\newcommand{\hatcurPPfluxavgeccenxxxxA}{\ensuremath{1.61_{-0.30}^{+0.43}}} 
\newcommand{\hatcurPPfluxavgdimeccenxxxxA}{\ensuremath{9}}          
\newcommand{\hatcurPPfluxavglogeccenxxxxA}{\ensuremath{9.21\pm0.11}} 
\newcommand{\hatcurXsecphaseeccenxxxxA}{\ensuremath{0.507\pm0.048}} 
\newcommand{\hatcurXsecondaryeccenxxxxA}{\ensuremath{2457244.36\pm0.22}} 
\newcommand{\hatcurXsecdureccenxxxxA}{\ensuremath{0.154\pm0.020}}   
\newcommand{\hatcurXsecingdureccenxxxxA}{\ensuremath{0.024\pm0.025}} 
\newcommand{\hatcurPPphiconjeccenxxxxA}{\ensuremath{0.07\pm0.30}}   
\newcommand{\hatcurPPperieccenxxxxA}{\ensuremath{2457241.7\pm1.4}}  
\newcommand{\hatcurPPaequiveccenxxxxA}{\ensuremath{0.0292\pm0.0035}} 
\newcommand{\hatcurPPtcirceccenxxxxA}{\ensuremath{180_{-97}^{+134}}} 
\newcommand{\hatcurPPtinfalleccenxxxxA}{\ensuremath{3400_{-1500}^{+2600}}} 
\newcommand{\hatcurXdisteccenxxxxA}{\ensuremath{760\pm120}}         
\newcommand{\hatcurXAveccenxxxxA}{\ensuremath{0.185\pm0.069}}       
\newcommand{\hatcurXdistredeccenxxxxA}{\ensuremath{760\pm120}}      
\newcommand{\hatcurXEBVeccenxxxxA}{\ensuremath{0.060\pm0.022}}      
\newcommand{\hatcurXmvisoredeccenxxxxA}{\ensuremath{12.750\pm0.020}} 
\newcommand{\hatcurXmiisoredeccenxxxxA}{\ensuremath{12.173\pm0.017}} 
\newcommand{\hatcurXmjisoredeccenxxxxA}{\ensuremath{11.857\pm0.014}} 
\newcommand{\hatcurXmhisoredeccenxxxxA}{\ensuremath{11.636\pm0.015}} 
\newcommand{\hatcurXmkisoredeccenxxxxA}{\ensuremath{11.583\pm0.016}} 
\newcommand{\hatcurXviisoredeccenxxxxA}{\ensuremath{0.577\pm0.020}} 
\newcommand{\hatcurXvkisoredeccenxxxxA}{\ensuremath{1.167\pm0.027}} 
\newcommand{\hatcurXjhisoredeccenxxxxA}{\ensuremath{0.2210\pm0.0097}} 
\newcommand{\hatcurXjkisoredeccenxxxxA}{\ensuremath{0.2740\pm0.0080}} 
\newcommand{\hatcurCCpmraeccenxxxxA}{\ensuremath{-0.2\pm1.0}}       
\newcommand{\hatcurCCpmdececcenxxxxA}{\ensuremath{-5.9\pm1.7}}      
\newcommand{\hatcurCCpmeccenxxxxA}{\ensuremath{5.9\pm2.0}}          
 \newcommand{\hatcurhtreccenxxxxB}{HATS601-056}                      
\newcommand{\hatcurfieldeccenxxxxB}{\ensuremath{string}}            
\newcommand{\hatcurCCraeccenxxxxB}{\ensuremath{06^{\mathrm h}42^{\mathrm m}17.10{\mathrm s}}}                     
\newcommand{\hatcurCCdececcenxxxxB}{\ensuremath{-29{\arcdeg}46{\arcmin}36.5{\arcsec}}}                    
\newcommand{\hatcurCCmageccenxxxxB}{13.377}                         
\newcommand{\hatcurCCtwomasseccenxxxxB}{2MASS~06421710-2946365}     
\newcommand{\hatcurCCgsceccenxxxxB}{GSC~6533-01514}                 
\newcommand{\hatcurCCtassmveccenxxxxB}{\ensuremath{13.377\pm0.030}} 
\newcommand{\hatcurCCtassmvshorteccenxxxxB}{\ensuremath{13.4}}      
\newcommand{\hatcurCCtassmBeccenxxxxB}{\ensuremath{13.866\pm0.020}} 
\newcommand{\hatcurCCtassmBshorteccenxxxxB}{\ensuremath{13.9}}      
\newcommand{\hatcurCCtassmIeccenxxxxB}{\ensuremath{nff\pmnff}}      
\newcommand{\hatcurCCtassmIshorteccenxxxxB}{\ensuremath{0.0}}       
\newcommand{\hatcurCCtassmgeccenxxxxB}{\ensuremath{13.557\pm0.020}} 
\newcommand{\hatcurCCtassmgshorteccenxxxxB}{\ensuremath{13.6}}      
\newcommand{\hatcurCCtassmreccenxxxxB}{\ensuremath{13.272\pm0.040}} 
\newcommand{\hatcurCCtassmrshorteccenxxxxB}{\ensuremath{13.3}}      
\newcommand{\hatcurCCtassmieccenxxxxB}{\ensuremath{13.197\pm0.040}} 
\newcommand{\hatcurCCtassmishorteccenxxxxB}{\ensuremath{13.2}}      
\newcommand{\hatcurCCtwomassJmageccenxxxxB}{\ensuremath{12.439\pm0.026}} 
\newcommand{\hatcurCCtwomassHmageccenxxxxB}{\ensuremath{12.242\pm0.032}} 
\newcommand{\hatcurCCtwomassKmageccenxxxxB}{\ensuremath{12.147\pm0.027}} 
\newcommand{\hatcurCCcitJmageccenxxxxB}{\ensuremath{12.460\pm0.026}} 
\newcommand{\hatcurCCcitHmageccenxxxxB}{\ensuremath{12.236\pm0.032}} 
\newcommand{\hatcurCCcitKmageccenxxxxB}{\ensuremath{12.171\pm0.027}} 
\newcommand{\hatcurCCbbJmageccenxxxxB}{\ensuremath{12.503\pm0.028}} 
\newcommand{\hatcurCCbbHmageccenxxxxB}{\ensuremath{12.258\pm0.033}} 
\newcommand{\hatcurCCbbKmageccenxxxxB}{\ensuremath{12.191\pm0.027}} 
\newcommand{\hatcurCCesoJmageccenxxxxB}{\ensuremath{12.504\pm0.029}} 
\newcommand{\hatcurCCesoHmageccenxxxxB}{\ensuremath{12.255\pm0.037}} 
\newcommand{\hatcurCCesoKmageccenxxxxB}{\ensuremath{12.190\pm0.028}} 
\newcommand{\hatcurCCesoJHmageccenxxxxB}{\ensuremath{0.249\pm0.045}} 
\newcommand{\hatcurCCesoJKmageccenxxxxB}{\ensuremath{0.314\pm0.040}} 
\newcommand{\hatcurCCesoHKmageccenxxxxB}{\ensuremath{0.065\pm0.047}} 
\newcommand{\hatcurLCdipeccenxxxxB}{\ensuremath{4.7}}               
\newcommand{\hatcurLCrprstareccenxxxxB}{\ensuremath{0.0718\pm0.0032}} 
\newcommand{\hatcurLCbsqeccenxxxxB}{\ensuremath{0.100_{-0.072}^{+0.104}}} 
\newcommand{\hatcurLCimpeccenxxxxB}{\ensuremath{0.32_{-0.15}^{+0.14}}} 
\newcommand{\hatcurLCzetaeccenxxxxB}{\ensuremath{9.76\pm0.11}}      
\newcommand{\hatcurLCdureccenxxxxB}{\ensuremath{0.2215\pm0.0029}}   
\newcommand{\hatcurLCdurshorteccenxxxxB}{\ensuremath{0.2215}}       
\newcommand{\hatcurLCdurhreccenxxxxB}{\ensuremath{5.316\pm0.070}}   
\newcommand{\hatcurLCdurhrshorteccenxxxxB}{\ensuremath{5.316}}      
\newcommand{\hatcurLCqeccenxxxxB}{\ensuremath{0.06790\pm0.00090}}   
\newcommand{\hatcurLCqshorteccenxxxxB}{\ensuremath{0.068}}          
\newcommand{\hatcurLCingdureccenxxxxB}{\ensuremath{0.0164\pm0.0020}} 
\newcommand{\hatcurLCPeccenxxxxB}{\ensuremath{3.2642729\pm0.0000058}} 
\newcommand{\hatcurLCPprececcenxxxxB}{\ensuremath{3.2642729}}       
\newcommand{\hatcurLCPshorteccenxxxxB}{\ensuremath{3.2643}}         
\newcommand{\hatcurLCTeccenxxxxB}{\ensuremath{2456949.6188\pm0.0011}} 
\newcommand{\hatcurLCTAeccenxxxxB}{\ensuremath{2455794.0662\pm0.0022}} 
\newcommand{\hatcurLCTBeccenxxxxB}{\ensuremath{2457393.5598\pm0.0014}} 
\newcommand{\hatcurLChatnetmAeccenxxxxB}{\ensuremath{13.317590\pm0.000070}} 
\newcommand{\hatcurLCiblendAeccenxxxxB}{\ensuremath{0.871\pm0.078}} 
\newcommand{\hatcurLChatnetmBeccenxxxxB}{\ensuremath{13.317530\pm0.000096}} 
\newcommand{\hatcurLCiblendBeccenxxxxB}{\ensuremath{0.539\pm0.073}} 
\newcommand{\hatcurLChatnetmCeccenxxxxB}{\ensuremath{13.31752\pm0.00028}} 
\newcommand{\hatcurLCiblendCeccenxxxxB}{\ensuremath{0.85\pm0.13}}   
\newcommand{\hatcurLCrhoeccenxxxxB}{\ensuremath{0.18\pm0.86}}       
\newcommand{\hatcurSMEiteffeccenxxxxB}{\ensuremath{6460\pm130}}     
\newcommand{\hatcurSMEizfeheccenxxxxB}{\ensuremath{0.010\pm0.077}}  
\newcommand{\hatcurSMEizfehshorteccenxxxxB}{\ensuremath{0.01}}      
\newcommand{\hatcurSMEiloggeccenxxxxB}{\ensuremath{3.90\pm0.24}}    
\newcommand{\hatcurSMEivsineccenxxxxB}{\ensuremath{9.52\pm0.25}}    
\newcommand{\hatcurSMEivmaceccenxxxxB}{\ensuremath{nff\pmnff}}      
\newcommand{\hatcurSMEivmiceccenxxxxB}{\ensuremath{nff\pmnff}}      
\newcommand{\hatcurLBizeccenxxxxB}{\ensuremath{0.1203}}             
\newcommand{\hatcurLBiizeccenxxxxB}{\ensuremath{0.3694}}            
\newcommand{\hatcurLBiieccenxxxxB}{\ensuremath{0.1695}}             
\newcommand{\hatcurLBiiieccenxxxxB}{\ensuremath{0.3788}}            
\newcommand{\hatcurLBiIeccenxxxxB}{\ensuremath{0.1507}}             
\newcommand{\hatcurLBiiIeccenxxxxB}{\ensuremath{0.3774}}            
\newcommand{\hatcurLBigeccenxxxxB}{\ensuremath{0.4123}}             
\newcommand{\hatcurLBiigeccenxxxxB}{\ensuremath{0.3323}}            
\newcommand{\hatcurLBireccenxxxxB}{\ensuremath{0.2457}}             
\newcommand{\hatcurLBiireccenxxxxB}{\ensuremath{0.3857}}            
\newcommand{\hatcurLBiReccenxxxxB}{\ensuremath{0.2239}}             
\newcommand{\hatcurLBiiReccenxxxxB}{\ensuremath{0.3855}}            
\newcommand{\hatcurLBikepeccenxxxxB}{\ensuremath{0.1000}}           
\newcommand{\hatcurLBiikepeccenxxxxB}{\ensuremath{0.1000}}          
\newcommand{\hatcurISOmeccenxxxxB}{\ensuremath{1.58\pm0.11}}        
\newcommand{\hatcurISOmshorteccenxxxxB}{\ensuremath{1.58}}          
\newcommand{\hatcurISOmlongeccenxxxxB}{\ensuremath{1.58\pm0.11}}    
\newcommand{\hatcurISOreccenxxxxB}{\ensuremath{2.33_{-0.22}^{+0.37}}} 
\newcommand{\hatcurISOrshorteccenxxxxB}{\ensuremath{2.33}}          
\newcommand{\hatcurISOrlongeccenxxxxB}{\ensuremath{2.33_{-0.22}^{+0.37}}} 
\newcommand{\hatcurISOrhoeccenxxxxB}{\ensuremath{0.176\pm0.071}}    
\newcommand{\hatcurISOrholongeccenxxxxB}{\ensuremath{0.176\pm0.071}} 
\newcommand{\hatcurISOloggeccenxxxxB}{\ensuremath{3.902\pm0.097}}   
\newcommand{\hatcurISOlumeccenxxxxB}{\ensuremath{8.5_{-1.8}^{+3.0}}} 
\newcommand{\hatcurISOlumshorteccenxxxxB}{\ensuremath{8.47}}        
\newcommand{\hatcurISOmveccenxxxxB}{\ensuremath{2.41\pm0.33}}       
\newcommand{\hatcurISOvieccenxxxxB}{\ensuremath{0.504\pm0.033}}     
\newcommand{\hatcurISOageeccenxxxxB}{\ensuremath{2.00\pm0.39}}      
\newcommand{\hatcurISOsigmaeccenxxxxB}{\ensuremath{0.00040\pm0.00013}} 
\newcommand{\hatcurISOMJeccenxxxxB}{\ensuremath{1.61\pm0.31}}       
\newcommand{\hatcurISOMHeccenxxxxB}{\ensuremath{1.39\pm0.31}}       
\newcommand{\hatcurISOMKeccenxxxxB}{\ensuremath{1.35\pm0.31}}       
\newcommand{\hatcurISOJKeccenxxxxB}{\ensuremath{0.260\pm0.030}}     
\newcommand{\hatcurISOspececcenxxxxB}{F}                            
\newcommand{\hatcurRVKeccenxxxxB}{\ensuremath{166\pm29}}            
\newcommand{\hatcurRVrkeccenxxxxB}{\ensuremath{0.08\pm0.20}}        
\newcommand{\hatcurRVrheccenxxxxB}{\ensuremath{0.08\pm0.25}}        
\newcommand{\hatcurRVkeccenxxxxB}{\ensuremath{0.019_{-0.056}^{+0.082}}} 
\newcommand{\hatcurRVheccenxxxxB}{\ensuremath{0.012_{-0.068}^{+0.140}}} 
\newcommand{\hatcurRVtroneeccenxxxxB}{\ensuremath{0\pm0}}           
\newcommand{\hatcurRVtrtwoeccenxxxxB}{\ensuremath{0\pm0}}           
\newcommand{\hatcurRVgammaeccenxxxxB}{\ensuremath{9196\pm18}}       
\newcommand{\hatcurRVjittereccenxxxxB}{\ensuremath{0\pm14}}         
\newcommand{\hatcurRVjittertwosiglimeccenxxxxB}{\ensuremath{<29.7}} 
\newcommand{\hatcurRVfitrmseccenxxxxB}{\ensuremath{.1fym}}          
\newcommand{\hatcurRVecceneccenxxxxB}{\ensuremath{0.086\pm0.099}}   
\newcommand{\hatcurRVeccentwosiglimeccenxxxxB}{\ensuremath{<0.312}} 
\newcommand{\hatcurRVomegaeccenxxxxB}{\ensuremath{100\pm120}}       
\newcommand{\hatcurPPieccenxxxxB}{\ensuremath{85.8\pm2.2}}          
\newcommand{\hatcurPPgeccenxxxxB}{\ensuremath{14.2_{-2.4}^{+4.6}}}  
\newcommand{\hatcurPPloggeccenxxxxB}{\ensuremath{3.152_{-0.082}^{+0.123}}} 
\newcommand{\hatcurPPareccenxxxxB}{\ensuremath{4.63_{-0.56}^{+0.41}}} 
\newcommand{\hatcurPPareleccenxxxxB}{\ensuremath{0.0502\pm0.0012}}  
\newcommand{\hatcurPPrhoeccenxxxxB}{\ensuremath{0.42_{-0.10}^{+0.22}}} 
\newcommand{\hatcurPPmeccenxxxxB}{\ensuremath{1.63\pm0.30}}         
\newcommand{\hatcurPPmshorteccenxxxxB}{\ensuremath{1.63}}           
\newcommand{\hatcurPPmlongeccenxxxxB}{\ensuremath{1.63\pm0.30}}     
\newcommand{\hatcurPPmeeccenxxxxB}{\ensuremath{517\pm94}}           
\newcommand{\hatcurPPmeshorteccenxxxxB}{\ensuremath{516.8}}         
\newcommand{\hatcurPPmelongeccenxxxxB}{\ensuremath{517\pm94}}       
\newcommand{\hatcurPPreccenxxxxB}{\ensuremath{1.64\pm0.26}}         
\newcommand{\hatcurPPrshorteccenxxxxB}{\ensuremath{1.64}}           
\newcommand{\hatcurPPrlongeccenxxxxB}{\ensuremath{1.64\pm0.26}}     
\newcommand{\hatcurPPreeccenxxxxB}{\ensuremath{18.4\pm2.9}}         
\newcommand{\hatcurPPreshorteccenxxxxB}{\ensuremath{18.4}}          
\newcommand{\hatcurPPrelongeccenxxxxB}{\ensuremath{18.4\pm2.9}}     
\newcommand{\hatcurPPmrcorreccenxxxxB}{\ensuremath{0.57}}           
\newcommand{\hatcurPPteffeccenxxxxB}{\ensuremath{2130_{-100}^{+150}}} 
\newcommand{\hatcurPPthetaeccenxxxxB}{\ensuremath{0.0616_{-0.0089}^{+0.0117}}} 
\newcommand{\hatcurPPfluxperieccenxxxxB}{\ensuremath{5.31_{-1.00}^{+3.53}}} 
\newcommand{\hatcurPPfluxperidimeccenxxxxB}{\ensuremath{9}}         
\newcommand{\hatcurPPfluxapeccenxxxxB}{\ensuremath{3.97\pm0.81}}    
\newcommand{\hatcurPPfluxapdimeccenxxxxB}{\ensuremath{9}}           
\newcommand{\hatcurPPfluxavgeccenxxxxB}{\ensuremath{4.61_{-0.84}^{+1.44}}} 
\newcommand{\hatcurPPfluxavgdimeccenxxxxB}{\ensuremath{9}}          
\newcommand{\hatcurPPfluxavglogeccenxxxxB}{\ensuremath{9.664_{-0.088}^{+0.118}}} 
\newcommand{\hatcurXsecphaseeccenxxxxB}{\ensuremath{0.512\pm0.052}} 
\newcommand{\hatcurXsecondaryeccenxxxxB}{\ensuremath{2456951.29\pm0.17}} 
\newcommand{\hatcurXsecdureccenxxxxB}{\ensuremath{0.228\pm0.048}}   
\newcommand{\hatcurXsecingdureccenxxxxB}{\ensuremath{0.0178\pm0.0093}} 
\newcommand{\hatcurPPphiconjeccenxxxxB}{\ensuremath{0.04_{-0.35}^{+0.19}}} 
\newcommand{\hatcurPPperieccenxxxxB}{\ensuremath{2456949.50\pm0.78}} 
\newcommand{\hatcurPPaequiveccenxxxxB}{\ensuremath{0.0172\pm0.0022}} 
\newcommand{\hatcurPPtcirceccenxxxxB}{\ensuremath{83_{-40}^{+70}}}  
\newcommand{\hatcurPPtinfalleccenxxxxB}{\ensuremath{65_{-33}^{+46}}} 
\newcommand{\hatcurXdisteccenxxxxB}{\ensuremath{1470_{-150}^{+240}}} 
\newcommand{\hatcurXAveccenxxxxB}{\ensuremath{0.121\pm0.096}}       
\newcommand{\hatcurXdistredeccenxxxxB}{\ensuremath{1470_{-140}^{+230}}} 
\newcommand{\hatcurXEBVeccenxxxxB}{\ensuremath{0.039\pm0.031}}      
\newcommand{\hatcurXmvisoredeccenxxxxB}{\ensuremath{13.385\pm0.031}} 
\newcommand{\hatcurXmiisoredeccenxxxxB}{\ensuremath{12.815\pm0.025}} 
\newcommand{\hatcurXmjisoredeccenxxxxB}{\ensuremath{12.485\pm0.017}} 
\newcommand{\hatcurXmhisoredeccenxxxxB}{\ensuremath{12.254\pm0.021}} 
\newcommand{\hatcurXmkisoredeccenxxxxB}{\ensuremath{12.201\pm0.021}} 
\newcommand{\hatcurXviisoredeccenxxxxB}{\ensuremath{0.569\pm0.025}} 
\newcommand{\hatcurXvkisoredeccenxxxxB}{\ensuremath{1.183\pm0.041}} 
\newcommand{\hatcurXjhisoredeccenxxxxB}{\ensuremath{0.230\pm0.017}} 
\newcommand{\hatcurXjkisoredeccenxxxxB}{\ensuremath{0.283\pm0.014}} 
\newcommand{\hatcurCCpmraeccenxxxxB}{\ensuremath{-4.6\pm2.1}}       
\newcommand{\hatcurCCpmdececcenxxxxB}{\ensuremath{5.6\pm2.3}}       
\newcommand{\hatcurCCpmeccenxxxxB}{\ensuremath{7.2\pm3.1}}          
 \newcommand{\hatcurhtreccenxxxxC}{HATS601-061}                      
\newcommand{\hatcurfieldeccenxxxxC}{\ensuremath{string}}            
\newcommand{\hatcurCCraeccenxxxxC}{\ensuremath{06^{\mathrm h}54^{\mathrm m}04.18{\mathrm s}}}                     
\newcommand{\hatcurCCdececcenxxxxC}{\ensuremath{-27{\arcdeg}03{\arcmin}01.4{\arcsec}}}                    
\newcommand{\hatcurCCmageccenxxxxC}{12.681}                         
\newcommand{\hatcurCCtwomasseccenxxxxC}{2MASS~06540416-2703013}     
\newcommand{\hatcurCCgsceccenxxxxC}{GSC~6530-01596}                 
\newcommand{\hatcurCCtassmveccenxxxxC}{\ensuremath{12.681\pm0.030}} 
\newcommand{\hatcurCCtassmvshorteccenxxxxC}{\ensuremath{12.7}}      
\newcommand{\hatcurCCtassmBeccenxxxxC}{\ensuremath{13.175\pm0.040}} 
\newcommand{\hatcurCCtassmBshorteccenxxxxC}{\ensuremath{13.2}}      
\newcommand{\hatcurCCtassmIeccenxxxxC}{\ensuremath{nff\pmnff}}      
\newcommand{\hatcurCCtassmIshorteccenxxxxC}{\ensuremath{0.0}}       
\newcommand{\hatcurCCtassmgeccenxxxxC}{\ensuremath{12.893\pm0.040}} 
\newcommand{\hatcurCCtassmgshorteccenxxxxC}{\ensuremath{12.9}}      
\newcommand{\hatcurCCtassmreccenxxxxC}{\ensuremath{12.555\pm0.020}} 
\newcommand{\hatcurCCtassmrshorteccenxxxxC}{\ensuremath{12.6}}      
\newcommand{\hatcurCCtassmieccenxxxxC}{\ensuremath{12.484\pm0.060}} 
\newcommand{\hatcurCCtassmishorteccenxxxxC}{\ensuremath{12.5}}      
\newcommand{\hatcurCCtwomassJmageccenxxxxC}{\ensuremath{11.765\pm0.024}} 
\newcommand{\hatcurCCtwomassHmageccenxxxxC}{\ensuremath{11.521\pm0.023}} 
\newcommand{\hatcurCCtwomassKmageccenxxxxC}{\ensuremath{11.498\pm0.023}} 
\newcommand{\hatcurCCcitJmageccenxxxxC}{\ensuremath{11.787\pm0.024}} 
\newcommand{\hatcurCCcitHmageccenxxxxC}{\ensuremath{11.517\pm0.024}} 
\newcommand{\hatcurCCcitKmageccenxxxxC}{\ensuremath{11.522\pm0.023}} 
\newcommand{\hatcurCCbbJmageccenxxxxC}{\ensuremath{11.828\pm0.026}} 
\newcommand{\hatcurCCbbHmageccenxxxxC}{\ensuremath{11.537\pm0.024}} 
\newcommand{\hatcurCCbbKmageccenxxxxC}{\ensuremath{11.542\pm0.023}} 
\newcommand{\hatcurCCesoJmageccenxxxxC}{\ensuremath{11.829\pm0.027}} 
\newcommand{\hatcurCCesoHmageccenxxxxC}{\ensuremath{11.530\pm0.025}} 
\newcommand{\hatcurCCesoKmageccenxxxxC}{\ensuremath{11.542\pm0.024}} 
\newcommand{\hatcurCCesoJHmageccenxxxxC}{\ensuremath{0.299\pm0.035}} 
\newcommand{\hatcurCCesoJKmageccenxxxxC}{\ensuremath{0.287\pm0.036}} 
\newcommand{\hatcurCCesoHKmageccenxxxxC}{\ensuremath{-0.0120\pm0.0070}} 
\newcommand{\hatcurLCdipeccenxxxxC}{\ensuremath{7.2}}               
\newcommand{\hatcurLCrprstareccenxxxxC}{\ensuremath{0.0798\pm0.0036}} 
\newcommand{\hatcurLCbsqeccenxxxxC}{\ensuremath{0.721_{-0.084}^{+0.050}}} 
\newcommand{\hatcurLCimpeccenxxxxC}{\ensuremath{0.849_{-0.051}^{+0.029}}} 
\newcommand{\hatcurLCzetaeccenxxxxC}{\ensuremath{27.55\pm0.89}}     
\newcommand{\hatcurLCdureccenxxxxC}{\ensuremath{0.0915\pm0.0044}}   
\newcommand{\hatcurLCdurshorteccenxxxxC}{\ensuremath{0.0915}}       
\newcommand{\hatcurLCdurhreccenxxxxC}{\ensuremath{2.20\pm0.10}}     
\newcommand{\hatcurLCdurhrshorteccenxxxxC}{\ensuremath{2.196}}      
\newcommand{\hatcurLCqeccenxxxxC}{\ensuremath{0.0218\pm0.0010}}     
\newcommand{\hatcurLCqshorteccenxxxxC}{\ensuremath{0.022}}          
\newcommand{\hatcurLCingdureccenxxxxC}{\ensuremath{0.0215\pm0.0065}} 
\newcommand{\hatcurLCPeccenxxxxC}{\ensuremath{4.193649\pm0.000013}} 
\newcommand{\hatcurLCPprececcenxxxxC}{\ensuremath{4.1936489}}       
\newcommand{\hatcurLCPshorteccenxxxxC}{\ensuremath{4.1936}}         
\newcommand{\hatcurLCTeccenxxxxC}{\ensuremath{2456795.7240\pm0.0014}} 
\newcommand{\hatcurLCTAeccenxxxxC}{\ensuremath{2455797.6356\pm0.0032}} 
\newcommand{\hatcurLCTBeccenxxxxC}{\ensuremath{2457319.9301\pm0.0023}} 
\newcommand{\hatcurLChatnetmeccenxxxxC}{\ensuremath{12.674930\pm0.000057}} 
\newcommand{\hatcurLCiblendeccenxxxxC}{\ensuremath{0.856\pm0.090}}  
\newcommand{\hatcurLCrhoeccenxxxxC}{\ensuremath{0.48_{-0.20}^{+0.37}}} 
\newcommand{\hatcurSMEiteffeccenxxxxC}{\ensuremath{6424\pm91}}      
\newcommand{\hatcurSMEizfeheccenxxxxC}{\ensuremath{0.470\pm0.041}}  
\newcommand{\hatcurSMEizfehshorteccenxxxxC}{\ensuremath{0.47}}      
\newcommand{\hatcurSMEiloggeccenxxxxC}{\ensuremath{4.10\pm0.22}}    
\newcommand{\hatcurSMEivsineccenxxxxC}{\ensuremath{19.21\pm0.23}}   
\newcommand{\hatcurSMEivmaceccenxxxxC}{\ensuremath{4.99\pm0.14}}    
\newcommand{\hatcurSMEivmiceccenxxxxC}{\ensuremath{1.66\pm0.12}}    
\newcommand{\hatcurLBizeccenxxxxC}{\ensuremath{0.1258}}             
\newcommand{\hatcurLBiizeccenxxxxC}{\ensuremath{0.3871}}            
\newcommand{\hatcurLBiieccenxxxxC}{\ensuremath{0.1795}}             
\newcommand{\hatcurLBiiieccenxxxxC}{\ensuremath{0.3965}}            
\newcommand{\hatcurLBiIeccenxxxxC}{\ensuremath{0.1592}}             
\newcommand{\hatcurLBiiIeccenxxxxC}{\ensuremath{0.3955}}            
\newcommand{\hatcurLBigeccenxxxxC}{\ensuremath{0.4468}}             
\newcommand{\hatcurLBiigeccenxxxxC}{\ensuremath{0.3221}}            
\newcommand{\hatcurLBireccenxxxxC}{\ensuremath{0.2634}}             
\newcommand{\hatcurLBiireccenxxxxC}{\ensuremath{0.3954}}            
\newcommand{\hatcurLBiReccenxxxxC}{\ensuremath{0.2393}}             
\newcommand{\hatcurLBiiReccenxxxxC}{\ensuremath{0.3976}}            
\newcommand{\hatcurLBikepeccenxxxxC}{\ensuremath{0.1000}}           
\newcommand{\hatcurLBiikepeccenxxxxC}{\ensuremath{0.1000}}          
\newcommand{\hatcurISOmeccenxxxxC}{\ensuremath{1.496_{-0.078}^{+0.115}}} 
\newcommand{\hatcurISOmshorteccenxxxxC}{\ensuremath{1.50}}          
\newcommand{\hatcurISOmlongeccenxxxxC}{\ensuremath{1.496_{-0.078}^{+0.115}}} 
\newcommand{\hatcurISOreccenxxxxC}{\ensuremath{1.71_{-0.24}^{+0.36}}} 
\newcommand{\hatcurISOrshorteccenxxxxC}{\ensuremath{1.71}}          
\newcommand{\hatcurISOrlongeccenxxxxC}{\ensuremath{1.71_{-0.24}^{+0.36}}} 
\newcommand{\hatcurISOrhoeccenxxxxC}{\ensuremath{0.42\pm0.17}}      
\newcommand{\hatcurISOrholongeccenxxxxC}{\ensuremath{0.42\pm0.17}}  
\newcommand{\hatcurISOloggeccenxxxxC}{\ensuremath{4.14\pm0.11}}     
\newcommand{\hatcurISOlumeccenxxxxC}{\ensuremath{4.5_{-1.2}^{+2.1}}} 
\newcommand{\hatcurISOlumshorteccenxxxxC}{\ensuremath{4.48}}        
\newcommand{\hatcurISOmveccenxxxxC}{\ensuremath{3.08\pm0.37}}       
\newcommand{\hatcurISOvieccenxxxxC}{\ensuremath{0.522\pm0.024}}     
\newcommand{\hatcurISOageeccenxxxxC}{\ensuremath{1.34_{-0.51}^{+0.31}}} 
\newcommand{\hatcurISOsigmaeccenxxxxC}{\ensuremath{0.0053\pm0.0024}} 
\newcommand{\hatcurISOMJeccenxxxxC}{\ensuremath{2.25\pm0.35}}       
\newcommand{\hatcurISOMHeccenxxxxC}{\ensuremath{2.04\pm0.35}}       
\newcommand{\hatcurISOMKeccenxxxxC}{\ensuremath{2.00\pm0.35}}       
\newcommand{\hatcurISOJKeccenxxxxC}{\ensuremath{0.240\pm0.060}}     
\newcommand{\hatcurISOspececcenxxxxC}{F}                            
\newcommand{\hatcurRVKeccenxxxxC}{\ensuremath{1080\pm190}}          
\newcommand{\hatcurRVrkeccenxxxxC}{\ensuremath{-0.43\pm0.12}}       
\newcommand{\hatcurRVrheccenxxxxC}{\ensuremath{0.42\pm0.16}}        
\newcommand{\hatcurRVkeccenxxxxC}{\ensuremath{-0.261\pm0.092}}      
\newcommand{\hatcurRVheccenxxxxC}{\ensuremath{0.25\pm0.12}}         
\newcommand{\hatcurRVtroneeccenxxxxC}{\ensuremath{0\pm0}}           
\newcommand{\hatcurRVtrtwoeccenxxxxC}{\ensuremath{0\pm0}}           
\newcommand{\hatcurRVgammaAeccenxxxxC}{\ensuremath{37080\pm160}}    
\newcommand{\hatcurRVjitterAeccenxxxxC}{\ensuremath{460\pm150}}     
\newcommand{\hatcurRVjittertwosiglimAeccenxxxxC}{\ensuremath{<773.3}} 
\newcommand{\hatcurRVfitrmsAeccenxxxxC}{\ensuremath{0.0}}           
\newcommand{\hatcurRVgammaBeccenxxxxC}{\ensuremath{37250\pm220}}    
\newcommand{\hatcurRVjitterBeccenxxxxC}{\ensuremath{350\pm210}}     
\newcommand{\hatcurRVjittertwosiglimBeccenxxxxC}{\ensuremath{<789.9}} 
\newcommand{\hatcurRVfitrmsBeccenxxxxC}{\ensuremath{0.0}}           
\newcommand{\hatcurRVgammaCeccenxxxxC}{\ensuremath{38000\pm200}}    
\newcommand{\hatcurRVjitterCeccenxxxxC}{\ensuremath{380\pm140}}     
\newcommand{\hatcurRVjittertwosiglimCeccenxxxxC}{\ensuremath{<678.2}} 
\newcommand{\hatcurRVfitrmsCeccenxxxxC}{\ensuremath{0.0}}           
\newcommand{\hatcurRVecceneccenxxxxC}{\ensuremath{0.38\pm0.11}}     
\newcommand{\hatcurRVeccentwosiglimeccenxxxxC}{\ensuremath{<0.561}} 
\newcommand{\hatcurRVomegaeccenxxxxC}{\ensuremath{136\pm18}}        
\newcommand{\hatcurPPieccenxxxxC}{\ensuremath{80.4_{-4.2}^{+2.3}}}  
\newcommand{\hatcurPPgeccenxxxxC}{\ensuremath{135\pm50}}            
\newcommand{\hatcurPPloggeccenxxxxC}{\ensuremath{4.13\pm0.16}}      
\newcommand{\hatcurPPareccenxxxxC}{\ensuremath{7.31\pm0.98}}        
\newcommand{\hatcurPPareleccenxxxxC}{\ensuremath{0.0583_{-0.0010}^{+0.0015}}} 
\newcommand{\hatcurPPrhoeccenxxxxC}{\ensuremath{5.1_{-2.3}^{+3.3}}} 
\newcommand{\hatcurPPmeccenxxxxC}{\ensuremath{9.7\pm1.6}}           
\newcommand{\hatcurPPmshorteccenxxxxC}{\ensuremath{9.74}}           
\newcommand{\hatcurPPmlongeccenxxxxC}{\ensuremath{9.7\pm1.6}}       
\newcommand{\hatcurPPmeeccenxxxxC}{\ensuremath{3100\pm520}}         
\newcommand{\hatcurPPmeshorteccenxxxxC}{\ensuremath{3095.3}}        
\newcommand{\hatcurPPmelongeccenxxxxC}{\ensuremath{3100\pm520}}     
\newcommand{\hatcurPPreccenxxxxC}{\ensuremath{1.33_{-0.20}^{+0.29}}} 
\newcommand{\hatcurPPrshorteccenxxxxC}{\ensuremath{1.33}}           
\newcommand{\hatcurPPrlongeccenxxxxC}{\ensuremath{1.33_{-0.20}^{+0.29}}} 
\newcommand{\hatcurPPreeccenxxxxC}{\ensuremath{15.0_{-2.3}^{+3.3}}} 
\newcommand{\hatcurPPreshorteccenxxxxC}{\ensuremath{15.0}}          
\newcommand{\hatcurPPrelongeccenxxxxC}{\ensuremath{15.0_{-2.3}^{+3.3}}} 
\newcommand{\hatcurPPmrcorreccenxxxxC}{\ensuremath{0.10}}           
\newcommand{\hatcurPPteffeccenxxxxC}{\ensuremath{1710_{-120}^{+170}}} 
\newcommand{\hatcurPPthetaeccenxxxxC}{\ensuremath{0.60\pm0.14}}     
\newcommand{\hatcurPPfluxperieccenxxxxC}{\ensuremath{4.6_{-1.8}^{+4.2}}} 
\newcommand{\hatcurPPfluxperidimeccenxxxxC}{\ensuremath{9}}         
\newcommand{\hatcurPPfluxapeccenxxxxC}{\ensuremath{9.8\pm2.3}}      
\newcommand{\hatcurPPfluxapdimeccenxxxxC}{\ensuremath{8}}           
\newcommand{\hatcurPPfluxavgeccenxxxxC}{\ensuremath{1.94_{-0.49}^{+0.88}}} 
\newcommand{\hatcurPPfluxavgdimeccenxxxxC}{\ensuremath{9}}          
\newcommand{\hatcurPPfluxavglogeccenxxxxC}{\ensuremath{9.29\pm0.14}} 
\newcommand{\hatcurXsecphaseeccenxxxxC}{\ensuremath{0.329\pm0.059}} 
\newcommand{\hatcurXsecondaryeccenxxxxC}{\ensuremath{2456797.10\pm0.25}} 
\newcommand{\hatcurXsecdureccenxxxxC}{\ensuremath{0.127\pm0.039}}   
\newcommand{\hatcurXsecingdureccenxxxxC}{\ensuremath{0.063\pm0.020}} 
\newcommand{\hatcurPPphiconjeccenxxxxC}{\ensuremath{-0.055_{-0.037}^{+0.026}}} 
\newcommand{\hatcurPPperieccenxxxxC}{\ensuremath{2456795.95\pm0.17}} 
\newcommand{\hatcurPPaequiveccenxxxxC}{\ensuremath{0.0276\pm0.0039}} 
\newcommand{\hatcurPPtcirceccenxxxxC}{\ensuremath{1700_{-1300}^{+3400}}} 
\newcommand{\hatcurPPtinfalleccenxxxxC}{\ensuremath{127_{-69}^{+116}}} 
\newcommand{\hatcurXdisteccenxxxxC}{\ensuremath{810_{-110}^{+170}}} 
\newcommand{\hatcurXAveccenxxxxC}{\ensuremath{0.081\pm0.073}}       
\newcommand{\hatcurXdistredeccenxxxxC}{\ensuremath{800_{-110}^{+170}}} 
\newcommand{\hatcurXEBVeccenxxxxC}{\ensuremath{0.026\pm0.023}}      
\newcommand{\hatcurXmvisoredeccenxxxxC}{\ensuremath{12.688\pm0.030}} 
\newcommand{\hatcurXmiisoredeccenxxxxC}{\ensuremath{12.120\pm0.022}} 
\newcommand{\hatcurXmjisoredeccenxxxxC}{\ensuremath{11.796\pm0.014}} 
\newcommand{\hatcurXmhisoredeccenxxxxC}{\ensuremath{11.573\pm0.016}} 
\newcommand{\hatcurXmkisoredeccenxxxxC}{\ensuremath{11.524\pm0.016}} 
\newcommand{\hatcurXviisoredeccenxxxxC}{\ensuremath{0.566\pm0.022}} 
\newcommand{\hatcurXvkisoredeccenxxxxC}{\ensuremath{1.164\pm0.036}} 
\newcommand{\hatcurXjhisoredeccenxxxxC}{\ensuremath{0.223\pm0.012}} 
\newcommand{\hatcurXjkisoredeccenxxxxC}{\ensuremath{0.272\pm0.011}} 
\newcommand{\hatcurCCpmraeccenxxxxC}{\ensuremath{0.60\pm0.90}}      
\newcommand{\hatcurCCpmdececcenxxxxC}{\ensuremath{-7.0\pm1.0}}      
\newcommand{\hatcurCCpmeccenxxxxC}{\ensuremath{7.0\pm1.3}}          
 \newcommand{\hatcurhtreccenxxxxD}{HATS602-002}                      
\newcommand{\hatcurfieldeccenxxxxD}{\ensuremath{string}}            
\newcommand{\hatcurCCraeccenxxxxD}{\ensuremath{07^{\mathrm h}13^{\mathrm m}48.58{\mathrm s}}}                     
\newcommand{\hatcurCCdececcenxxxxD}{\ensuremath{-33{\arcdeg}26{\arcmin}14.4{\arcsec}}}                    
\newcommand{\hatcurCCmageccenxxxxD}{13.617}                         
\newcommand{\hatcurCCtwomasseccenxxxxD}{2MASS~07134857-3326143}     
\newcommand{\hatcurCCgsceccenxxxxD}{GSC~7107-03973}                 
\newcommand{\hatcurCCtassmveccenxxxxD}{\ensuremath{13.617\pm0.010}} 
\newcommand{\hatcurCCtassmvshorteccenxxxxD}{\ensuremath{13.6}}      
\newcommand{\hatcurCCtassmBeccenxxxxD}{\ensuremath{14.243\pm0.010}} 
\newcommand{\hatcurCCtassmBshorteccenxxxxD}{\ensuremath{14.2}}      
\newcommand{\hatcurCCtassmIeccenxxxxD}{\ensuremath{nff\pmnff}}      
\newcommand{\hatcurCCtassmIshorteccenxxxxD}{\ensuremath{0.0}}       
\newcommand{\hatcurCCtassmgeccenxxxxD}{\ensuremath{13.922\pm0.050}} 
\newcommand{\hatcurCCtassmgshorteccenxxxxD}{\ensuremath{13.9}}      
\newcommand{\hatcurCCtassmreccenxxxxD}{\ensuremath{13.498\pm0.010}} 
\newcommand{\hatcurCCtassmrshorteccenxxxxD}{\ensuremath{13.5}}      
\newcommand{\hatcurCCtassmieccenxxxxD}{\ensuremath{13.396\pm0.050}} 
\newcommand{\hatcurCCtassmishorteccenxxxxD}{\ensuremath{13.4}}      
\newcommand{\hatcurCCtwomassJmageccenxxxxD}{\ensuremath{12.543\pm0.024}} 
\newcommand{\hatcurCCtwomassHmageccenxxxxD}{\ensuremath{12.281\pm0.026}} 
\newcommand{\hatcurCCtwomassKmageccenxxxxD}{\ensuremath{12.245\pm0.029}} 
\newcommand{\hatcurCCcitJmageccenxxxxD}{\ensuremath{12.564\pm0.024}} 
\newcommand{\hatcurCCcitHmageccenxxxxD}{\ensuremath{12.277\pm0.026}} 
\newcommand{\hatcurCCcitKmageccenxxxxD}{\ensuremath{12.269\pm0.029}} 
\newcommand{\hatcurCCbbJmageccenxxxxD}{\ensuremath{12.607\pm0.026}} 
\newcommand{\hatcurCCbbHmageccenxxxxD}{\ensuremath{12.297\pm0.027}} 
\newcommand{\hatcurCCbbKmageccenxxxxD}{\ensuremath{12.289\pm0.029}} 
\newcommand{\hatcurCCesoJmageccenxxxxD}{\ensuremath{12.608\pm0.027}} 
\newcommand{\hatcurCCesoHmageccenxxxxD}{\ensuremath{12.291\pm0.030}} 
\newcommand{\hatcurCCesoKmageccenxxxxD}{\ensuremath{12.288\pm0.030}} 
\newcommand{\hatcurCCesoJHmageccenxxxxD}{\ensuremath{0.317\pm0.038}} 
\newcommand{\hatcurCCesoJKmageccenxxxxD}{\ensuremath{0.321\pm0.040}} 
\newcommand{\hatcurCCesoHKmageccenxxxxD}{\ensuremath{0.003\pm0.043}} 
\newcommand{\hatcurLCdipeccenxxxxD}{\ensuremath{10.5}}              
\newcommand{\hatcurLCrprstareccenxxxxD}{\ensuremath{0.0980\pm0.0038}} 
\newcommand{\hatcurLCbsqeccenxxxxD}{\ensuremath{0.22_{-0.14}^{+0.15}}} 
\newcommand{\hatcurLCimpeccenxxxxD}{\ensuremath{0.47_{-0.18}^{+0.14}}} 
\newcommand{\hatcurLCzetaeccenxxxxD}{\ensuremath{16.53\pm0.21}}     
\newcommand{\hatcurLCdureccenxxxxD}{\ensuremath{0.1365\pm0.0033}}   
\newcommand{\hatcurLCdurshorteccenxxxxD}{\ensuremath{0.1365}}       
\newcommand{\hatcurLCdurhreccenxxxxD}{\ensuremath{3.276\pm0.079}}   
\newcommand{\hatcurLCdurhrshorteccenxxxxD}{\ensuremath{3.276}}      
\newcommand{\hatcurLCqeccenxxxxD}{\ensuremath{0.0595\pm0.0014}}     
\newcommand{\hatcurLCqshorteccenxxxxD}{\ensuremath{0.059}}          
\newcommand{\hatcurLCingdureccenxxxxD}{\ensuremath{0.0153\pm0.0038}} 
\newcommand{\hatcurLCPeccenxxxxD}{\ensuremath{2.2921020\pm0.0000020}} 
\newcommand{\hatcurLCPprececcenxxxxD}{\ensuremath{2.2921020}}       
\newcommand{\hatcurLCPshorteccenxxxxD}{\ensuremath{2.2921}}         
\newcommand{\hatcurLCTeccenxxxxD}{\ensuremath{2456656.29436\pm0.00068}} 
\newcommand{\hatcurLCTAeccenxxxxD}{\ensuremath{2455794.4640\pm0.0010}} 
\newcommand{\hatcurLCTBeccenxxxxD}{\ensuremath{2457396.64326\pm0.00093}} 
\newcommand{\hatcurLChatnetmAeccenxxxxD}{\ensuremath{13.483310\pm0.000096}} 
\newcommand{\hatcurLCiblendAeccenxxxxD}{\ensuremath{0.923\pm0.057}} 
\newcommand{\hatcurLChatnetmBeccenxxxxD}{\ensuremath{13.483100\pm0.000094}} 
\newcommand{\hatcurLCiblendBeccenxxxxD}{\ensuremath{0.873\pm0.060}} 
\newcommand{\hatcurLCrhoeccenxxxxD}{\ensuremath{0.60_{-0.20}^{+0.44}}} 
\newcommand{\hatcurSMEiteffeccenxxxxD}{\ensuremath{6320\pm170}}     
\newcommand{\hatcurSMEizfeheccenxxxxD}{\ensuremath{0.320\pm0.076}}  
\newcommand{\hatcurSMEizfehshorteccenxxxxD}{\ensuremath{0.32}}      
\newcommand{\hatcurSMEiloggeccenxxxxD}{\ensuremath{4.55\pm0.21}}    
\newcommand{\hatcurSMEivsineccenxxxxD}{\ensuremath{5.86\pm0.49}}    
\newcommand{\hatcurSMEivmaceccenxxxxD}{\ensuremath{4.82\pm0.26}}    
\newcommand{\hatcurSMEivmiceccenxxxxD}{\ensuremath{1.53\pm0.19}}    
\newcommand{\hatcurSMEiiteffeccenxxxxD}{\ensuremath{6060\pm120}}    
\newcommand{\hatcurSMEiizfeheccenxxxxD}{\ensuremath{0.220\pm0.070}} 
\newcommand{\hatcurSMEiizfehshorteccenxxxxD}{\ensuremath{0.22}}     
\newcommand{\hatcurSMEiiloggeccenxxxxD}{\ensuremath{4.199\pm0.070}} 
\newcommand{\hatcurSMEiivsineccenxxxxD}{\ensuremath{6.04\pm0.36}}   
\newcommand{\hatcurSMEiivmaceccenxxxxD}{\ensuremath{4.42\pm0.19}}   
\newcommand{\hatcurSMEiivmiceccenxxxxD}{\ensuremath{1.28\pm0.10}}   
\newcommand{\hatcurLBizeccenxxxxD}{\ensuremath{0.1719}}             
\newcommand{\hatcurLBiizeccenxxxxD}{\ensuremath{0.3525}}            
\newcommand{\hatcurLBiieccenxxxxD}{\ensuremath{0.2292}}             
\newcommand{\hatcurLBiiieccenxxxxD}{\ensuremath{0.3581}}            
\newcommand{\hatcurLBiIeccenxxxxD}{\ensuremath{0.2090}}             
\newcommand{\hatcurLBiiIeccenxxxxD}{\ensuremath{0.3575}}            
\newcommand{\hatcurLBigeccenxxxxD}{\ensuremath{0.5019}}             
\newcommand{\hatcurLBiigeccenxxxxD}{\ensuremath{0.2762}}            
\newcommand{\hatcurLBireccenxxxxD}{\ensuremath{0.3138}}             
\newcommand{\hatcurLBiireccenxxxxD}{\ensuremath{0.3565}}            
\newcommand{\hatcurLBiReccenxxxxD}{\ensuremath{0.2901}}             
\newcommand{\hatcurLBiiReccenxxxxD}{\ensuremath{0.3582}}            
\newcommand{\hatcurLBikepeccenxxxxD}{\ensuremath{0.1000}}           
\newcommand{\hatcurLBiikepeccenxxxxD}{\ensuremath{0.1000}}          
\newcommand{\hatcurISOmeccenxxxxD}{\ensuremath{1.271\pm0.082}}      
\newcommand{\hatcurISOmshorteccenxxxxD}{\ensuremath{1.27}}          
\newcommand{\hatcurISOmlongeccenxxxxD}{\ensuremath{1.271\pm0.082}}  
\newcommand{\hatcurISOreccenxxxxD}{\ensuremath{1.48\pm0.22}}        
\newcommand{\hatcurISOrshorteccenxxxxD}{\ensuremath{1.48}}          
\newcommand{\hatcurISOrlongeccenxxxxD}{\ensuremath{1.48\pm0.22}}    
\newcommand{\hatcurISOrhoeccenxxxxD}{\ensuremath{0.54_{-0.17}^{+0.24}}} 
\newcommand{\hatcurISOrholongeccenxxxxD}{\ensuremath{0.54_{-0.17}^{+0.24}}} 
\newcommand{\hatcurISOloggeccenxxxxD}{\ensuremath{4.20\pm0.10}}     
\newcommand{\hatcurISOlumeccenxxxxD}{\ensuremath{2.66_{-0.67}^{+0.96}}} 
\newcommand{\hatcurISOlumshorteccenxxxxD}{\ensuremath{2.66}}        
\newcommand{\hatcurISOmveccenxxxxD}{\ensuremath{3.72\pm0.34}}       
\newcommand{\hatcurISOvieccenxxxxD}{\ensuremath{0.617\pm0.036}}     
\newcommand{\hatcurISOageeccenxxxxD}{\ensuremath{3.16\pm0.94}}      
\newcommand{\hatcurISOsigmaeccenxxxxD}{\ensuremath{0.00120\pm0.00066}} 
\newcommand{\hatcurISOMJeccenxxxxD}{\ensuremath{2.70\pm0.32}}       
\newcommand{\hatcurISOMHeccenxxxxD}{\ensuremath{2.41\pm0.31}}       
\newcommand{\hatcurISOMKeccenxxxxD}{\ensuremath{2.36\pm0.31}}       
\newcommand{\hatcurISOJKeccenxxxxD}{\ensuremath{0.340\pm0.020}}     
\newcommand{\hatcurISOspececcenxxxxD}{F}                            
\newcommand{\hatcurRVKeccenxxxxD}{\ensuremath{243\pm20}}            
\newcommand{\hatcurRVrkeccenxxxxD}{\ensuremath{-0.08\pm0.10}}       
\newcommand{\hatcurRVrheccenxxxxD}{\ensuremath{-0.03\pm0.24}}       
\newcommand{\hatcurRVkeccenxxxxD}{\ensuremath{-0.016_{-0.026}^{+0.019}}} 
\newcommand{\hatcurRVheccenxxxxD}{\ensuremath{-0.004_{-0.089}^{+0.048}}} 
\newcommand{\hatcurRVtroneeccenxxxxD}{\ensuremath{0\pm0}}           
\newcommand{\hatcurRVtrtwoeccenxxxxD}{\ensuremath{0\pm0}}           
\newcommand{\hatcurRVgammaAeccenxxxxD}{\ensuremath{8129\pm64}}      
\newcommand{\hatcurRVjitterAeccenxxxxD}{\ensuremath{102\pm61}}      
\newcommand{\hatcurRVjittertwosiglimAeccenxxxxD}{\ensuremath{<218.6}} 
\newcommand{\hatcurRVfitrmsAeccenxxxxD}{\ensuremath{0.0}}           
\newcommand{\hatcurRVgammaBeccenxxxxD}{\ensuremath{8158\pm19}}      
\newcommand{\hatcurRVjitterBeccenxxxxD}{\ensuremath{0\pm17}}        
\newcommand{\hatcurRVjittertwosiglimBeccenxxxxD}{\ensuremath{<42.3}} 
\newcommand{\hatcurRVfitrmsBeccenxxxxD}{\ensuremath{0.0}}           
\newcommand{\hatcurRVecceneccenxxxxD}{\ensuremath{0.053\pm0.074}}   
\newcommand{\hatcurRVeccentwosiglimeccenxxxxD}{\ensuremath{<0.229}} 
\newcommand{\hatcurRVomegaeccenxxxxD}{\ensuremath{194\pm73}}        
\newcommand{\hatcurPPieccenxxxxD}{\ensuremath{84.9\pm2.3}}          
\newcommand{\hatcurPPgeccenxxxxD}{\ensuremath{22.6_{-5.6}^{+7.4}}}  
\newcommand{\hatcurPPloggeccenxxxxD}{\ensuremath{3.35\pm0.13}}      
\newcommand{\hatcurPPareccenxxxxD}{\ensuremath{5.33\pm0.66}}        
\newcommand{\hatcurPPareleccenxxxxD}{\ensuremath{0.03687\pm0.00078}} 
\newcommand{\hatcurPPrhoeccenxxxxD}{\ensuremath{0.80_{-0.29}^{+0.44}}} 
\newcommand{\hatcurPPmeccenxxxxD}{\ensuremath{1.85\pm0.17}}         
\newcommand{\hatcurPPmshorteccenxxxxD}{\ensuremath{1.85}}           
\newcommand{\hatcurPPmlongeccenxxxxD}{\ensuremath{1.85\pm0.17}}     
\newcommand{\hatcurPPmeeccenxxxxD}{\ensuremath{588\pm54}}           
\newcommand{\hatcurPPmeshorteccenxxxxD}{\ensuremath{588.4}}         
\newcommand{\hatcurPPmelongeccenxxxxD}{\ensuremath{588\pm54}}       
\newcommand{\hatcurPPreccenxxxxD}{\ensuremath{1.42_{-0.19}^{+0.26}}} 
\newcommand{\hatcurPPrshorteccenxxxxD}{\ensuremath{1.42}}           
\newcommand{\hatcurPPrlongeccenxxxxD}{\ensuremath{1.42_{-0.19}^{+0.26}}} 
\newcommand{\hatcurPPreeccenxxxxD}{\ensuremath{15.9_{-2.2}^{+2.9}}} 
\newcommand{\hatcurPPreshorteccenxxxxD}{\ensuremath{15.9}}          
\newcommand{\hatcurPPrelongeccenxxxxD}{\ensuremath{15.9_{-2.2}^{+2.9}}} 
\newcommand{\hatcurPPmrcorreccenxxxxD}{\ensuremath{0.48}}           
\newcommand{\hatcurPPteffeccenxxxxD}{\ensuremath{1860\pm120}}       
\newcommand{\hatcurPPthetaeccenxxxxD}{\ensuremath{0.075\pm0.013}}   
\newcommand{\hatcurPPfluxperieccenxxxxD}{\ensuremath{3.04_{-0.61}^{+0.98}}} 
\newcommand{\hatcurPPfluxperidimeccenxxxxD}{\ensuremath{9}}         
\newcommand{\hatcurPPfluxapeccenxxxxD}{\ensuremath{2.42\pm0.71}}    
\newcommand{\hatcurPPfluxapdimeccenxxxxD}{\ensuremath{9}}           
\newcommand{\hatcurPPfluxavgeccenxxxxD}{\ensuremath{2.68_{-0.59}^{+0.82}}} 
\newcommand{\hatcurPPfluxavgdimeccenxxxxD}{\ensuremath{9}}          
\newcommand{\hatcurPPfluxavglogeccenxxxxD}{\ensuremath{9.43\pm0.11}} 
\newcommand{\hatcurXsecphaseeccenxxxxD}{\ensuremath{0.490\pm0.020}} 
\newcommand{\hatcurXsecondaryeccenxxxxD}{\ensuremath{2456657.417\pm0.047}} 
\newcommand{\hatcurXsecdureccenxxxxD}{\ensuremath{0.136\pm0.018}}   
\newcommand{\hatcurXsecingdureccenxxxxD}{\ensuremath{0.0149\pm0.0084}} 
\newcommand{\hatcurPPphiconjeccenxxxxD}{\ensuremath{-0.20\pm0.27}}  
\newcommand{\hatcurPPperieccenxxxxD}{\ensuremath{2456656.74\pm0.62}} 
\newcommand{\hatcurPPaequiveccenxxxxD}{\ensuremath{0.0226\pm0.0030}} 
\newcommand{\hatcurPPtcirceccenxxxxD}{\ensuremath{40_{-21}^{+35}}}  
\newcommand{\hatcurPPtinfalleccenxxxxD}{\ensuremath{65_{-31}^{+52}}} 
\newcommand{\hatcurXdisteccenxxxxD}{\ensuremath{970\pm140}}         
\newcommand{\hatcurXAveccenxxxxD}{\ensuremath{0.000\pm0.064}}       
\newcommand{\hatcurXdistredeccenxxxxD}{\ensuremath{940\pm140}}      
\newcommand{\hatcurXEBVeccenxxxxD}{\ensuremath{0.000\pm0.021}}      
\newcommand{\hatcurXmvisoredeccenxxxxD}{\ensuremath{13.624_{-0.012}^{+0.016}}} 
\newcommand{\hatcurXmiisoredeccenxxxxD}{\ensuremath{12.992\pm0.015}} 
\newcommand{\hatcurXmjisoredeccenxxxxD}{\ensuremath{12.589_{-0.028}^{+0.018}}} 
\newcommand{\hatcurXmhisoredeccenxxxxD}{\ensuremath{12.300_{-0.048}^{+0.027}}} 
\newcommand{\hatcurXmkisoredeccenxxxxD}{\ensuremath{12.247_{-0.047}^{+0.027}}} 
\newcommand{\hatcurXviisoredeccenxxxxD}{\ensuremath{0.631_{-0.016}^{+0.025}}} 
\newcommand{\hatcurXvkisoredeccenxxxxD}{\ensuremath{1.375_{-0.032}^{+0.062}}} 
\newcommand{\hatcurXjhisoredeccenxxxxD}{\ensuremath{0.292\pm0.016}} 
\newcommand{\hatcurXjkisoredeccenxxxxD}{\ensuremath{0.343_{-0.012}^{+0.019}}} 
\newcommand{\hatcurCCpmraeccenxxxxD}{\ensuremath{0.9\pm1.3}}        
\newcommand{\hatcurCCpmdececcenxxxxD}{\ensuremath{1.7\pm1.3}}       
\newcommand{\hatcurCCpmeccenxxxxD}{\ensuremath{1.9\pm1.8}}          
 \newcommand{\hatcurCCbbHmageccen}[1]{\ifnum#1=39 
\hatcurCCbbHmageccenxxxxA
\else
\ifnum#1=40 
\hatcurCCbbHmageccenxxxxB
\else
\ifnum#1=41 
\hatcurCCbbHmageccenxxxxC
\else
\ifnum#1=42 
\hatcurCCbbHmageccenxxxxD
\else
??????\fi
\fi
\fi
\fi
}
\newcommand{\hatcurCCbbJmageccen}[1]{\ifnum#1=39 
\hatcurCCbbJmageccenxxxxA
\else
\ifnum#1=40 
\hatcurCCbbJmageccenxxxxB
\else
\ifnum#1=41 
\hatcurCCbbJmageccenxxxxC
\else
\ifnum#1=42 
\hatcurCCbbJmageccenxxxxD
\else
??????\fi
\fi
\fi
\fi
}
\newcommand{\hatcurCCbbKmageccen}[1]{\ifnum#1=39 
\hatcurCCbbKmageccenxxxxA
\else
\ifnum#1=40 
\hatcurCCbbKmageccenxxxxB
\else
\ifnum#1=41 
\hatcurCCbbKmageccenxxxxC
\else
\ifnum#1=42 
\hatcurCCbbKmageccenxxxxD
\else
??????\fi
\fi
\fi
\fi
}
\newcommand{\hatcurCCcitHmageccen}[1]{\ifnum#1=39 
\hatcurCCcitHmageccenxxxxA
\else
\ifnum#1=40 
\hatcurCCcitHmageccenxxxxB
\else
\ifnum#1=41 
\hatcurCCcitHmageccenxxxxC
\else
\ifnum#1=42 
\hatcurCCcitHmageccenxxxxD
\else
??????\fi
\fi
\fi
\fi
}
\newcommand{\hatcurCCcitJmageccen}[1]{\ifnum#1=39 
\hatcurCCcitJmageccenxxxxA
\else
\ifnum#1=40 
\hatcurCCcitJmageccenxxxxB
\else
\ifnum#1=41 
\hatcurCCcitJmageccenxxxxC
\else
\ifnum#1=42 
\hatcurCCcitJmageccenxxxxD
\else
??????\fi
\fi
\fi
\fi
}
\newcommand{\hatcurCCcitKmageccen}[1]{\ifnum#1=39 
\hatcurCCcitKmageccenxxxxA
\else
\ifnum#1=40 
\hatcurCCcitKmageccenxxxxB
\else
\ifnum#1=41 
\hatcurCCcitKmageccenxxxxC
\else
\ifnum#1=42 
\hatcurCCcitKmageccenxxxxD
\else
??????\fi
\fi
\fi
\fi
}
\newcommand{\hatcurCCdececcen}[1]{\ifnum#1=39 
\hatcurCCdececcenxxxxA
\else
\ifnum#1=40 
\hatcurCCdececcenxxxxB
\else
\ifnum#1=41 
\hatcurCCdececcenxxxxC
\else
\ifnum#1=42 
\hatcurCCdececcenxxxxD
\else
??????\fi
\fi
\fi
\fi
}
\newcommand{\hatcurCCesoHKmageccen}[1]{\ifnum#1=39 
\hatcurCCesoHKmageccenxxxxA
\else
\ifnum#1=40 
\hatcurCCesoHKmageccenxxxxB
\else
\ifnum#1=41 
\hatcurCCesoHKmageccenxxxxC
\else
\ifnum#1=42 
\hatcurCCesoHKmageccenxxxxD
\else
??????\fi
\fi
\fi
\fi
}
\newcommand{\hatcurCCesoHmageccen}[1]{\ifnum#1=39 
\hatcurCCesoHmageccenxxxxA
\else
\ifnum#1=40 
\hatcurCCesoHmageccenxxxxB
\else
\ifnum#1=41 
\hatcurCCesoHmageccenxxxxC
\else
\ifnum#1=42 
\hatcurCCesoHmageccenxxxxD
\else
??????\fi
\fi
\fi
\fi
}
\newcommand{\hatcurCCesoJHmageccen}[1]{\ifnum#1=39 
\hatcurCCesoJHmageccenxxxxA
\else
\ifnum#1=40 
\hatcurCCesoJHmageccenxxxxB
\else
\ifnum#1=41 
\hatcurCCesoJHmageccenxxxxC
\else
\ifnum#1=42 
\hatcurCCesoJHmageccenxxxxD
\else
??????\fi
\fi
\fi
\fi
}
\newcommand{\hatcurCCesoJKmageccen}[1]{\ifnum#1=39 
\hatcurCCesoJKmageccenxxxxA
\else
\ifnum#1=40 
\hatcurCCesoJKmageccenxxxxB
\else
\ifnum#1=41 
\hatcurCCesoJKmageccenxxxxC
\else
\ifnum#1=42 
\hatcurCCesoJKmageccenxxxxD
\else
??????\fi
\fi
\fi
\fi
}
\newcommand{\hatcurCCesoJmageccen}[1]{\ifnum#1=39 
\hatcurCCesoJmageccenxxxxA
\else
\ifnum#1=40 
\hatcurCCesoJmageccenxxxxB
\else
\ifnum#1=41 
\hatcurCCesoJmageccenxxxxC
\else
\ifnum#1=42 
\hatcurCCesoJmageccenxxxxD
\else
??????\fi
\fi
\fi
\fi
}
\newcommand{\hatcurCCesoKmageccen}[1]{\ifnum#1=39 
\hatcurCCesoKmageccenxxxxA
\else
\ifnum#1=40 
\hatcurCCesoKmageccenxxxxB
\else
\ifnum#1=41 
\hatcurCCesoKmageccenxxxxC
\else
\ifnum#1=42 
\hatcurCCesoKmageccenxxxxD
\else
??????\fi
\fi
\fi
\fi
}
\newcommand{\hatcurCCgsceccen}[1]{\ifnum#1=39 
\hatcurCCgsceccenxxxxA
\else
\ifnum#1=40 
\hatcurCCgsceccenxxxxB
\else
\ifnum#1=41 
\hatcurCCgsceccenxxxxC
\else
\ifnum#1=42 
\hatcurCCgsceccenxxxxD
\else
??????\fi
\fi
\fi
\fi
}
\newcommand{\hatcurCCmageccen}[1]{\ifnum#1=39 
\hatcurCCmageccenxxxxA
\else
\ifnum#1=40 
\hatcurCCmageccenxxxxB
\else
\ifnum#1=41 
\hatcurCCmageccenxxxxC
\else
\ifnum#1=42 
\hatcurCCmageccenxxxxD
\else
??????\fi
\fi
\fi
\fi
}
\newcommand{\hatcurCCpmdececcen}[1]{\ifnum#1=39 
\hatcurCCpmdececcenxxxxA
\else
\ifnum#1=40 
\hatcurCCpmdececcenxxxxB
\else
\ifnum#1=41 
\hatcurCCpmdececcenxxxxC
\else
\ifnum#1=42 
\hatcurCCpmdececcenxxxxD
\else
??????\fi
\fi
\fi
\fi
}
\newcommand{\hatcurCCpmeccen}[1]{\ifnum#1=39 
\hatcurCCpmeccenxxxxA
\else
\ifnum#1=40 
\hatcurCCpmeccenxxxxB
\else
\ifnum#1=41 
\hatcurCCpmeccenxxxxC
\else
\ifnum#1=42 
\hatcurCCpmeccenxxxxD
\else
??????\fi
\fi
\fi
\fi
}
\newcommand{\hatcurCCpmraeccen}[1]{\ifnum#1=39 
\hatcurCCpmraeccenxxxxA
\else
\ifnum#1=40 
\hatcurCCpmraeccenxxxxB
\else
\ifnum#1=41 
\hatcurCCpmraeccenxxxxC
\else
\ifnum#1=42 
\hatcurCCpmraeccenxxxxD
\else
??????\fi
\fi
\fi
\fi
}
\newcommand{\hatcurCCraeccen}[1]{\ifnum#1=39 
\hatcurCCraeccenxxxxA
\else
\ifnum#1=40 
\hatcurCCraeccenxxxxB
\else
\ifnum#1=41 
\hatcurCCraeccenxxxxC
\else
\ifnum#1=42 
\hatcurCCraeccenxxxxD
\else
??????\fi
\fi
\fi
\fi
}
\newcommand{\hatcurCCtassmBeccen}[1]{\ifnum#1=39 
\hatcurCCtassmBeccenxxxxA
\else
\ifnum#1=40 
\hatcurCCtassmBeccenxxxxB
\else
\ifnum#1=41 
\hatcurCCtassmBeccenxxxxC
\else
\ifnum#1=42 
\hatcurCCtassmBeccenxxxxD
\else
??????\fi
\fi
\fi
\fi
}
\newcommand{\hatcurCCtassmBshorteccen}[1]{\ifnum#1=39 
\hatcurCCtassmBshorteccenxxxxA
\else
\ifnum#1=40 
\hatcurCCtassmBshorteccenxxxxB
\else
\ifnum#1=41 
\hatcurCCtassmBshorteccenxxxxC
\else
\ifnum#1=42 
\hatcurCCtassmBshorteccenxxxxD
\else
??????\fi
\fi
\fi
\fi
}
\newcommand{\hatcurCCtassmgeccen}[1]{\ifnum#1=39 
\hatcurCCtassmgeccenxxxxA
\else
\ifnum#1=40 
\hatcurCCtassmgeccenxxxxB
\else
\ifnum#1=41 
\hatcurCCtassmgeccenxxxxC
\else
\ifnum#1=42 
\hatcurCCtassmgeccenxxxxD
\else
??????\fi
\fi
\fi
\fi
}
\newcommand{\hatcurCCtassmgshorteccen}[1]{\ifnum#1=39 
\hatcurCCtassmgshorteccenxxxxA
\else
\ifnum#1=40 
\hatcurCCtassmgshorteccenxxxxB
\else
\ifnum#1=41 
\hatcurCCtassmgshorteccenxxxxC
\else
\ifnum#1=42 
\hatcurCCtassmgshorteccenxxxxD
\else
??????\fi
\fi
\fi
\fi
}
\newcommand{\hatcurCCtassmieccen}[1]{\ifnum#1=39 
\hatcurCCtassmieccenxxxxA
\else
\ifnum#1=40 
\hatcurCCtassmieccenxxxxB
\else
\ifnum#1=41 
\hatcurCCtassmieccenxxxxC
\else
\ifnum#1=42 
\hatcurCCtassmieccenxxxxD
\else
??????\fi
\fi
\fi
\fi
}
\newcommand{\hatcurCCtassmIeccen}[1]{\ifnum#1=39 
\hatcurCCtassmIeccenxxxxA
\else
\ifnum#1=40 
\hatcurCCtassmIeccenxxxxB
\else
\ifnum#1=41 
\hatcurCCtassmIeccenxxxxC
\else
\ifnum#1=42 
\hatcurCCtassmIeccenxxxxD
\else
??????\fi
\fi
\fi
\fi
}
\newcommand{\hatcurCCtassmishorteccen}[1]{\ifnum#1=39 
\hatcurCCtassmishorteccenxxxxA
\else
\ifnum#1=40 
\hatcurCCtassmishorteccenxxxxB
\else
\ifnum#1=41 
\hatcurCCtassmishorteccenxxxxC
\else
\ifnum#1=42 
\hatcurCCtassmishorteccenxxxxD
\else
??????\fi
\fi
\fi
\fi
}
\newcommand{\hatcurCCtassmIshorteccen}[1]{\ifnum#1=39 
\hatcurCCtassmIshorteccenxxxxA
\else
\ifnum#1=40 
\hatcurCCtassmIshorteccenxxxxB
\else
\ifnum#1=41 
\hatcurCCtassmIshorteccenxxxxC
\else
\ifnum#1=42 
\hatcurCCtassmIshorteccenxxxxD
\else
??????\fi
\fi
\fi
\fi
}
\newcommand{\hatcurCCtassmreccen}[1]{\ifnum#1=39 
\hatcurCCtassmreccenxxxxA
\else
\ifnum#1=40 
\hatcurCCtassmreccenxxxxB
\else
\ifnum#1=41 
\hatcurCCtassmreccenxxxxC
\else
\ifnum#1=42 
\hatcurCCtassmreccenxxxxD
\else
??????\fi
\fi
\fi
\fi
}
\newcommand{\hatcurCCtassmrshorteccen}[1]{\ifnum#1=39 
\hatcurCCtassmrshorteccenxxxxA
\else
\ifnum#1=40 
\hatcurCCtassmrshorteccenxxxxB
\else
\ifnum#1=41 
\hatcurCCtassmrshorteccenxxxxC
\else
\ifnum#1=42 
\hatcurCCtassmrshorteccenxxxxD
\else
??????\fi
\fi
\fi
\fi
}
\newcommand{\hatcurCCtassmveccen}[1]{\ifnum#1=39 
\hatcurCCtassmveccenxxxxA
\else
\ifnum#1=40 
\hatcurCCtassmveccenxxxxB
\else
\ifnum#1=41 
\hatcurCCtassmveccenxxxxC
\else
\ifnum#1=42 
\hatcurCCtassmveccenxxxxD
\else
??????\fi
\fi
\fi
\fi
}
\newcommand{\hatcurCCtassmvshorteccen}[1]{\ifnum#1=39 
\hatcurCCtassmvshorteccenxxxxA
\else
\ifnum#1=40 
\hatcurCCtassmvshorteccenxxxxB
\else
\ifnum#1=41 
\hatcurCCtassmvshorteccenxxxxC
\else
\ifnum#1=42 
\hatcurCCtassmvshorteccenxxxxD
\else
??????\fi
\fi
\fi
\fi
}
\newcommand{\hatcurCCtwomasseccen}[1]{\ifnum#1=39 
\hatcurCCtwomasseccenxxxxA
\else
\ifnum#1=40 
\hatcurCCtwomasseccenxxxxB
\else
\ifnum#1=41 
\hatcurCCtwomasseccenxxxxC
\else
\ifnum#1=42 
\hatcurCCtwomasseccenxxxxD
\else
??????\fi
\fi
\fi
\fi
}
\newcommand{\hatcurCCtwomassHmageccen}[1]{\ifnum#1=39 
\hatcurCCtwomassHmageccenxxxxA
\else
\ifnum#1=40 
\hatcurCCtwomassHmageccenxxxxB
\else
\ifnum#1=41 
\hatcurCCtwomassHmageccenxxxxC
\else
\ifnum#1=42 
\hatcurCCtwomassHmageccenxxxxD
\else
??????\fi
\fi
\fi
\fi
}
\newcommand{\hatcurCCtwomassJmageccen}[1]{\ifnum#1=39 
\hatcurCCtwomassJmageccenxxxxA
\else
\ifnum#1=40 
\hatcurCCtwomassJmageccenxxxxB
\else
\ifnum#1=41 
\hatcurCCtwomassJmageccenxxxxC
\else
\ifnum#1=42 
\hatcurCCtwomassJmageccenxxxxD
\else
??????\fi
\fi
\fi
\fi
}
\newcommand{\hatcurCCtwomassKmageccen}[1]{\ifnum#1=39 
\hatcurCCtwomassKmageccenxxxxA
\else
\ifnum#1=40 
\hatcurCCtwomassKmageccenxxxxB
\else
\ifnum#1=41 
\hatcurCCtwomassKmageccenxxxxC
\else
\ifnum#1=42 
\hatcurCCtwomassKmageccenxxxxD
\else
??????\fi
\fi
\fi
\fi
}
\newcommand{\hatcurfieldeccen}[1]{\ifnum#1=39 
\hatcurfieldeccenxxxxA
\else
\ifnum#1=40 
\hatcurfieldeccenxxxxB
\else
\ifnum#1=41 
\hatcurfieldeccenxxxxC
\else
\ifnum#1=42 
\hatcurfieldeccenxxxxD
\else
??????\fi
\fi
\fi
\fi
}
\newcommand{\hatcurhtreccen}[1]{\ifnum#1=39 
\hatcurhtreccenxxxxA
\else
\ifnum#1=40 
\hatcurhtreccenxxxxB
\else
\ifnum#1=41 
\hatcurhtreccenxxxxC
\else
\ifnum#1=42 
\hatcurhtreccenxxxxD
\else
??????\fi
\fi
\fi
\fi
}
\newcommand{\hatcurISOageeccen}[1]{\ifnum#1=39 
\hatcurISOageeccenxxxxA
\else
\ifnum#1=40 
\hatcurISOageeccenxxxxB
\else
\ifnum#1=41 
\hatcurISOageeccenxxxxC
\else
\ifnum#1=42 
\hatcurISOageeccenxxxxD
\else
??????\fi
\fi
\fi
\fi
}
\newcommand{\hatcurISOJKeccen}[1]{\ifnum#1=39 
\hatcurISOJKeccenxxxxA
\else
\ifnum#1=40 
\hatcurISOJKeccenxxxxB
\else
\ifnum#1=41 
\hatcurISOJKeccenxxxxC
\else
\ifnum#1=42 
\hatcurISOJKeccenxxxxD
\else
??????\fi
\fi
\fi
\fi
}
\newcommand{\hatcurISOloggeccen}[1]{\ifnum#1=39 
\hatcurISOloggeccenxxxxA
\else
\ifnum#1=40 
\hatcurISOloggeccenxxxxB
\else
\ifnum#1=41 
\hatcurISOloggeccenxxxxC
\else
\ifnum#1=42 
\hatcurISOloggeccenxxxxD
\else
??????\fi
\fi
\fi
\fi
}
\newcommand{\hatcurISOlumeccen}[1]{\ifnum#1=39 
\hatcurISOlumeccenxxxxA
\else
\ifnum#1=40 
\hatcurISOlumeccenxxxxB
\else
\ifnum#1=41 
\hatcurISOlumeccenxxxxC
\else
\ifnum#1=42 
\hatcurISOlumeccenxxxxD
\else
??????\fi
\fi
\fi
\fi
}
\newcommand{\hatcurISOlumshorteccen}[1]{\ifnum#1=39 
\hatcurISOlumshorteccenxxxxA
\else
\ifnum#1=40 
\hatcurISOlumshorteccenxxxxB
\else
\ifnum#1=41 
\hatcurISOlumshorteccenxxxxC
\else
\ifnum#1=42 
\hatcurISOlumshorteccenxxxxD
\else
??????\fi
\fi
\fi
\fi
}
\newcommand{\hatcurISOmeccen}[1]{\ifnum#1=39 
\hatcurISOmeccenxxxxA
\else
\ifnum#1=40 
\hatcurISOmeccenxxxxB
\else
\ifnum#1=41 
\hatcurISOmeccenxxxxC
\else
\ifnum#1=42 
\hatcurISOmeccenxxxxD
\else
??????\fi
\fi
\fi
\fi
}
\newcommand{\hatcurISOMHeccen}[1]{\ifnum#1=39 
\hatcurISOMHeccenxxxxA
\else
\ifnum#1=40 
\hatcurISOMHeccenxxxxB
\else
\ifnum#1=41 
\hatcurISOMHeccenxxxxC
\else
\ifnum#1=42 
\hatcurISOMHeccenxxxxD
\else
??????\fi
\fi
\fi
\fi
}
\newcommand{\hatcurISOMJeccen}[1]{\ifnum#1=39 
\hatcurISOMJeccenxxxxA
\else
\ifnum#1=40 
\hatcurISOMJeccenxxxxB
\else
\ifnum#1=41 
\hatcurISOMJeccenxxxxC
\else
\ifnum#1=42 
\hatcurISOMJeccenxxxxD
\else
??????\fi
\fi
\fi
\fi
}
\newcommand{\hatcurISOMKeccen}[1]{\ifnum#1=39 
\hatcurISOMKeccenxxxxA
\else
\ifnum#1=40 
\hatcurISOMKeccenxxxxB
\else
\ifnum#1=41 
\hatcurISOMKeccenxxxxC
\else
\ifnum#1=42 
\hatcurISOMKeccenxxxxD
\else
??????\fi
\fi
\fi
\fi
}
\newcommand{\hatcurISOmlongeccen}[1]{\ifnum#1=39 
\hatcurISOmlongeccenxxxxA
\else
\ifnum#1=40 
\hatcurISOmlongeccenxxxxB
\else
\ifnum#1=41 
\hatcurISOmlongeccenxxxxC
\else
\ifnum#1=42 
\hatcurISOmlongeccenxxxxD
\else
??????\fi
\fi
\fi
\fi
}
\newcommand{\hatcurISOmshorteccen}[1]{\ifnum#1=39 
\hatcurISOmshorteccenxxxxA
\else
\ifnum#1=40 
\hatcurISOmshorteccenxxxxB
\else
\ifnum#1=41 
\hatcurISOmshorteccenxxxxC
\else
\ifnum#1=42 
\hatcurISOmshorteccenxxxxD
\else
??????\fi
\fi
\fi
\fi
}
\newcommand{\hatcurISOmveccen}[1]{\ifnum#1=39 
\hatcurISOmveccenxxxxA
\else
\ifnum#1=40 
\hatcurISOmveccenxxxxB
\else
\ifnum#1=41 
\hatcurISOmveccenxxxxC
\else
\ifnum#1=42 
\hatcurISOmveccenxxxxD
\else
??????\fi
\fi
\fi
\fi
}
\newcommand{\hatcurISOreccen}[1]{\ifnum#1=39 
\hatcurISOreccenxxxxA
\else
\ifnum#1=40 
\hatcurISOreccenxxxxB
\else
\ifnum#1=41 
\hatcurISOreccenxxxxC
\else
\ifnum#1=42 
\hatcurISOreccenxxxxD
\else
??????\fi
\fi
\fi
\fi
}
\newcommand{\hatcurISOrhoeccen}[1]{\ifnum#1=39 
\hatcurISOrhoeccenxxxxA
\else
\ifnum#1=40 
\hatcurISOrhoeccenxxxxB
\else
\ifnum#1=41 
\hatcurISOrhoeccenxxxxC
\else
\ifnum#1=42 
\hatcurISOrhoeccenxxxxD
\else
??????\fi
\fi
\fi
\fi
}
\newcommand{\hatcurISOrholongeccen}[1]{\ifnum#1=39 
\hatcurISOrholongeccenxxxxA
\else
\ifnum#1=40 
\hatcurISOrholongeccenxxxxB
\else
\ifnum#1=41 
\hatcurISOrholongeccenxxxxC
\else
\ifnum#1=42 
\hatcurISOrholongeccenxxxxD
\else
??????\fi
\fi
\fi
\fi
}
\newcommand{\hatcurISOrlongeccen}[1]{\ifnum#1=39 
\hatcurISOrlongeccenxxxxA
\else
\ifnum#1=40 
\hatcurISOrlongeccenxxxxB
\else
\ifnum#1=41 
\hatcurISOrlongeccenxxxxC
\else
\ifnum#1=42 
\hatcurISOrlongeccenxxxxD
\else
??????\fi
\fi
\fi
\fi
}
\newcommand{\hatcurISOrshorteccen}[1]{\ifnum#1=39 
\hatcurISOrshorteccenxxxxA
\else
\ifnum#1=40 
\hatcurISOrshorteccenxxxxB
\else
\ifnum#1=41 
\hatcurISOrshorteccenxxxxC
\else
\ifnum#1=42 
\hatcurISOrshorteccenxxxxD
\else
??????\fi
\fi
\fi
\fi
}
\newcommand{\hatcurISOsigmaeccen}[1]{\ifnum#1=39 
\hatcurISOsigmaeccenxxxxA
\else
\ifnum#1=40 
\hatcurISOsigmaeccenxxxxB
\else
\ifnum#1=41 
\hatcurISOsigmaeccenxxxxC
\else
\ifnum#1=42 
\hatcurISOsigmaeccenxxxxD
\else
??????\fi
\fi
\fi
\fi
}
\newcommand{\hatcurISOspececcen}[1]{\ifnum#1=39 
\hatcurISOspececcenxxxxA
\else
\ifnum#1=40 
\hatcurISOspececcenxxxxB
\else
\ifnum#1=41 
\hatcurISOspececcenxxxxC
\else
\ifnum#1=42 
\hatcurISOspececcenxxxxD
\else
??????\fi
\fi
\fi
\fi
}
\newcommand{\hatcurISOvieccen}[1]{\ifnum#1=39 
\hatcurISOvieccenxxxxA
\else
\ifnum#1=40 
\hatcurISOvieccenxxxxB
\else
\ifnum#1=41 
\hatcurISOvieccenxxxxC
\else
\ifnum#1=42 
\hatcurISOvieccenxxxxD
\else
??????\fi
\fi
\fi
\fi
}
\newcommand{\hatcurLBigeccen}[1]{\ifnum#1=39 
\hatcurLBigeccenxxxxA
\else
\ifnum#1=40 
\hatcurLBigeccenxxxxB
\else
\ifnum#1=41 
\hatcurLBigeccenxxxxC
\else
\ifnum#1=42 
\hatcurLBigeccenxxxxD
\else
??????\fi
\fi
\fi
\fi
}
\newcommand{\hatcurLBiieccen}[1]{\ifnum#1=39 
\hatcurLBiieccenxxxxA
\else
\ifnum#1=40 
\hatcurLBiieccenxxxxB
\else
\ifnum#1=41 
\hatcurLBiieccenxxxxC
\else
\ifnum#1=42 
\hatcurLBiieccenxxxxD
\else
??????\fi
\fi
\fi
\fi
}
\newcommand{\hatcurLBiIeccen}[1]{\ifnum#1=39 
\hatcurLBiIeccenxxxxA
\else
\ifnum#1=40 
\hatcurLBiIeccenxxxxB
\else
\ifnum#1=41 
\hatcurLBiIeccenxxxxC
\else
\ifnum#1=42 
\hatcurLBiIeccenxxxxD
\else
??????\fi
\fi
\fi
\fi
}
\newcommand{\hatcurLBiigeccen}[1]{\ifnum#1=39 
\hatcurLBiigeccenxxxxA
\else
\ifnum#1=40 
\hatcurLBiigeccenxxxxB
\else
\ifnum#1=41 
\hatcurLBiigeccenxxxxC
\else
\ifnum#1=42 
\hatcurLBiigeccenxxxxD
\else
??????\fi
\fi
\fi
\fi
}
\newcommand{\hatcurLBiiieccen}[1]{\ifnum#1=39 
\hatcurLBiiieccenxxxxA
\else
\ifnum#1=40 
\hatcurLBiiieccenxxxxB
\else
\ifnum#1=41 
\hatcurLBiiieccenxxxxC
\else
\ifnum#1=42 
\hatcurLBiiieccenxxxxD
\else
??????\fi
\fi
\fi
\fi
}
\newcommand{\hatcurLBiiIeccen}[1]{\ifnum#1=39 
\hatcurLBiiIeccenxxxxA
\else
\ifnum#1=40 
\hatcurLBiiIeccenxxxxB
\else
\ifnum#1=41 
\hatcurLBiiIeccenxxxxC
\else
\ifnum#1=42 
\hatcurLBiiIeccenxxxxD
\else
??????\fi
\fi
\fi
\fi
}
\newcommand{\hatcurLBiikepeccen}[1]{\ifnum#1=39 
\hatcurLBiikepeccenxxxxA
\else
\ifnum#1=40 
\hatcurLBiikepeccenxxxxB
\else
\ifnum#1=41 
\hatcurLBiikepeccenxxxxC
\else
\ifnum#1=42 
\hatcurLBiikepeccenxxxxD
\else
??????\fi
\fi
\fi
\fi
}
\newcommand{\hatcurLBiireccen}[1]{\ifnum#1=39 
\hatcurLBiireccenxxxxA
\else
\ifnum#1=40 
\hatcurLBiireccenxxxxB
\else
\ifnum#1=41 
\hatcurLBiireccenxxxxC
\else
\ifnum#1=42 
\hatcurLBiireccenxxxxD
\else
??????\fi
\fi
\fi
\fi
}
\newcommand{\hatcurLBiiReccen}[1]{\ifnum#1=39 
\hatcurLBiiReccenxxxxA
\else
\ifnum#1=40 
\hatcurLBiiReccenxxxxB
\else
\ifnum#1=41 
\hatcurLBiiReccenxxxxC
\else
\ifnum#1=42 
\hatcurLBiiReccenxxxxD
\else
??????\fi
\fi
\fi
\fi
}
\newcommand{\hatcurLBiizeccen}[1]{\ifnum#1=39 
\hatcurLBiizeccenxxxxA
\else
\ifnum#1=40 
\hatcurLBiizeccenxxxxB
\else
\ifnum#1=41 
\hatcurLBiizeccenxxxxC
\else
\ifnum#1=42 
\hatcurLBiizeccenxxxxD
\else
??????\fi
\fi
\fi
\fi
}
\newcommand{\hatcurLBikepeccen}[1]{\ifnum#1=39 
\hatcurLBikepeccenxxxxA
\else
\ifnum#1=40 
\hatcurLBikepeccenxxxxB
\else
\ifnum#1=41 
\hatcurLBikepeccenxxxxC
\else
\ifnum#1=42 
\hatcurLBikepeccenxxxxD
\else
??????\fi
\fi
\fi
\fi
}
\newcommand{\hatcurLBireccen}[1]{\ifnum#1=39 
\hatcurLBireccenxxxxA
\else
\ifnum#1=40 
\hatcurLBireccenxxxxB
\else
\ifnum#1=41 
\hatcurLBireccenxxxxC
\else
\ifnum#1=42 
\hatcurLBireccenxxxxD
\else
??????\fi
\fi
\fi
\fi
}
\newcommand{\hatcurLBiReccen}[1]{\ifnum#1=39 
\hatcurLBiReccenxxxxA
\else
\ifnum#1=40 
\hatcurLBiReccenxxxxB
\else
\ifnum#1=41 
\hatcurLBiReccenxxxxC
\else
\ifnum#1=42 
\hatcurLBiReccenxxxxD
\else
??????\fi
\fi
\fi
\fi
}
\newcommand{\hatcurLBizeccen}[1]{\ifnum#1=39 
\hatcurLBizeccenxxxxA
\else
\ifnum#1=40 
\hatcurLBizeccenxxxxB
\else
\ifnum#1=41 
\hatcurLBizeccenxxxxC
\else
\ifnum#1=42 
\hatcurLBizeccenxxxxD
\else
??????\fi
\fi
\fi
\fi
}
\newcommand{\hatcurLCbsqeccen}[1]{\ifnum#1=39 
\hatcurLCbsqeccenxxxxA
\else
\ifnum#1=40 
\hatcurLCbsqeccenxxxxB
\else
\ifnum#1=41 
\hatcurLCbsqeccenxxxxC
\else
\ifnum#1=42 
\hatcurLCbsqeccenxxxxD
\else
??????\fi
\fi
\fi
\fi
}
\newcommand{\hatcurLCdipeccen}[1]{\ifnum#1=39 
\hatcurLCdipeccenxxxxA
\else
\ifnum#1=40 
\hatcurLCdipeccenxxxxB
\else
\ifnum#1=41 
\hatcurLCdipeccenxxxxC
\else
\ifnum#1=42 
\hatcurLCdipeccenxxxxD
\else
??????\fi
\fi
\fi
\fi
}
\newcommand{\hatcurLCdureccen}[1]{\ifnum#1=39 
\hatcurLCdureccenxxxxA
\else
\ifnum#1=40 
\hatcurLCdureccenxxxxB
\else
\ifnum#1=41 
\hatcurLCdureccenxxxxC
\else
\ifnum#1=42 
\hatcurLCdureccenxxxxD
\else
??????\fi
\fi
\fi
\fi
}
\newcommand{\hatcurLCdurhreccen}[1]{\ifnum#1=39 
\hatcurLCdurhreccenxxxxA
\else
\ifnum#1=40 
\hatcurLCdurhreccenxxxxB
\else
\ifnum#1=41 
\hatcurLCdurhreccenxxxxC
\else
\ifnum#1=42 
\hatcurLCdurhreccenxxxxD
\else
??????\fi
\fi
\fi
\fi
}
\newcommand{\hatcurLCdurhrshorteccen}[1]{\ifnum#1=39 
\hatcurLCdurhrshorteccenxxxxA
\else
\ifnum#1=40 
\hatcurLCdurhrshorteccenxxxxB
\else
\ifnum#1=41 
\hatcurLCdurhrshorteccenxxxxC
\else
\ifnum#1=42 
\hatcurLCdurhrshorteccenxxxxD
\else
??????\fi
\fi
\fi
\fi
}
\newcommand{\hatcurLCdurshorteccen}[1]{\ifnum#1=39 
\hatcurLCdurshorteccenxxxxA
\else
\ifnum#1=40 
\hatcurLCdurshorteccenxxxxB
\else
\ifnum#1=41 
\hatcurLCdurshorteccenxxxxC
\else
\ifnum#1=42 
\hatcurLCdurshorteccenxxxxD
\else
??????\fi
\fi
\fi
\fi
}
\newcommand{\hatcurLChatnetmAeccen}[1]{\ifnum#1=39 
\hatcurLChatnetmAeccenxxxxA
\else
\ifnum#1=40 
\hatcurLChatnetmAeccenxxxxB
\else
\ifnum#1=42 
\hatcurLChatnetmAeccenxxxxD
\else
??????\fi
\fi
\fi
}
\newcommand{\hatcurLChatnetmBeccen}[1]{\ifnum#1=39 
\hatcurLChatnetmBeccenxxxxA
\else
\ifnum#1=40 
\hatcurLChatnetmBeccenxxxxB
\else
\ifnum#1=42 
\hatcurLChatnetmBeccenxxxxD
\else
??????\fi
\fi
\fi
}
\newcommand{\hatcurLChatnetmCeccen}[1]{\ifnum#1=40 
\hatcurLChatnetmCeccenxxxxB
\else
??????\fi
}
\newcommand{\hatcurLChatnetmeccen}[1]{\ifnum#1=41 
\hatcurLChatnetmeccenxxxxC
\else
??????\fi
}
\newcommand{\hatcurLCiblendAeccen}[1]{\ifnum#1=39 
\hatcurLCiblendAeccenxxxxA
\else
\ifnum#1=40 
\hatcurLCiblendAeccenxxxxB
\else
\ifnum#1=42 
\hatcurLCiblendAeccenxxxxD
\else
??????\fi
\fi
\fi
}
\newcommand{\hatcurLCiblendBeccen}[1]{\ifnum#1=39 
\hatcurLCiblendBeccenxxxxA
\else
\ifnum#1=40 
\hatcurLCiblendBeccenxxxxB
\else
\ifnum#1=42 
\hatcurLCiblendBeccenxxxxD
\else
??????\fi
\fi
\fi
}
\newcommand{\hatcurLCiblendCeccen}[1]{\ifnum#1=40 
\hatcurLCiblendCeccenxxxxB
\else
??????\fi
}
\newcommand{\hatcurLCiblendeccen}[1]{\ifnum#1=41 
\hatcurLCiblendeccenxxxxC
\else
??????\fi
}
\newcommand{\hatcurLCimpeccen}[1]{\ifnum#1=39 
\hatcurLCimpeccenxxxxA
\else
\ifnum#1=40 
\hatcurLCimpeccenxxxxB
\else
\ifnum#1=41 
\hatcurLCimpeccenxxxxC
\else
\ifnum#1=42 
\hatcurLCimpeccenxxxxD
\else
??????\fi
\fi
\fi
\fi
}
\newcommand{\hatcurLCingdureccen}[1]{\ifnum#1=39 
\hatcurLCingdureccenxxxxA
\else
\ifnum#1=40 
\hatcurLCingdureccenxxxxB
\else
\ifnum#1=41 
\hatcurLCingdureccenxxxxC
\else
\ifnum#1=42 
\hatcurLCingdureccenxxxxD
\else
??????\fi
\fi
\fi
\fi
}
\newcommand{\hatcurLCPeccen}[1]{\ifnum#1=39 
\hatcurLCPeccenxxxxA
\else
\ifnum#1=40 
\hatcurLCPeccenxxxxB
\else
\ifnum#1=41 
\hatcurLCPeccenxxxxC
\else
\ifnum#1=42 
\hatcurLCPeccenxxxxD
\else
??????\fi
\fi
\fi
\fi
}
\newcommand{\hatcurLCPprececcen}[1]{\ifnum#1=39 
\hatcurLCPprececcenxxxxA
\else
\ifnum#1=40 
\hatcurLCPprececcenxxxxB
\else
\ifnum#1=41 
\hatcurLCPprececcenxxxxC
\else
\ifnum#1=42 
\hatcurLCPprececcenxxxxD
\else
??????\fi
\fi
\fi
\fi
}
\newcommand{\hatcurLCPshorteccen}[1]{\ifnum#1=39 
\hatcurLCPshorteccenxxxxA
\else
\ifnum#1=40 
\hatcurLCPshorteccenxxxxB
\else
\ifnum#1=41 
\hatcurLCPshorteccenxxxxC
\else
\ifnum#1=42 
\hatcurLCPshorteccenxxxxD
\else
??????\fi
\fi
\fi
\fi
}
\newcommand{\hatcurLCqeccen}[1]{\ifnum#1=39 
\hatcurLCqeccenxxxxA
\else
\ifnum#1=40 
\hatcurLCqeccenxxxxB
\else
\ifnum#1=41 
\hatcurLCqeccenxxxxC
\else
\ifnum#1=42 
\hatcurLCqeccenxxxxD
\else
??????\fi
\fi
\fi
\fi
}
\newcommand{\hatcurLCqshorteccen}[1]{\ifnum#1=39 
\hatcurLCqshorteccenxxxxA
\else
\ifnum#1=40 
\hatcurLCqshorteccenxxxxB
\else
\ifnum#1=41 
\hatcurLCqshorteccenxxxxC
\else
\ifnum#1=42 
\hatcurLCqshorteccenxxxxD
\else
??????\fi
\fi
\fi
\fi
}
\newcommand{\hatcurLCrhoeccen}[1]{\ifnum#1=39 
\hatcurLCrhoeccenxxxxA
\else
\ifnum#1=40 
\hatcurLCrhoeccenxxxxB
\else
\ifnum#1=41 
\hatcurLCrhoeccenxxxxC
\else
\ifnum#1=42 
\hatcurLCrhoeccenxxxxD
\else
??????\fi
\fi
\fi
\fi
}
\newcommand{\hatcurLCrprstareccen}[1]{\ifnum#1=39 
\hatcurLCrprstareccenxxxxA
\else
\ifnum#1=40 
\hatcurLCrprstareccenxxxxB
\else
\ifnum#1=41 
\hatcurLCrprstareccenxxxxC
\else
\ifnum#1=42 
\hatcurLCrprstareccenxxxxD
\else
??????\fi
\fi
\fi
\fi
}
\newcommand{\hatcurLCTAeccen}[1]{\ifnum#1=39 
\hatcurLCTAeccenxxxxA
\else
\ifnum#1=40 
\hatcurLCTAeccenxxxxB
\else
\ifnum#1=41 
\hatcurLCTAeccenxxxxC
\else
\ifnum#1=42 
\hatcurLCTAeccenxxxxD
\else
??????\fi
\fi
\fi
\fi
}
\newcommand{\hatcurLCTBeccen}[1]{\ifnum#1=39 
\hatcurLCTBeccenxxxxA
\else
\ifnum#1=40 
\hatcurLCTBeccenxxxxB
\else
\ifnum#1=41 
\hatcurLCTBeccenxxxxC
\else
\ifnum#1=42 
\hatcurLCTBeccenxxxxD
\else
??????\fi
\fi
\fi
\fi
}
\newcommand{\hatcurLCTeccen}[1]{\ifnum#1=39 
\hatcurLCTeccenxxxxA
\else
\ifnum#1=40 
\hatcurLCTeccenxxxxB
\else
\ifnum#1=41 
\hatcurLCTeccenxxxxC
\else
\ifnum#1=42 
\hatcurLCTeccenxxxxD
\else
??????\fi
\fi
\fi
\fi
}
\newcommand{\hatcurLCzetaeccen}[1]{\ifnum#1=39 
\hatcurLCzetaeccenxxxxA
\else
\ifnum#1=40 
\hatcurLCzetaeccenxxxxB
\else
\ifnum#1=41 
\hatcurLCzetaeccenxxxxC
\else
\ifnum#1=42 
\hatcurLCzetaeccenxxxxD
\else
??????\fi
\fi
\fi
\fi
}
\newcommand{\hatcurPPaequiveccen}[1]{\ifnum#1=39 
\hatcurPPaequiveccenxxxxA
\else
\ifnum#1=40 
\hatcurPPaequiveccenxxxxB
\else
\ifnum#1=41 
\hatcurPPaequiveccenxxxxC
\else
\ifnum#1=42 
\hatcurPPaequiveccenxxxxD
\else
??????\fi
\fi
\fi
\fi
}
\newcommand{\hatcurPPareccen}[1]{\ifnum#1=39 
\hatcurPPareccenxxxxA
\else
\ifnum#1=40 
\hatcurPPareccenxxxxB
\else
\ifnum#1=41 
\hatcurPPareccenxxxxC
\else
\ifnum#1=42 
\hatcurPPareccenxxxxD
\else
??????\fi
\fi
\fi
\fi
}
\newcommand{\hatcurPPareleccen}[1]{\ifnum#1=39 
\hatcurPPareleccenxxxxA
\else
\ifnum#1=40 
\hatcurPPareleccenxxxxB
\else
\ifnum#1=41 
\hatcurPPareleccenxxxxC
\else
\ifnum#1=42 
\hatcurPPareleccenxxxxD
\else
??????\fi
\fi
\fi
\fi
}
\newcommand{\hatcurPPfluxapdimeccen}[1]{\ifnum#1=39 
\hatcurPPfluxapdimeccenxxxxA
\else
\ifnum#1=40 
\hatcurPPfluxapdimeccenxxxxB
\else
\ifnum#1=41 
\hatcurPPfluxapdimeccenxxxxC
\else
\ifnum#1=42 
\hatcurPPfluxapdimeccenxxxxD
\else
??????\fi
\fi
\fi
\fi
}
\newcommand{\hatcurPPfluxapeccen}[1]{\ifnum#1=39 
\hatcurPPfluxapeccenxxxxA
\else
\ifnum#1=40 
\hatcurPPfluxapeccenxxxxB
\else
\ifnum#1=41 
\hatcurPPfluxapeccenxxxxC
\else
\ifnum#1=42 
\hatcurPPfluxapeccenxxxxD
\else
??????\fi
\fi
\fi
\fi
}
\newcommand{\hatcurPPfluxavgdimeccen}[1]{\ifnum#1=39 
\hatcurPPfluxavgdimeccenxxxxA
\else
\ifnum#1=40 
\hatcurPPfluxavgdimeccenxxxxB
\else
\ifnum#1=41 
\hatcurPPfluxavgdimeccenxxxxC
\else
\ifnum#1=42 
\hatcurPPfluxavgdimeccenxxxxD
\else
??????\fi
\fi
\fi
\fi
}
\newcommand{\hatcurPPfluxavgeccen}[1]{\ifnum#1=39 
\hatcurPPfluxavgeccenxxxxA
\else
\ifnum#1=40 
\hatcurPPfluxavgeccenxxxxB
\else
\ifnum#1=41 
\hatcurPPfluxavgeccenxxxxC
\else
\ifnum#1=42 
\hatcurPPfluxavgeccenxxxxD
\else
??????\fi
\fi
\fi
\fi
}
\newcommand{\hatcurPPfluxavglogeccen}[1]{\ifnum#1=39 
\hatcurPPfluxavglogeccenxxxxA
\else
\ifnum#1=40 
\hatcurPPfluxavglogeccenxxxxB
\else
\ifnum#1=41 
\hatcurPPfluxavglogeccenxxxxC
\else
\ifnum#1=42 
\hatcurPPfluxavglogeccenxxxxD
\else
??????\fi
\fi
\fi
\fi
}
\newcommand{\hatcurPPfluxperidimeccen}[1]{\ifnum#1=39 
\hatcurPPfluxperidimeccenxxxxA
\else
\ifnum#1=40 
\hatcurPPfluxperidimeccenxxxxB
\else
\ifnum#1=41 
\hatcurPPfluxperidimeccenxxxxC
\else
\ifnum#1=42 
\hatcurPPfluxperidimeccenxxxxD
\else
??????\fi
\fi
\fi
\fi
}
\newcommand{\hatcurPPfluxperieccen}[1]{\ifnum#1=39 
\hatcurPPfluxperieccenxxxxA
\else
\ifnum#1=40 
\hatcurPPfluxperieccenxxxxB
\else
\ifnum#1=41 
\hatcurPPfluxperieccenxxxxC
\else
\ifnum#1=42 
\hatcurPPfluxperieccenxxxxD
\else
??????\fi
\fi
\fi
\fi
}
\newcommand{\hatcurPPgeccen}[1]{\ifnum#1=39 
\hatcurPPgeccenxxxxA
\else
\ifnum#1=40 
\hatcurPPgeccenxxxxB
\else
\ifnum#1=41 
\hatcurPPgeccenxxxxC
\else
\ifnum#1=42 
\hatcurPPgeccenxxxxD
\else
??????\fi
\fi
\fi
\fi
}
\newcommand{\hatcurPPieccen}[1]{\ifnum#1=39 
\hatcurPPieccenxxxxA
\else
\ifnum#1=40 
\hatcurPPieccenxxxxB
\else
\ifnum#1=41 
\hatcurPPieccenxxxxC
\else
\ifnum#1=42 
\hatcurPPieccenxxxxD
\else
??????\fi
\fi
\fi
\fi
}
\newcommand{\hatcurPPloggeccen}[1]{\ifnum#1=39 
\hatcurPPloggeccenxxxxA
\else
\ifnum#1=40 
\hatcurPPloggeccenxxxxB
\else
\ifnum#1=41 
\hatcurPPloggeccenxxxxC
\else
\ifnum#1=42 
\hatcurPPloggeccenxxxxD
\else
??????\fi
\fi
\fi
\fi
}
\newcommand{\hatcurPPmeccen}[1]{\ifnum#1=39 
\hatcurPPmeccenxxxxA
\else
\ifnum#1=40 
\hatcurPPmeccenxxxxB
\else
\ifnum#1=41 
\hatcurPPmeccenxxxxC
\else
\ifnum#1=42 
\hatcurPPmeccenxxxxD
\else
??????\fi
\fi
\fi
\fi
}
\newcommand{\hatcurPPmeeccen}[1]{\ifnum#1=39 
\hatcurPPmeeccenxxxxA
\else
\ifnum#1=40 
\hatcurPPmeeccenxxxxB
\else
\ifnum#1=41 
\hatcurPPmeeccenxxxxC
\else
\ifnum#1=42 
\hatcurPPmeeccenxxxxD
\else
??????\fi
\fi
\fi
\fi
}
\newcommand{\hatcurPPmelongeccen}[1]{\ifnum#1=39 
\hatcurPPmelongeccenxxxxA
\else
\ifnum#1=40 
\hatcurPPmelongeccenxxxxB
\else
\ifnum#1=41 
\hatcurPPmelongeccenxxxxC
\else
\ifnum#1=42 
\hatcurPPmelongeccenxxxxD
\else
??????\fi
\fi
\fi
\fi
}
\newcommand{\hatcurPPmeshorteccen}[1]{\ifnum#1=39 
\hatcurPPmeshorteccenxxxxA
\else
\ifnum#1=40 
\hatcurPPmeshorteccenxxxxB
\else
\ifnum#1=41 
\hatcurPPmeshorteccenxxxxC
\else
\ifnum#1=42 
\hatcurPPmeshorteccenxxxxD
\else
??????\fi
\fi
\fi
\fi
}
\newcommand{\hatcurPPmlongeccen}[1]{\ifnum#1=39 
\hatcurPPmlongeccenxxxxA
\else
\ifnum#1=40 
\hatcurPPmlongeccenxxxxB
\else
\ifnum#1=41 
\hatcurPPmlongeccenxxxxC
\else
\ifnum#1=42 
\hatcurPPmlongeccenxxxxD
\else
??????\fi
\fi
\fi
\fi
}
\newcommand{\hatcurPPmrcorreccen}[1]{\ifnum#1=39 
\hatcurPPmrcorreccenxxxxA
\else
\ifnum#1=40 
\hatcurPPmrcorreccenxxxxB
\else
\ifnum#1=41 
\hatcurPPmrcorreccenxxxxC
\else
\ifnum#1=42 
\hatcurPPmrcorreccenxxxxD
\else
??????\fi
\fi
\fi
\fi
}
\newcommand{\hatcurPPmshorteccen}[1]{\ifnum#1=39 
\hatcurPPmshorteccenxxxxA
\else
\ifnum#1=40 
\hatcurPPmshorteccenxxxxB
\else
\ifnum#1=41 
\hatcurPPmshorteccenxxxxC
\else
\ifnum#1=42 
\hatcurPPmshorteccenxxxxD
\else
??????\fi
\fi
\fi
\fi
}
\newcommand{\hatcurPPperieccen}[1]{\ifnum#1=39 
\hatcurPPperieccenxxxxA
\else
\ifnum#1=40 
\hatcurPPperieccenxxxxB
\else
\ifnum#1=41 
\hatcurPPperieccenxxxxC
\else
\ifnum#1=42 
\hatcurPPperieccenxxxxD
\else
??????\fi
\fi
\fi
\fi
}
\newcommand{\hatcurPPphiconjeccen}[1]{\ifnum#1=39 
\hatcurPPphiconjeccenxxxxA
\else
\ifnum#1=40 
\hatcurPPphiconjeccenxxxxB
\else
\ifnum#1=41 
\hatcurPPphiconjeccenxxxxC
\else
\ifnum#1=42 
\hatcurPPphiconjeccenxxxxD
\else
??????\fi
\fi
\fi
\fi
}
\newcommand{\hatcurPPreccen}[1]{\ifnum#1=39 
\hatcurPPreccenxxxxA
\else
\ifnum#1=40 
\hatcurPPreccenxxxxB
\else
\ifnum#1=41 
\hatcurPPreccenxxxxC
\else
\ifnum#1=42 
\hatcurPPreccenxxxxD
\else
??????\fi
\fi
\fi
\fi
}
\newcommand{\hatcurPPreeccen}[1]{\ifnum#1=39 
\hatcurPPreeccenxxxxA
\else
\ifnum#1=40 
\hatcurPPreeccenxxxxB
\else
\ifnum#1=41 
\hatcurPPreeccenxxxxC
\else
\ifnum#1=42 
\hatcurPPreeccenxxxxD
\else
??????\fi
\fi
\fi
\fi
}
\newcommand{\hatcurPPrelongeccen}[1]{\ifnum#1=39 
\hatcurPPrelongeccenxxxxA
\else
\ifnum#1=40 
\hatcurPPrelongeccenxxxxB
\else
\ifnum#1=41 
\hatcurPPrelongeccenxxxxC
\else
\ifnum#1=42 
\hatcurPPrelongeccenxxxxD
\else
??????\fi
\fi
\fi
\fi
}
\newcommand{\hatcurPPreshorteccen}[1]{\ifnum#1=39 
\hatcurPPreshorteccenxxxxA
\else
\ifnum#1=40 
\hatcurPPreshorteccenxxxxB
\else
\ifnum#1=41 
\hatcurPPreshorteccenxxxxC
\else
\ifnum#1=42 
\hatcurPPreshorteccenxxxxD
\else
??????\fi
\fi
\fi
\fi
}
\newcommand{\hatcurPPrhoeccen}[1]{\ifnum#1=39 
\hatcurPPrhoeccenxxxxA
\else
\ifnum#1=40 
\hatcurPPrhoeccenxxxxB
\else
\ifnum#1=41 
\hatcurPPrhoeccenxxxxC
\else
\ifnum#1=42 
\hatcurPPrhoeccenxxxxD
\else
??????\fi
\fi
\fi
\fi
}
\newcommand{\hatcurPPrlongeccen}[1]{\ifnum#1=39 
\hatcurPPrlongeccenxxxxA
\else
\ifnum#1=40 
\hatcurPPrlongeccenxxxxB
\else
\ifnum#1=41 
\hatcurPPrlongeccenxxxxC
\else
\ifnum#1=42 
\hatcurPPrlongeccenxxxxD
\else
??????\fi
\fi
\fi
\fi
}
\newcommand{\hatcurPPrshorteccen}[1]{\ifnum#1=39 
\hatcurPPrshorteccenxxxxA
\else
\ifnum#1=40 
\hatcurPPrshorteccenxxxxB
\else
\ifnum#1=41 
\hatcurPPrshorteccenxxxxC
\else
\ifnum#1=42 
\hatcurPPrshorteccenxxxxD
\else
??????\fi
\fi
\fi
\fi
}
\newcommand{\hatcurPPtcirceccen}[1]{\ifnum#1=39 
\hatcurPPtcirceccenxxxxA
\else
\ifnum#1=40 
\hatcurPPtcirceccenxxxxB
\else
\ifnum#1=41 
\hatcurPPtcirceccenxxxxC
\else
\ifnum#1=42 
\hatcurPPtcirceccenxxxxD
\else
??????\fi
\fi
\fi
\fi
}
\newcommand{\hatcurPPteffeccen}[1]{\ifnum#1=39 
\hatcurPPteffeccenxxxxA
\else
\ifnum#1=40 
\hatcurPPteffeccenxxxxB
\else
\ifnum#1=41 
\hatcurPPteffeccenxxxxC
\else
\ifnum#1=42 
\hatcurPPteffeccenxxxxD
\else
??????\fi
\fi
\fi
\fi
}
\newcommand{\hatcurPPthetaeccen}[1]{\ifnum#1=39 
\hatcurPPthetaeccenxxxxA
\else
\ifnum#1=40 
\hatcurPPthetaeccenxxxxB
\else
\ifnum#1=41 
\hatcurPPthetaeccenxxxxC
\else
\ifnum#1=42 
\hatcurPPthetaeccenxxxxD
\else
??????\fi
\fi
\fi
\fi
}
\newcommand{\hatcurPPtinfalleccen}[1]{\ifnum#1=39 
\hatcurPPtinfalleccenxxxxA
\else
\ifnum#1=40 
\hatcurPPtinfalleccenxxxxB
\else
\ifnum#1=41 
\hatcurPPtinfalleccenxxxxC
\else
\ifnum#1=42 
\hatcurPPtinfalleccenxxxxD
\else
??????\fi
\fi
\fi
\fi
}
\newcommand{\hatcurRVecceneccen}[1]{\ifnum#1=39 
\hatcurRVecceneccenxxxxA
\else
\ifnum#1=40 
\hatcurRVecceneccenxxxxB
\else
\ifnum#1=41 
\hatcurRVecceneccenxxxxC
\else
\ifnum#1=42 
\hatcurRVecceneccenxxxxD
\else
??????\fi
\fi
\fi
\fi
}
\newcommand{\hatcurRVeccentwosiglimeccen}[1]{\ifnum#1=39 
\hatcurRVeccentwosiglimeccenxxxxA
\else
\ifnum#1=40 
\hatcurRVeccentwosiglimeccenxxxxB
\else
\ifnum#1=41 
\hatcurRVeccentwosiglimeccenxxxxC
\else
\ifnum#1=42 
\hatcurRVeccentwosiglimeccenxxxxD
\else
??????\fi
\fi
\fi
\fi
}
\newcommand{\hatcurRVfitrmsAeccen}[1]{\ifnum#1=39 
\hatcurRVfitrmsAeccenxxxxA
\else
\ifnum#1=41 
\hatcurRVfitrmsAeccenxxxxC
\else
\ifnum#1=42 
\hatcurRVfitrmsAeccenxxxxD
\else
??????\fi
\fi
\fi
}
\newcommand{\hatcurRVfitrmsBeccen}[1]{\ifnum#1=39 
\hatcurRVfitrmsBeccenxxxxA
\else
\ifnum#1=41 
\hatcurRVfitrmsBeccenxxxxC
\else
\ifnum#1=42 
\hatcurRVfitrmsBeccenxxxxD
\else
??????\fi
\fi
\fi
}
\newcommand{\hatcurRVfitrmsCeccen}[1]{\ifnum#1=41 
\hatcurRVfitrmsCeccenxxxxC
\else
??????\fi
}
\newcommand{\hatcurRVfitrmseccen}[1]{\ifnum#1=40 
\hatcurRVfitrmseccenxxxxB
\else
??????\fi
}
\newcommand{\hatcurRVgammaAeccen}[1]{\ifnum#1=39 
\hatcurRVgammaAeccenxxxxA
\else
\ifnum#1=41 
\hatcurRVgammaAeccenxxxxC
\else
\ifnum#1=42 
\hatcurRVgammaAeccenxxxxD
\else
??????\fi
\fi
\fi
}
\newcommand{\hatcurRVgammaBeccen}[1]{\ifnum#1=39 
\hatcurRVgammaBeccenxxxxA
\else
\ifnum#1=41 
\hatcurRVgammaBeccenxxxxC
\else
\ifnum#1=42 
\hatcurRVgammaBeccenxxxxD
\else
??????\fi
\fi
\fi
}
\newcommand{\hatcurRVgammaCeccen}[1]{\ifnum#1=41 
\hatcurRVgammaCeccenxxxxC
\else
??????\fi
}
\newcommand{\hatcurRVgammaeccen}[1]{\ifnum#1=40 
\hatcurRVgammaeccenxxxxB
\else
??????\fi
}
\newcommand{\hatcurRVheccen}[1]{\ifnum#1=39 
\hatcurRVheccenxxxxA
\else
\ifnum#1=40 
\hatcurRVheccenxxxxB
\else
\ifnum#1=41 
\hatcurRVheccenxxxxC
\else
\ifnum#1=42 
\hatcurRVheccenxxxxD
\else
??????\fi
\fi
\fi
\fi
}
\newcommand{\hatcurRVjitterAeccen}[1]{\ifnum#1=39 
\hatcurRVjitterAeccenxxxxA
\else
\ifnum#1=41 
\hatcurRVjitterAeccenxxxxC
\else
\ifnum#1=42 
\hatcurRVjitterAeccenxxxxD
\else
??????\fi
\fi
\fi
}
\newcommand{\hatcurRVjitterBeccen}[1]{\ifnum#1=39 
\hatcurRVjitterBeccenxxxxA
\else
\ifnum#1=41 
\hatcurRVjitterBeccenxxxxC
\else
\ifnum#1=42 
\hatcurRVjitterBeccenxxxxD
\else
??????\fi
\fi
\fi
}
\newcommand{\hatcurRVjitterCeccen}[1]{\ifnum#1=41 
\hatcurRVjitterCeccenxxxxC
\else
??????\fi
}
\newcommand{\hatcurRVjittereccen}[1]{\ifnum#1=40 
\hatcurRVjittereccenxxxxB
\else
??????\fi
}
\newcommand{\hatcurRVjittertwosiglimAeccen}[1]{\ifnum#1=39 
\hatcurRVjittertwosiglimAeccenxxxxA
\else
\ifnum#1=41 
\hatcurRVjittertwosiglimAeccenxxxxC
\else
\ifnum#1=42 
\hatcurRVjittertwosiglimAeccenxxxxD
\else
??????\fi
\fi
\fi
}
\newcommand{\hatcurRVjittertwosiglimBeccen}[1]{\ifnum#1=39 
\hatcurRVjittertwosiglimBeccenxxxxA
\else
\ifnum#1=41 
\hatcurRVjittertwosiglimBeccenxxxxC
\else
\ifnum#1=42 
\hatcurRVjittertwosiglimBeccenxxxxD
\else
??????\fi
\fi
\fi
}
\newcommand{\hatcurRVjittertwosiglimCeccen}[1]{\ifnum#1=41 
\hatcurRVjittertwosiglimCeccenxxxxC
\else
??????\fi
}
\newcommand{\hatcurRVjittertwosiglimeccen}[1]{\ifnum#1=40 
\hatcurRVjittertwosiglimeccenxxxxB
\else
??????\fi
}
\newcommand{\hatcurRVkeccen}[1]{\ifnum#1=39 
\hatcurRVkeccenxxxxA
\else
\ifnum#1=40 
\hatcurRVkeccenxxxxB
\else
\ifnum#1=41 
\hatcurRVkeccenxxxxC
\else
\ifnum#1=42 
\hatcurRVkeccenxxxxD
\else
??????\fi
\fi
\fi
\fi
}
\newcommand{\hatcurRVKeccen}[1]{\ifnum#1=39 
\hatcurRVKeccenxxxxA
\else
\ifnum#1=40 
\hatcurRVKeccenxxxxB
\else
\ifnum#1=41 
\hatcurRVKeccenxxxxC
\else
\ifnum#1=42 
\hatcurRVKeccenxxxxD
\else
??????\fi
\fi
\fi
\fi
}
\newcommand{\hatcurRVomegaeccen}[1]{\ifnum#1=39 
\hatcurRVomegaeccenxxxxA
\else
\ifnum#1=40 
\hatcurRVomegaeccenxxxxB
\else
\ifnum#1=41 
\hatcurRVomegaeccenxxxxC
\else
\ifnum#1=42 
\hatcurRVomegaeccenxxxxD
\else
??????\fi
\fi
\fi
\fi
}
\newcommand{\hatcurRVrheccen}[1]{\ifnum#1=39 
\hatcurRVrheccenxxxxA
\else
\ifnum#1=40 
\hatcurRVrheccenxxxxB
\else
\ifnum#1=41 
\hatcurRVrheccenxxxxC
\else
\ifnum#1=42 
\hatcurRVrheccenxxxxD
\else
??????\fi
\fi
\fi
\fi
}
\newcommand{\hatcurRVrkeccen}[1]{\ifnum#1=39 
\hatcurRVrkeccenxxxxA
\else
\ifnum#1=40 
\hatcurRVrkeccenxxxxB
\else
\ifnum#1=41 
\hatcurRVrkeccenxxxxC
\else
\ifnum#1=42 
\hatcurRVrkeccenxxxxD
\else
??????\fi
\fi
\fi
\fi
}
\newcommand{\hatcurRVtroneeccen}[1]{\ifnum#1=39 
\hatcurRVtroneeccenxxxxA
\else
\ifnum#1=40 
\hatcurRVtroneeccenxxxxB
\else
\ifnum#1=41 
\hatcurRVtroneeccenxxxxC
\else
\ifnum#1=42 
\hatcurRVtroneeccenxxxxD
\else
??????\fi
\fi
\fi
\fi
}
\newcommand{\hatcurRVtrtwoeccen}[1]{\ifnum#1=39 
\hatcurRVtrtwoeccenxxxxA
\else
\ifnum#1=40 
\hatcurRVtrtwoeccenxxxxB
\else
\ifnum#1=41 
\hatcurRVtrtwoeccenxxxxC
\else
\ifnum#1=42 
\hatcurRVtrtwoeccenxxxxD
\else
??????\fi
\fi
\fi
\fi
}
\newcommand{\hatcurSMEiiloggeccen}[1]{\ifnum#1=39 
\hatcurSMEiiloggeccenxxxxA
\else
\ifnum#1=42 
\hatcurSMEiiloggeccenxxxxD
\else
??????\fi
\fi
}
\newcommand{\hatcurSMEiiteffeccen}[1]{\ifnum#1=39 
\hatcurSMEiiteffeccenxxxxA
\else
\ifnum#1=42 
\hatcurSMEiiteffeccenxxxxD
\else
??????\fi
\fi
}
\newcommand{\hatcurSMEiivmaceccen}[1]{\ifnum#1=39 
\hatcurSMEiivmaceccenxxxxA
\else
\ifnum#1=42 
\hatcurSMEiivmaceccenxxxxD
\else
??????\fi
\fi
}
\newcommand{\hatcurSMEiivmiceccen}[1]{\ifnum#1=39 
\hatcurSMEiivmiceccenxxxxA
\else
\ifnum#1=42 
\hatcurSMEiivmiceccenxxxxD
\else
??????\fi
\fi
}
\newcommand{\hatcurSMEiivsineccen}[1]{\ifnum#1=39 
\hatcurSMEiivsineccenxxxxA
\else
\ifnum#1=42 
\hatcurSMEiivsineccenxxxxD
\else
??????\fi
\fi
}
\newcommand{\hatcurSMEiizfeheccen}[1]{\ifnum#1=39 
\hatcurSMEiizfeheccenxxxxA
\else
\ifnum#1=42 
\hatcurSMEiizfeheccenxxxxD
\else
??????\fi
\fi
}
\newcommand{\hatcurSMEiizfehshorteccen}[1]{\ifnum#1=39 
\hatcurSMEiizfehshorteccenxxxxA
\else
\ifnum#1=42 
\hatcurSMEiizfehshorteccenxxxxD
\else
??????\fi
\fi
}
\newcommand{\hatcurSMEiloggeccen}[1]{\ifnum#1=39 
\hatcurSMEiloggeccenxxxxA
\else
\ifnum#1=40 
\hatcurSMEiloggeccenxxxxB
\else
\ifnum#1=41 
\hatcurSMEiloggeccenxxxxC
\else
\ifnum#1=42 
\hatcurSMEiloggeccenxxxxD
\else
??????\fi
\fi
\fi
\fi
}
\newcommand{\hatcurSMEiteffeccen}[1]{\ifnum#1=39 
\hatcurSMEiteffeccenxxxxA
\else
\ifnum#1=40 
\hatcurSMEiteffeccenxxxxB
\else
\ifnum#1=41 
\hatcurSMEiteffeccenxxxxC
\else
\ifnum#1=42 
\hatcurSMEiteffeccenxxxxD
\else
??????\fi
\fi
\fi
\fi
}
\newcommand{\hatcurSMEivmaceccen}[1]{\ifnum#1=39 
\hatcurSMEivmaceccenxxxxA
\else
\ifnum#1=40 
\hatcurSMEivmaceccenxxxxB
\else
\ifnum#1=41 
\hatcurSMEivmaceccenxxxxC
\else
\ifnum#1=42 
\hatcurSMEivmaceccenxxxxD
\else
??????\fi
\fi
\fi
\fi
}
\newcommand{\hatcurSMEivmiceccen}[1]{\ifnum#1=39 
\hatcurSMEivmiceccenxxxxA
\else
\ifnum#1=40 
\hatcurSMEivmiceccenxxxxB
\else
\ifnum#1=41 
\hatcurSMEivmiceccenxxxxC
\else
\ifnum#1=42 
\hatcurSMEivmiceccenxxxxD
\else
??????\fi
\fi
\fi
\fi
}
\newcommand{\hatcurSMEivsineccen}[1]{\ifnum#1=39 
\hatcurSMEivsineccenxxxxA
\else
\ifnum#1=40 
\hatcurSMEivsineccenxxxxB
\else
\ifnum#1=41 
\hatcurSMEivsineccenxxxxC
\else
\ifnum#1=42 
\hatcurSMEivsineccenxxxxD
\else
??????\fi
\fi
\fi
\fi
}
\newcommand{\hatcurSMEizfeheccen}[1]{\ifnum#1=39 
\hatcurSMEizfeheccenxxxxA
\else
\ifnum#1=40 
\hatcurSMEizfeheccenxxxxB
\else
\ifnum#1=41 
\hatcurSMEizfeheccenxxxxC
\else
\ifnum#1=42 
\hatcurSMEizfeheccenxxxxD
\else
??????\fi
\fi
\fi
\fi
}
\newcommand{\hatcurSMEizfehshorteccen}[1]{\ifnum#1=39 
\hatcurSMEizfehshorteccenxxxxA
\else
\ifnum#1=40 
\hatcurSMEizfehshorteccenxxxxB
\else
\ifnum#1=41 
\hatcurSMEizfehshorteccenxxxxC
\else
\ifnum#1=42 
\hatcurSMEizfehshorteccenxxxxD
\else
??????\fi
\fi
\fi
\fi
}
\newcommand{\hatcurXAveccen}[1]{\ifnum#1=39 
\hatcurXAveccenxxxxA
\else
\ifnum#1=40 
\hatcurXAveccenxxxxB
\else
\ifnum#1=41 
\hatcurXAveccenxxxxC
\else
\ifnum#1=42 
\hatcurXAveccenxxxxD
\else
??????\fi
\fi
\fi
\fi
}
\newcommand{\hatcurXdisteccen}[1]{\ifnum#1=39 
\hatcurXdisteccenxxxxA
\else
\ifnum#1=40 
\hatcurXdisteccenxxxxB
\else
\ifnum#1=41 
\hatcurXdisteccenxxxxC
\else
\ifnum#1=42 
\hatcurXdisteccenxxxxD
\else
??????\fi
\fi
\fi
\fi
}
\newcommand{\hatcurXdistredeccen}[1]{\ifnum#1=39 
\hatcurXdistredeccenxxxxA
\else
\ifnum#1=40 
\hatcurXdistredeccenxxxxB
\else
\ifnum#1=41 
\hatcurXdistredeccenxxxxC
\else
\ifnum#1=42 
\hatcurXdistredeccenxxxxD
\else
??????\fi
\fi
\fi
\fi
}
\newcommand{\hatcurXEBVeccen}[1]{\ifnum#1=39 
\hatcurXEBVeccenxxxxA
\else
\ifnum#1=40 
\hatcurXEBVeccenxxxxB
\else
\ifnum#1=41 
\hatcurXEBVeccenxxxxC
\else
\ifnum#1=42 
\hatcurXEBVeccenxxxxD
\else
??????\fi
\fi
\fi
\fi
}
\newcommand{\hatcurXjhisoredeccen}[1]{\ifnum#1=39 
\hatcurXjhisoredeccenxxxxA
\else
\ifnum#1=40 
\hatcurXjhisoredeccenxxxxB
\else
\ifnum#1=41 
\hatcurXjhisoredeccenxxxxC
\else
\ifnum#1=42 
\hatcurXjhisoredeccenxxxxD
\else
??????\fi
\fi
\fi
\fi
}
\newcommand{\hatcurXjkisoredeccen}[1]{\ifnum#1=39 
\hatcurXjkisoredeccenxxxxA
\else
\ifnum#1=40 
\hatcurXjkisoredeccenxxxxB
\else
\ifnum#1=41 
\hatcurXjkisoredeccenxxxxC
\else
\ifnum#1=42 
\hatcurXjkisoredeccenxxxxD
\else
??????\fi
\fi
\fi
\fi
}
\newcommand{\hatcurXmhisoredeccen}[1]{\ifnum#1=39 
\hatcurXmhisoredeccenxxxxA
\else
\ifnum#1=40 
\hatcurXmhisoredeccenxxxxB
\else
\ifnum#1=41 
\hatcurXmhisoredeccenxxxxC
\else
\ifnum#1=42 
\hatcurXmhisoredeccenxxxxD
\else
??????\fi
\fi
\fi
\fi
}
\newcommand{\hatcurXmiisoredeccen}[1]{\ifnum#1=39 
\hatcurXmiisoredeccenxxxxA
\else
\ifnum#1=40 
\hatcurXmiisoredeccenxxxxB
\else
\ifnum#1=41 
\hatcurXmiisoredeccenxxxxC
\else
\ifnum#1=42 
\hatcurXmiisoredeccenxxxxD
\else
??????\fi
\fi
\fi
\fi
}
\newcommand{\hatcurXmjisoredeccen}[1]{\ifnum#1=39 
\hatcurXmjisoredeccenxxxxA
\else
\ifnum#1=40 
\hatcurXmjisoredeccenxxxxB
\else
\ifnum#1=41 
\hatcurXmjisoredeccenxxxxC
\else
\ifnum#1=42 
\hatcurXmjisoredeccenxxxxD
\else
??????\fi
\fi
\fi
\fi
}
\newcommand{\hatcurXmkisoredeccen}[1]{\ifnum#1=39 
\hatcurXmkisoredeccenxxxxA
\else
\ifnum#1=40 
\hatcurXmkisoredeccenxxxxB
\else
\ifnum#1=41 
\hatcurXmkisoredeccenxxxxC
\else
\ifnum#1=42 
\hatcurXmkisoredeccenxxxxD
\else
??????\fi
\fi
\fi
\fi
}
\newcommand{\hatcurXmvisoredeccen}[1]{\ifnum#1=39 
\hatcurXmvisoredeccenxxxxA
\else
\ifnum#1=40 
\hatcurXmvisoredeccenxxxxB
\else
\ifnum#1=41 
\hatcurXmvisoredeccenxxxxC
\else
\ifnum#1=42 
\hatcurXmvisoredeccenxxxxD
\else
??????\fi
\fi
\fi
\fi
}
\newcommand{\hatcurXsecdureccen}[1]{\ifnum#1=39 
\hatcurXsecdureccenxxxxA
\else
\ifnum#1=40 
\hatcurXsecdureccenxxxxB
\else
\ifnum#1=41 
\hatcurXsecdureccenxxxxC
\else
\ifnum#1=42 
\hatcurXsecdureccenxxxxD
\else
??????\fi
\fi
\fi
\fi
}
\newcommand{\hatcurXsecingdureccen}[1]{\ifnum#1=39 
\hatcurXsecingdureccenxxxxA
\else
\ifnum#1=40 
\hatcurXsecingdureccenxxxxB
\else
\ifnum#1=41 
\hatcurXsecingdureccenxxxxC
\else
\ifnum#1=42 
\hatcurXsecingdureccenxxxxD
\else
??????\fi
\fi
\fi
\fi
}
\newcommand{\hatcurXsecondaryeccen}[1]{\ifnum#1=39 
\hatcurXsecondaryeccenxxxxA
\else
\ifnum#1=40 
\hatcurXsecondaryeccenxxxxB
\else
\ifnum#1=41 
\hatcurXsecondaryeccenxxxxC
\else
\ifnum#1=42 
\hatcurXsecondaryeccenxxxxD
\else
??????\fi
\fi
\fi
\fi
}
\newcommand{\hatcurXsecphaseeccen}[1]{\ifnum#1=39 
\hatcurXsecphaseeccenxxxxA
\else
\ifnum#1=40 
\hatcurXsecphaseeccenxxxxB
\else
\ifnum#1=41 
\hatcurXsecphaseeccenxxxxC
\else
\ifnum#1=42 
\hatcurXsecphaseeccenxxxxD
\else
??????\fi
\fi
\fi
\fi
}
\newcommand{\hatcurXviisoredeccen}[1]{\ifnum#1=39 
\hatcurXviisoredeccenxxxxA
\else
\ifnum#1=40 
\hatcurXviisoredeccenxxxxB
\else
\ifnum#1=41 
\hatcurXviisoredeccenxxxxC
\else
\ifnum#1=42 
\hatcurXviisoredeccenxxxxD
\else
??????\fi
\fi
\fi
\fi
}
\newcommand{\hatcurXvkisoredeccen}[1]{\ifnum#1=39 
\hatcurXvkisoredeccenxxxxA
\else
\ifnum#1=40 
\hatcurXvkisoredeccenxxxxB
\else
\ifnum#1=41 
\hatcurXvkisoredeccenxxxxC
\else
\ifnum#1=42 
\hatcurXvkisoredeccenxxxxD
\else
??????\fi
\fi
\fi
\fi
}

\newcommand{\hatcurxxxxA}{HATS-39}
\newcommand{\hatcurbxxxxA}{HATS-39b}
\newcommand{\hatcurcxxxxA}{HATS-39c}

\newcommand{\hatcurplanetnumxxxxA}{39}

\newcommand{\hatcurCCtwomassshortxxxxA}{07294061-2956163}

\newcommand{\hatcurRVgammaabsxxxxA}{\hatcurRVgammaA{\hatcurplanetnumxxxxA}}                           

\newcommand{\hatcurRVgammarelxxxxA}{\hatcurRVgammaA{\hatcurplanetnumxxxxA}}                           

\newcommand{\hatcurCCtassvixxxxA}{\ensuremath{NULL\pm NULL}}                  

\newcommand{\hatcurSMEversionxxxxA}{ii}                                       

\newcommand{\hatcurisoshortxxxxA}{YY}
\newcommand{\hatcurisofullxxxxA}{Yonsei-Yale (YY)}
\newcommand{\hatcurisocitexxxxA}{yi:2001}

\newcommand{\hatcurlumindxxxxA}{\arstar}

\newcommand{\hatcurjhkfilsetxxxxA}{ESO}

\newcommand{\hatcurSMEteffxxxxA}{\ifthenelse{\equal{\hatcurSMEversionxxxxA}{i}}{\hatcurSMEiteff{\hatcurplanetnumxxxxA}}{\hatcurSMEiiteff{\hatcurplanetnumxxxxA}}}
\newcommand{\hatcurSMEzfehxxxxA}{\ifthenelse{\equal{\hatcurSMEversionxxxxA}{i}}{\hatcurSMEizfeh{\hatcurplanetnumxxxxA}}{\hatcurSMEiizfeh{\hatcurplanetnumxxxxA}}}
\newcommand{\hatcurSMEzfehshortxxxxA}{\ifthenelse{\equal{\hatcurSMEversionxxxxA}{i}}{\hatcurSMEizfehshort{\hatcurplanetnumxxxxA}}{\hatcurSMEiizfehshort{\hatcurplanetnumxxxxA}}}
\newcommand{\hatcurSMEloggxxxxA}{\ifthenelse{\equal{\hatcurSMEversionxxxxA}{i}}{\hatcurSMEilogg{\hatcurplanetnumxxxxA}}{\hatcurSMEiilogg{\hatcurplanetnumxxxxA}}}
\newcommand{\hatcurSMEvsinxxxxA}{\ifthenelse{\equal{\hatcurSMEversionxxxxA}{i}}{\hatcurSMEivsin{\hatcurplanetnumxxxxA}}{\hatcurSMEiivsin{\hatcurplanetnumxxxxA}}}
\newcommand{\hatcurSMEvmacxxxxA}{\ifthenelse{\equal{\hatcurSMEversionxxxxA}{i}}{\hatcurSMEivmac{\hatcurplanetnumxxxxA}}{\hatcurSMEiivmac{\hatcurplanetnumxxxxA}}}
\newcommand{\hatcurSMEvmicxxxxA}{\ifthenelse{\equal{\hatcurSMEversionxxxxA}{i}}{\hatcurSMEivmic{\hatcurplanetnumxxxxA}}{\hatcurSMEiivmic{\hatcurplanetnumxxxxA}}}

\newcommand{\hatcurxxxxB}{HATS-40}
\newcommand{\hatcurbxxxxB}{HATS-40b}
\newcommand{\hatcurcxxxxB}{HATS-40c}

\newcommand{\hatcurplanetnumxxxxB}{40}

\newcommand{\hatcurCCtwomassshortxxxxB}{06421710-2946365}

\newcommand{\hatcurRVgammaabsxxxxB}{\hatcurRVgamma{\hatcurplanetnumxxxxB}}                           

\newcommand{\hatcurRVgammarelxxxxB}{\hatcurRVgamma{\hatcurplanetnumxxxxB}}                           

\newcommand{\hatcurCCtassvixxxxB}{\ensuremath{NULL\pm NULL}}                  

\newcommand{\hatcurSMEversionxxxxB}{i}                                       

\newcommand{\hatcurisoshortxxxxB}{YY}
\newcommand{\hatcurisofullxxxxB}{Yonsei-Yale (YY)}
\newcommand{\hatcurisocitexxxxB}{yi:2001}

\newcommand{\hatcurlumindxxxxB}{\arstar}

\newcommand{\hatcurjhkfilsetxxxxB}{ESO}

\newcommand{\hatcurSMEteffxxxxB}{\ifthenelse{\equal{\hatcurSMEversionxxxxB}{i}}{\hatcurSMEiteff{\hatcurplanetnumxxxxB}}{\hatcurSMEiiteff{\hatcurplanetnumxxxxB}}}
\newcommand{\hatcurSMEzfehxxxxB}{\ifthenelse{\equal{\hatcurSMEversionxxxxB}{i}}{\hatcurSMEizfeh{\hatcurplanetnumxxxxB}}{\hatcurSMEiizfeh{\hatcurplanetnumxxxxB}}}
\newcommand{\hatcurSMEzfehshortxxxxB}{\ifthenelse{\equal{\hatcurSMEversionxxxxB}{i}}{\hatcurSMEizfehshort{\hatcurplanetnumxxxxB}}{\hatcurSMEiizfehshort{\hatcurplanetnumxxxxB}}}
\newcommand{\hatcurSMEloggxxxxB}{\ifthenelse{\equal{\hatcurSMEversionxxxxB}{i}}{\hatcurSMEilogg{\hatcurplanetnumxxxxB}}{\hatcurSMEiilogg{\hatcurplanetnumxxxxB}}}
\newcommand{\hatcurSMEvsinxxxxB}{\ifthenelse{\equal{\hatcurSMEversionxxxxB}{i}}{\hatcurSMEivsin{\hatcurplanetnumxxxxB}}{\hatcurSMEiivsin{\hatcurplanetnumxxxxB}}}
\newcommand{\hatcurSMEvmacxxxxB}{\ifthenelse{\equal{\hatcurSMEversionxxxxB}{i}}{\hatcurSMEivmac{\hatcurplanetnumxxxxB}}{\hatcurSMEiivmac{\hatcurplanetnumxxxxB}}}
\newcommand{\hatcurSMEvmicxxxxB}{\ifthenelse{\equal{\hatcurSMEversionxxxxB}{i}}{\hatcurSMEivmic{\hatcurplanetnumxxxxB}}{\hatcurSMEiivmic{\hatcurplanetnumxxxxB}}}

\newcommand{\hatcurxxxxC}{HATS-41}
\newcommand{\hatcurbxxxxC}{HATS-41b}
\newcommand{\hatcurcxxxxC}{HATS-41c}

\newcommand{\hatcurplanetnumxxxxC}{41}

\newcommand{\hatcurCCtwomassshortxxxxC}{06540416-2703013}

\newcommand{\hatcurRVgammaabsxxxxC}{\hatcurRVgammaBeccen{\hatcurplanetnumxxxxC}}                           

\newcommand{\hatcurRVgammarelxxxxC}{\hatcurRVgammaBeccen{\hatcurplanetnumxxxxC}}                           

\newcommand{\hatcurCCtassvixxxxC}{\ensuremath{NULL\pm NULL}}                  

\newcommand{\hatcurSMEversionxxxxC}{i}                                       

\newcommand{\hatcurisoshortxxxxC}{YY}
\newcommand{\hatcurisofullxxxxC}{Yonsei-Yale (YY)}
\newcommand{\hatcurisocitexxxxC}{yi:2001}

\newcommand{\hatcurlumindxxxxC}{\arstar}

\newcommand{\hatcurjhkfilsetxxxxC}{ESO}

\newcommand{\hatcurSMEteffxxxxC}{\ifthenelse{\equal{\hatcurSMEversionxxxxC}{i}}{\hatcurSMEiteffeccen{\hatcurplanetnumxxxxC}}{\hatcurSMEiiteffeccen{\hatcurplanetnumxxxxC}}}
\newcommand{\hatcurSMEzfehxxxxC}{\ifthenelse{\equal{\hatcurSMEversionxxxxC}{i}}{\hatcurSMEizfeheccen{\hatcurplanetnumxxxxC}}{\hatcurSMEiizfeheccen{\hatcurplanetnumxxxxC}}}
\newcommand{\hatcurSMEzfehshortxxxxC}{\ifthenelse{\equal{\hatcurSMEversionxxxxC}{i}}{\hatcurSMEizfehshorteccen{\hatcurplanetnumxxxxC}}{\hatcurSMEiizfehshorteccen{\hatcurplanetnumxxxxC}}}
\newcommand{\hatcurSMEloggxxxxC}{\ifthenelse{\equal{\hatcurSMEversionxxxxC}{i}}{\hatcurSMEiloggeccen{\hatcurplanetnumxxxxC}}{\hatcurSMEiiloggeccen{\hatcurplanetnumxxxxC}}}
\newcommand{\hatcurSMEvsinxxxxC}{\ifthenelse{\equal{\hatcurSMEversionxxxxC}{i}}{\hatcurSMEivsineccen{\hatcurplanetnumxxxxC}}{\hatcurSMEiivsineccen{\hatcurplanetnumxxxxC}}}
\newcommand{\hatcurSMEvmacxxxxC}{\ifthenelse{\equal{\hatcurSMEversionxxxxC}{i}}{\hatcurSMEivmaceccen{\hatcurplanetnumxxxxC}}{\hatcurSMEiivmaceccen{\hatcurplanetnumxxxxC}}}
\newcommand{\hatcurSMEvmicxxxxC}{\ifthenelse{\equal{\hatcurSMEversionxxxxC}{i}}{\hatcurSMEivmiceccen{\hatcurplanetnumxxxxC}}{\hatcurSMEiivmiceccen{\hatcurplanetnumxxxxC}}}

\newcommand{\hatcurxxxxD}{HATS-42}
\newcommand{\hatcurbxxxxD}{HATS-42b}
\newcommand{\hatcurcxxxxD}{HATS-42c}

\newcommand{\hatcurplanetnumxxxxD}{42}

\newcommand{\hatcurCCtwomassshortxxxxD}{07134857-3326143}

\newcommand{\hatcurRVgammaabsxxxxD}{\hatcurRVgammaB{\hatcurplanetnumxxxxD}}                           

\newcommand{\hatcurRVgammarelxxxxD}{\hatcurRVgammaB{\hatcurplanetnumxxxxD}}                           

\newcommand{\hatcurCCtassvixxxxD}{\ensuremath{NULL\pm NULL}}                  

\newcommand{\hatcurSMEversionxxxxD}{ii}                                       

\newcommand{\hatcurisoshortxxxxD}{YY}
\newcommand{\hatcurisofullxxxxD}{Yonsei-Yale (YY)}
\newcommand{\hatcurisocitexxxxD}{yi:2001}

\newcommand{\hatcurlumindxxxxD}{\arstar}

\newcommand{\hatcurjhkfilsetxxxxD}{ESO}

\newcommand{\hatcurSMEteffxxxxD}{\ifthenelse{\equal{\hatcurSMEversionxxxxD}{i}}{\hatcurSMEiteff{\hatcurplanetnumxxxxD}}{\hatcurSMEiiteff{\hatcurplanetnumxxxxD}}}
\newcommand{\hatcurSMEzfehxxxxD}{\ifthenelse{\equal{\hatcurSMEversionxxxxD}{i}}{\hatcurSMEizfeh{\hatcurplanetnumxxxxD}}{\hatcurSMEiizfeh{\hatcurplanetnumxxxxD}}}
\newcommand{\hatcurSMEzfehshortxxxxD}{\ifthenelse{\equal{\hatcurSMEversionxxxxD}{i}}{\hatcurSMEizfehshort{\hatcurplanetnumxxxxD}}{\hatcurSMEiizfehshort{\hatcurplanetnumxxxxD}}}
\newcommand{\hatcurSMEloggxxxxD}{\ifthenelse{\equal{\hatcurSMEversionxxxxD}{i}}{\hatcurSMEilogg{\hatcurplanetnumxxxxD}}{\hatcurSMEiilogg{\hatcurplanetnumxxxxD}}}
\newcommand{\hatcurSMEvsinxxxxD}{\ifthenelse{\equal{\hatcurSMEversionxxxxD}{i}}{\hatcurSMEivsin{\hatcurplanetnumxxxxD}}{\hatcurSMEiivsin{\hatcurplanetnumxxxxD}}}
\newcommand{\hatcurSMEvmacxxxxD}{\ifthenelse{\equal{\hatcurSMEversionxxxxD}{i}}{\hatcurSMEivmac{\hatcurplanetnumxxxxD}}{\hatcurSMEiivmac{\hatcurplanetnumxxxxD}}}
\newcommand{\hatcurSMEvmicxxxxD}{\ifthenelse{\equal{\hatcurSMEversionxxxxD}{i}}{\hatcurSMEivmic{\hatcurplanetnumxxxxD}}{\hatcurSMEiivmic{\hatcurplanetnumxxxxD}}}
 \newcommand{\hatcur}[1]{\ifnum#1=39 
\hatcurxxxxA
\else
\ifnum#1=40 
\hatcurxxxxB
\else
\ifnum#1=41 
\hatcurxxxxC
\else
\ifnum#1=42 
\hatcurxxxxD
\else
??????\fi
\fi
\fi
\fi
}
\newcommand{\hatcurb}[1]{\ifnum#1=39 
\hatcurbxxxxA
\else
\ifnum#1=40 
\hatcurbxxxxB
\else
\ifnum#1=41 
\hatcurbxxxxC
\else
\ifnum#1=42 
\hatcurbxxxxD
\else
??????\fi
\fi
\fi
\fi
}
\newcommand{\hatcurc}[1]{\ifnum#1=39 
\hatcurcxxxxA
\else
\ifnum#1=40 
\hatcurcxxxxB
\else
\ifnum#1=41 
\hatcurcxxxxC
\else
\ifnum#1=42 
\hatcurcxxxxD
\else
??????\fi
\fi
\fi
\fi
}
\newcommand{\hatcurCCtassvi}[1]{\ifnum#1=39 
\hatcurCCtassvixxxxA
\else
\ifnum#1=40 
\hatcurCCtassvixxxxB
\else
\ifnum#1=41 
\hatcurCCtassvixxxxC
\else
\ifnum#1=42 
\hatcurCCtassvixxxxD
\else
??????\fi
\fi
\fi
\fi
}
\newcommand{\hatcurCCtwomassshort}[1]{\ifnum#1=39 
\hatcurCCtwomassshortxxxxA
\else
\ifnum#1=40 
\hatcurCCtwomassshortxxxxB
\else
\ifnum#1=41 
\hatcurCCtwomassshortxxxxC
\else
\ifnum#1=42 
\hatcurCCtwomassshortxxxxD
\else
??????\fi
\fi
\fi
\fi
}
\newcommand{\hatcurisocite}[1]{\ifnum#1=39 
\hatcurisocitexxxxA
\else
\ifnum#1=40 
\hatcurisocitexxxxB
\else
\ifnum#1=41 
\hatcurisocitexxxxC
\else
\ifnum#1=42 
\hatcurisocitexxxxD
\else
??????\fi
\fi
\fi
\fi
}
\newcommand{\hatcurisofull}[1]{\ifnum#1=39 
\hatcurisofullxxxxA
\else
\ifnum#1=40 
\hatcurisofullxxxxB
\else
\ifnum#1=41 
\hatcurisofullxxxxC
\else
\ifnum#1=42 
\hatcurisofullxxxxD
\else
??????\fi
\fi
\fi
\fi
}
\newcommand{\hatcurisoshort}[1]{\ifnum#1=39 
\hatcurisoshortxxxxA
\else
\ifnum#1=40 
\hatcurisoshortxxxxB
\else
\ifnum#1=41 
\hatcurisoshortxxxxC
\else
\ifnum#1=42 
\hatcurisoshortxxxxD
\else
??????\fi
\fi
\fi
\fi
}
\newcommand{\hatcurjhkfilset}[1]{\ifnum#1=39 
\hatcurjhkfilsetxxxxA
\else
\ifnum#1=40 
\hatcurjhkfilsetxxxxB
\else
\ifnum#1=41 
\hatcurjhkfilsetxxxxC
\else
\ifnum#1=42 
\hatcurjhkfilsetxxxxD
\else
??????\fi
\fi
\fi
\fi
}
\newcommand{\hatcurlumind}[1]{\ifnum#1=39 
\hatcurlumindxxxxA
\else
\ifnum#1=40 
\hatcurlumindxxxxB
\else
\ifnum#1=41 
\hatcurlumindxxxxC
\else
\ifnum#1=42 
\hatcurlumindxxxxD
\else
??????\fi
\fi
\fi
\fi
}
\newcommand{\hatcurplanetnum}[1]{\ifnum#1=39 
\hatcurplanetnumxxxxA
\else
\ifnum#1=40 
\hatcurplanetnumxxxxB
\else
\ifnum#1=41 
\hatcurplanetnumxxxxC
\else
\ifnum#1=42 
\hatcurplanetnumxxxxD
\else
??????\fi
\fi
\fi
\fi
}
\newcommand{\hatcurRVgammaabs}[1]{\ifnum#1=39 
\hatcurRVgammaabsxxxxA
\else
\ifnum#1=40 
\hatcurRVgammaabsxxxxB
\else
\ifnum#1=41 
\hatcurRVgammaabsxxxxC
\else
\ifnum#1=42 
\hatcurRVgammaabsxxxxD
\else
??????\fi
\fi
\fi
\fi
}
\newcommand{\hatcurRVgammarel}[1]{\ifnum#1=39 
\hatcurRVgammarelxxxxA
\else
\ifnum#1=40 
\hatcurRVgammarelxxxxB
\else
\ifnum#1=41 
\hatcurRVgammarelxxxxC
\else
\ifnum#1=42 
\hatcurRVgammarelxxxxD
\else
??????\fi
\fi
\fi
\fi
}
\newcommand{\hatcurSMElogg}[1]{\ifnum#1=39 
\hatcurSMEloggxxxxA
\else
\ifnum#1=40 
\hatcurSMEloggxxxxB
\else
\ifnum#1=41 
\hatcurSMEloggxxxxC
\else
\ifnum#1=42 
\hatcurSMEloggxxxxD
\else
??????\fi
\fi
\fi
\fi
}
\newcommand{\hatcurSMEteff}[1]{\ifnum#1=39 
\hatcurSMEteffxxxxA
\else
\ifnum#1=40 
\hatcurSMEteffxxxxB
\else
\ifnum#1=41 
\hatcurSMEteffxxxxC
\else
\ifnum#1=42 
\hatcurSMEteffxxxxD
\else
??????\fi
\fi
\fi
\fi
}
\newcommand{\hatcurSMEversion}[1]{\ifnum#1=39 
\hatcurSMEversionxxxxA
\else
\ifnum#1=40 
\hatcurSMEversionxxxxB
\else
\ifnum#1=41 
\hatcurSMEversionxxxxC
\else
\ifnum#1=42 
\hatcurSMEversionxxxxD
\else
??????\fi
\fi
\fi
\fi
}
\newcommand{\hatcurSMEvmac}[1]{\ifnum#1=39 
\hatcurSMEvmacxxxxA
\else
\ifnum#1=40 
\hatcurSMEvmacxxxxB
\else
\ifnum#1=41 
\hatcurSMEvmacxxxxC
\else
\ifnum#1=42 
\hatcurSMEvmacxxxxD
\else
??????\fi
\fi
\fi
\fi
}
\newcommand{\hatcurSMEvmic}[1]{\ifnum#1=39 
\hatcurSMEvmicxxxxA
\else
\ifnum#1=40 
\hatcurSMEvmicxxxxB
\else
\ifnum#1=41 
\hatcurSMEvmicxxxxC
\else
\ifnum#1=42 
\hatcurSMEvmicxxxxD
\else
??????\fi
\fi
\fi
\fi
}
\newcommand{\hatcurSMEvsin}[1]{\ifnum#1=39 
\hatcurSMEvsinxxxxA
\else
\ifnum#1=40 
\hatcurSMEvsinxxxxB
\else
\ifnum#1=41 
\hatcurSMEvsinxxxxC
\else
\ifnum#1=42 
\hatcurSMEvsinxxxxD
\else
??????\fi
\fi
\fi
\fi
}
\newcommand{\hatcurSMEzfeh}[1]{\ifnum#1=39 
\hatcurSMEzfehxxxxA
\else
\ifnum#1=40 
\hatcurSMEzfehxxxxB
\else
\ifnum#1=41 
\hatcurSMEzfehxxxxC
\else
\ifnum#1=42 
\hatcurSMEzfehxxxxD
\else
??????\fi
\fi
\fi
\fi
}
\newcommand{\hatcurSMEzfehshort}[1]{\ifnum#1=39 
\hatcurSMEzfehshortxxxxA
\else
\ifnum#1=40 
\hatcurSMEzfehshortxxxxB
\else
\ifnum#1=41 
\hatcurSMEzfehshortxxxxC
\else
\ifnum#1=42 
\hatcurSMEzfehshortxxxxD
\else
??????\fi
\fi
\fi
\fi
}
 
\newcounter{planetcounter}

\newboolean{emulateapj}
\setboolean{emulateapj}{true}

\newboolean{rvtablelong}
\setboolean{rvtablelong}{true}

\newboolean{astroph}
\setboolean{astroph}{true}

\shortauthors{Bento et al.}
\shorttitle{\hatcur{39}\lowercase{b}--\hatcur{42}\lowercase{b}}
\ifthenelse{\boolean{emulateapj}}{
    
}{
    
}

\begin{document}

\title{\hatcur{39}\lowercase{b}, \hatcur{40}\lowercase{b}, \hatcur{41}\lowercase{b}, and \hatcur{42}\lowercase{b}: Three Inflated Hot Jupiters and a Super-Jupiter Transiting F Stars
}

\author{J.~Bento}
\affiliation{Research School of Astronomy and Astrophysics, Mount Stromlo Observatory, Australian National University, Cotter Road, Weston, ACT 2611, Australia. E-mail: joao.bento@anu.edu.au}
\author{J.~D.~Hartman}
\author{G.~\'A. Bakos}
\author{W.~Bhatti}
\author{Z.~Csubry}
\affiliation{Department of Astrophysical Sciences, 4 Ivy Ln., Princeton, NJ 08544}
\author{K.~Penev}
\affiliation{Department of Astrophysical Sciences, 4 Ivy Ln., Princeton, NJ 08544}
\affiliation{Physics Department, University of Texas at Dallas, 800 W Campbell Rd. MS WT15, Richardson, TX 75080, USA}
\author{D.~Bayliss}
\affiliation{Department of Physics, University of Warwick, Coventry CV4 7AL, UK}
\author{M.~de~Val-Borro}
\affiliation{Astrochemistry Laboratory, Goddard Space Flight Center, NASA, 8800 Greenbelt Rd, Greenbelt, MD 20771, USA}
\author{G.~Zhou}
\affiliation{Harvard-Smithsonian Center for Astrophysics, Cambridge, MA 02138, USA}
\author{R.~Brahm}
\affiliation{Millennium Institute for Astrophysics, Santiago, Chile}
\affiliation{Instituto de Astrof\'isica, Facultad de F\'isica, Pontificia Universidad Cat\'olica de Chile, Vicu\~{n}a Mackenna 4860, 7820436 Macul, Santiago, Chile}
\author{N.~Espinoza}
\affiliation{Max Planck Institute for Astronomy, K\"{o}nigstuhl 17, 69117 Heidelberg, Germany}
\author{M.~Rabus}
\affiliation{Instituto de Astrof\'isica, Facultad de F\'isica, Pontificia Universidad Cat\'olica de Chile, Vicu\~{n}a Mackenna 4860, 7820436 Macul, Santiago, Chile}
\affiliation{Max Planck Institute for Astronomy, K\"{o}nigstuhl 17, 69117 Heidelberg, Germany}
\author{A.~Jord\'an}
\affiliation{Instituto de Astrof\'isica, Facultad de F\'isica, Pontificia Universidad Cat\'olica de Chile, Vicu\~{n}a Mackenna 4860, 7820436 Macul, Santiago, Chile}
\affiliation{Millennium Institute for Astrophysics, Santiago, Chile}
\affiliation{Max Planck Institute for Astronomy, K\"{o}nigstuhl 17, 69117 Heidelberg, Germany}
\author{V.~Suc}
\affiliation{Instituto de Astrof\'isica, Facultad de F\'isica, Pontificia Universidad Cat\'olica de Chile, Vicu\~{n}a Mackenna 4860, 7820436 Macul, Santiago, Chile}
\author{S.~Ciceri}
\author{P.~Sarkis}
\author{T.~Henning}
\affiliation{Max Planck Institute for Astronomy, K\"{o}nigstuhl 17, 69117 Heidelberg, Germany}
\author{L.~Mancini}
\affiliation{Department of Physics, University of Rome Tor Vergata, Via della
Ricerca Scientifica 1, 00133 Roma, Italy}
\affiliation{Max Planck Institute for Astronomy, K\"{o}nigstuhl 17, 69117 Heidelberg, Germany}
\affiliation{INAF -- Astrophysical Observatory of Turin, Via Osservatorio 20,
10025 -- Pino Torinese, Italy}
\author{C.~G.~Tinney}
\author{D.~J.~Wright}
\affiliation{Exoplanetary Science at UNSW, School of Physics
UNSW Sydney, NSW 2052, Australia}
\affiliation{Australian Centre for Astrobiology, School of Physics,
UNSW Sydney, NSW 2052, Australia}
\author{S.~Durkan}
\affiliation{Astrophysics Research Centre, Queens University, Belfast, Belfast, Northern Ireland, UK}
\author{T.~G.~Tan}
\affiliation{Perth Exoplanet Survey Telescope, Perth, Australia}
\author{J.~L\'az\'ar}
\author{I.~Papp}
\author{P.~S\'ari}
\affiliation{Hungarian Astronomical Association, 1451 Budapest, Hungary}

\begin{abstract}

\setcounter{footnote}{10}
  We report the discovery of four transiting \hjs\ from the HATSouth survey: \hatcurb{39}, \hatcurb{40}, \hatcurb{41} and \hatcurb{42}. These discoveries add to the growing number of transiting planets orbiting moderately bright ($12.5 \lesssim V \lesssim 13.7$) F dwarf stars on short (2-5 day) periods. 
The planets have similar radii, ranging from \hatcurPPrlongeccen{41}\,\rjup\, for \hatcurb{41} to \hatcurPPrlong{40}\,\rjup\, for \hatcurb{40}.
Their masses and bulk densities, however, span more than an order of magnitude. \hatcurb{39} has a mass of \hatcurPPmlong{39}\,\mjup,  and an inflated radius of \hatcurPPrlong{39}\,\rjup, making it a good target for future transmission spectroscopic studies. \hatcurb{41} is a very massive \hatcurPPmlongeccen{41}\,\mjup\ planet and one of only a few \hjs\ found to date with a mass over 5 \mjup. This planet orbits the highest metallicity star ($\feh = 0.470 \pm 0.010$) known to host a transiting planet and is also likely on an eccentric orbit. The high mass, coupled with a relatively young age (\hatcurISOageeccen{41}\,Gyr) for the host star, are factors that may explain why this planet's orbit has not yet circularised.

\setcounter{footnote}{0}
\end{abstract}

\keywords{
    planetary systems ---
    stars: individual (
\setcounter{planetcounter}{1}
\hatcur{39},
\hatcurCCgsc{39}\loopcommanoperiod
\setcounter{planetcounter}{2}
\hatcur{40},
\hatcurCCgsc{40}\loopcommanoperiod
\setcounter{planetcounter}{3}
\hatcur{41},
\hatcurCCgsceccen{41}\loopcommanoperiod
\setcounter{planetcounter}{3}
\hatcur{42},
\hatcurCCgsc{42}\loopcommanoperiod
\setcounter{planetcounter}{4}
) 
    techniques: spectroscopic, photometric
}

\section{Introduction}
\label{sec:introduction}

Planets that transit their host star are key for understanding the details of planet formation, structure and evolution. These systems not only provide a unique opportunity for further studies of atmospheric and surface conditions \citep[e.g.][]{bento:2014,jordan:2013,sing:2011,zhou:2014:sec,desert:2011,louden:2015}, 
but are also the exoplanets for which two complementary measurement techniques (i.e. transit photometry and host-star radial velocity) can be combined to deliver both  planet mass and planet radius, so yielding a measurement of bulk planet density. The vast majority of well-characterised exoplanets to date have been discovered using wide-field photometric surveys, either from the ground (e.g. HATNet, \citet{bakos:2004:hatnet} and SuperWASP, \citet{pollacco:2006}) or space \citep[e.g. Kepler and K2,][]{borucki:2010}.

In particular, \hjs\, (broadly defined as Jupiter-mass planets orbiting close to their host stars with orbital periods less than $\sim 10$ days) are still challenging models of planetary formation and evolution, despite over twenty years of study. The general consensus is that these planets are formed at large separations and migrate inwards to their current positions. 
There is, however, no consensus yet as to how these planets migrate, with a variety of mechanisms having been proposed \citep[e.g.][and references therein]{chambers:2009,ford:2008,wu:2003,petrovich:2015}. 
The increasing number of discoveries is now allowing studies that can statistically test the significance of these mechanisms -- for example investigating the dependence of eccentricity on mass and orbital separation \citep{mazeh:1997,southworth:2009,pont:2011}, to determine which migration mechanism (if any) is dominant. If planet-planet scattering dominates over disk migration, then it would be reasonable to expect eccentric planets at large separations in young systems. This drives the need to discover larger samples of planets spanning a larger range of ages and orbital separations. 

In this paper we report the discovery and characterisation of four new transiting \hjs\, from the HATSouth survey: \hatcurb{39}, \hatcurb{40}, \hatcurb{41}  and \hatcurb{42}. 
In Section \ref{sec:obs} we describe the photometric and spectroscopic observations undertaken for all four targets. Section \ref{sec:analysis} contains a description of the global data analysis and presents the modelled stellar and planetary parameters. We also describe the methods employed to reject false positive scenarios. Our findings are finally discussed in Section \ref{sec:discussion}.

\section{Observations}
\label{sec:obs}

A range of astrophysical events can mimic the photometric transit events for an exoplanet in a wide-field survey. These include grazing eclipses in binary systems, transiting late-M dwarfs, eclipses by dwarf star companions of evolved primary stars and eclipsing binary star systems whose light is blended with a third unresolved star. Substantial follow-up campaigns are required to obtain the additional photometric and spectroscopic observations required to reject these contaminants and confirm the planetary nature of the candidates found by the survey.

\subsection{Photometric detection}
\label{sec:detection}

The HATSouth project is an ongoing effort by a number of collaborating institutions\footnote{The HATSouth network is operated by a collaboration consisting of
Princeton University (PU), the Max Planck Institute f\"ur Astronomie
(MPIA), the Australian National University (ANU), and the Pontificia
Universidad Cat\'olica de Chile (PUC).  The station at Las Campanas
Observatory (LCO) of the Carnegie Institute is operated by PU in
conjunction with PUC, the station at the High Energy Spectroscopic
Survey (H.E.S.S.) site is operated in conjunction with MPIA, and the
station at Siding Spring Observatory (SSO) is operated jointly with
ANU.} aimed at discovering transiting planets orbiting moderately bright stars visible from the southern hemisphere \citep{bakos:2013:hatsouth}. It is composed of three identical facilities at Las Campanas Observatory in Chile, the High Energy Spectroscopic Survey (HESS) site in Namibia, and Siding Spring Observatory, Australia. The longitudinal coverage of these sites means that together they can 
continuously monitor 128 sq degree fields in the southern sky. This is highlighted by the discovery of HATS-17b \citep{brahm:2016}, the longest period transiting exoplanet found to date by a wide-field ground-based survey. A full list of discovered planets along with corresponding discovery \lcs\, can be found at \url{https://hatsouth.org/}.

Table \ref{tab:photobs} shows a summary of the HATSouth photometric
observations for the four new exoplanetary systems described in the present
work (along with observing details for subsequent follow-up observations with
the Las Cumbres Observatory Global Telescope (LCOGT) and the 0.3\,m PEST
telescope in Western Australia -- see Section 2.3). For HATSouth data, we list
the HATSouth unit, CCD and field name from which the observations were taken.
The detection of all targets relied on data from all HATSouth telescopes. The
HATSouth data for these targets spans a period of just under two years, from
August 2011 to April 2013, resulting in a total of 16,488 data points for
\hatcur{39}, 27,476 for \hatcur{40}, 11,938 for \hatcur{41}, and 21,210 for
\hatcur{42}. 

All HATSouth observations are obtained through a Sloan \emph{r} filter with a typical cadence of 4 minutes. The data were reduced with a custom pipeline described by \cite{penev:2013:hats1}, and  \lcs{} were de-trended using an External Parameter Decorrelation method \citep{bakos:2010:hat11}, followed by the application of a Trend Filtering Algorithm \citep[TFA,][]{kovacs:2005:TFA}. The Box-fitting Least-Squares algorithm \citep[BLS; see][]{kovacs:2002:BLS} was then used to search for periodic transit-like signals. The resulting discovery \lcs\ are shown in Figure \ref{fig:hatsouth}, phase-folded to the highest likelihood periods. This figure contains both the full phase \lcs\ for all four systems, as well as an expanded section around the transit, binned data points and the best fit model. Clear transit signals are readily visible. We highlight the case of \hatcur{40} where the apparent transit depth is $\sim$\hatcurLCdip{40}\,mmag, which is comparable to the smallest transit depths of previous HATSouth discovered planets \citep[HATS-9b, HATS-12b and HATS-17b][respectively]{brahm:2015,rabus:2016,brahm:2016}. HATSouth is able to consistently detect transit signals of a few mmag depth for its target magnitude range down to $V=15$. 

After having removed the best fit Box Least Squares model corresponding to the hot-Jupiter transit signal from the \lcs, we searched for additional periodic signals in an attempt to identify other transiting planets or potential stellar photometric activity. None of the \lcs\, revealed any other significant signals, where ``significant'' is defined by the formal false alarm probability (assuming Gaussian white noise, of less than 0.1\%) on a second BLS pass of the residuals. Additionally, a Generalised Lomb Scargle \citep[GLS,][]{zechmeister:2009} search for sinusoidal patterns related to stellar activity (either in the form of spots or pulsations) detected no significant periodic signals. We conclude there is no evidence for additional transiting planets in the systems, or clear evidence of photometric activity in the host stars. We note, additionally, that three of our targets were present in overlapping regions for multiple cameras on the same site, and therefore were observed by multiple cameras simultaneously. This further adds to a robust photometric signal where some systematic errors are averaged out by data combination from multiple sources. Depending on the characteristics and sampling of the light curves under analysis, the process of applying the TFA algorithm occasionally removes astrophysical signals that may have an impact on the conclusions regarding each system. We therefore looked for periodic signals in the pre-TFA light curves. A sinusoidal signal with a period of 29.04 days is detected with a false alarm probability of $10^{-12}$ in the light curve for \hatcur{39}. The false alarm probability is based on bootstrap simulations. This signal is most likely an instrumental artifact associated with systematic variations in the sky background corresponding the lunar orbital period. No other significant periodic signals are found in the light curves of the remaining targets.

\startlongtable
\ifthenelse{\boolean{emulateapj}}{
    \begin{deluxetable*}{llrrrr}
}{
    \begin{deluxetable}{llrrrr}
}
\tablewidth{0pc}
\tabletypesize{\scriptsize}
\tablecaption{
    Summary of photometric observations
    \label{tab:photobs}
}
\tablehead{
    \multicolumn{1}{c}{Instrument/Field\tablenotemark{a}} &
    \multicolumn{1}{c}{Date(s)} &
    \multicolumn{1}{c}{\# Images} &
    \multicolumn{1}{c}{Cadence\tablenotemark{b}} &
    \multicolumn{1}{c}{Filter} &
    \multicolumn{1}{c}{Precision\tablenotemark{c}} \\
    \multicolumn{1}{c}{} &
    \multicolumn{1}{c}{} &
    \multicolumn{1}{c}{} &
    \multicolumn{1}{c}{(sec)} &
    \multicolumn{1}{c}{} &
    \multicolumn{1}{c}{(mmag)}
}
\startdata
\sidehead{\textbf{\hatcur{39}}}
~~~~HS-2.3/G602 & 2011 Aug--2012 Feb & 4942 & 295 & $r$ & 9.0 \\
~~~~HS-4.3/G602 & 2011 Aug--2012 Feb & 1835 & 304 & $r$ & 8.9 \\
~~~~HS-6.3/G602 & 2011 Oct--2012 Feb & 1362 & 302 & $r$ & 10.4 \\
~~~~HS-2.4/G602 & 2011 Aug--2012 Feb & 4098 & 295 & $r$ & 11.2 \\
~~~~HS-4.4/G602 & 2011 Aug--2012 Feb & 3044 & 296 & $r$ & 9.2 \\
~~~~HS-6.4/G602 & 2011 Oct--2012 Feb & 1207 & 303 & $r$ & 10.3 \\
~~~~LCOGT~1\,m+CTIO/sinistro & 2015 Oct 23 & 67 & 159 & $i$ & 1.6 \\
~~~~LCOGT~1\,m+SSO/SBIG & 2015 Nov 11 & 53 & 132 & $i$ & 2.1 \\
~~~~LCOGT~1\,m+SAAO/SBIG & 2015 Dec 31 & 80 & 134 & $i$ & 2.4 \\
~~~~LCOGT~1\,m+CTIO/sinistro & 2016 Jan 09 & 122 & 159 & $i$ & 1.1 \\
~~~~Swope~1\,m/e2v & 2016 Jan 09 & 449 & 59 & $i$ & 2.5 \\
\sidehead{\textbf{\hatcur{40}}}
~~~~HS-2.3/G600 & 2012 Sep--2013 Apr & 7339 & 281 & $r$ & 11.3 \\
~~~~HS-4.3/G600 & 2012 Sep--2013 Apr & 2908 & 291 & $r$ & 11.1 \\
~~~~HS-6.3/G600 & 2012 Sep--2013 Jan & 2954 & 289 & $r$ & 10.8 \\
~~~~HS-4.4/G600 & 2012 Sep--2013 Feb & 2313 & 291 & $r$ & 12.7 \\
~~~~HS-6.4/G600 & 2012 Sep & 3 & $\cdots$ & $r$ & 17.9 \\
~~~~HS-1.2/G601 & 2011 Aug--2012 Jan & 4806 & 296 & $r$ & 8.3 \\
~~~~HS-3.2/G601 & 2011 Aug--2012 Jan & 4072 & 296 & $r$ & 8.5 \\
~~~~HS-5.2/G601 & 2011 Aug--2012 Jan & 3081 & 290 & $r$ & 9.0 \\
~~~~PEST~0.3\,m & 2014 Nov 07 & 88 & 133 & $R_{C}$ & 4.2 \\
~~~~LCOGT~1\,m+SSO/SBIG & 2015 Mar 08 & 69 & 196 & $i$ & 3.5 \\
~~~~LCOGT~1\,m+SSO/SBIG & 2015 Apr 13 & 54 & 195 & $i$ & 3.2 \\
~~~~LCOGT~1\,m+CTIO/sinistro & 2015 Sep 06 & 11 & 222 & $i$ & 2.7 \\
~~~~LCOGT~1\,m+SAAO/SBIG & 2015 Sep 13 & 28 & 194 & $i$ & 2.7 \\
~~~~LCOGT~1\,m+SAAO/SBIG & 2015 Oct 06 & 61 & 192 & $i$ & 2.3 \\
~~~~LCOGT~1\,m+SAAO/SBIG & 2015 Nov 04 & 27 & 192 & $i$ & 2.0 \\
~~~~LCOGT~1\,m+SSO/SBIG & 2015 Dec 07 & 79 & 192 & $i$ & 3.1 \\
~~~~LCOGT~1\,m+CTIO/sinistro & 2015 Dec 13 & 95 & 219 & $i$ & 1.3 \\
~~~~LCOGT~1\,m+CTIO/SBIG & 2015 Dec 23 & 107 & 192 & $i$ & 2.8 \\
~~~~LCOGT~1\,m+SSO/SBIG & 2015 Dec 30 & 8 & 193 & $i$ & 1.9 \\
~~~~LCOGT~1\,m+SAAO/SBIG & 2016 Jan 02 & 62 & 195 & $i$ & 2.7 \\
~~~~LCOGT~1\,m+SAAO/SBIG & 2016 Jan 05 & 55 & 192 & $i$ & 4.2 \\
\sidehead{\textbf{\hatcur{41}}}
~~~~HS-1.2/G601 & 2011 Aug--2012 Jan & 4790 & 296 & $r$ & 6.5 \\
~~~~HS-3.2/G601 & 2011 Aug--2012 Jan & 4059 & 296 & $r$ & 7.1 \\
~~~~HS-5.2/G601 & 2011 Aug--2012 Jan & 3089 & 290 & $r$ & 6.3 \\
~~~~LCOGT~1\,m+CTIO/sinistro & 2014 Nov 30 & 55 & 229 & $i$ & 1.1 \\
~~~~Swope~1\,m/e2v & 2014 Nov 30 & 171 & 99 & $i$ & 1.8 \\
~~~~LCOGT~1\,m+CTIO/sinistro & 2015 Sep 07 & 35 & 162 & $i$ & 1.7 \\
~~~~LCOGT~1\,m+SAAO/SBIG & 2015 Oct 15 & 65 & 137 & $i$ & 1.9 \\
\sidehead{\textbf{\hatcur{42}}}
~~~~HS-1.4/G601 & 2011 Aug--2012 Jan & 4840 & 296 & $r$ & 10.2 \\
~~~~HS-3.4/G601 & 2011 Aug--2012 Jan & 4033 & 296 & $r$ & 10.8 \\
~~~~HS-5.4/G601 & 2011 Aug--2012 Jan & 3075 & 290 & $r$ & 10.2 \\
~~~~HS-2.1/G602 & 2011 Aug--2012 Feb & 5247 & 295 & $r$ & 8.8 \\
~~~~HS-4.1/G602 & 2011 Aug--2012 Feb & 2621 & 297 & $r$ & 9.9 \\
~~~~HS-6.1/G602 & 2011 Oct--2012 Feb & 1394 & 303 & $r$ & 9.5 \\
~~~~Swope~1\,m/e2v & 2016 Jan 08 & 181 & 99 & $i$ & 2.3 \\
\enddata
\tablenotetext{a}{
    For HATSouth data we list the HATSouth unit, CCD and field name
    from which the observations are taken. HS-1 and -2 are located at
    Las Campanas Observatory in Chile, HS-3 and -4 are located at the
    HESS site in Namibia, and HS-5 and -6 are located at Siding
    Spring Observatory in Australia. Each unit has 4 CCDs. Each field
    corresponds to one of 838 fixed pointings used to cover the full
    4$\pi$ celestial sphere. All data from a given HATSouth field and
    CCD number are reduced together, while detrending through External
    Parameter Decorrelation (EPD) is done independently for each
    unique unit+CCD+field combination.
}
\tablenotetext{b}{
    The median time between consecutive images rounded to the nearest
    second. Due to factors such as weather, the day--night cycle,
    guiding and focus corrections the cadence is only approximately
    uniform over short timescales.
}
\tablenotetext{c}{
    The RMS of the residuals from the best-fit model.
} \ifthenelse{\boolean{emulateapj}}{
    \end{deluxetable*}
}{
    \end{deluxetable}
}

\ifthenelse{\boolean{emulateapj}}{
    \begin{figure*}[!ht]
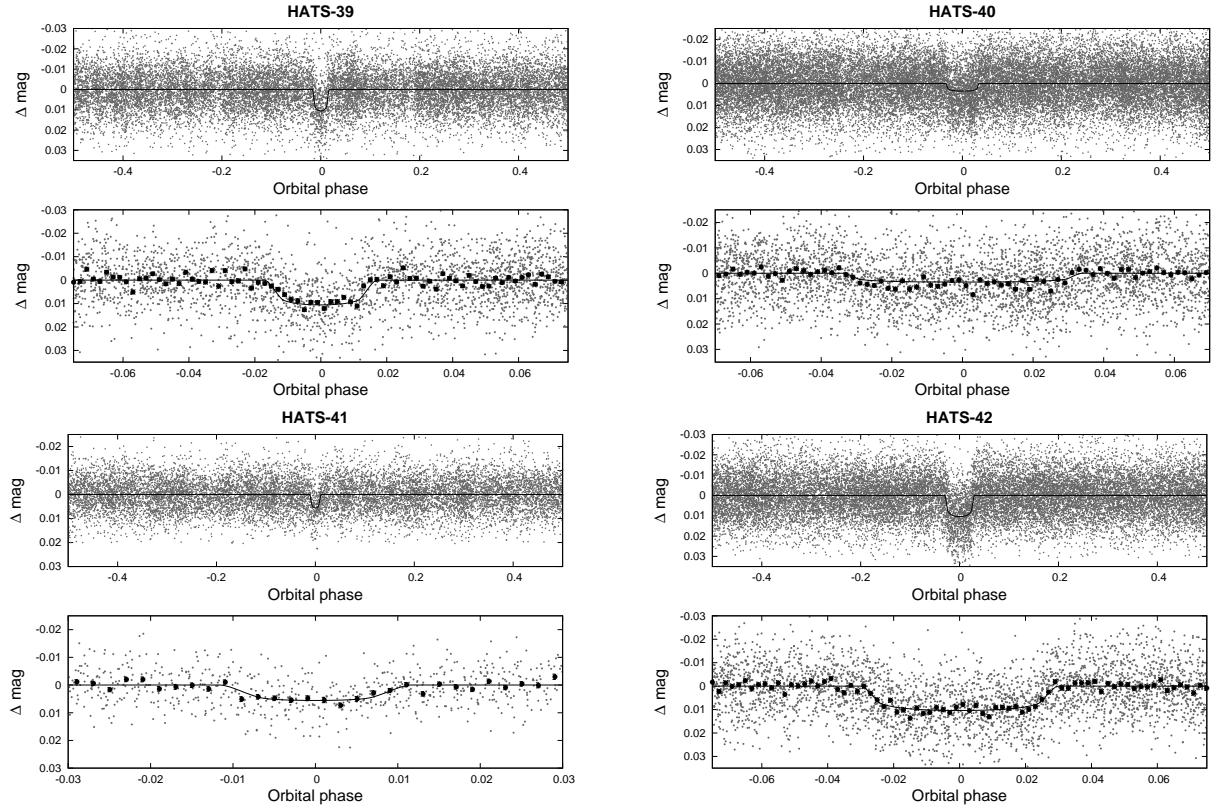

}{
    \begin{figure}[!ht]
}
\plottwo{img/\hatcurhtr{39}-hs.eps}{img/\hatcurhtr{40}-hs.eps}
\plottwo{img/\hatcurhtreccen{41}-eccen-hs.eps}{img/\hatcurhtr{42}-hs.eps}
\caption{
    Phase-folded unbinned HATSouth light curves for \hatcur{39} (upper left), \hatcur{40} (upper right), \hatcur{41} (lower left) and \hatcur{42} (lower right). In each case we show two panels. The
    top panel shows the full light curve, while the bottom panel shows
    the light curve zoomed-in on the transit. The solid lines show the
    model fits to the light curves. The dark filled circles in the
    bottom panels show the light curves binned in phase with a bin
    size of 0.002.
\label{fig:hatsouth}}
\ifthenelse{\boolean{emulateapj}}{
    \end{figure*}
}{
    \end{figure}
}

\subsection{Spectroscopic Observations}
\label{sec:obsspec}

In this section we describe our spectroscopic follow-up observations, from initial candidate vetting, through to orbital characterisation.

\subsubsection{Reconnaissance spectroscopic observations}
\label{sec:recspec}

The initial follow-up phase for HATSouth planet candidates utilised spectra acquired with the WiFeS instrument on the 2.3m ANU telescope at Siding Spring Observatory \citep[SSO;][]{Dopita:2007}. For our targets this combination delivers low-resolution spectra, over a wide wavelength range at high speed -- upwards of 60 targets per night can be easily observed. The purpose of these reconnaissance observations is to quickly eliminate those systems whose detectable transits are clearly not from planets. Observations at $R \equiv \Delta \lambda / \lambda \approx 3000$ using the blue arm of the spectrograph are used to determine the stellar type of the host star. We estimate three key stellar properties, the effective temperature $\teff$, $\loggstar$ and $\feh$ by performing a $\chi ^2$ minimisation grid search  between each the observed, normalised spectrum and synthetic templates from the MARCS model atmospheres \citep{gustafsson:2008}. 2MASS J--K colors are used to restrict the $\teff$ parameter space and extinction correction is applied using the method of \citet{cardelli:1989}. A detailed description of the observing and data reduction procedure is described in \citet{bayliss:2013:hats3}. These data identify giant host stars, for which the observed dip in the \lc\, could only have been caused by a stellar companion, and to identify stars not suitable for precise radial velocity follow-up due to high \teff\, or large \vsini. 

\ifthenelse{\boolean{emulateapj}}{
    \begin{deluxetable*}{llrrrrr}
}{
    \begin{deluxetable}{llrrrrrrrr}
}
\tablewidth{0pc}
\tabletypesize{\scriptsize}
\tablecaption{
    Summary of spectroscopic observations
    \label{tab:specobs}
}
\tablehead{
    \multicolumn{1}{c}{Instrument}          &
    \multicolumn{1}{c}{UT Date(s)}             &
    \multicolumn{1}{c}{\# Spec.}   &
    \multicolumn{1}{c}{Res.}          &
    \multicolumn{1}{c}{S/N Range\tablenotemark{a}}           &
    \multicolumn{1}{c}{$\gamma_{\rm RV}$\tablenotemark{b}} &
    \multicolumn{1}{c}{Precision\tablenotemark{c}} \\
    &
    &
    &
    \multicolumn{1}{c}{$\Delta \lambda$/$\lambda$/1000} &
    &
    \multicolumn{1}{c}{(\kms)}              &
    \multicolumn{1}{c}{(\ms)}
}
\startdata
\sidehead{\textbf{\hatcur{39}}}\\
ANU~2.3\,m/WiFeS & 2014 Feb 23 & 1 & 3 & 65 & $\cdots$ & $\cdots$ \\
ANU~2.3\,m/WiFeS & 2014 Jun--Dec & 3 & 7 & 7--82 & 1.2 & 4000 \\
ESO~3.6\,m/HARPS & 2015 Feb--2016 Apr & 17 & 115 & 15--31 & 2.916 & 39 \\
AAT~3.9\,m/CYCLOPS2+UCLES & 2015 March 1 & 3 & 70 & 15-17 & 3.092 & 89 \\
\sidehead{\textbf{\hatcur{40}}}\\
ANU~2.3\,m/WiFeS & 2014 Oct 7 & 1 & 3 & 42 & $\cdots$ & $\cdots$ \\
ANU~2.3\,m/WiFeS & 2014 Oct 8--11 & 2 & 7 & 14--24 & 9.8 & 4000 \\
Euler~1.2\,m/Coralie & 2014 Oct--2015 Oct & 6 & 60 & 10--14 & 9.30 & 460 \\
ESO~3.6\,m/HARPS & 2015 Feb--Nov & 10 & 115 & 6--17 & 9.194 & 49 \\
\sidehead{\textbf{\hatcur{41}}}\\
ANU~2.3\,m/WiFeS & 2014 Oct 7 & 1 & 3 & 54 & $\cdots$ & $\cdots$ \\
ANU~2.3\,m/WiFeS & 2014 Oct 8--10 & 2 & 7 & 51--62 & 33.3 & 4000 \\
Euler~1.2\,m/Coralie & 2014 Oct--2016 Jan & 11 & 60 & 14--29 & 37.08 & 440 \\
AAT~3.9\,m/CYCLOPS2+UCLES & 2015 Feb--May & 8 & 70 & 13-17 & 38.00 & 375 \\
ESO~3.6\,m/HARPS & 2015 Nov--2016 Mar & 5 & 115 & 13--25 & 37.25 & 275 \\
\sidehead{\textbf{\hatcur{42}}}\\
ANU~2.3\,m/WiFeS & 2014 Jun 4 & 1 & 3 & 40 & $\cdots$ & $\cdots$ \\
ANU~2.3\,m/WiFeS & 2014 Dec 11--12 & 2 & 7 & 34--46 & 7.4 & 4000 \\
ESO~3.6\,m/HARPS & 2015 Apr--Nov & 5 & 115 & 11--18 & 8.163 & 34 \\
MPG~2.2\,m/FEROS & 2016 Jan 16--21 & 4 & 48 & 41--47 & 8.131 & 90 \\
\enddata 
\tablenotetext{a}{
    S/N per resolution element near 5180\,\AA.
}
\tablenotetext{b}{
    For high-precision radial velocity observations included in the orbit determination this is the zero-point radial velocity from the best-fit orbit. For other instruments it is the mean value. We do not provide this quantity for the lower resolution WiFeS observations which were only used to measure stellar atmospheric parameters.
}
\tablenotetext{c}{
    For high-precision radial velocity observations included in the orbit
    determination this is the scatter in the radial velocity residuals from the
    best-fit orbit (which may include astrophysical jitter), for other
    instruments this is either an estimate of the precision (not
    including jitter), or the measured standard deviation. We do not
    provide this quantity for low-resolution observations from the
    ANU~2.3\,m/WiFeS.
}
\ifthenelse{\boolean{emulateapj}}{
    \end{deluxetable*}
}{
    \end{deluxetable}
}

Targets not eliminated by these data are observed at predicted quadrature phases using a WiFES higher resolving power grating ($R \sim 7,000$) to obtain radial velocity measurements with $\sim 2 \kms$ precision (the true precision varies depending on stellar type and signal-to-noise of each individual target). 
Radial velocities are measured by cross-correlation against velocity standards observed every night, calibrated using bracketed NeAr exposures and a selection of telluric lines.  This allows the detection of radial velocity variations with amplitudes above $\sim 5 \kms$, which indicate that the transiting companion is a star. The results of these initial vetting observations for our four targets are:

\begin{itemize}

\item \hatcur{39} has an effective temperature of $6460 \pm 300 K$, $\loggstar$ of $3.9 \pm 0.3$ and metallicity of $\feh = -0.5 \pm 0.5$, leading to the conclusion that this is a F-dwarf host star. Two radial velocity measurements at each quadrature showed no significant variation.

\item \hatcur{40} has an effective temperature of $6720 \pm 300 K$, $\loggstar$ of $4.0 \pm 0.3$ and metallicity of $\feh = 0.0 \pm 0.5$. We conclude that the host star is an F dwarf. Two radial velocity measurements showed no significant variation, though they were both obtained near the same quadrature phase.

\item \hatcur{41} has an effective temperature of $6327 \pm 300 K$, $\loggstar$ of $3.9 \pm 0.3$  and metallicity of $\feh = 0.0 \pm 0.5$. We conclude that the target is an F dwarf. The two radial velocity measurements taken at either quadrature phase showed no significant variation.

\item \hatcur{42} was measured to have an effective temperature of $6249 \pm 300 K$, $\loggstar$ of $4.0 \pm 0.3$  and metallicity of $\feh = 0.0 \pm 0.5$. Based on this we conclude that the target is a G- or F-dwarf. Two radial velocity measurements taken at either quadrature phase also showed no significant variation.

\end{itemize}

This initial vetting excluded these targets as giant host stars, and except for \hatcur{40}, as eclipsing binaries, and thus all were then promoted to the next phase in the follow-up campaign, leading to further higher
radial-velocity-precision spectroscopy and photometric follow-up. 
In the case of \hatcur{40} we began collecting higher precision RV observations before both quadrature phases were covered by WiFeS, and it became clear from these data that this object is not an eclipsing binary, and further WiFeS observations were not needed.

\subsubsection{High-precision spectroscopic observations}
\label{sec:highspec}

A full radial velocity characterisation covering a wide portion of the orbital phase of all of our targets is required in order to determine fundamental parameters such as the planetary masses and orbital eccentricities. As such, observations were performed with a range of facilities capable of high precision radial velocity measurements on single visits. Exposures were taken with the High Accuracy Radial Velocity Planet Searcher \cite[HARPS]{mayor:2003}, fed by the ESO 3.6m telescope at a resolving power of $R \sim 115,000$, the FEROS spectrograph \citep[][ $R \sim 48,000$]{kaufer:1998} fed by the MPG 2.2m telescope, and spectra at $R \sim 60,000$ were also taken with the CORALIE spectrograph \citep{queloz:2001} fed by the 1.2m Euler telescope, all located at La Silla Observatory, Chile. The data reduction for all these spectra was performed using the method described in \citet{jordan:2014:hats4} and \citet{brahm:2017:ceres}. Additionally, eleven spectra of \hatcur{39} and \hatcur{41} were also obtained with the CYCLOPS2 fibre-feed and the UCLES spectrograph on the 3.9m Anglo-Australian telescope (AAT) at SSO at a resolving power of $R \sim 70,000$. These data were reduced using the methods described in \cite{addison:2013}. Further details about these observations can be found in Table \ref{tab:specobs}. The resulting data sets for all targets can be found in Table \ref{tab:rvs}, and are shown in Figure \ref{fig:rvbis}, which includes radial velocity curves, best-fit models and bisector span (BS) \citep{queloz:2001} estimates shown in the bottom panels for each target. All systems clearly show a radial velocity variation consistent with the detected transit ephemeris from the photometric \lcs\, and no clear correlation between the radial velocity measurements and the bisector-spans, indicating the systems are likely bona fide transiting planets (see Section \ref{sec:blend}). We note that bisector span measurements from CYCLOPS2+UCLES are not available as the pipeline does not have the facility to measure these at this time.

\startlongtable
\tabletypesize{\scriptsize}
\ifthenelse{\boolean{emulateapj}}{
    \begin{deluxetable*}{lrrrrrl}
}{
    \begin{deluxetable}{lrrrrrl}
}
\tablewidth{0pc}
\tablecaption{
    Relative radial velocities (RV) and bisector spans (BS) for \hatcur{39}--\hatcur{42}.
    \label{tab:rvs}
}
\tablehead{
    \colhead{BJD} &
    \colhead{RV\tablenotemark{a}} &
    \colhead{\ensuremath{\sigma_{\rm RV}}\tablenotemark{b}} &
    \colhead{BS\tablenotemark{c}} &
    \colhead{\ensuremath{\sigma_{\rm BS}}} &
    \colhead{Phase} &
    \colhead{Instrument}\\
    \colhead{\hbox{(2,450,000$+$)}} &
    \colhead{(\ms)} &
    \colhead{(\ms)} &
    \colhead{(\ms)} &
    \colhead{(\ms)} &
    \colhead{} &
    \colhead{}
}
\startdata
\multicolumn{7}{c}{\bf HATS-39} \\
\hline\\
    $ 7067.62167 $ & $   -25.89 $ & $    22.00 $ & $   95.0 $ & $   19.0 $ & $   0.897 $ & HARPS \\
$ 7068.65140 $ & $  -102.89 $ & $    33.00 $ & $   97.0 $ & $   30.0 $ & $   0.122 $ & HARPS \\
$ 7069.66662 $ & $   -76.89 $ & $    23.00 $ & $   -2.0 $ & $   21.0 $ & $   0.344 $ & HARPS \\
$ 7070.67868 $ & $   -22.89 $ & $    20.00 $ & $    7.0 $ & $   17.0 $ & $   0.565 $ & HARPS \\
$ 7071.64200 $ & $    75.11 $ & $    17.00 $ & $    9.0 $ & $   14.0 $ & $   0.776 $ & HARPS \\
$ 7072.64113 $ & $    13.11 $ & $    26.00 $ & $    9.0 $ & $   21.0 $ & $   0.994 $ & HARPS \\
$ 7083.02827 $ & $    16.80 $ & $    17.10 $ & \nodata      & \nodata      & $   0.263 $ & CYCLOPS \\
$ 7083.04429 $ & $  -140.40 $ & $    16.80 $ & \nodata      & \nodata      & $   0.267 $ & CYCLOPS \\
$ 7083.06038 $ & $  -134.10 $ & $    39.50 $ & \nodata      & \nodata      & $   0.270 $ & CYCLOPS \\
$ 7118.58681 $ & $    -8.89 $ & $    22.00 $ & $   -7.0 $ & $   19.0 $ & $   0.031 $ & HARPS \\
$ 7119.52550 $ & $     3.11 $ & $    25.00 $ & $  -51.0 $ & $   21.0 $ & $   0.236 $ & HARPS \\
$ 7120.57144 $ & $    10.11 $ & $    33.00 $ & $  146.0 $ & $   26.0 $ & $   0.464 $ & HARPS \\
$ 7329.85316 $ & $  -106.89 $ & $    35.00 $ & $   37.0 $ & $   26.0 $ & $   0.183 $ & HARPS \\
$ 7331.84959 $ & $    72.11 $ & $    21.00 $ & $  -49.0 $ & $   17.0 $ & $   0.619 $ & HARPS \\
$ 7331.86410 $ & $    82.11 $ & $    21.00 $ & $  -28.0 $ & $   17.0 $ & $   0.622 $ & HARPS \\
$ 7332.83972 $ & $    73.11 $ & $    21.00 $ & $    9.0 $ & $   19.0 $ & $   0.835 $ & HARPS \\
$ 7466.55350 $ & $    29.11 $ & $    32.00 $ & $   16.0 $ & $   21.0 $ & $   0.045 $ & HARPS \\
$ 7467.54434 $ & $   -33.89 $ & $    18.00 $ & $   -1.0 $ & $   11.0 $ & $   0.262 $ & HARPS \\
$ 7468.55331 $ & $     2.11 $ & $    18.00 $ & $  -25.0 $ & $   11.0 $ & $   0.482 $ & HARPS \\
$ 7495.58378 $ & $   -40.89 $ & $    20.00 $ & $   24.0 $ & $   14.0 $ & $   0.387 $ & HARPS \\
 \cutinhead{\bf HATS-40}
    $ 7067.56779 $ & $   -78.13 $ & $    44.00 $ & \nodata      & \nodata      & $   0.133 $ & HARPS \\
$ 7069.58400 $ & $   216.87 $ & $    41.00 $ & $   60.0 $ & $   29.0 $ & $   0.751 $ & HARPS \\
$ 7070.56482 $ & $  -113.13 $ & $    51.00 $ & \nodata      & \nodata      & $   0.051 $ & HARPS \\
$ 7071.57968 $ & $  -148.13 $ & $    39.00 $ & \nodata      & \nodata      & $   0.362 $ & HARPS \\
$ 7072.59059 $ & $   127.87 $ & $    58.00 $ & \nodata      & \nodata      & $   0.672 $ & HARPS \\
$ 7118.55616 $ & $    60.87 $ & $    93.00 $ & $ -420.0 $ & $   94.0 $ & $   0.753 $ & HARPS \\
$ 7119.49884 $ & $   -51.13 $ & $    48.00 $ & $   59.0 $ & $   34.0 $ & $   0.042 $ & HARPS \\
$ 7120.49030 $ & $  -109.13 $ & $    36.00 $ & $  -58.0 $ & $   26.0 $ & $   0.346 $ & HARPS \\
$ 7329.74873 $ & $   -12.13 $ & $    79.00 $ & $ -298.0 $ & $   47.0 $ & $   0.452 $ & HARPS \\
$ 7332.77781 $ & $   -66.13 $ & $    38.00 $ & $  -19.0 $ & $   26.0 $ & $   0.380 $ & HARPS \\
 \cutinhead{\bf HATS-41}
    $ 6939.87473 $ & $   237.04 $ & $   112.00 $ & $ -404.0 $ & $   21.0 $ & $   0.373 $ & Coralie \\
$ 6969.82238 $ & $   107.04 $ & $   117.00 $ & $ -399.0 $ & $   22.0 $ & $   0.515 $ & Coralie \\
$ 6971.81055 $ & $  -693.96 $ & $   146.00 $ & $ -617.0 $ & $   35.0 $ & $   0.989 $ & Coralie \\
$ 7080.00880 $ & $   666.54 $ & $   118.60 $ & \nodata      & \nodata      & $   0.789 $ & CYCLOPS \\
$ 7080.07261 $ & $   431.24 $ & $   223.90 $ & \nodata      & \nodata      & $   0.804 $ & CYCLOPS \\
$ 7080.08861 $ & $  1289.54 $ & $    31.60 $ & \nodata      & \nodata      & $   0.808 $ & CYCLOPS \\
$ 7082.92936 $ & $  -171.56 $ & $    90.80 $ & \nodata      & \nodata      & $   0.486 $ & CYCLOPS \\
$ 7082.94532 $ & $    26.94 $ & $    59.80 $ & \nodata      & \nodata      & $   0.489 $ & CYCLOPS \\
$ 7082.96128 $ & $   157.44 $ & $    74.90 $ & \nodata      & \nodata      & $   0.493 $ & CYCLOPS \\
$ 7109.59087 $ & $   -68.96 $ & $   166.00 $ & $ -654.0 $ & $   32.0 $ & $   0.843 $ & Coralie \\
$ 7150.86119 $ & $  1139.44 $ & $   111.50 $ & \nodata      & \nodata      & $   0.684 $ & CYCLOPS \\
$ 7150.87728 $ & $   985.24 $ & $    87.40 $ & \nodata      & \nodata      & $   0.688 $ & CYCLOPS \\
$ 7282.89965 $ & $ -1104.96 $ & $   117.00 $ & $ -296.0 $ & $   29.0 $ & $   0.170 $ & Coralie \\
$ 7312.79578 $ & $  -656.96 $ & $   107.00 $ & $ -166.0 $ & $   26.0 $ & $   0.299 $ & Coralie \\
$ 7318.75358 $ & $   336.04 $ & $   120.00 $ & $  239.0 $ & $   26.0 $ & $   0.719 $ & Coralie \\
$ 7329.76516 $ & $  -584.98 $ & $    83.00 $ & $ -121.0 $ & $   49.0 $ & $   0.345 $ & HARPS \\
$ 7331.78101 $ & $   744.02 $ & $    52.00 $ & $ -238.0 $ & $   26.0 $ & $   0.826 $ & HARPS \\
$ 7332.79161 $ & $ -1197.98 $ & $    57.00 $ & $  -72.0 $ & $   31.0 $ & $   0.067 $ & HARPS \\
$ 7408.60163 $ & $ -1167.96 $ & $    82.00 $ & $  -44.0 $ & $   22.0 $ & $   0.144 $ & Coralie \\
$ 7409.56696 $ & $   367.04 $ & $    93.00 $ & $ -679.0 $ & $   25.0 $ & $   0.374 $ & Coralie \\
$ 7410.69245 $ & $  1233.04 $ & $    79.00 $ & $ -307.0 $ & $   21.0 $ & $   0.643 $ & Coralie \\
$ 7411.56574 $ & $   912.04 $ & $   137.00 $ & $ -10178.0 $ & $   26.0 $ & $   0.851 $ & Coralie \\
$ 7467.50487 $ & $  -918.98 $ & $    69.00 $ & $ -134.0 $ & $   24.0 $ & $   0.190 $ & HARPS \\
$ 7468.49670 $ & $   426.02 $ & $    81.00 $ & $   11.0 $ & $   28.0 $ & $   0.426 $ & HARPS \\
 \cutinhead{\bf HATS-42}
    $ 7119.55661 $ & $  -222.35 $ & $    35.00 $ & $    7.0 $ & $   38.0 $ & $   0.113 $ & HARPS \\
$ 7120.51754 $ & $    49.65 $ & $    20.00 $ & $ -100.0 $ & $   27.0 $ & $   0.532 $ & HARPS \\
$ 7330.82381 $ & $  -241.35 $ & $    31.00 $ & $  -20.0 $ & $   38.0 $ & $   0.285 $ & HARPS \\
$ 7331.76070 $ & $   223.65 $ & $    17.00 $ & $   37.0 $ & $   21.0 $ & $   0.693 $ & HARPS \\
$ 7332.80517 $ & $  -194.35 $ & $    29.00 $ & $   18.0 $ & $   32.0 $ & $   0.149 $ & HARPS \\
$ 7403.81476 $ & $  -112.92 $ & $    15.00 $ & $   -1.0 $ & $   15.0 $ & $   0.129 $ & FEROS \\
$ 7404.83808 $ & $   185.08 $ & $    17.00 $ & $  126.0 $ & $   17.0 $ & $   0.576 $ & FEROS \\
$ 7407.54463 $ & $   113.08 $ & $    16.00 $ & $ -143.0 $ & $   16.0 $ & $   0.756 $ & FEROS \\
$ 7408.68903 $ & $  -271.92 $ & $    16.00 $ & $  -80.0 $ & $   15.0 $ & $   0.256 $ & FEROS \\
 \enddata
\tablenotetext{a}{
    The zero-point of these velocities is arbitrary. An overall offset
    $\gamma_{\rm rel}$ fitted independently to the velocities from
    each instrument has been subtracted.
}
\tablenotetext{b}{
    Internal errors excluding the component of astrophysical jitter
    considered in \refsecl{globmod}.
}
\tablenotetext{c}{
    Bisector span measurements are only shown for observations in which the automated routines in the individual instrument pipelines were able to determine them. For cases where the peak of the cross-correlated function was too low to obtain a reliable measurement, these values are not presented.
    }
%\tablenotetext{c}{
%    These observations were excluded from the analysis because the observations were (partially) obtained with the planet in transit, and thus may be affected by the Rossiter-McLaughlin effect. 
%}
% \ifthenelse{\boolean{rvtablelong}}{
%     \tablecomments{
%     }
% }{
%     \tablecomments{
%     }
% } 
\ifthenelse{\boolean{emulateapj}}{
    \end{deluxetable*}
}{
    \end{deluxetable}
}

\setcounter{planetcounter}{1}
\ifthenelse{\boolean{emulateapj}}{
    \begin{figure*} [ht]
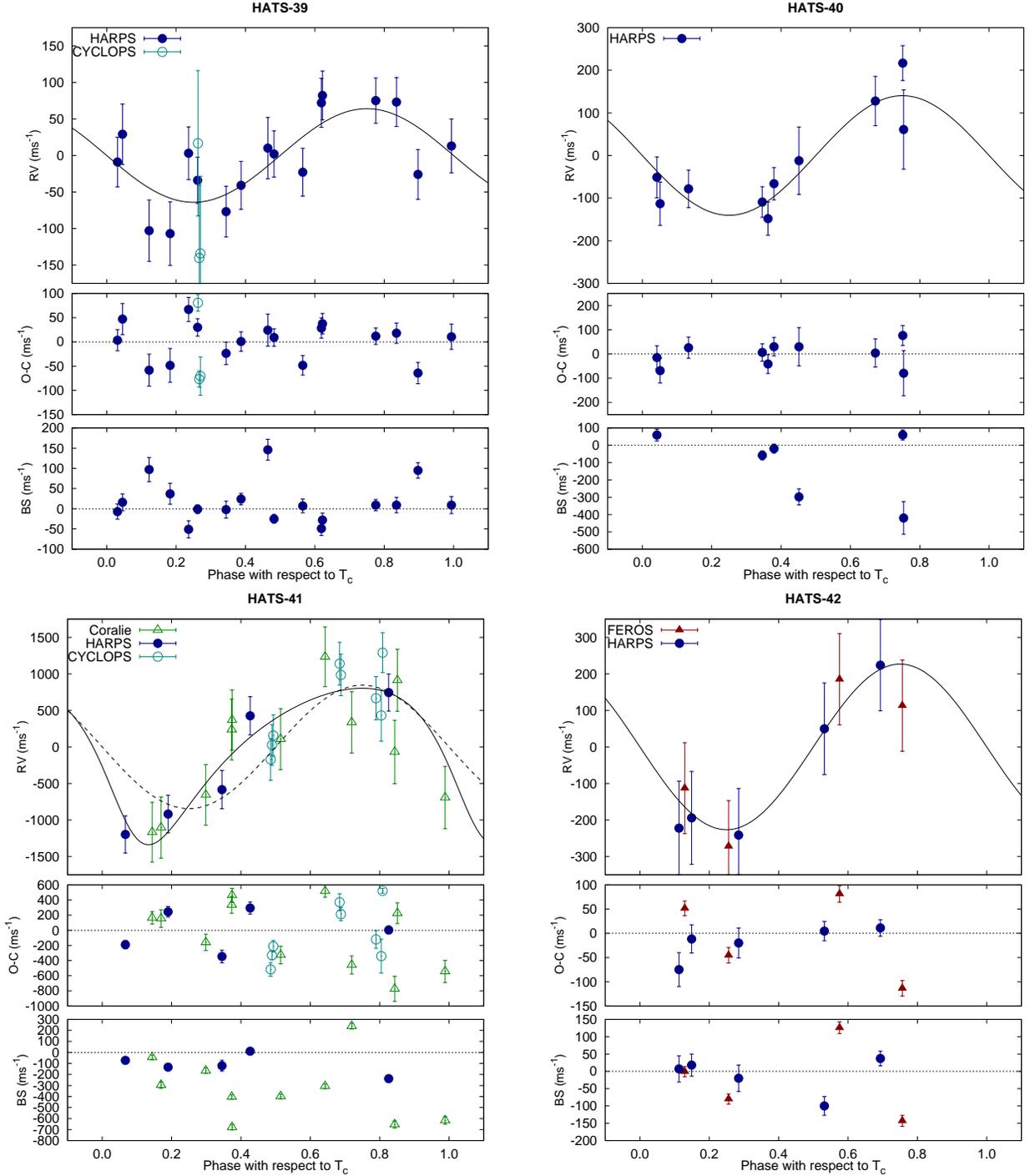

}{
    \begin{figure}[ht]
}
\plottwo{img/\hatcurhtr{39}-rv.eps}{img/\hatcurhtr{40}-rv.eps}
\plottwo{img/\hatcurhtreccen{41}-rv.eps}{img/\hatcurhtr{42}-rv.eps}
\caption{
    Phased high-precision radial velocity measurements for \hbox{\hatcur{39}{}} (upper left), \hbox{\hatcur{40}{}} (upper right), \hbox{\hatcur{41}{}} (lower left) and \hbox{\hatcur{42}{}} (lower right). The instruments used are labeled in the plots. In each case we show three panels. The top panel shows the phased measurements together with our best-fit model (see \reftabl{planetparam}) for each system. Zero-phase corresponds to the time of mid-transit. The center-of-mass velocity has been subtracted. The second panel shows the velocity $O\!-\!C$ residuals from the best fit. The error bars in the middle panel correspond to the formal uncertainties only and, on the top panel, we show final uncertainties with the jitter terms listed in \reftabl{planetparam} added in quadrature. The third panel shows the bisector spans (BS). Note the different vertical scales of the panels. We include the zero eccentricity model fit for the case of \hatcur{41} for reference. 
}
\label{fig:rvbis}
\ifthenelse{\boolean{emulateapj}}{
    \end{figure*}
}{
    \end{figure}
}

\subsection{Photometric follow-up observations}
\label{sec:phot}

\begin{figure}[!ht]
\epsscale{1.2}
\plotone{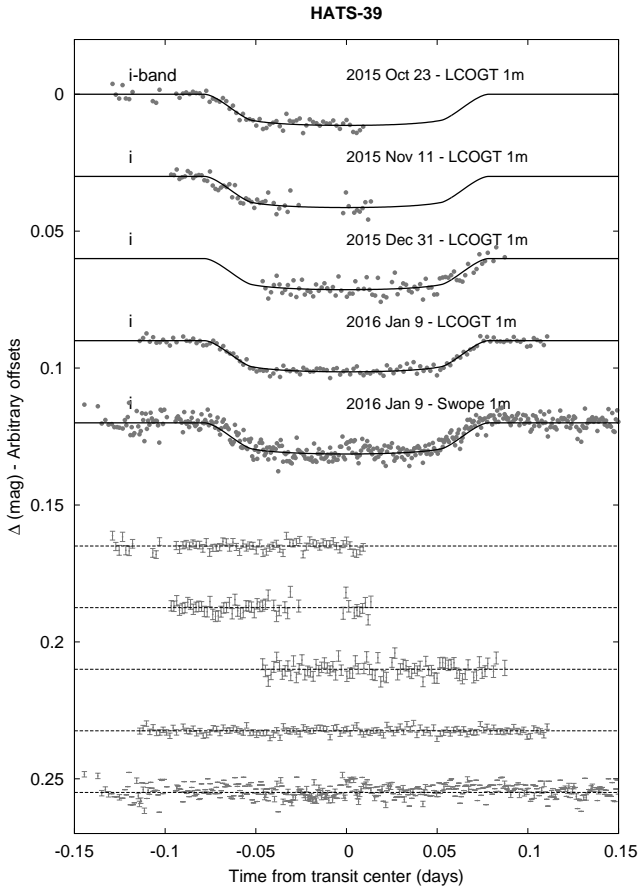}
\caption{
    Unbinned transit \lcs{} for \hatcur{39}.  The light curves have been
    corrected for quadratic trends in time, and linear trends with up
    to three parameters characterizing the shape of the PSF, fitted
    simultaneously with the transit model.
    The dates of the events, filters and instruments used are
    indicated.  Light curves following the first are displaced
    vertically for clarity.  Our best fit from the global modeling
    described in \refsecl{globmod} is shown by the solid lines. The
    residuals from the best-fit model are shown below in the same
    order as the original light curves.  The error bars represent the
    photon and background shot noise, plus the readout noise.
}
\label{fig:lc:39}
\end{figure}

\begin{figure*}[!ht]
\epsscale{0.95}
\plotone{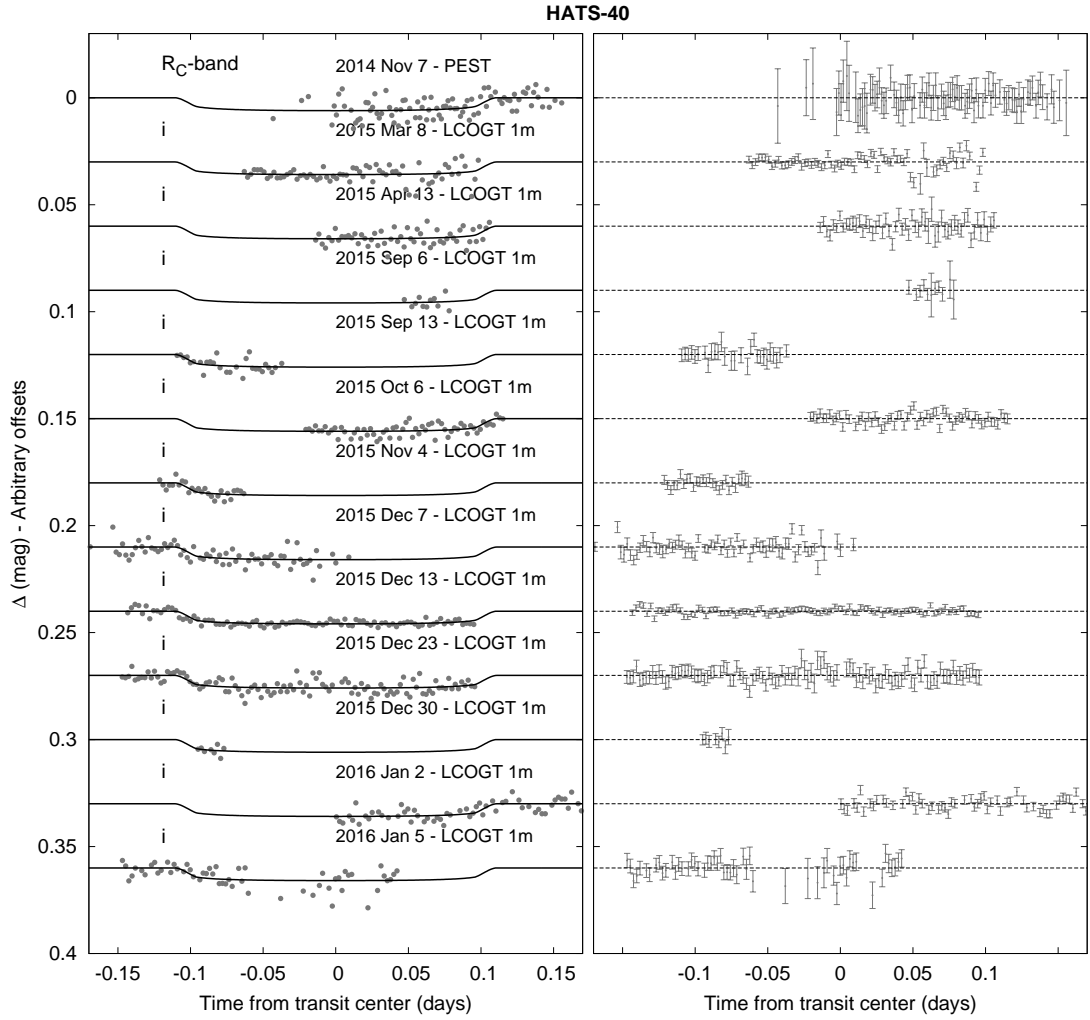}
\caption{
    Similar to Fig.~\ref{fig:lc:39}, here we show \lcs{} for \hatcur{40}. In this case the residuals are plotted on the right-hand-side of the figure, in the same order as the original light curves on the left-hand-side.
}
\label{fig:lc:40}
\end{figure*}

Photometric follow-up is also undertaken to both confirm the transit signal and improve \lcs\ parameter estimates for each system. All four candidates were observed  with the LCOGT network \citep{brown:2013:lcogt} -- specifically using the 1m aperture telescopes of this network in the \emph{i} band, which obtained several full- and partial-transits for \hatcur{39}, \hatcur{40} and \hatcur{41}. Additionally, a partial transit of \hatcur{40} was observed with the PEST 0.3m telescope in Western Australia in the $R_C$ band. Full transits of \hatcur{39} and \hatcur{42}, as well as a partial transit of \hatcur{41}, were also observed with the Swope 1m telescope in Las Campanas, Chile in the \emph{i} band. This data set includes a full transit of \hatcur{39} observed simultaneously with both Swope and LCOGT on Jan 9th 2016. The photometric data were acquired using the same strategy, and reduced using a customisable pipeline and the methods described in \cite{penev:2013:hats1}, with details of setup in \citet{bayliss:2015}. This pipeline uses standard photometric reduction frames (master bias, darks, twilight flats) and the DAOPHOT aperture photometry package for flux extraction of target and comparison stars. A quadratic trend in time, as well as variations correlated with point-spread-function shape, were fitted simultaneously with the transit shape to compensate for variable seeing and differential refraction. We assume an ellipsoidal Gaussian PSF parameterized by 

\begin{equation}
e^{-\frac{1}{2}(S(x^2+y^2) + D(x^2-y^2) + K(2xy))}, 
\end{equation}

where the coefficients $S$, $D$ and $K$ are allowed to vary freely and can be mapped to FWHM, elongation and position angle. These photometric follow-up observations are summarized in \reftabl{photobs}, and all the resulting photometric data are available in electronic format in Table \ref{tab:phfu}. The full set of photometric follow-up \lcs\ are shown in Figures~\ref{fig:lc:39}, \ref{fig:lc:40}, \ref{fig:lc:41}, and \ref{fig:lc:42}, for \hatcur{39}, \hatcur{40}, \hatcur{41}, and \hatcur{42}, with the data plotted along with the best fit models and residuals plotted underneath.

\setcounter{planetcounter}{1}

\begin{figure*}[!ht]
\epsscale{0.85}
\plotone{img/\hatcurhtreccen{41}-eccen-lc.eps}
\caption{
    Same as Fig.~\ref{fig:lc:40}, here we show \lcs{} for \hatcur{41}.
}
\label{fig:lc:41}
\end{figure*}

\begin{figure}[!ht]
\epsscale{1.2}
\plotone{img/\hatcurhtr{42}-lc.eps}
\caption{
    Same as Fig.~\ref{fig:lc:41}, here we show \lc{} for \hatcur{42}.
}
\label{fig:lc:42}
\end{figure}

%\clearpage

\ifthenelse{\boolean{emulateapj}}{
    \begin{deluxetable*}{llrrrrl}
}{
    \begin{deluxetable}{llrrrrl}
}
\tablewidth{0pc}
\tablecaption{
    Light curve data for \hatcur{39}, \hatcur{40}, \hatcur{41} and \hatcur{42}\label{tab:phfu}.
}
\tablehead{
    \colhead{Object\tablenotemark{a}} &
    \colhead{BJD\tablenotemark{b}} & 
    \colhead{Mag\tablenotemark{c}} & 
    \colhead{\ensuremath{\sigma_{\rm Mag}}} &
    \colhead{Mag(orig)\tablenotemark{d}} & 
    \colhead{Filter} &
    \colhead{Instrument} \\
    \colhead{} &
    \colhead{\hbox{~~~~(2,400,000$+$)~~~~}} & 
    \colhead{} & 
    \colhead{} &
    \colhead{} & 
    \colhead{} &
    \colhead{}
}
\startdata
   HATS-39 & $ 55939.70502 $ & $  -0.01394 $ & $   0.00416 $ & $ \cdots $ & $ r$ &  HS/G602.3\\
   HATS-39 & $ 55916.81716 $ & $  -0.00641 $ & $   0.00364 $ & $ \cdots $ & $ r$ &  HS/G602.3\\
   HATS-39 & $ 55948.86125 $ & $  -0.00034 $ & $   0.00414 $ & $ \cdots $ & $ r$ &  HS/G602.3\\
   HATS-39 & $ 55962.59487 $ & $  -0.00385 $ & $   0.00494 $ & $ \cdots $ & $ r$ &  HS/G602.3\\
   HATS-39 & $ 55880.19892 $ & $   0.01622 $ & $   0.00511 $ & $ \cdots $ & $ r$ &  HS/G602.3\\
   HATS-39 & $ 55971.75158 $ & $   0.00319 $ & $   0.00420 $ & $ \cdots $ & $ r$ &  HS/G602.3\\
   HATS-39 & $ 55843.57819 $ & $   0.02082 $ & $   0.00564 $ & $ \cdots $ & $ r$ &  HS/G602.3\\
   HATS-39 & $ 55939.70882 $ & $  -0.02094 $ & $   0.00429 $ & $ \cdots $ & $ r$ &  HS/G602.3\\
   HATS-39 & $ 55916.82201 $ & $   0.00981 $ & $   0.00372 $ & $ \cdots $ & $ r$ &  HS/G602.3\\
   HATS-39 & $ 55880.20276 $ & $   0.01193 $ & $   0.00527 $ & $ \cdots $ & $ r$ &  HS/G602.3\\
 \enddata
\tablenotetext{a}{
    Either \hatcur{39}, \hatcur{40}, \hatcur{41} or \hatcur{42}.
}
\tablenotetext{b}{
    Barycentric Julian Date is computed directly from the UTC time
    without correction for leap seconds.
}
\tablenotetext{c}{
    The out-of-transit level has been subtracted. For observations
    made with the HATSouth instruments (identified by ``HS'' in the
    ``Instrument'' column) these magnitudes have been corrected for
    trends using the EPD and TFA procedures applied {\em prior} to
    fitting the transit model. This procedure may lead to an
    artificial dilution in the transit depths. The blend factors for
    the HATSouth light curves are listed in
    Table~\ref{tab:planetparam}. For
    observations made with follow-up instruments (anything other than
    ``HS'' in the ``Instrument'' column), the magnitudes have been
    corrected for a quadratic trend in time, and for variations
    correlated with up to three PSF shape parameters, fit simultaneously
    with the transit.
}
\tablenotetext{d}{
    Raw magnitude values without correction for the quadratic trend in
    time, or for trends correlated with the seeing. These are only
    reported for the follow-up observations.
}
\tablecomments{
    This table is available in a machine-readable form in the online
    journal.  A portion is shown here for guidance regarding its form
    and content.
}
\ifthenelse{\boolean{emulateapj}}{
    \end{deluxetable*}
}{
    \end{deluxetable}
}

\subsection{Lucky Imaging}
\label{sec:luckyimaging}

\ifthenelse{\boolean{emulateapj}}{
    \begin{figure*}[!ht]
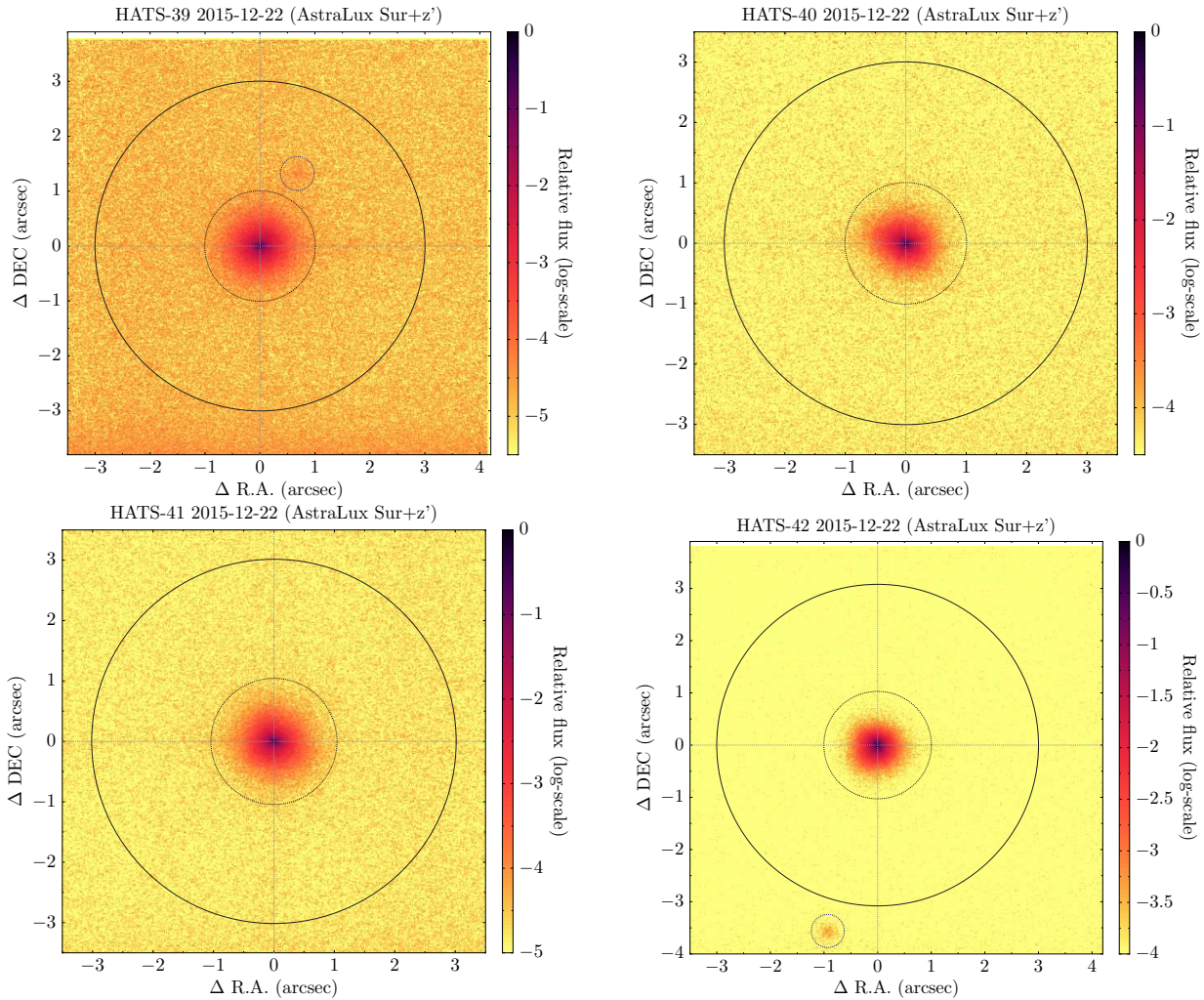

}{
    \begin{figure}[!ht]
}
\plottwo{img/\hatcur{39}_astralux.eps}{img/\hatcur{40}_astralux.eps}
\plottwo{img/\hatcur{41}_astralux.eps}{img/\hatcur{42}_astralux.eps}
\caption{
    Astralux lucky images of \hatcur{39} ({\em top left}), \hatcur{40} ({\em top right}), \hatcur{41} ({\em bottom left}), and \hatcur{42} ({\em bottom right}). A neighboring source to \hatcur{39}, indicated with a red circle, is detected at $2\sigma$ confidence at a separation of $\sim 2\arcsec$ in Declination and $\sim -1\arcsec$ in Right Ascension. If real, it has $\Delta z^{\prime} = 5.65 \pm 0.35$\,mag relative to \hatcur{39}. No neighboring sources are detected in the observations of \hatcur{40}, or \hatcur{41}. \hatcur{42} has a real companion at $\sim -3.5\arcsec$ in Declination and $\sim -1\arcsec$ in Right Ascension also detected by the Gaia space observatory, but too faint to affect our results. 
\label{fig:luckyimages}}
\ifthenelse{\boolean{emulateapj}}{
    \end{figure*}
}{
    \end{figure}
}

\ifthenelse{\boolean{emulateapj}}{
    \begin{figure*}[!ht]
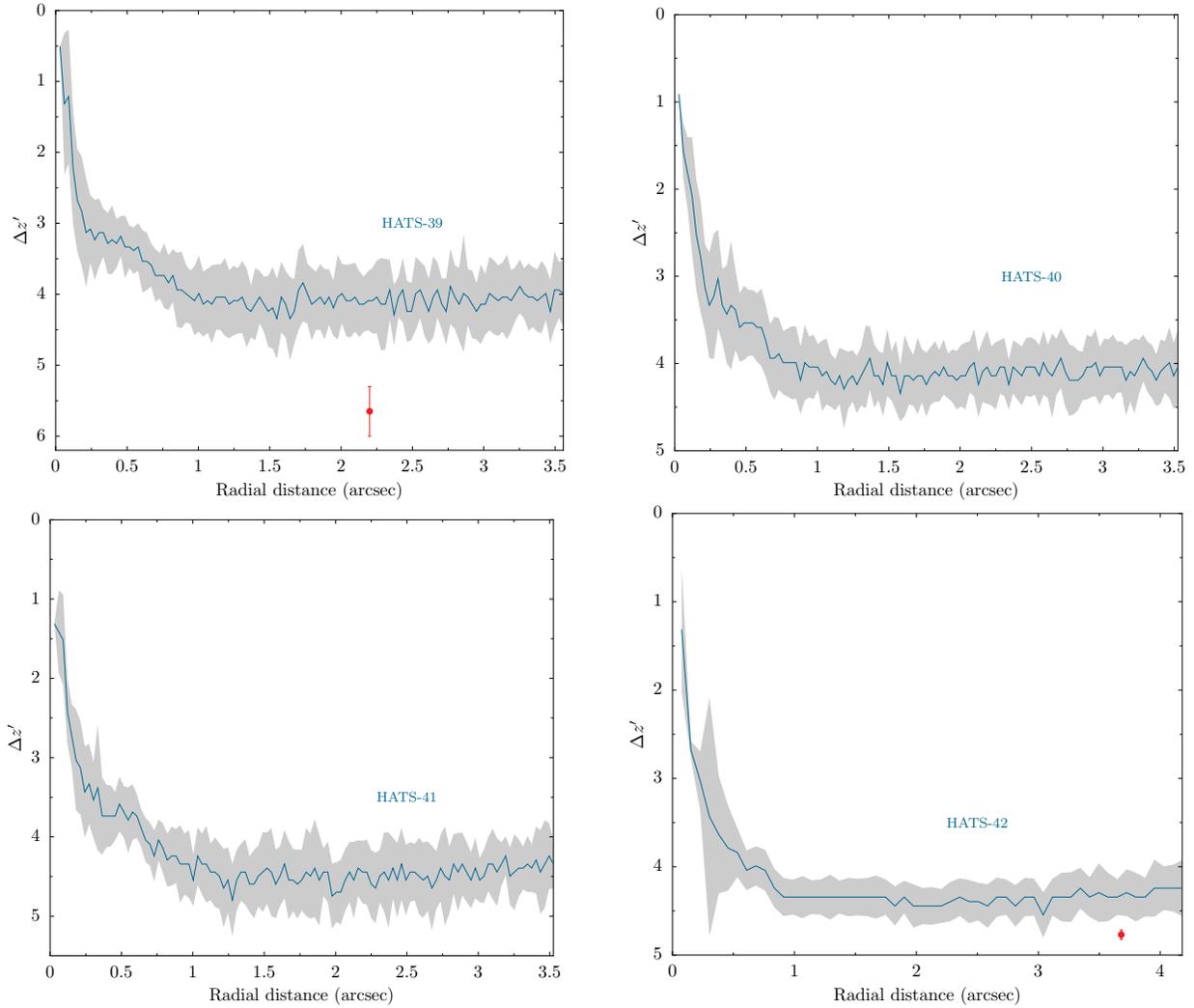

}{
    \begin{figure}[!ht]
}
\plottwo{img/contrast_curve_\hatcur{39}.eps}{img/contrast_curve_\hatcur{40}.eps}
\plottwo{img/contrast_curve_\hatcur{41}.eps}{img/contrast_curve_\hatcur{42}.eps}
\caption{
    $5\sigma$ contrast curves for \hatcur{39} ({\em upper left}), \hatcur{40} ({\em upper right}), \hatcur{41} ({\em lower left}) and \hatcur{42} ({\em lower right}) based on our AstraLux Sur $z^{\prime}-band$ observations. The candidate neighboring source to \hatcur{39} detected with $2\sigma$ confidence is indicated by the filled circle and errorbar, and the same is done for the companion of \hatcur{42}. Additional details on these companions can be found in the main text and on the footnotes of Table \ref{tab:stellar}. Gray bands show the uncertainty given by the scatter in the contrast in the azimuthal direction at a given radius.
\label{fig:luckyimagecontrastcurves}}
\ifthenelse{\boolean{emulateapj}}{
    \end{figure*}
}{
    \end{figure}
}

Lucky imaging observations were obtained through a $z^{\prime}$ filter
for all four systems using the Astralux Sur camera
\citep{hippler:2009} on the New Technology Telescope (NTT) at La Silla
Observatory in Chile on the night of 2015 December 22. Observations
with this facility were carried out and reduced following
\citet{espinoza:2016:hats25hats30}, but a plate scale of $\rm 15.20\, mas\, pixel^{-1}$ was used, derived in the work of \cite{janson:2017}. Figure~\ref{fig:luckyimages} shows
the reduced final images for each system, while
Figure~\ref{fig:luckyimagecontrastcurves} shows the $5\sigma$ contrast
curves based on these images produced using the technique and software
described in \citet{espinoza:2016:hats25hats30}. 

For \hatcur{39} we achieve an effective FWHM for the final image of $0\farcs0368 \pm 0\farcs0046$, equivalent to $2.42 \pm 0.30$ pixels. For this object a neighboring source is detected at $1\farcs32 \pm 0\farcs02$ in Declination and $0\farcs68 \pm 0\farcs02$ in RA (i.e., at a distance of $2\farcs2$ from the target; errors on RA an DEC are obtained as the effective FWHM divided by 2.355) from the target at $\sim 2\sigma$ confidence. The apparent source, if real, has $\Delta z^{\prime} = 5.65 \pm 0.35$ mag relative to \hatcur{39}, and cannot be responsible for the transits. This candidate neighbour also has a negligible impact on the inferred parameters of the \hatcurb{39} system. 

For \hatcur{40} we obtained an effective FWHM of $2.92 \pm 0.35$ pixels, or $0\farcs0444 \pm 0\farcs0053$ and no companion sources were detected. Similarly, for \hatcur{41} we obtained an effective FWHM of $2.64 \pm 0.34$ pixels, or $0\farcs0401 \pm 0\farcs0052$ and no companions were detected.

In the case of \hatcur{42} (effective FWHM of $5.01 \pm 0.32$ pixels, equivalent to $0\farcs0761 \pm 0\farcs0049$) a nearby source is also detected at a $\sim 2\sigma$ level. The target is at $-3\farcs56 \pm 0\farcs03$ in Dec and $-0\farcs93 \pm 0\farcs03$ in RA (i.e., at a distance of $3\farcs68$ from the target). The magnitude difference for these two stars is $4.769 \pm 0.052$. This nearby target is also detected by the Gaia space observatory \citep{lindegren:2016} at a separation of $-3\farcs6187 \pm 0\farcs0003$ in Dec and $-0\farcs9104 \pm 0.0002$ in RA, numbers which are in perfect agreement with our values; they also find a magnitude difference in the \emph{g} band of 3.553 magnitudes, confirming the existence of this target as real. This source is, however, not able to be responsible for the observed transits at this brightness. 

\section{Analysis}
\label{sec:analysis}
\begin{comment}
\end{comment}

\subsection{Properties of the parent star}
\label{sec:stelparam}

We used the Zonal Atmospheric Stellar Parameter Estimator \citep[ZASPE;][]{brahm:2016:ZASPE} to model the stellar parameters of all four host stars. ZASPE is capable of precise stellar atmospheric parameter estimation from high-resolution echelle spectra of FGK-type stars. It compares the observed continuum-normalized spectrum with a grid of synthetic spectra by a least squares minimisation in the most sensitive regions of the stellar spectrum. The complete FGK-type star parameter space is searched using this method. To take into account the microturbulence dependence of the line widths, we use an empirical relation between the microturbulence and the stellar parameters. In particular, we used the stellar parameters provided by the SweetCat catalogue \citep{santos:2013} define a polynomial that delivers the microturbulence as function of effective temperature and \logg. Then, the microturbulence value used in the synthesis of each spectrum was obtained using that empirical function. More details on this method can be found in \citet{brahm:2016:ZASPE}. We performed this analysis on the combined HARPS spectra for \hatcur{39}, \hatcur{40} and \hatcur{41}, and on the FEROS spectra for \hatcur{42}.

We calculate an initial estimate of the effective temperature (\teff), the surface gravity (\logg), metallicity (\feh) and projected stellar rotational velocity of the stars (\vsini). Following \citet{sozzetti:2007}, we used the stellar density $\rhostar$, which was determined from the modeling described in Section~\ref{sec:globmod}, together with \teff\ and \feh\ to determine the other physical parameters of the host star through a comparison with the Yonsei-Yale \citep[Y2;][]{yi:2001} isochrones.
If the value of \loggstar\ from the stellar evolution modeling is discrepant from the value determined in the initial ZASPE analysis of the spectrum by more than 1$\sigma$, we perform a second iteration of ZASPE using \loggstar\ determined from the isochrones, followed by a second iteration of the analysis in Section~\ref{sec:globmod} and comparison to the Y2 isochrones. This was done to improve the results for \hatcur{39} and \hatcur{42}. The second iteration was not needed for the other two candidates. We present the adopted results and an extensive set of host star parameters from several sources in Table \ref{tab:stellar}.

All host stars were found to be F-type dwarf stars.
We find \hatcur{39} to be a solar metallicity star with $\teff = \hatcurSMEteff{39}$\,K,  mass $\mstar = \hatcurISOmlongeccen{39}$ \msun\, and radius $\rstar = \hatcurISOrlongeccen{39}$ \rsun. \hatcur{40} is also solar metallicity, but more massive and larger ($\mstar = \hatcurISOmlongeccen{40}$ \msun\, and $\rstar = \hatcurISOrlongeccen{40}$ \rsun\,). 
\hatcur{41} and \hatcur{42} have metallicities above the solar one with $\feh = \hatcurSMEzfeh{41} $ and $\feh = \hatcurSMEzfeh{42}$ (respectively). Both stars are also somewhat above solar mass, and have similar radii. We refer the reader to Table \ref{tab:stellar} for further details. 

Distances to these stars were determined by comparing the measured broad-band photometry listed in Table \ref{tab:stellar} to the predicted magnitudes in each filter from the isochrones. We assumed a $R_{V} = 3.1$ extinction law from \citet{cardelli:1989} to determine the extinction and find these to be consistent within their uncertainties to reddening maps available on the NASA/IPAC infrared science archive\footnote{Publicly available at \\\url{http://irsa.ipac.caltech.edu/applications/DUST/}}. The locations of each star on an $\teffstar$--$\rhostar$ diagram (similar to a Hertzsprung-Russell diagram) are shown in \reffigl{iso}. 
%{\em CGT: $\rhostar$ is never defined. I can work out what it is, but its not defined. Also, while $\rhostar$ is "like" a H-R diagram, it really doesn't get used much, and it isn't explained why its being used here. I've seen logg vs Teff diagramas far more frequently. In any case in the absence of an observed main sequence and/or giant branch, Fig. 9 really doesn't seem to convery much information. Especially as the isochrones are unlabelled! Use the power of plotting pcdes to write the age on each isochrone!}

\ifthenelse{\boolean{emulateapj}}{
    \begin{figure*}[!ht]
}{
    \begin{figure}[!ht]
}
\plottwo{img/\hatcurhtr{39}-iso-rho.eps}{img/\hatcurhtr{40}-iso-rho.eps}
\plottwo{img/\hatcurhtreccen{41}-iso-rho.eps}{img/\hatcurhtr{42}-iso-rho.eps}
\caption{Plots of stellar density \rhostar\ as a function of effective temperature $\teffstar$  for the four new exoplanet discoveries. Model isochrones from \cite{\hatcurisocite{39}} for the measured metallicities of \hatcur{39} (upper left), \hatcur{40} (upper right), \hatcur{41} (lower left) and \hatcur{42} (lower right) are also shown as the black solid lines. These models have been chosen for a starting age of 0.2\,Gyr, and then a range from 1.0 to 14.0\,Gyr in 1\,Gyr increments (ages increasing from left to right). We also plot in green dashed lines the evolutionary tracks for stars with masses listed in Solar units. The adopted values of $\teffstar$ and \rhostar\ are shown in the filled circle together with their 1$\sigma$ and 2$\sigma$ confidence ellipsoids. The initial values of \teffstar\ and \rhostar\ from the first ZASPE and \lc\ analyses of \hatcur{39} and \hatcur{42} are represented with open triangles. The other two candidates did not require a second iteration.
}
\label{fig:iso}
\ifthenelse{\boolean{emulateapj}}{
    \end{figure*}
}{
    \end{figure}
}

\ifthenelse{\boolean{emulateapj}}{
    \begin{deluxetable*}{lccccl}
}{
    \begin{deluxetable}{lccccl}
}
\tablewidth{0pc}
\tabletypesize{\scriptsize}
\tablecaption{
    Stellar parameters for \hatcur{39}--\hatcur{42}
    \label{tab:stellar}
}
\tablehead{
    \multicolumn{1}{c}{} &
    \multicolumn{1}{c}{\bf HATS-39} &
    \multicolumn{1}{c}{\bf HATS-40} &
    \multicolumn{1}{c}{\bf HATS-41} &
    \multicolumn{1}{c}{\bf HATS-42} &
    \multicolumn{1}{c}{} \\
    \multicolumn{1}{c}{~~~~~~~~Parameter~~~~~~~~} &
    \multicolumn{1}{c}{Value}                     &
    \multicolumn{1}{c}{Value}                     &
    \multicolumn{1}{c}{Value}                     &
    \multicolumn{1}{c}{Value}                     &
    \multicolumn{1}{c}{Source}
}
\startdata
\noalign{\vskip -3pt}
\sidehead{Astrometric properties and cross-identifications}
~~~~2MASS-ID\dotfill               & \hatcurCCtwomassshort{39}  & \hatcurCCtwomassshort{40} & \hatcurCCtwomassshort{41} & \hatcurCCtwomassshort{42} & \\
~~~~GSC-ID\dotfill                 & \hatcurCCgsc{39}      & \hatcurCCgsc{40}     & \hatcurCCgsceccen{41}     & \hatcurCCgsc{42}     & \\
~~~~R.A. (J2000)\dotfill            & \hatcurCCra{39}       & \hatcurCCra{40}    & \hatcurCCraeccen{41}    & \hatcurCCra{42}    & 2MASS\\
~~~~Dec. (J2000)\dotfill            & \hatcurCCdec{39}      & \hatcurCCdec{40}   & \hatcurCCdececcen{41}   & \hatcurCCdec{42}   & 2MASS\\
~~~~$\mu_{\rm R.A.}$ (\masy)              & \hatcurCCpmra{39}     & \hatcurCCpmra{40} & \hatcurCCpmraeccen{41} & \hatcurCCpmra{42} & UCAC4\\
~~~~$\mu_{\rm Dec.}$ (\masy)              & \hatcurCCpmdec{39}    & \hatcurCCpmdec{40} & \hatcurCCpmdececcen{41} & \hatcurCCpmdec{42} & UCAC4\\
\sidehead{Spectroscopic properties}
~~~~$\teffstar$ (K)\dotfill         &  \hatcurSMEteff{39}   & \hatcurSMEteff{40} & \hatcurSMEteff{41} & \hatcurSMEteff{42} & ZASPE\tablenotemark{a}\\
~~~~$\feh$\dotfill                  &  \hatcurSMEzfeh{39}   & \hatcurSMEzfeh{40} & \hatcurSMEzfeh{41} & \hatcurSMEzfeh{42} & ZASPE               \\
~~~~$\vsini$ (\kms)\dotfill         &  \hatcurSMEvsin{39}   & \hatcurSMEvsin{40} & \hatcurSMEvsin{41} & \hatcurSMEvsin{42} & ZASPE                \\
~~~~$\vmac$ (\kms)\dotfill          &  $\hatcurSMEvmac{39}$   & $\hatcurSMEvmac{40}$ & $\hatcurSMEvmac{41}$ & $\hatcurSMEvmac{42}$ & Assumed              \\
~~~~$\vmic$ (\kms)\dotfill          &  $\hatcurSMEvmic{39}$   & $\hatcurSMEvmic{40}$ & $\hatcurSMEvmic{41}$ & $\hatcurSMEvmic{42}$ & Assumed              \\
~~~~$\gamma_{\rm RV}$ (\ms)\dotfill&  \hatcurRVgammaabs{39}  & \hatcurRVgammaabs{40} & \hatcurRVgammaabs{41} & \hatcurRVgammaabs{42} & HARPS\tablenotemark{b}  \\
\sidehead{Photometric properties}
~~~~$G$ (mag)\dotfill               &  12.58  & 13.22 & 12.52 & 13.48 & GAIA DR1\tablenotemark{c} \\
~~~~$B$ (mag)\dotfill               &  \hatcurCCtassmB{39}  & \hatcurCCtassmB{40} & \hatcurCCtassmBeccen{41} & \hatcurCCtassmB{42} & APASS\tablenotemark{d} \\
~~~~$V$ (mag)\dotfill               &  \hatcurCCtassmv{39}  & \hatcurCCtassmv{40} & \hatcurCCtassmveccen{41} & \hatcurCCtassmv{42} & APASS\tablenotemark{d} \\
~~~~$g$ (mag)\dotfill               &  \hatcurCCtassmg{39}  & \hatcurCCtassmg{40} & \hatcurCCtassmgeccen{41} & \hatcurCCtassmg{42} & APASS\tablenotemark{d} \\
~~~~$r$ (mag)\dotfill               &  \hatcurCCtassmr{39}  & \hatcurCCtassmr{40} & \hatcurCCtassmreccen{41} & \hatcurCCtassmr{42} & APASS\tablenotemark{d} \\
~~~~$i$ (mag)\dotfill               &  \hatcurCCtassmi{39}  & \hatcurCCtassmi{40} & \hatcurCCtassmieccen{41} & \hatcurCCtassmi{42} & APASS\tablenotemark{d} \\
~~~~$J$ (mag)\dotfill               &  \hatcurCCtwomassJmag{39} & \hatcurCCtwomassJmag{40} & \hatcurCCtwomassJmageccen{41} & \hatcurCCtwomassJmag{42} & 2MASS           \\
~~~~$H$ (mag)\dotfill               &  \hatcurCCtwomassHmag{39} & \hatcurCCtwomassHmag{40} & \hatcurCCtwomassHmageccen{41} & \hatcurCCtwomassHmag{42} & 2MASS           \\
~~~~$K_s$ (mag)\dotfill             &  \hatcurCCtwomassKmag{39} & \hatcurCCtwomassKmag{40} & \hatcurCCtwomassKmageccen{41} & \hatcurCCtwomassKmag{42} & 2MASS           \\
\sidehead{Derived properties}
~~~~$\mstar$ ($\msun$)\dotfill      &  \hatcurISOmlong{39}   & \hatcurISOmlong{40} & \hatcurISOmlongeccen{41} & \hatcurISOmlong{42} & YY+$\rhostar$+ZASPE \tablenotemark{e}\\
~~~~$\rstar$ ($\rsun$)\dotfill      &  \hatcurISOrlong{39}   & \hatcurISOrlong{40} & \hatcurISOrlongeccen{41} & \hatcurISOrlong{42} & YY+$\rhostar$+ZASPE         \\
~~~~$\loggstar$ (cgs)\dotfill       &  \hatcurISOlogg{39}    & \hatcurISOlogg{40} & \hatcurISOloggeccen{41} & \hatcurISOlogg{42} & YY+$\rhostar$+ZASPE         \\
~~~~$\rhostar$ (\gcmc) \tablenotemark{f}\dotfill       &  \hatcurLCrho{39}    & \hatcurLCrho{40} & \hatcurLCrhoeccen{41} & \hatcurLCrho{42} & Light curves         \\
~~~~$\rhostar$ (\gcmc) \tablenotemark{f}\dotfill       &  \hatcurISOrho{39}    & \hatcurISOrho{40} & \hatcurISOrhoeccen{41} & \hatcurISOrho{42} & YY+Light curves+ZASPE         \\
~~~~$\lstar$ ($\lsun$)\dotfill      &  \hatcurISOlum{39}     & \hatcurISOlum{40} & \hatcurISOlumeccen{41} & \hatcurISOlum{42} & YY+$\rhostar$+ZASPE         \\
~~~~$M_V$ (mag)\dotfill             &  \hatcurISOmv{39}      & \hatcurISOmv{40} & \hatcurISOmveccen{41} & \hatcurISOmv{42} & YY+$\rhostar$+ZASPE         \\
~~~~$M_K$ (mag,\hatcurjhkfilset{39})\dotfill &  \hatcurISOMK{39} & \hatcurISOMK{40} & \hatcurISOMKeccen{41} & \hatcurISOMK{42} & YY+$\rhostar$+ZASPE         \\
~~~~Age (Gyr)\dotfill               &  \hatcurISOage{39}     & \hatcurISOage{40} & \hatcurISOageeccen{41} & \hatcurISOage{42} & YY+$\rhostar$+ZASPE         \\
~~~~$A_{V}$ (mag)\dotfill               &  \hatcurXAv{39}     & \hatcurXAv{40} & \hatcurXAveccen{41} & \hatcurXAv{42} & YY+$\rhostar$+ZASPE         \\
~~~~Distance (pc)\dotfill           &  \hatcurXdistred{39}\phn  & \hatcurXdistred{40} & \hatcurXdistredeccen{41} & \hatcurXdistred{42} & YY+$\rhostar$+ZASPE\\
\enddata
\tablecomments{
For \hatcurb{41} we adopt a model in which the eccentricity is allowed to vary. For the other three systems we adopt a model in which the orbit is assumed to be circular. See the discussion in Section~\ref{sec:globmod}.
}
\tablenotetext{a}{
    ZASPE = Zonal Atmospherical Stellar Parameter Estimator routine
    for the analysis of high-resolution spectra
    \citep{brahm:2016:ZASPE}, applied to the HARPS spectra of \hatcur{39}, \hatcur{40}, and \hatcur{41}, and to the FEROS spectra of \hatcur{42}. These parameters rely primarily on ZASPE, but have a small
    dependence also on the iterative analysis incorporating the
    isochrone search and global modelling of the data.
}
\tablenotetext{b}{
    The error on $\gamma_{\rm RV}$ is determined from the orbital fit
    to the radial velocity measurements, and does not include the systematic
    uncertainty in transforming the velocities to the IAU standard
    system. The velocities have not been corrected for gravitational
    redshifts.
} \tablenotetext{c}{
    From GAIA Data Release 1 \citep{lindegren:2016}. \hatcur{39} has a neighbour detected $4.35\arcsec$ away with a magnitude of $G=18.27$ ($\Delta G = 5.69$). \hatcur{42} has a detected nearby source at $3.61\arcsec$ distance and with $G=17.04$ ($\Delta G = 3.56$).
} \tablenotetext{d}{
    From APASS DR6 \citep{henden:2014} for as
    listed in the UCAC 4 catalog \citep{zacharias:2012:ucac4}.  
}
\tablenotetext{e}{
    \hatcurisoshort{39}+\rhostar+ZASPE = Based on the \hatcurisoshort{39}
    isochrones \citep{\hatcurisocite{39}}, \rhostar\ as a luminosity
    indicator, and the ZASPE results.
}
\tablenotetext{f}{
    In the case of $\rhostar$ we list two values. The first value is
    determined from the global fit to the light curves and radial velocity data,
    without imposing a constraint that the parameters match the
    stellar evolution models. The second value results from
    restricting the posterior distribution to combinations of
    $\rhostar$+$\teffstar$+$\feh$ that match to a \hatcurisoshort{39}
    stellar model.
}
\ifthenelse{\boolean{emulateapj}}{
    \end{deluxetable*}
}{
    \end{deluxetable}
}

\subsection{Excluding blend scenarios}
\label{sec:blend}
\begin{comment}
\end{comment}

In order to exclude blend scenarios we carried out an analysis
following \citet{hartman:2012:hat39hat41}. We attempt to model the
available photometric data (including light curves and catalog
broad-band photometric measurements) for each object as a blend
between an eclipsing binary star system and a third star along the
line of sight. The physical properties of the stars are constrained
using the Padova isochrones \citep{girardi:2000}, while we also
require that the brightest of the three stars in the blend have
atmospheric parameters consistent with those measured with ZASPE. We
also simulate composite cross-correlation functions and use
them to predict radial velocities and bisector spans for each blend scenario considered. The results for each system are as follows:
\begin{itemize}
\item {\em \hatcur{39}} -- all blend scenarios tested give a poorer fit to the photometric data than a model consisting of a single star with a planet, though for the best-fit blend models the difference in $\chi^2$ compared to the best-fit planet model is not statistically significant. The simulated bisector spans and radial velocities for all blend models that cannot be ruled out by the photometry (i.e., those that cannot be rejected with greater than $5\sigma$ confidence) show variations in excess of 100\,\ms, and in most cases in excess of 1\,\kms.  This contrasts with the measured HARPS velocities which have a sinusoidal variation with an amplitude of $K = \hatcurRVK{39}$\,\ms, and a standard deviation (not subtracting the Keplerian orbit) of 59\,\ms. Likewise the measured HARPS bisector spans have a standard deviation of 52\,\ms, which is significantly less than the simulated values for all blend models that cannot be ruled out by the photometry. Based on this, we reject the hypothesis that \hatcur{39} is a blended stellar eclipsing binary object rather than a transiting planet system.
\item {\em \hatcur{40}} -- in this case we find that all blend scenarios tested provide a much poorer fit to the photometric data than a single star with a planet. In fact, all blend models can be rejected with a confidence greater than $4.6\sigma$, based on the photometry alone. We conclude that \hatcur{40} is not a blended stellar eclipsing binary object, but is a transiting planet system.
\item {\em \hatcur{41}} -- there exist blend models that provide slightly better fits to the photometric data than a single star with a planet. We find that the best-fit blend model (a hierarchical triple system with a bright third star having $M_3 = 1.453$\,\msun, and an eclipsing binary with $M_1 = 1.05$\,\msun, and $M_2  = 0.28$\,\msun) has a value of $\chi^2$ (based on all of the photometric data) that is 7.6 less than the value of $\chi^2$ for the best-fit model consisting of a single star with a planet. Based on Monte Carlo simulations of photometric data with pink noise properties comparable to what is observed in the light curves, this corresponds to a $1.5\sigma$ confidence difference, and is thus not a large enough difference to be statistically significant. However, we find that none of the blend models that provide a reasonable fit to the photometric data are able to simultaneously reproduce both the observed radial velocity variation with $K = \hatcurRVK{41}$\,\ms, and the measured $91$\,\ms\ scatter in the HARPS bisector span values. In general the simulated bisector span values have a correlated variation that is comparable in amplitude to the simulated radial velocity values. Blend scenarios that produce radial velocity variations at an amplitude above 1\,\kms, also result in large BS variations at amplitude above 1\,\kms, while blend scenarios that produce simulated BS variations with an amplitude below 100\,\ms, produce similarly low amplitude radial velocity variations.  We conclude that blend scenarios cannot account for all of the photometric and spectroscopic observations of \hatcur{41}, and furthermore conclude that \hatcur{41} is a transiting planet system.
\item {\em \hatcur{42}} -- like \hatcur{40}, all blend models tested can be rejected with a confidence greater than $4\sigma$, based solely on the photometry. We conclude that \hatcur{42} is a transiting planet system, and not a blended stellar eclipsing binary object.
\end{itemize}

\subsection{Global modeling of the data}
\label{sec:globmod}

We modeled the full available data for each target (initial photometry, follow-up photometry and spectroscopy) following the same method described in previous discoveries 
\citet{pal:2008:hat7,bakos:2010:hat11,hartman:2012:hat39hat41}. We fit
\citet{mandel:2002} transit models to all light curves, allowing for the possible dilution of the HATSouth transit depths as a result of blending from
neighboring stars and over-correction by the trend-filtering method. 
To correct for systematic errors in the follow-up light curves, such as airmass and pointing errors, 
we include in our model for each event a quadratic trend in time, and
linear trends with up to three parameters describing the shape of the
PSF. This ensures that seeing changes and centroiding errors are minimised. We then fit Keplerian orbits to the radial velocity curves allowing the zero-point
for each instrument to vary independently in the fit, and allowing for
radial velocity jitter, which is also allowed to vary for each
instrument. A Differential Evolution Markov Chain Monte Carlo
procedure is then performed to explore the fitness landscape and to determine the
posterior distribution of the parameters.

Note that we tried fitting both fixed circular orbits and
free-eccentricity models to the data for all 4 systems, and then use
the method of \citet{weinberg:2013} to estimate the Bayesian evidence
for each scenario. We find eccentricities consistent with zero for \hatcur{39}, \hatcur{40} and
\hatcur{42}, in which the
Bayesian evidence for the fixed circular orbit models are higher. For
these three systems we adopt the parameters from the fixed circular
orbit model solutions. For \hatcur{41} the free eccentricity model
yields a marginally significant eccentricity of $e = \hatcurRVecceneccen{41}$, with $\Delta \chi^2 = -14$ between the best-fit
free eccentricity model, and the best-fit fixed circular orbit
model. The Bayesian evidence for the fixed circular model is slightly
higher by a factor of $6.7$, but the best-fit circular orbit model yields a stellar density of \hatcurLCrho{41}\,\gcmc, which is higher than allowed by the stellar evolution models at $\teffstar = \hatcurSMEteff{41}$\,K. The free
eccentricity model, on the other hand, yields a stellar density that
falls within the range allowed by the stellar evolution models. For
\hatcur{41} we adopt the parameters from a model where the
eccentricity is allowed to vary in the fit, and include the zero eccentricity radial velocity solution in figure \ref{fig:rvbis} for comparison. The high planet-to-star mass ratio leads to an estimate of 0.5 -- 1.5 Gyr for a tidal circularisation timescale assuming present orbital characteristics and depending on assumptions on the quality factor $Q_P$ between $1-3 \times 10^5$, typical values assumed for Jovian and dense Jovian planets \citep{pont:2011}. This value is consistent with the determined age for this system, and therefore some eccentricity is not unexpected.

The resulting parameters for each system are listed in
\reftabl{planetparam}.

\section{Discussion}
\label{sec:discussion}

We report the discovery of four transiting \hjs\ orbiting F-type stars by the HATSouth survey: \hatcurb{39}, \hatcurb{40}, \hatcurb{41} and \hatcurb{42}. Among these is the particularly interesting case of \hatcurb{41} which is one of the most massive \hjs\ found to date and orbits the highest metallicity star to host a transiting planet, making it particularly important in the context of exoplanet discoveries to date. These add to the growing number of well-characterised exoplanets and provide further evidence of the diversity of these exotic worlds.

\ifthenelse{\boolean{emulateapj}}{
    \begin{figure*}[!ht]
}{
    \begin{figure}[!ht]
}
\centering
\includegraphics[width={0.8\linewidth}]{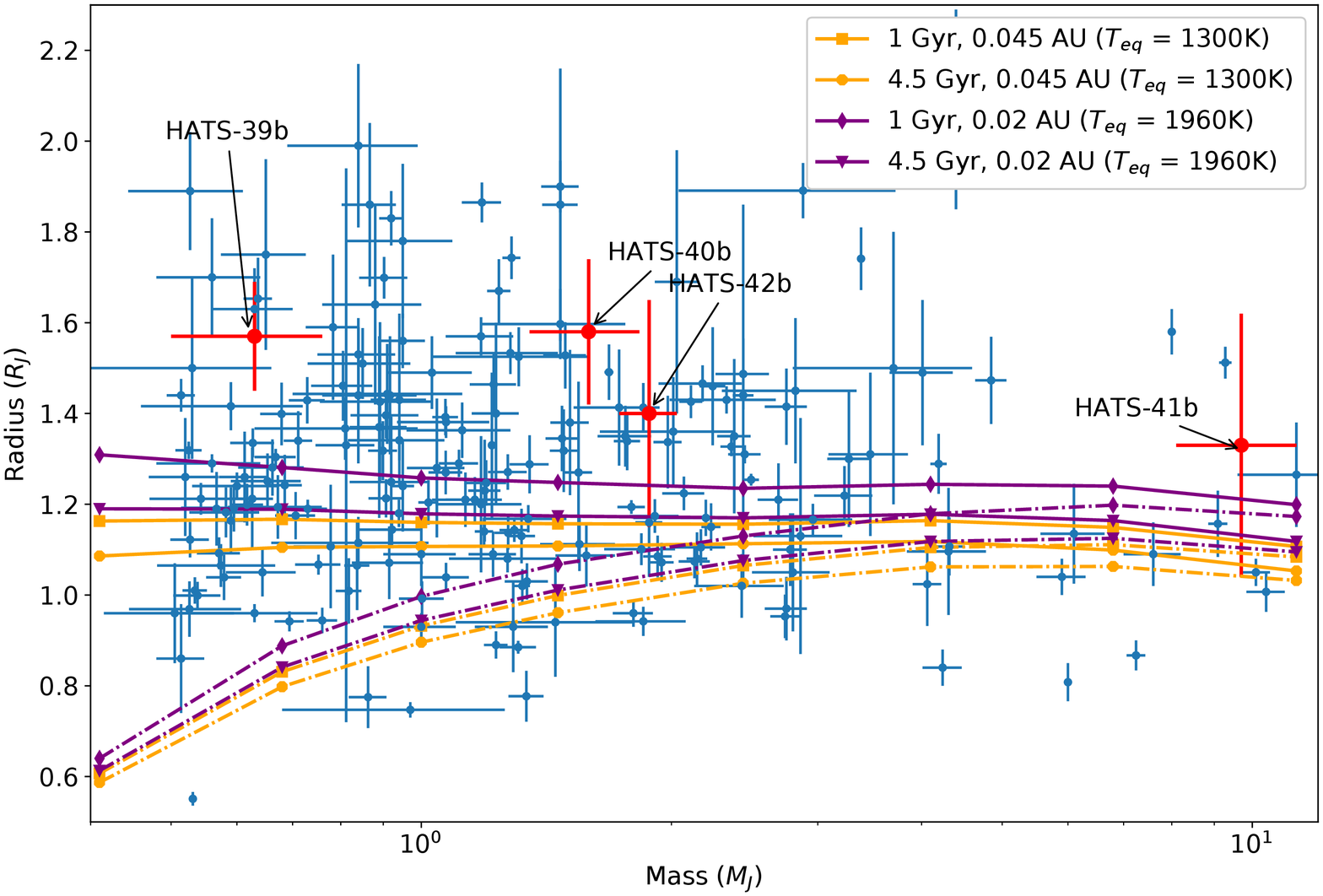}
\caption{
     Mass-radius relation for \hjs, defined as those planets with masses higher than $0.5 \mjup$ and periods shorter than 10 days. We show theoretical models for planet structures from \cite{fortney:2007} for both no core (dashed lines) and $100 \mearth$ core (solid lines) scenarios. The new HATSouth planets are indicated. We present models at 1 Gyr and 4.5 Gyr for planets at a separation of 0.045 AU (orange lines), consistent with the separations of our new discoveries, and models at a separation of 0.02 AU that predict an equilibrium surface temperature of 1960K, closer to that of the four HATSouth targets. A colour version of this plot is available in the online version of this article. 
}
\label{fig:massradius}
\ifthenelse{\boolean{emulateapj}}{
    \end{figure*}
}{
    \end{figure}
}

\begin{figure}
{
\centering
\includegraphics[width={\linewidth}]{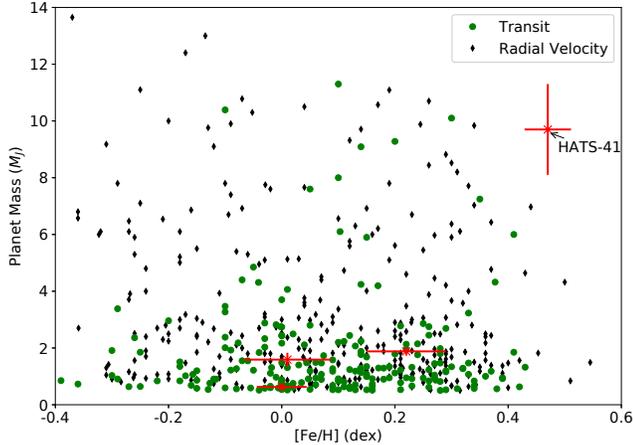}
}
\caption{
     Planet Mass (or \mplsini) as a function of stellar metallicity for known exoplanets above $\rm 0.5 \mjup$ mass.  We show two data sets; the first consists of planets originally discovered via the transit method (green circles) and which have reliable measured masses and radii. The second, corresponds to those discovered by Radial Velocity only (black diamonds) and, thus, the minimum mass value (\mplsini) is shown. We also show our new discovered planets, highlighting the case of \hatcur{41} as the highest metallicity host star with a known transiting planet to date.
     }
\label{fig:metmass}
\end{figure}

In Figure \ref{fig:massradius} we show these discoveries in the context of all other known \hjs, which we define as planets with masses higher than $0.5 \mjup$ and orbital periods less than 10 days\footnote{Previously known planets shown in Figures \ref{fig:massradius} and \ref{fig:metmass} taken from the NASA Exoplanet Archive at \url{http://exoplanetarchive.ipac.caltech.edu/} \citep{akeson:2013:exoplanets}}. 

In addition to previously known planets, we plot a selection of predicted mass--radius relations from \cite{fortney:2007} relevant for each of our planets. We have selected models for planets orbiting solar twins at 1 Gyr and 4.5 Gyr at a separation of 0.045 AU, and we plot two extreme values of the core masses -- 0\,\mearth{} (solid lines) and 100\,\mearth{} (dashed lines). While the orbital distances of our new planets are mostly consistent with 0.045\,AU, the nature of the host stars leads to a higher equilibrium temperature. Hence, we also show models for orbital separations of 0.02 AU, which correspond to an equilibrium temperature of 1960\,K, more closely matching that of the four highlighted targets. While it is premature to make statements regarding the composition of these planets, \hatcurb{39} and \hatcurb{40} seem to be inflated with respect to these predicted models. In particular, \hatcurb{39} is likely to be a good candidate for future transmission spectroscopy follow-up studies. Assuming a mean molecular mass similar to that of Jupiter, the scale height for this planets is approximately 970 km, which corresponds to a transmission signal during transit of 170 p.p.m. For a star of this magnitude ($J$ = \hatcurCCtwomassJmag{39}) this signal is expected to fall within the detection limits of JWST and would result in a more than 3$\sigma$ detection \citep{pepe:2014}. 

Of particular note is the case of \hatcurb{41}. This very high mass planet is found to orbit the highest metallicity star to host a transiting planet to date. While there is a known relation between the stellar metallicity and giant planet frequency for low mass stars \citep{santos:2004,fisher:2005}, the recent work by \citet{santos:2017} suggests that perhaps there are, in fact, two distinct planet populations represented by those with masses above and below $\approx 4 \mjup$. The majority of higher mass planets are also found around higher mass (and sometimes evolved) host stars. The authors explore this in further detail and conclude that the dependence on host star metallicity found for lower mass planets is not present in those planets with masses higher than $4 \mjup$. Furthermore, on average, high mass giant planets are found orbiting hosts with slightly lower metallicity than their lower mass counterparts and therefore consistent with the metallicity distribution of average field stars with similar masses. These factors could be interpreted as these two populations of planets forming by different mechanisms, where lower mass planets are formed via a core-accretion process \citep{perri:1974, mizuno:1980, kennedy:2008} and the higher mass planets via another process where disk instability plays a role, as proposed by \citet{cameron:1978, boss:1998} and later revised by \citet{rafikov:2005} and \citet{nayakshin:2017}.

\startlongtable
\ifthenelse{\boolean{emulateapj}}{
    \begin{deluxetable*}{lcccc}
}{
    \begin{deluxetable}{lcccc}
}
\tabletypesize{\scriptsize}
\tablecaption{Orbital and planetary parameters for \hatcurb{39}--\hatcurb{42}\label{tab:planetparam}}
\tablehead{
    \multicolumn{1}{c}{} &
    \multicolumn{1}{c}{\bf HATS-39b} &
    \multicolumn{1}{c}{\bf HATS-40b} &
    \multicolumn{1}{c}{\bf HATS-41b} &
    \multicolumn{1}{c}{\bf HATS-42b} \\ 
    \multicolumn{1}{c}{~~~~~~~~~~~~~~~Parameter~~~~~~~~~~~~~~~} &
    \multicolumn{1}{c}{Value} &
    \multicolumn{1}{c}{Value} &
    \multicolumn{1}{c}{Value} &
    \multicolumn{1}{c}{Value}
}
\startdata
\noalign{\vskip -3pt}
\sidehead{\Lc{} parameters}
~~~$P$ (days)             \dotfill    & $\hatcurLCP{39}$ & $\hatcurLCP{40}$ & $\hatcurLCPeccen{41}$ & $\hatcurLCP{42}$ \\
~~~$T_c$ (${\rm BJD}$)    
      \tablenotemark{a}   \dotfill    & $\hatcurLCT{39}$ & $\hatcurLCT{40}$ & $\hatcurLCTeccen{41}$ & $\hatcurLCT{42}$ \\
~~~$T_{14}$ (days)
      \tablenotemark{a}   \dotfill    & $\hatcurLCdur{39}$ & $\hatcurLCdur{40}$ & $\hatcurLCdureccen{41}$ & $\hatcurLCdur{42}$ \\
~~~$T_{12} = T_{34}$ (days)
      \tablenotemark{a}   \dotfill    & $\hatcurLCingdur{39}$ & $\hatcurLCingdur{40}$ & $\hatcurLCingdureccen{41}$ & $\hatcurLCingdur{42}$ \\
~~~$\arstar$              \dotfill    & $\hatcurPPar{39}$ & $\hatcurPPar{40}$ & $\hatcurPPareccen{41}$ & $\hatcurPPar{42}$ \\
~~~$\zrstar$ \tablenotemark{b}             \dotfill    & $\hatcurLCzeta{39}$\phn & $\hatcurLCzeta{40}$\phn & $\hatcurLCzetaeccen{41}$\phn & $\hatcurLCzeta{42}$\phn \\
~~~$\rpl/\rstar$          \dotfill    & $\hatcurLCrprstar{39}$ & $\hatcurLCrprstar{40}$ & $\hatcurLCrprstareccen{41}$ & $\hatcurLCrprstar{42}$ \\
~~~$b^2$                  \dotfill    & $\hatcurLCbsq{39}$ & $\hatcurLCbsq{40}$ & $\hatcurLCbsqeccen{41}$ & $\hatcurLCbsq{42}$ \\
~~~$b \equiv a \cos i/\rstar$
                          \dotfill    & $\hatcurLCimp{39}$ & $\hatcurLCimp{40}$ & $\hatcurLCimpeccen{41}$ & $\hatcurLCimp{42}$ \\
~~~$i$ (deg)              \dotfill    & $\hatcurPPi{39}$\phn & $\hatcurPPi{40}$\phn & $\hatcurPPieccen{41}$\phn & $\hatcurPPi{42}$\phn \\
\sidehead{HATSouth dilution factors \tablenotemark{c}}
~~~Dilution factor 1 \dotfill & \hatcurLCiblendA{39} & \hatcurLCiblendA{40} & $\hatcurLCiblendeccen{41}$ & \hatcurLCiblendA{42} \\
~~~Dilution factor 2 \dotfill & \hatcurLCiblendB{39} & \hatcurLCiblendB{40} & $\cdots$ & \hatcurLCiblendB{42} \\
~~~Dilution factor 3 \dotfill & $\cdots$ & \hatcurLCiblendC{40} & $\cdots$ & $\cdots$ \\
\sidehead{Limb-darkening coefficients \tablenotemark{d}}
~~~$c_1,r$                  \dotfill    & $\hatcurLBir{39}$ & $\hatcurLBir{40}$ & $\hatcurLBireccen{41}$ & $\hatcurLBir{42}$ \\
~~~$c_2,r$                  \dotfill    & $\hatcurLBiir{39}$ & $\hatcurLBiir{40}$ & $\hatcurLBiireccen{41}$ & $\hatcurLBiir{42}$ \\
~~~$c_1,R$                  \dotfill    & $\cdots$ & $\cdots$ & $\hatcurLBiReccen{41}$ & $\cdots$ \\
~~~$c_2,R$                  \dotfill    & $\cdots$ & $\cdots$ & $\hatcurLBiiReccen{41}$ & $\cdots$ \\
~~~$c_1,i$                  \dotfill    & $\hatcurLBii{39}$ & $\hatcurLBii{40}$ & $\hatcurLBiieccen{41}$ & $\hatcurLBii{42}$ \\
~~~$c_2,i$                  \dotfill    & $\hatcurLBiii{39}$ & $\hatcurLBiii{40}$ & $\hatcurLBiiieccen{41}$ & $\hatcurLBiii{42}$ \\
\sidehead{Radial Velocity parameters}
~~~$K$ (\ms)              \dotfill    & $\hatcurRVK{39}$\phn\phn & $\hatcurRVK{40}$\phn\phn & $\hatcurRVKeccen{41}$\phn\phn & $\hatcurRVK{42}$\phn\phn \\
~~~$e$ \tablenotemark{e}               \dotfill    & $\hatcurRVeccentwosiglimeccen{39}$ & $\hatcurRVeccentwosiglimeccen{40}$ & $\hatcurRVecceneccen{41}$ & $\hatcurRVeccentwosiglimeccen{42}$ \\
~~~$\omega$ (deg) \dotfill    & $\cdots$ & $\cdots$ & $\hatcurRVomegaeccen{41}$ & $\cdots$ \\
~~~$\sqrt{e}\cos\omega$               \dotfill    & $\cdots$ & $\cdots$ & $\hatcurRVrkeccen{41}$ & $\cdots$ \\
~~~$\sqrt{e}\sin\omega$               \dotfill    & $\cdots$ & $\cdots$ & $\hatcurRVrheccen{41}$ & $\cdots$ \\
~~~ jitter HARPS (\ms)        \dotfill    & \hatcurRVjitterA{39} & \hatcurRVjittertwosiglim{40} & \hatcurRVjitterBeccen{41} & \hatcurRVjittertwosiglimB{42} \\
~~~ jitter FEROS (\ms) \tablenotemark{f}       \dotfill    & $\cdots$ & $\cdots$ & $\cdots$ & \hatcurRVjitterA{42} \\
~~~ jitter CYCLOPS (\ms)        \dotfill    & \hatcurRVjittertwosiglimB{39} & $\cdots$ & \hatcurRVjitterCeccen{41} & $\cdots$ \\
~~~ jitter Coralie (\ms)        \dotfill    & $\cdots$ & $\cdots$ & \hatcurRVjitterAeccen{41} & $\cdots$ \\
\sidehead{Planetary parameters}
~~~$\mpl$ ($\mjup$)       \dotfill    & $\hatcurPPmlong{39}$ & $\hatcurPPmlong{40}$ & $\hatcurPPmlongeccen{41}$ & $\hatcurPPmlong{42}$ \\
~~~$\rpl$ ($\rjup$)       \dotfill    & $\hatcurPPrlong{39}$ & $\hatcurPPrlong{40}$ & $\hatcurPPrlongeccen{41}$ & $\hatcurPPrlong{42}$ \\
~~~$C(\mpl,\rpl)$
    \tablenotemark{g}     \dotfill    & $\hatcurPPmrcorr{39}$ & $\hatcurPPmrcorr{40}$ & $\hatcurPPmrcorreccen{41}$ & $\hatcurPPmrcorr{42}$ \\
~~~$\rhopl$ (\gcmc)       \dotfill    & $\hatcurPPrho{39}$ & $\hatcurPPrho{40}$ & $\hatcurPPrhoeccen{41}$ & $\hatcurPPrho{42}$ \\
~~~$\log g_p$ (cgs)       \dotfill    & $\hatcurPPlogg{39}$ & $\hatcurPPlogg{40}$ & $\hatcurPPloggeccen{41}$ & $\hatcurPPlogg{42}$ \\
~~~$a$ (AU)               \dotfill    & $\hatcurPParel{39}$ & $\hatcurPParel{40}$ & $\hatcurPPareleccen{41}$ & $\hatcurPParel{42}$ \\
~~~$T_{\rm eq}$ (K)        \dotfill   & $\hatcurPPteff{39}$ & $\hatcurPPteff{40}$ & $\hatcurPPteffeccen{41}$ & $\hatcurPPteff{42}$ \\
~~~$\Theta$ \tablenotemark{h} \dotfill & $\hatcurPPtheta{39}$ & $\hatcurPPtheta{40}$ & $\hatcurPPthetaeccen{41}$ & $\hatcurPPtheta{42}$ \\
~~~$\log_{10}\langle F \rangle$ (cgs) \tablenotemark{i}
                          \dotfill    & $\hatcurPPfluxavglog{39}$ & $\hatcurPPfluxavglog{40}$ & $\hatcurPPfluxavglogeccen{41}$ & $\hatcurPPfluxavglog{42}$ \\
\enddata
\tablecomments{
For \hatcurb{41} we adopt a model in which the eccentricity is allowed to vary. For the other three systems we adopt a model in which the orbit is assumed to be circular. See the discussion in Section~\ref{sec:globmod}.
}
\tablenotetext{a}{
    Times are in Barycentric Julian Date calculated directly from UTC {\em without} correction for leap seconds.
    \ensuremath{T_c}: Reference epoch of
    mid transit that minimizes the correlation with the orbital
    period.
    \ensuremath{T_{12}}: total transit duration, time
    between first to last contact;
    \ensuremath{T_{12}=T_{34}}: ingress/egress time, time between first
    and second, or third and fourth contact.
}
\tablenotetext{b}{
   Reciprocal of the half duration of the transit used as a jump parameter in our MCMC analysis in place of $\arstar$. It is related to $\arstar$ by the expression $\zrstar = \arstar(2\pi(1+e\sin\omega))/(P\sqrt{1-b^2}\sqrt{1-e^2})$ \citep{bakos:2010:hat11}.
}
\tablenotetext{c}{
    Scaling factor applied to the model transit that is fit to the HATSouth light curves. This factor accounts for dilution of the transit due to blending from neighboring stars and over-filtering of the light curve.  These factors are varied in the fit, with independent values adopted for each HATSouth light curve. The factors listed for \hatcur{39} are for the G602.3 and G602.4 light curves, respectively. For \hatcur{40}, we list the factors for G601.2, G600.3, and G600.4, respectively. For \hatcur{41} the listed factor is for G601.2. For \hatcur{42}, the listed factors are for G602.1 and G601.4, respectively.
}
\tablenotetext{d}{
    Values for a quadratic law, adopted from the tabulations by
    \cite{claret:2004} according to the spectroscopic (ZASPE) parameters
    listed in \reftabl{stellar}.
}
\tablenotetext{e}{
    For \hatcur{39}, \hatcur{40} and \hatcur{42} we list
    the 95\% confidence upper limit on the eccentricity determined
    when $\sqrt{e}\cos\omega$ and $\sqrt{e}\sin\omega$ are allowed to
    vary in the fit.
}
\tablenotetext{f}{
    Term added in quadrature to the formal radial velocity uncertainties for each
    instrument. This is treated as a free parameter in the fitting
    routine. In cases where the jitter is consistent with zero, we
    list its 95\% confidence upper limit.
}
\tablenotetext{g}{
    Correlation coefficient between the planetary mass \mpl\ and radius
    \rpl\ estimated from the posterior parameter distribution.
}
\tablenotetext{h}{
    The Safronov number is given by $\Theta = \frac{1}{2}(V_{\rm
    esc}/V_{\rm orb})^2 = (a/\rpl)(\mpl / \mstar )$
    \citep[see][]{hansen:2007}.
}
\tablenotetext{i}{
    Incoming flux per unit surface area, averaged over the orbit.
}
\ifthenelse{\boolean{emulateapj}}{
    \end{deluxetable*}
}{
    \end{deluxetable}
}

While the authors focus on a sample of planets with orbital periods above 10 days to deliberately reject \hjs, they note that their conclusion regarding the potential existence of two separate populations still stands if those planets are included, and therefore we can place our new discoveries in this context. In Figure \ref{fig:metmass} we show a plot of planet mass as a function of stellar host metallicity for known exoplanets in which we have distinguished those discovered by the transit method that have measured masses (green circles) and those discovered by radial velocity only (black diamonds). For those planets with no detected transits, we plot the minimum mass (\mplsini) instead. In this plot we also show our four new discovered planets, highlighting the position of \hatcur{41} as the highest metallicity star hosting a planet with well characterised mass and radius. The clustering of planets below $4\mjup$ masses in the above solar metallicity regime is clearly seen, despite the existence of a significant number of low metallicity stars hosting low mass giant planets. However, for planets above this mass threshold, most transiting planets are still found to orbit stars with higher than solar metallicity, and the top-left region of the plot is dominated by planets found by radial velocity. Given the observational bias on the discovery of transiting planets favoring short period orbits, typically less than 20 days, these two samples are, in fact, somewhat different in nature as they represent planets with very different orbital periods. Therefore, while the statement is still true that planets above $4\mjup$ masses have less of a dependence on metallicity, hot Jupiters are still found to follow the previous known relation. This suggests that the relation between giant planet mass and host star metallicity may also depend on the orbital period of the planet and that the inward migration process for giant planets that results in the known sample of \hjs\ may be dependent on the host star properties. This is further evidence that the known sample of \hjs\ is indeed distinct from the remaining planet population. A larger number of well characterised planets is required to further address this issue, and the advent of the next generation of instruments will improve our understanding of these processes.

Despite the fact that these new four exoplanets reside in relatively similar environments in terms of stellar host type and orbital separation (and thus similar equilibrium temperatures within a $\sim 500K$ range), they span effectively the entire mass range of known \hjs, as show in Figure \ref{fig:massradius}. We note that, given the large uncertainties in the radii of \hatcurb{41} and \hatcurb{42} (likely related to a combination of the limited precision on the stellar radius estimates and lack of extensive photometric follow-up), further observations are required to obtain improved estimates of this parameter for these targets. We therefore find them to be consistent with models of planetary radii under comparable equilibrium temperatures.

\acknowledgements

Development of the HATSouth project was funded by NSF MRI grant
NSF/AST-0723074, operations have been supported by NASA grants NNX09AB29G, NNX12AH91H, and NNX17AB61G, and
follow-up observations have received partial support from grant
NSF/AST-1108686.
J.H. acknowledges support from NASA grant NNX14AE87G.
A.J.\ acknowledges support from FONDECYT project 1171208, BASAL CATA PFB-06, and project IC120009 ``Millennium Institute of Astrophysics (MAS)'' of the Millenium Science Initiative, Chilean Ministry of Economy. R.B.\ and N.E.\ acknowledge  support from project IC120009 ``Millenium Institute of Astrophysics  (MAS)'' of the Millennium Science Initiative, Chilean Ministry of Economy. 
V.S.\ acknowledges support form BASAL CATA PFB-06. 
This work also uses observations obtained
with facilities of the Las Cumbres Observatory Global
Telescope (LCOGT). 
This research has made use of the NASA Exoplanet Archive, which is operated by the California Institute of Technology, under contract with the National Aeronautics and Space Administration under the Exoplanet Exploration Program.
This work is based on observations made with ESO Telescopes at the La
Silla Observatory.
Work at the Australian National University is supported by ARC Laureate
Fellowship Grant FL0992131.
We acknowledge the use of the AAVSO Photometric All-Sky Survey (APASS),
funded by the Robert Martin Ayers Sciences Fund, and the SIMBAD
database, operated at CDS, Strasbourg, France.
Operations at the MPG~2.2\,m Telescope are jointly performed by the
Max Planck Gesellschaft and the European Southern Observatory. 
We are grateful to P.Sackett for her help in the early phase of the
HATSouth project.
This research has made use of the NASA/ IPAC Infrared Science Archive, 
which is operated by the Jet Propulsion Laboratory, California Institute 
of Technology, under contract with the National Aeronautics and Space Administration.
This work is based on observations
collected with HARPS and FEROS at the European Organisation for
Astronomical Research in the Southern Hemisphere under ESO programmes 095.C-0367, CN2013A-171,
CN2013B-55, CN2014A-104, CN2014B-57, CN2015A-51 and ESO 096.C-0544.
This work has made use of data from the European Space Agency (ESA)
mission {\it Gaia} (\url{https://www.cosmos.esa.int/gaia}), processed by
the {\it Gaia} Data Processing and Analysis Consortium (DPAC,
\url{https://www.cosmos.esa.int/web/gaia/dpac/consortium}). Funding
for the DPAC has been provided by national institutions, in particular
the institutions participating in the {\it Gaia} Multilateral Agreement.

%\clearpage
\bibliographystyle{aasjournal}
\bibliography{hatsbib}

\begin{thebibliography}{}
\expandafter\ifx\csname natexlab\endcsname\relax\def\natexlab#1{#1}\fi
\providecommand{\url}[1]{\href{#1}{#1}}

\bibitem[{{Addison} {et~al.}(2013){Addison}, {Tinney}, {Wright}, {Bayliss},
  {Zhou}, {Hartman}, {Bakos}, \& {Schmidt}}]{addison:2013}
{Addison}, B.~C., {Tinney}, C.~G., {Wright}, D.~J., {et~al.} 2013, \apjl, 774,
  L9

\bibitem[{{Akeson} {et~al.}(2013){Akeson}, {Chen}, {Ciardi}, {Crane}, {Good},
  {Harbut}, {Jackson}, {Kane}, {Laity}, {Leifer}, {Lynn}, {McElroy}, {Papin},
  {Plavchan}, {Ram{\'{\i}}rez}, {Rey}, {von Braun}, {Wittman}, {Abajian},
  {Ali}, {Beichman}, {Beekley}, {Berriman}, {Berukoff}, {Bryden}, {Chan},
  {Groom}, {Lau}, {Payne}, {Regelson}, {Saucedo}, {Schmitz}, {Stauffer},
  {Wyatt}, \& {Zhang}}]{akeson:2013:exoplanets}
{Akeson}, R.~L., {Chen}, X., {Ciardi}, D., {et~al.} 2013, \pasp, 125, 989

\bibitem[{{Bakos} {et~al.}(2004){Bakos}, {Noyes}, {Kov{\'a}cs}, {Stanek},
  {Sasselov}, \& {Domsa}}]{bakos:2004:hatnet}
{Bakos}, G., {Noyes}, R.~W., {Kov{\'a}cs}, G., {et~al.} 2004, \pasp, 116, 266

\bibitem[{{Bakos} {et~al.}(2010){Bakos}, {Torres}, {P{\'a}l}, {Hartman},
  {Kov{\'a}cs}, {Noyes}, {Latham}, {Sasselov}, {Sip{\H o}cz}, {Esquerdo},
  {Fischer}, {Johnson}, {Marcy}, {Butler}, {Isaacson}, {Howard}, {Vogt},
  {Kov{\'a}cs}, {Fernandez}, {Mo{\'o}r}, {Stefanik}, {L{\'a}z{\'a}r}, {Papp},
  \& {S{\'a}ri}}]{bakos:2010:hat11}
{Bakos}, G.~{\'A}., {Torres}, G., {P{\'a}l}, A., {et~al.} 2010, \apj, 710, 1724

\bibitem[{{Bakos} {et~al.}(2013){Bakos}, {Csubry}, {Penev}, {Bayliss},
  {Jord{\'a}n}, {Afonso}, {Hartman}, {Henning}, {Kov{\'a}cs}, {Noyes},
  {B{\'e}ky}, {Suc}, {Cs{\'a}k}, {Rabus}, {L{\'a}z{\'a}r}, {Papp}, {S{\'a}ri},
  {Conroy}, {Zhou}, {Sackett}, {Schmidt}, {Mancini}, {Sasselov}, \&
  {Ueltzhoeffer}}]{bakos:2013:hatsouth}
{Bakos}, G.~{\'A}., {Csubry}, Z., {Penev}, K., {et~al.} 2013, \pasp, 125, 154

\bibitem[{{Bayliss} {et~al.}(2013){Bayliss}, {Zhou}, {Penev}, {Bakos},
  {Hartman}, {Jord{\'a}n}, {Mancini}, {Mohler-Fischer}, {Suc}, {Rabus},
  {B{\'e}ky}, {Csubry}, {Buchhave}, {Henning}, {Nikolov}, {Cs{\'a}k}, {Brahm},
  {Espinoza}, {Noyes}, {Schmidt}, {Conroy}, {Wright}, {Tinney}, {Addison},
  {Sackett}, {Sasselov}, {L{\'a}z{\'a}r}, {Papp}, \&
  {S{\'a}ri}}]{bayliss:2013:hats3}
{Bayliss}, D., {Zhou}, G., {Penev}, K., {et~al.} 2013, \aj, 146, 113

\bibitem[{{Bayliss} {et~al.}(2015){Bayliss}, {Hartman}, {Bakos}, {Penev},
  {Zhou}, {Brahm}, {Rabus}, {Jord{\'a}n}, {Mancini}, {de Val-Borro}, {Bhatti},
  {Espinoza}, {Csubry}, {Howard}, {Fulton}, {Buchhave}, {Henning}, {Schmidt},
  {Ciceri}, {Noyes}, {Isaacson}, {Marcy}, {Suc}, {L{\'a}z{\'a}r}, {Papp}, \&
  {S{\'a}ri}}]{bayliss:2015}
{Bayliss}, D., {Hartman}, J.~D., {Bakos}, G.~{\'A}., {et~al.} 2015, \aj, 150,
  49

\bibitem[{{Bento} {et~al.}(2014){Bento}, {Wheatley}, {Copperwheat}, {Fortney},
  {Dhillon}, {Hickman}, {Littlefair}, {Marsh}, {Parsons}, \&
  {Southworth}}]{bento:2014}
{Bento}, J., {Wheatley}, P.~J., {Copperwheat}, C.~M., {et~al.} 2014, \mnras,
  437, 1511

\bibitem[{{Borucki} {et~al.}(2010){Borucki}, {Koch}, {Basri}, {Batalha},
  {Brown}, {Caldwell}, {Caldwell}, {Christensen-Dalsgaard}, {Cochran},
  {DeVore}, {Dunham}, {Dupree}, {Gautier}, {Geary}, {Gilliland}, {Gould},
  {Howell}, {Jenkins}, {Kondo}, {Latham}, {Marcy}, {Meibom}, {Kjeldsen},
  {Lissauer}, {Monet}, {Morrison}, {Sasselov}, {Tarter}, {Boss}, {Brownlee},
  {Owen}, {Buzasi}, {Charbonneau}, {Doyle}, {Fortney}, {Ford}, {Holman},
  {Seager}, {Steffen}, {Welsh}, {Rowe}, {Anderson}, {Buchhave}, {Ciardi},
  {Walkowicz}, {Sherry}, {Horch}, {Isaacson}, {Everett}, {Fischer}, {Torres},
  {Johnson}, {Endl}, {MacQueen}, {Bryson}, {Dotson}, {Haas}, {Kolodziejczak},
  {Van Cleve}, {Chandrasekaran}, {Twicken}, {Quintana}, {Clarke}, {Allen},
  {Li}, {Wu}, {Tenenbaum}, {Verner}, {Bruhweiler}, {Barnes}, \&
  {Prsa}}]{borucki:2010}
{Borucki}, W.~J., {Koch}, D., {Basri}, G., {et~al.} 2010, Science, 327, 977

\bibitem[{{Boss}(1998)}]{boss:1998}
{Boss}, A.~P. 1998, \apj, 503, 923

\bibitem[{{Brahm} {et~al.}(2017{\natexlab{a}}){Brahm}, {Jord{\'a}n}, \&
  {Espinoza}}]{brahm:2017:ceres}
{Brahm}, R., {Jord{\'a}n}, A., \& {Espinoza}, N. 2017{\natexlab{a}}, \pasp,
  129, 034002

\bibitem[{{Brahm} {et~al.}(2017{\natexlab{b}}){Brahm}, {Jord{\'a}n}, {Hartman},
  \& {Bakos}}]{brahm:2016:ZASPE}
{Brahm}, R., {Jord{\'a}n}, A., {Hartman}, J., \& {Bakos}, G.
  2017{\natexlab{b}}, \mnras, 467, 971

\bibitem[{{Brahm} {et~al.}(2015){Brahm}, {Jord{\'a}n}, {Hartman}, {Bakos},
  {Bayliss}, {Penev}, {Zhou}, {Ciceri}, {Rabus}, {Espinoza}, {Mancini}, {de
  Val-Borro}, {Bhatti}, {Sato}, {Tan}, {Csubry}, {Buchhave}, {Henning},
  {Schmidt}, {Suc}, {Noyes}, {Papp}, {L{\'a}z{\'a}r}, \&
  {S{\'a}ri}}]{brahm:2015}
{Brahm}, R., {Jord{\'a}n}, A., {Hartman}, J.~D., {et~al.} 2015, \aj, 150, 33

\bibitem[{{Brahm} {et~al.}(2016){Brahm}, {Jord{\'a}n}, {Bakos}, {Penev},
  {Espinoza}, {Rabus}, {Hartman}, {Bayliss}, {Ciceri}, {Zhou}, {Mancini},
  {Tan}, {de Val-Borro}, {Bhatti}, {Csubry}, {Bento}, {Henning}, {Schmidt},
  {Rojas}, {Suc}, {L{\'a}z{\'a}r}, {Papp}, \& {S{\'a}ri}}]{brahm:2016}
{Brahm}, R., {Jord{\'a}n}, A., {Bakos}, G.~{\'A}., {et~al.} 2016, \aj, 151, 89

\bibitem[{{Brown} {et~al.}(2013){Brown}, {Baliber}, {Bianco}, {Bowman},
  {Burleson}, {Conway}, {Crellin}, {Depagne}, {De Vera}, {Dilday}, {Dragomir},
  {Dubberley}, {Eastman}, {Elphick}, {Falarski}, {Foale}, {Ford}, {Fulton},
  {Garza}, {Gomez}, {Graham}, {Greene}, {Haldeman}, {Hawkins}, {Haworth},
  {Haynes}, {Hidas}, {Hjelstrom}, {Howell}, {Hygelund}, {Lister}, {Lobdill},
  {Martinez}, {Mullins}, {Norbury}, {Parrent}, {Paulson}, {Petry}, {Pickles},
  {Posner}, {Rosing}, {Ross}, {Sand}, {Saunders}, {Shobbrook}, {Shporer},
  {Street}, {Thomas}, {Tsapras}, {Tufts}, {Valenti}, {Vander Horst}, {Walker},
  {White}, \& {Willis}}]{brown:2013:lcogt}
{Brown}, T.~M., {Baliber}, N., {Bianco}, F.~B., {et~al.} 2013, \pasp, 125, 1031

\bibitem[{{Cameron}(1978)}]{cameron:1978}
{Cameron}, A.~G.~W. 1978, in IAU Colloq. 52: Protostars and Planets, ed.
  T.~{Gehrels}, 453--487

\bibitem[{{Cardelli} {et~al.}(1989){Cardelli}, {Clayton}, \&
  {Mathis}}]{cardelli:1989}
{Cardelli}, J.~A., {Clayton}, G.~C., \& {Mathis}, J.~S. 1989, \apj, 345, 245

\bibitem[{{Chambers}(2009)}]{chambers:2009}
{Chambers}, J.~E. 2009, \apj, 705, 1206

\bibitem[{{Claret}(2004)}]{claret:2004}
{Claret}, A. 2004, \aap, 428, 1001

\bibitem[{{D{\'e}sert} {et~al.}(2011){D{\'e}sert}, {Charbonneau}, {Fortney},
  {Madhusudhan}, {Knutson}, {Fressin}, {Deming}, {Borucki}, {Brown},
  {Caldwell}, {Ford}, {Gilliland}, {Latham}, {Marcy}, \&
  {Seager}}]{desert:2011}
{D{\'e}sert}, J.-M., {Charbonneau}, D., {Fortney}, J.~J., {et~al.} 2011, \apjs,
  197, 11

\bibitem[{{Dopita} {et~al.}(2007){Dopita}, {Hart}, {McGregor}, {Oates},
  {Bloxham}, \& {Jones}}]{Dopita:2007}
{Dopita}, M., {Hart}, J., {McGregor}, P., {et~al.} 2007, \apss, 310, 255

\bibitem[{{Espinoza} {et~al.}(2016){Espinoza}, {Bayliss}, {Hartman}, {Bakos},
  {Jord{\'a}n}, {Zhou}, {Mancini}, {Brahm}, {Ciceri}, {Bhatti}, {Csubry},
  {Rabus}, {Penev}, {Bento}, {de Val-Borro}, {Henning}, {Schmidt}, {Suc},
  {Wright}, {Tinney}, {Tan}, \& {Noyes}}]{espinoza:2016:hats25hats30}
{Espinoza}, N., {Bayliss}, D., {Hartman}, J.~D., {et~al.} 2016, \aj, 152, 108

\bibitem[{{Fischer} \& {Valenti}(2005)}]{fisher:2005}
{Fischer}, D.~A., \& {Valenti}, J. 2005, \apj, 622, 1102

\bibitem[{{Ford} \& {Rasio}(2008)}]{ford:2008}
{Ford}, E.~B., \& {Rasio}, F.~A. 2008, \apj, 686, 621

\bibitem[{{Fortney} {et~al.}(2007){Fortney}, {Marley}, \&
  {Barnes}}]{fortney:2007}
{Fortney}, J.~J., {Marley}, M.~S., \& {Barnes}, J.~W. 2007, \apj, 659, 1661

\bibitem[{{Girardi} {et~al.}(2000){Girardi}, {Bressan}, {Bertelli}, \&
  {Chiosi}}]{girardi:2000}
{Girardi}, L., {Bressan}, A., {Bertelli}, G., \& {Chiosi}, C. 2000, \aaps, 141,
  371

\bibitem[{{Gustafsson} {et~al.}(2008){Gustafsson}, {Edvardsson}, {Eriksson},
  {J{\o}rgensen}, {Nordlund}, \& {Plez}}]{gustafsson:2008}
{Gustafsson}, B., {Edvardsson}, B., {Eriksson}, K., {et~al.} 2008, \aap, 486,
  951

\bibitem[{{Hansen} \& {Barman}(2007)}]{hansen:2007}
{Hansen}, B.~M.~S., \& {Barman}, T. 2007, \apj, 671, 861

\bibitem[{{Hartman} {et~al.}(2012){Hartman}, {Bakos}, {B{\'e}ky}, {Torres},
  {Latham}, {Csubry}, {Penev}, {Shporer}, {Fulton}, {Buchhave}, {Johnson},
  {Howard}, {Marcy}, {Fischer}, {Kov{\'a}cs}, {Noyes}, {Esquerdo}, {Everett},
  {Szklen{\'a}r}, {Quinn}, {Bieryla}, {Knox}, {Hinz}, {Sasselov}, {F{\H
  u}r{\'e}sz}, {Stefanik}, {L{\'a}z{\'a}r}, {Papp}, \&
  {S{\'a}ri}}]{hartman:2012:hat39hat41}
{Hartman}, J.~D., {Bakos}, G.~{\'A}., {B{\'e}ky}, B., {et~al.} 2012, \aj, 144,
  139

\bibitem[{{Henden} \& {Munari}(2014)}]{henden:2014}
{Henden}, A., \& {Munari}, U. 2014, Contributions of the Astronomical
  Observatory Skalnate Pleso, 43, 518

\bibitem[{{Hippler} {et~al.}(2009){Hippler}, {Bergfors}, {Brandner Wolfgang},
  {Daemgen}, {Henning}, {Hormuth}, {Huber}, {Janson}, {Rochau}, {Rohloff}, \&
  {Wagner}}]{hippler:2009}
{Hippler}, S., {Bergfors}, C., {Brandner Wolfgang}, {et~al.} 2009, The
  Messenger, 137, 14

\bibitem[{{Janson} {et~al.}(2017){Janson}, {Durkan}, {Hippler}, {Dai},
  {Brandner}, {Schlieder}, {Bonnefoy}, \& {Henning}}]{janson:2017}
{Janson}, M., {Durkan}, S., {Hippler}, S., {et~al.} 2017, \aap, 599, A70

\bibitem[{{Jord{\'a}n} {et~al.}(2013){Jord{\'a}n}, {Espinoza}, {Rabus},
  {Eyheramendy}, {Sing}, {D{\'e}sert}, {Bakos}, {Fortney}, {L{\'o}pez-Morales},
  {Maxted}, {Triaud}, \& {Szentgyorgyi}}]{jordan:2013}
{Jord{\'a}n}, A., {Espinoza}, N., {Rabus}, M., {et~al.} 2013, \apj, 778, 184

\bibitem[{{Jord{\'a}n} {et~al.}(2014){Jord{\'a}n}, {Brahm}, {Bakos}, {Bayliss},
  {Penev}, {Hartman}, {Zhou}, {Mancini}, {Mohler-Fischer}, {Ciceri}, {Sato},
  {Csubry}, {Rabus}, {Suc}, {Espinoza}, {Bhatti}, {Borro}, {Buchhave},
  {Cs{\'a}k}, {Henning}, {Schmidt}, {Tan}, {Noyes}, {B{\'e}ky}, {Butler},
  {Shectman}, {Crane}, {Thompson}, {Williams}, {Martin}, {Contreras},
  {L{\'a}z{\'a}r}, {Papp}, \& {S{\'a}ri}}]{jordan:2014:hats4}
{Jord{\'a}n}, A., {Brahm}, R., {Bakos}, G.~{\'A}., {et~al.} 2014, \aj, 148, 29

\bibitem[{{Kaufer} \& {Pasquini}(1998)}]{kaufer:1998}
{Kaufer}, A., \& {Pasquini}, L. 1998, in Society of Photo-Optical
  Instrumentation Engineers (SPIE) Conference Series, Vol. 3355, Optical
  Astronomical Instrumentation, ed. S.~{D'Odorico}, 844--854

\bibitem[{{Kennedy} \& {Kenyon}(2008)}]{kennedy:2008}
{Kennedy}, G.~M., \& {Kenyon}, S.~J. 2008, \apj, 673, 502

\bibitem[{{Kov{\'a}cs} {et~al.}(2005){Kov{\'a}cs}, {Bakos}, \&
  {Noyes}}]{kovacs:2005:TFA}
{Kov{\'a}cs}, G., {Bakos}, G., \& {Noyes}, R.~W. 2005, \mnras, 356, 557

\bibitem[{{Kov{\'a}cs} {et~al.}(2002){Kov{\'a}cs}, {Zucker}, \&
  {Mazeh}}]{kovacs:2002:BLS}
{Kov{\'a}cs}, G., {Zucker}, S., \& {Mazeh}, T. 2002, \aap, 391, 369

\bibitem[{{Lindegren} {et~al.}(2016){Lindegren}, {Lammers}, {Bastian},
  {Hern{\'a}ndez}, {Klioner}, {Hobbs}, {Bombrun}, {Michalik}, {Ramos-Lerate},
  {Butkevich}, {Comoretto}, {Joliet}, {Holl}, {Hutton}, {Parsons},
  {Steidelm{\"u}ller}, {Abbas}, {Altmann}, {Andrei}, {Anton}, {Bach},
  {Barache}, {Becciani}, {Berthier}, {Bianchi}, {Biermann}, {Bouquillon},
  {Bourda}, {Br{\"u}semeister}, {Bucciarelli}, {Busonero}, {Carlucci},
  {Casta{\~n}eda}, {Charlot}, {Clotet}, {Crosta}, {Davidson}, {de Felice},
  {Drimmel}, {Fabricius}, {Fienga}, {Figueras}, {Fraile}, {Gai}, {Garralda},
  {Geyer}, {Gonz{\'a}lez-Vidal}, {Guerra}, {Hambly}, {Hauser}, {Jordan},
  {Lattanzi}, {Lenhardt}, {Liao}, {L{\"o}ffler}, {McMillan}, {Mignard}, {Mora},
  {Morbidelli}, {Portell}, {Riva}, {Sarasso}, {Serraller}, {Siddiqui}, {Smart},
  {Spagna}, {Stampa}, {Steele}, {Taris}, {Torra}, {van Reeven}, {Vecchiato},
  {Zschocke}, {de Bruijne}, {Gracia}, {Raison}, {Lister}, {Marchant},
  {Messineo}, {Soffel}, {Osorio}, {de Torres}, \& {O'Mullane}}]{lindegren:2016}
{Lindegren}, L., {Lammers}, U., {Bastian}, U., {et~al.} 2016, \aap, 595, A4

\bibitem[{{Louden} \& {Wheatley}(2015)}]{louden:2015}
{Louden}, T., \& {Wheatley}, P.~J. 2015, \apjl, 814, L24

\bibitem[{{Mandel} \& {Agol}(2002)}]{mandel:2002}
{Mandel}, K., \& {Agol}, E. 2002, \apjl, 580, L171

\bibitem[{{Mayor} {et~al.}(2003){Mayor}, {Pepe}, {Queloz}, {Bouchy},
  {Rupprecht}, {Lo Curto}, {Avila}, {Benz}, {Bertaux}, {Bonfils}, {Dall},
  {Dekker}, {Delabre}, {Eckert}, {Fleury}, {Gilliotte}, {Gojak}, {Guzman},
  {Kohler}, {Lizon}, {Longinotti}, {Lovis}, {Megevand}, {Pasquini}, {Reyes},
  {Sivan}, {Sosnowska}, {Soto}, {Udry}, {van Kesteren}, {Weber}, \&
  {Weilenmann}}]{mayor:2003}
{Mayor}, M., {Pepe}, F., {Queloz}, D., {et~al.} 2003, The Messenger, 114, 20

\bibitem[{{Mazeh} {et~al.}(1997){Mazeh}, {Mayor}, \& {Latham}}]{mazeh:1997}
{Mazeh}, T., {Mayor}, M., \& {Latham}, D.~W. 1997, \apj, 478, 367

\bibitem[{{Mizuno}(1980)}]{mizuno:1980}
{Mizuno}, H. 1980, Progress of Theoretical Physics, 64, 544

\bibitem[{{Nayakshin}(2017)}]{nayakshin:2017}
{Nayakshin}, S. 2017, \mnras, 470, 2387

\bibitem[{{P{\'a}l} {et~al.}(2008){P{\'a}l}, {Bakos}, {Torres}, {Noyes},
  {Latham}, {Kov{\'a}cs}, {Marcy}, {Fischer}, {Butler}, {Sasselov}, {Sip{\H
  o}cz}, {Esquerdo}, {Kov{\'a}cs}, {Stefanik}, {L{\'a}z{\'a}r}, {Papp}, \&
  {S{\'a}ri}}]{pal:2008:hat7}
{P{\'a}l}, A., {Bakos}, G.~{\'A}., {Torres}, G., {et~al.} 2008, \apj, 680, 1450

\bibitem[{{Penev} {et~al.}(2013){Penev}, {Bakos}, {Bayliss}, {Jord{\'a}n},
  {Mohler}, {Zhou}, {Suc}, {Rabus}, {Hartman}, {Mancini}, {B{\'e}ky}, {Csubry},
  {Buchhave}, {Henning}, {Nikolov}, {Cs{\'a}k}, {Brahm}, {Espinoza}, {Conroy},
  {Noyes}, {Sasselov}, {Schmidt}, {Wright}, {Tinney}, {Addison},
  {L{\'a}z{\'a}r}, {Papp}, \& {S{\'a}ri}}]{penev:2013:hats1}
{Penev}, K., {Bakos}, G.~{\'A}., {Bayliss}, D., {et~al.} 2013, \aj, 145, 5

\bibitem[{{Pepe} {et~al.}(2014){Pepe}, {Ehrenreich}, \& {Meyer}}]{pepe:2014}
{Pepe}, F., {Ehrenreich}, D., \& {Meyer}, M.~R. 2014, \nat, 513, 358

\bibitem[{{Perri} \& {Cameron}(1974)}]{perri:1974}
{Perri}, F., \& {Cameron}, A.~G.~W. 1974, \icarus, 22, 416

\bibitem[{{Petrovich}(2015)}]{petrovich:2015}
{Petrovich}, C. 2015, \apj, 805, 75

\bibitem[{{Pollacco} {et~al.}(2006){Pollacco}, {Skillen}, {Collier Cameron},
  {Christian}, {Hellier}, {Irwin}, {Lister}, {Street}, {West}, {Anderson},
  {Clarkson}, {Deeg}, {Enoch}, {Evans}, {Fitzsimmons}, {Haswell}, {Hodgkin},
  {Horne}, {Kane}, {Keenan}, {Maxted}, {Norton}, {Osborne}, {Parley}, {Ryans},
  {Smalley}, {Wheatley}, \& {Wilson}}]{pollacco:2006}
{Pollacco}, D.~L., {Skillen}, I., {Collier Cameron}, A., {et~al.} 2006, \pasp,
  118, 1407

\bibitem[{{Pont} {et~al.}(2011){Pont}, {Husnoo}, {Mazeh}, \&
  {Fabrycky}}]{pont:2011}
{Pont}, F., {Husnoo}, N., {Mazeh}, T., \& {Fabrycky}, D. 2011, \mnras, 414,
  1278

\bibitem[{{Queloz} {et~al.}(2001){Queloz}, {Mayor}, {Udry}, {Burnet},
  {Carrier}, {Eggenberger}, {Naef}, {Santos}, {Pepe}, {Rupprecht}, {Avila},
  {Baeza}, {Benz}, {Bertaux}, {Bouchy}, {Cavadore}, {Delabre}, {Eckert},
  {Fischer}, {Fleury}, {Gilliotte}, {Goyak}, {Guzman}, {Kohler}, {Lacroix},
  {Lizon}, {Megevand}, {Sivan}, {Sosnowska}, \& {Weilenmann}}]{queloz:2001}
{Queloz}, D., {Mayor}, M., {Udry}, S., {et~al.} 2001, The Messenger, 105, 1

\bibitem[{{Rabus} {et~al.}(2016){Rabus}, {Jord{\'a}n}, {Hartman}, {Bakos},
  {Espinoza}, {Brahm}, {Penev}, {Ciceri}, {Zhou}, {Bayliss}, {Mancini},
  {Bhatti}, {de Val-Borro}, {Csbury}, {Sato}, {Tan}, {Henning}, {Schmidt},
  {Bento}, {Suc}, {Noyes}, {L{\'a}z{\'a}r}, {Papp}, \& {S{\'a}ri}}]{rabus:2016}
{Rabus}, M., {Jord{\'a}n}, A., {Hartman}, J.~D., {et~al.} 2016, \aj, 152, 88

\bibitem[{{Rafikov}(2005)}]{rafikov:2005}
{Rafikov}, R.~R. 2005, \apjl, 621, L69

\bibitem[{{Santos} {et~al.}(2004){Santos}, {Israelian}, \&
  {Mayor}}]{santos:2004}
{Santos}, N.~C., {Israelian}, G., \& {Mayor}, M. 2004, \aap, 415, 1153

\bibitem[{{Santos} {et~al.}(2013){Santos}, {Sousa}, {Mortier}, {Neves},
  {Adibekyan}, {Tsantaki}, {Delgado Mena}, {Bonfils}, {Israelian}, {Mayor}, \&
  {Udry}}]{santos:2013}
{Santos}, N.~C., {Sousa}, S.~G., {Mortier}, A., {et~al.} 2013, \aap, 556, A150

\bibitem[{{Santos} {et~al.}(2017){Santos}, {Adibekyan}, {Figueira},
  {Andreasen}, {Barros}, {Delgado-Mena}, {Demangeon}, {Faria}, {Oshagh},
  {Sousa}, {Viana}, \& {Ferreira}}]{santos:2017}
{Santos}, N.~C., {Adibekyan}, V., {Figueira}, P., {et~al.} 2017, \aap, 603, A30

\bibitem[{{Sing} {et~al.}(2011){Sing}, {Pont}, {Aigrain}, {Charbonneau},
  {D{\'e}sert}, {Gibson}, {Gilliland}, {Hayek}, {Henry}, {Knutson}, {Lecavelier
  Des Etangs}, {Mazeh}, \& {Shporer}}]{sing:2011}
{Sing}, D.~K., {Pont}, F., {Aigrain}, S., {et~al.} 2011, \mnras, 416, 1443

\bibitem[{{Southworth} {et~al.}(2009){Southworth}, {Hinse}, {Dominik},
  {Glitrup}, {J{\o}rgensen}, {Liebig}, {Mathiasen}, {Anderson}, {Bozza},
  {Browne}, {Burgdorf}, {Calchi Novati}, {Dreizler}, {Finet}, {Harps{\o}e},
  {Hessman}, {Hundertmark}, {Maier}, {Mancini}, {Maxted}, {Rahvar}, {Ricci},
  {Scarpetta}, {Skottfelt}, {Snodgrass}, {Surdej}, \&
  {Zimmer}}]{southworth:2009}
{Southworth}, J., {Hinse}, T.~C., {Dominik}, M., {et~al.} 2009, \apj, 707, 167

\bibitem[{{Sozzetti} {et~al.}(2007){Sozzetti}, {Torres}, {Charbonneau},
  {Latham}, {Holman}, {Winn}, {Laird}, \& {O'Donovan}}]{sozzetti:2007}
{Sozzetti}, A., {Torres}, G., {Charbonneau}, D., {et~al.} 2007, \apj, 664, 1190

\bibitem[{{Weinberg} {et~al.}(2013){Weinberg}, {Yoon}, \&
  {Katz}}]{weinberg:2013}
{Weinberg}, M.~D., {Yoon}, I., \& {Katz}, N. 2013, ArXiv e-prints,
  arXiv:1301.3156

\bibitem[{{Wu} \& {Murray}(2003)}]{wu:2003}
{Wu}, Y., \& {Murray}, N. 2003, \apj, 589, 605

\bibitem[{{Yi} {et~al.}(2001){Yi}, {Demarque}, {Kim}, {Lee}, {Ree}, {Lejeune},
  \& {Barnes}}]{yi:2001}
{Yi}, S., {Demarque}, P., {Kim}, Y.-C., {et~al.} 2001, \apjs, 136, 417

\bibitem[{{Zacharias} {et~al.}(2012){Zacharias}, {Finch}, {Girard}, {Henden},
  {Bartlett}, {Monet}, \& {Zacharias}}]{zacharias:2012:ucac4}
{Zacharias}, N., {Finch}, C.~T., {Girard}, T.~M., {et~al.} 2012, VizieR Online
  Data Catalog, 1322, 0

\bibitem[{{Zechmeister} \& {K{\"u}rster}(2009)}]{zechmeister:2009}
{Zechmeister}, M., \& {K{\"u}rster}, M. 2009, \aap, 496, 577

\bibitem[{{Zhou} {et~al.}(2014){Zhou}, {Bayliss}, {Kedziora-Chudczer},
  {Salter}, {Tinney}, \& {Bailey}}]{zhou:2014:sec}
{Zhou}, G., {Bayliss}, D.~D.~R., {Kedziora-Chudczer}, L., {et~al.} 2014,
  \mnras, 445, 2746

\end{thebibliography}

\clearpage

\end{document}